\pdfoutput=1
\documentclass[usenatbib]{mnras}
\usepackage[figuresright]{rotating}
\usepackage{times}
\usepackage{txfonts}

\usepackage[T1]{fontenc}
\usepackage{ae,aecompl}

\usepackage{caption}
\usepackage{array,footnote,color,}
\usepackage{lscape}

\usepackage{graphicx}	

\newcommand\nodata{ ~$\cdots$~ }
\newcolumntype{x}[1]{>{\centering\let\newline\\\arraybackslash\hspace{0pt}}p{#1}}
\newcommand{\atlas}{A{\small TLAS}$^{\rm 3D}$}

\title[Radio LLAGNs in ETGs]{The ATLAS$^{\mathrm{3D}}$ Project -XXXI.  Nuclear Radio Emission in Nearby Early-Type Galaxies}

\author[Kristina Nyland et al.]{\parbox{\textwidth}{
Kristina Nyland$^{1,2}$\thanks{E-mail: knyland@nrao.edu}, 
Lisa M. Young$^{3}$, 
Joan M. Wrobel$^{4}$,
Marc Sarzi$^{5}$,
Raffaella Morganti$^{2,6}$,
Katherine Alatalo$^{7,8}$\thanks{Hubble fellow},
Leo Blitz$^{9}$,
Fr\'ed\'eric Bournaud$^{10}$,
Martin Bureau$^{11}$,
Michele Cappellari$^{11}$,
Alison F. Crocker$^{12}$,
Roger L. Davies$^{11}$,
Timothy A. Davis$^{13}$,
P. T. de Zeeuw$^{14,15}$,
Pierre-Alain Duc$^{10}$,
Eric Emsellem$^{14,16}$,
Sadegh Khochfar$^{17}$,
Davor Krajnovi\'c$^{18}$,
Harald Kuntschner$^{14}$,
Richard M. McDermid$^{19,20}$,
Thorsten Naab$^{21}$,
Tom Oosterloo$^{2,6}$,
Nicholas Scott$^{22}$,
Paolo Serra$^{23}$,
and Anne-Marie Weijmans$^{24}$}
\vspace{0.4cm}
\\ 
\parbox{\textwidth}{
$^{1}$National Radio Astronomy Observatory, Charlottesville, VA 22903, USA\\
$^{2}$Netherlands Institute for Radio Astronomy (ASTRON), Postbus 2, 7990 AA Dwingeloo, The Netherlands\\
$^{3}$Physics Department, New Mexico Institute of Mining and Technology, Socorro, NM 87801, USA\\
$^{4}$National Radio Astronomy Observatory, Socorro, NM 87801, USA\\
$^{5}$Centre for Astrophysics Research, University of Hertfordshire, Hatfield, Herts AL1 9AB, UK\\
$^{6}$Kapteyn Astronomical Institute, University of Groningen, Postbus 800, 9700 AV Groningen, The Netherlands\\
$^{7}$ Infrared Processing and Analysis Center, California Institute of Technology, Pasadena, California 91125, USA \\
$^{8}$Observatories of the Carnegie Institution of Washington, 813 Santa Barbara Street, Pasadena, CA 91101, USA\\
$^{9}$Department of Astronomy, Campbell Hall, University of California, Berkeley, CA 94720, USA\\
$^{10}$Laboratoire AIM Paris-Saclay, CEA/IRFU/SAp -- CNRS -- Universit\'e Paris Diderot, 91191 Gif-sur-Yvette Cedex, France\\
$^{11}$Sub-department of Astrophysics, Department of Physics, University of Oxford, Denys Wilkinson Building, Keble Road, Oxford, OX1 3RH, UK\\
$^{12}$Ritter Astrophysical Observatory, University of Toledo, Toledo, OH 43606, USA\\
$^{13}$School of Physics \&\ Astronomy, Cardiff University, Queens Buildings, The Parade, Cardiff, CF24 3AA, UK\\
$^{14}$European Southern Observatory, Karl-Schwarzschild-Str. 2, 85748 Garching, Germany\\
$^{15}$Sterrewacht Leiden, Leiden University, Postbus 9513, 2300 RA Leiden, the Netherlands\\
$^{16}$Universit\'e Lyon 1, Observatoire de Lyon, Centre de Recherche Astrophysique de Lyon and Ecole Normale Sup\'erieure de Lyon, 9 avenue Charles Andr\'e, F-69230 Saint-Genis Laval, France\\
$^{17}$Royal Observatory Edinburgh, Blackford Hill, Edinburgh, EH9 3HJ, UK\\
$^{18}$Leibniz-Institut f\"ur Astrophysik Potsdam (AIP), An der Sternwarte 16, D-14482 Potsdam, Germany\\
$^{19}$Department of Physics and Astronomy, Macquarie University, Sydney NSW 2109, Australia\\
$^{20}$Australian Gemini Office, Australian Astronomical Observatory, PO Box 915, Sydney, NSW 1670, Australia \\
$^{21}$Max-Planck-Institut f\"ur Astrophysik, Karl-Schwarzschild-Str. 1, 85741 Garching, Germany\\
$^{22}$Sydney Institute for Astronomy (SIfA), School of Physics, The University of Sydney, NSW 2006, Australia\\
$^{23}$CSIRO Astronomy \& Space Science, PO Box 76, Epping, NSW 1710, Australia \\
$^{24}$School of Physics and Astronomy, University of St Andrews, North Haugh, St Andrews KY16 9SS, UK}
}

\date{Accepted YEAR MONTH DAY.  Received YEAR MONTH DAY; in original form YEAR MONTH DAY}

\pagerange{\pageref{firstpage}--\pageref{lastpage}} 
\pubyear{2016}

\begin{document}
\label{firstpage}
\maketitle
\clearpage

\begin{abstract}
We present the results of a high-resolution, 5~GHz, Karl G. Jansky Very Large Array study of the nuclear radio emission in a representative subset of the \atlas\ survey of early-type galaxies (ETGs).  We find that $51 \pm 4$\% of the ETGs in our sample contain nuclear radio emission with luminosities as low as 10$^{18}$ W~Hz$^{-1}$.  Most of the nuclear radio sources have compact ($\lesssim 25-110$~pc) morphologies, although $\sim$10\% display multi-component core+jet or extended jet/lobe structures.  Based on the radio continuum properties, as well as optical emission line diagnostics and the nuclear X-ray properties, we conclude that the majority of the central 5~GHz sources detected in the \atlas\ galaxies are associated with the presence of an active galactic nucleus (AGN).  However, even at sub-arcsecond spatial resolution, the nuclear radio emission in some cases appears to arise from low-level nuclear star formation rather than an AGN, particularly when molecular gas and a young central stellar population is present.  This is in contrast to popular assumptions in the literature that the presence of a compact, unresolved, nuclear radio continuum source universally signifies the presence of an AGN.  Additionally, we examine the relationships between the 5~GHz luminosity and various galaxy properties including the molecular gas mass and - for the first time - the global kinematic state.  We discuss implications for the growth, triggering, and fueling of radio AGNs, as well as AGN-driven feedback in the continued evolution of nearby ETGs.
\end{abstract}



\defcitealias{cappellari+11a}{Paper~I}
\defcitealias{krajnovic+11}{Paper~II}
\defcitealias{emsellem+11}{Paper~III}
\defcitealias{young+11}{Paper~IV}
\defcitealias{bois+11}{Paper~VI}
\defcitealias{cappellari+11b}{Paper~VII}
\defcitealias{khochfar+11}{Paper~VIII}
\defcitealias{duc+11}{Paper~IX}
\defcitealias{davis+11}{Paper~X}
\defcitealias{crocker+12}{Paper~XI}
\defcitealias{lablanche+12}{Paper~XII}
\defcitealias{serra+12}{Paper~XIII}
\defcitealias{davis+13}{Paper~XIV}
\defcitealias{cappellari+13a}{Paper~XV}
\defcitealias{bayet+13}{Paper~XVI}
\defcitealias{krajnovic+13a}{Paper~XVII}
\defcitealias{alatalo+13}{Paper~XVIII}
\defcitealias{sarzi+13}{Paper~XIX}
\defcitealias{cappellari+13b}{Paper~XX}
\defcitealias{scott+13}{Paper~XXI}
\defcitealias{martig+13}{Paper~XXII}
\defcitealias{krajnovic+13b}{Paper~XXIII}
\defcitealias{weijmans+14}{Paper~XXIV}
\defcitealias{naab+14}{Paper~XXV}
\defcitealias{serra+14}{Paper~XXVI}
\defcitealias{young+14}{Paper~XXVII}
\defcitealias{davis+14}{Paper~XXVIII}
\defcitealias{duc+15}{Paper~XXIX}
\defcitealias{mcdermid+15}{Paper~XXX}

\section{Introduction and Motivation}
One of the most pressing issues in current models of galaxy formation and evolution is the uncertain role of accreting supermassive black holes (SMBHs) in shaping the characteristics of their host galaxies.  The importance of improving our understanding of the properties of galaxy nuclei is highlighted by the growing body of evidence suggesting that the evolution of galaxies and their SMBHs are intricately linked \citep{heckman+14, kormendy+13}.  This symbiotic relationship may be at the root of the observed scaling relations between SMBH and host galaxy properties, active galactic nucleus (AGN)-driven outflows, and the regulation of star formation (SF).  Although the bulk of rapid SMBH growth, SF, and galaxy mergers are believed to occur at higher redshifts (e.g., z $\sim$ 1-3; \citealt{genzel+14}), studies of the less extreme versions of these processes in low-redshift, nearby galaxies approaching their evolutionary endpoints nevertheless offer detailed insights into the primary drivers of galaxy evolution, such as the mechanisms responsible for AGN triggering and the importance of AGN feedback in regulating SF.

Recent studies indicate the existence of two main channels (Figures 15 and 16 in \citealt{cappellari+13b}, hereafter \citetalias{cappellari+13b}; \citealt{vandokkum+15}) for the late-time assembly of early-type galaxies (ETGs), where the most massive and slowly rotating galaxies proceed through a series of dry minor mergers, while gas-rich mergers (minor or major) are responsible for the production of less-massive but more rapidly rotating systems (\citealt{bois+11}, hereafter \citetalias{bois+11}; \citealt{khochfar+11}, hereafter \citetalias{khochfar+11}; \citealt{naab+14}, hereafter \citetalias{naab+14}).
Given that the formation of these galaxies is likely subject to feedback from processes such as stellar winds and AGNs (e.g., \citealt{kaviraj+11, heckman+14}), studying the final stages of their evolution is particularly relevant to our understanding of galaxy evolution as a whole.  For instance, information on the dominant AGN fueling mode (external cold gas accretion vs. the accretion of hot gas associated with stellar winds or X-ray halos) and the relative importance of various AGN triggering mechanisms (minor vs. major mergers, or secular processes) may help further refine our knowledge of the formation histories of ETGs.  Sensitive observations that have been optimized for studying the AGN emission in a large sample of ETGs with constraints on the content and kinematics of their cold gas reservoirs are thus an excellent means of distinguishing between different galaxy evolution scenarios.  

Radio continuum emission offers an extinction-free tracer of emission from even weak AGNs in nearby ETGs (e.g., \citealt{nagar+05, ho+08}).  Recent technological advances at observatories such as the NRAO\footnote{The National Radio Astronomy Observatory is a facility of the National Science Foundation operated under cooperative agreement by Associated Universities, Inc.} Karl G. Jansky Very Large Array (VLA) have allowed interferometric radio observations to reach impressively deep sensitivities over relatively short timescales, making radio continuum data well-suited for studies of low-level radio emission in large samples.  In addition to their superb sensitivities, radio interferometers with maximum baseline lengths of a few tens of kilometers provide spatial resolutions better than 1$^{\prime \prime}$ at frequencies of a few GHz, a regime in which non-thermal synchrotron emission is both bright and the dominant radio emission mechanism.  

\begin{table*}
\begin{minipage}{14cm}
\caption{New High-Resolution 5~GHz VLA Observations}
\label{tab:projects}
\begin{tabular*}{14cm}{ccccccc}
\hline
\hline
 Project ID & Observing Dates$^{\dagger}$ & Time     & Galaxies & BW      & Spws & Frequency \\
                    &                                & (hours)  &                 & (MHz)  &            & (GHz)         \\
      (1)        &                   (2)         &      (3)     &      (4)      &     (5)    &   (6)    &    (7)          \\
\hline 
 11A-226 &  May 27, 2011 - August 29, 2011 & 12.5 & 17 &  256 & 2 & 4.94\\
 12B-281$^{*}$ &  October 5, 2012 - January 14, 2013 & 16.0 & 108 &  2048 & 16 & 5.49\\
\hline
\hline
\end{tabular*}
 
\medskip
{\bf Notes.} Column 1: VLA project code.  Column 2:  observing dates.  Column 3: number of completed observing hours.  Column 4:  number of galaxies.  Column 5:  total bandwidth.  Column 6:  total number of spectral windows.  Column 7: central observing frequency.

\medskip
$^{\dagger}$ Both projects were observed in the VLA $C$-band during the A-configuration, yielding a typical spatial resolution of $\theta_{\mathrm{FWHM}} \sim 0.5^{\prime \prime}$.

\medskip
$^{*}$ Although the regular A-configuration observing period during semester 2012B ended on January 6, 2013, only 6.25 hours of the 16 hours allocated to project 12B-281 had been observed.  We therefore requested to use the VLA during the move from the A configuration to the D configuration for the remaining 9.75 hours of project 12B-281.

\end{minipage} 
\end{table*}

Over the past few decades there have been a number of radio surveys of optically-selected samples of ETGs at a variety of sensitivities, frequencies, and resolutions.  The two most comprehensive ETG continuum studies thus far are the VLA 5~GHz surveys of \citet{sadler+89} and \citet{wrobel+91b} with spatial resolutions of a few arcseconds.  More recently, studies of ETGs in dense environments such as the Virgo \citep{balmaverde+06, capetti+09, kharb+12} and Coma \citep{miller+09} clusters have also been published.  In addition, large galaxy studies including substantial fractions of ETGs, such as the radio continuum follow-up studies to the Palomar Spectroscopic Survey \citep{ho+97} with instruments such as the VLA \citep{ho+01, ulvestad+01} and Very Long Baseline Array (VLBA; \citealt{nagar+05}), have provided critical information on the properties of low-luminosity AGNs (LLAGNs; for a review see \citealt{ho+08}) in ETGs.  Although these studies have provided many new insights, an assessment of the nuclear radio continuum emission in a well-defined sample of nearby ETGs spanning a variety of environments, kinematic properties, and cold gas characteristics has been lacking.

Unlike powerful radio galaxies or radio-loud quasars, the radio emission in nearby ETGs harboring LLAGNs is often minuscule compared to contamination from stellar processes that also produce centimeter-wave continuum emission (e.g., residual SF or supernovae remnants) and may dominate on extranuclear spatial scales (e.g., \citealt{ho+01b}).  This is especially important for studies of ETGs in light of recent evidence for on-going SF in ellipticals and lenticulars at low-redshifts (e.g., \citealt{yi+05, kaviraj+07, shapiro+10, ford+13}), which may produce radio continuum emission on kiloparsec-scales.  Thus, we emphasize that high angular resolution interferometric continuum observations are essential for extracting the {\it nuclear} component of the radio emission, which is more likely to be associated with the LLAGN, from the rest of the galaxy.  Such high-resolution radio studies are of critical importance for characterizing the local population of LLAGNs residing in ETGs so that they can be meaningfully linked to studies of more distant sources at earlier epochs of galaxy assembly.  

In this work, we present new, high-resolution, VLA 5~GHz observations of a sample of ETGs drawn from the \atlas\ survey (\citealt{cappellari+11a}, hereafter \citetalias{cappellari+11a}).  The \atlas\ survey provides new information not available to previous studies including: (i) classification of the global kinematic state (fast and slow rotators) using two-dimensional stellar kinematics (\citealt{krajnovic+11}, hereafter \citetalias{krajnovic+11}; \citealt{emsellem+11}, \citetalias{emsellem+11}), (ii) an inventory of the molecular (\citealt{young+11}, hereafter \citetalias{young+11}) and atomic (\citealt{serra+12}, hereafter \citetalias{serra+12}; \citealt{serra+14}, hereafter \citetalias{serra+14}) cold gas content, (iii) measurements of stellar kinematic misalignment \citepalias{krajnovic+11}, and (iv) dynamical stellar mass measurements (\citealt{cappellari+13a}, hereafter \citetalias{cappellari+13a}).
The rich multiwavelength database of the \atlas\ survey, described in Section~\ref{sec:a3d}, is an essential tool for the interpretation of our nuclear radio continuum observations in a broader evolutionary context.  In Section~\ref{sec:sample}, we explain the selection of our 5~GHz VLA sample.  We describe our VLA observations, data reduction procedure, and basic results in Sections~\ref{sec:obs} and \ref{sec:results}.  We discuss the origin of the 5~GHz sources detected in our sample of ETGs, which may be either LLAGN emission or circumnuclear SF, in Section~\ref{sec:radio_origin}.  In Section~\ref{sec:galaxy_properties}, we investigate the relationships between the nuclear radio properties and a variety of host galaxy properties, with an emphasis on the global kinematic state and the presence/absence (i.e., CO detection or upper limit) of a molecular gas reservoir.  We discuss our results in the broader context of galaxy evolution in Section~\ref{sec:discussion}.  Our results are summarized in Section~\ref{sec:summary}.

\begin{figure*}
\includegraphics[clip=true, trim=2.25cm 0.3cm 3.5cm 1.5cm, scale=0.48]{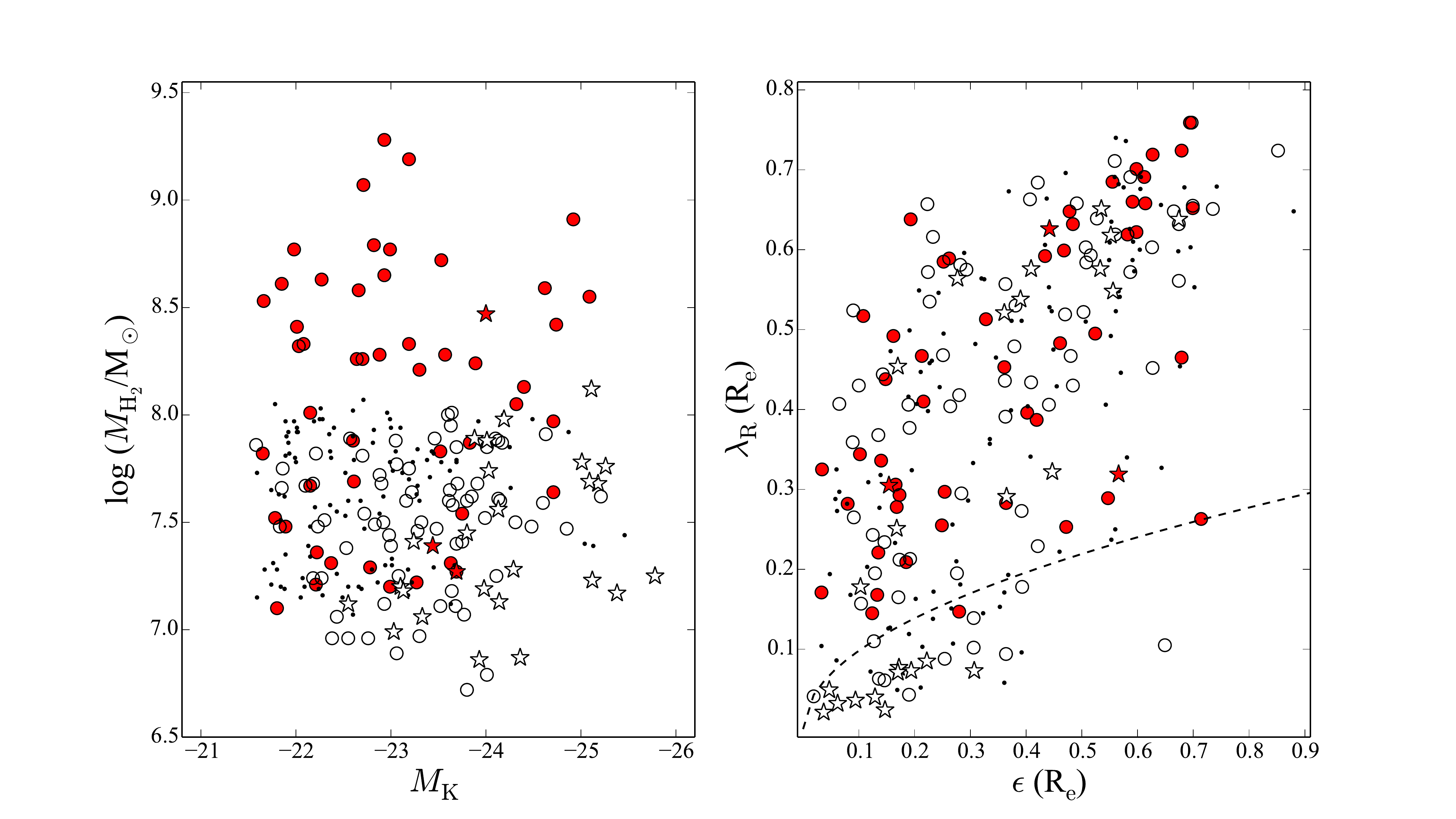}
\caption{Properties of the 5~GHz ETG sample.  
The 125 ETGs included in our new 5~GHz VLA observations are shown as circles.  The ETGs with archival radio continuum measurements are shown as stars.  (We note that in subsequent figures, we include ETGs with both new and archival 5~GHz measurements and make no distinction between them on plots.)  The remaining \atlas\ ETGs without high-resolution radio observations are shown as black points and are all CO upper limits by construction.  Symbols filled in red represent the \atlas\ IRAM single-dish CO detections \citepalias{young+11}.  Unfilled symbols represent CO non-detections (upper limits).  {\bf (Left:)} Molecular hydrogen mass plotted as a function of the optical $K$-band magnitude (from 2MASS; \citealt{skrutskie+06}), a proxy for stellar mass, at the distances adopted in \citetalias{cappellari+11a}.  {\bf (Right:)}  Specific angular momentum parameter ($\lambda_{\mathrm{R}}$; \citetalias{emsellem+11}) plotted against apparent ellipticity ($\epsilon$).  Both $\lambda_{\mathrm{R}}$ and $\epsilon$ are measured within one effective radius ($R_{\mathrm{e}}$).  The dashed line is $\lambda_\mathrm{R}(R_e)=0.31\times\sqrt{\epsilon(R_e)}$  \citepalias{emsellem+11}, and it separates the fast and slow rotators \citep{emsellem+07} above and below the line, respectively.}
\label{fig:samp_props}
\end{figure*}

\section{The \atlas\ Survey}
\label{sec:a3d}
\subsection{Overview}
For the first time, a statistical study of a sample of ETGs probing the photometric, kinemetric, and dynamical properties of their stellar populations and gas in the atomic, molecular, and ionized phases is available.  The {\atlas\ survey of 260 morphologically-selected ETGs was drawn from a volume- and magnitude-limited ($D < 42$ Mpc and $M_{\mathrm{K}} < -21.5$) parent sample of 871 galaxies \citepalias{cappellari+11a}.  {\atlas\ ETGs were selected on a morphological basis (i.e., the absence of spiral arms) and are thus not biased by any color selections.  A variety of environments (field, group, and the Virgo cluster; \citealt{cappellari+11b}, hereafter \citetalias{cappellari+11b}), kinematics \citepalias{krajnovic+11}, and stellar populations (\citealt{mcdermid+15}, hereafter \citetalias{mcdermid+15}) are also represented.  The survey combines multiwavelength data \citepalias{cappellari+11a} and theoretical models \citepalias{bois+11, khochfar+11, naab+14} with the aim of characterizing the local population of ETGs and exploring their formation and evolutionary histories.  

\subsection{Multiwavelength Data}
Available data for the full sample include integral-field spectroscopic maps \citepalias{cappellari+11a} with the {\tt SAURON} instrument \citep{bacon+01} on the William Herschel Telescope, optical imaging from the Sloan Digital Sky Survey (SDSS; \citealt{york+00}) or Isaac Newton Telescope (INT; \citealt{scott+13}, hereafter \citetalias{scott+13}), extremely deep optical imaging with the MegaCam instrument at the Canada-France-Hawaii Telescope (\citealt{duc+11}, hereafter \citetalias{duc+11}; \citealt{duc+15}, hereafter \citetalias{duc+15}), and single-dish $^{12}$CO(1-0) and (2-1) observations with the Institut de Radioastronomie Millim\'{e}trique (IRAM) 30-m telescope \citepalias{young+11}.  The CO observations have a detection rate of 22 $\pm$ 3\%} (56/259) and represent the first large, statistical search for molecular gas in ETGs.  Additional cold gas tracers for subsets of the \atlas\ survey include H{\tt I} data from the Westerbork Radio Synthesis Telescope (WRST) available for 166 galaxies \citepalias{serra+12, serra+14} and interferometric $^{12}$CO(1-0) maps from the Combined Array for Research in Millimeter Astronomy (CARMA) available for 40 of the brightest single-dish CO detections (\citealt{alatalo+13}, hereafter \citetalias{alatalo+13}; \citealt{davis+13}, hereafter \citetalias{davis+13}).

\subsection{Kinematic Classification}
The \atlas\ survey has compiled a number of parameters for assessing the kinematic states and evolutionary histories of ETGs.  The single most important parameter, the specific angular momentum ($\lambda_{\mathrm{R}}$; \citealt{emsellem+07, cappellari+07}), is derived from two-dimensional integral-field spectroscopic measurements of the stellar kinematics.  The $\lambda_{\mathrm{R}}$ parameter was originally defined and studied through the course of the SAURON survey \citep{dezeeuw+02}, which provided integral-field spectroscopic measurements for a representative sample of 48 nearby ETGs.  One of the key results of the SAURON survey was that ETGs fall into two distinct kinematic classes based on $\lambda_{\mathrm{R}}$ \citep{emsellem+07}: slow and fast rotators.  Slow rotators (SRs) are generally massive ellipticals, display little ordered stellar rotation, and often have complicated stellar velocity structures (e.g., kinematically distinct cores).  Fast rotators (FRs) are a class of lenticulars and less-massive ellipticals and are characterized by regular rotation in their stellar velocity fields.  In addition, FRs sometimes contain cold gas \citepalias{young+11}.  The \atlas\ team has refined the kinematic classification of ETGs into SRs and FRs by taking into account $\lambda_{\mathrm{R}}$ as well as the ellipticity ($\epsilon$; \citetalias{emsellem+11}).  
We compare the radio continuum properties of our sample galaxies with other galaxy parameters in terms of their kinematic classification, as well as molecular gas content, in Sections~\ref{sec:radio_origin} and \ref{sec:galaxy_properties}.

\section{Sample Selection}
\label{sec:sample}
\subsection{New VLA Observations}
The 125 ETGs included in our new high-resolution 5~GHz VLA observations are selected from the \atlas\ survey \citepalias{cappellari+11a}.  A list of these ETGs is provided in Table \ref{tab:radio_parms}.  Since one of our goals is to probe the relationship between LLAGN-driven radio emission and molecular gas content, we included as many of the 56 IRAM single-dish CO detections as possible.  We excluded\footnote{We emphasize that ETGs identified in this section as being ``excluded'' from our new VLA observations are in fact included in all subsequent analyses presented in this paper whenever archival data are available.} only 4 CO-detected ETGs from our new VLA observations (NGC3245, NGC4203, NGC4476, and NGC5866).  NGC4476 was excluded due to its proximity  (12.6$^{\prime}$) to the bright radio source hosted by NGC4486 (M87).  The other 3 CO-detected sources (NGC3245, NGC4203, and NGC5866) were not included in our new observations based on the availability of archival VLA data with similar properties.

In addition to the 52 CO-detected galaxies, we also include 73 of the CO non-detections in our new VLA observations as a ``control sample\footnote{We use the term ``control sample" in a general sense.  Our intention is to highlight the fact that we designed our study to allow us to investigate differences in the nuclear radio properties of ETGs with and without molecular gas.}" for comparison.  Although we would have ideally selected a random subset of the CO non-detections for the control sample, a number of observing complications hindered this goal.  For the galaxies observed during project 12B-281 (108/125; see Table~\ref{tab:projects}), the observations coincided with extensive daytime commissioning activities related to the Expanded VLA (EVLA) project \citep{perley+11}.  As a consequence, daytime observations during project 12B-281 while galaxies in the Virgo cluster were primarily observable were limited.  Thus, we were only able to obtain new high-resolution 5~GHz VLA observations for 17 Virgo Cluster galaxies. 

We also excluded galaxies from the control sample that are known to host extremely bright radio sources.  Aside from the challenges of high dynamic range imaging, bright radio sources (e.g., $S$ $\gtrsim$ 1~Jy) have generally been previously studied in great detail by the VLA and measurements are available in the literature.  Thus, we explicitly excluded NGC4486 from our VLA 5~GHz sample.  In addition, we also excluded two galaxies from the control group of CO non-detections located within 13$^{\prime}$ of NGC4486 (NGC4486A and NGC4478).  The closest galaxy to NGC4486 that was actually observed is NGC4435 at an angular distance of $\sim62^{\prime}$.

\subsection{Archival Data}
When available, we incorporate archival high-resolution radio data in our analysis.  We formally required archival radio data to have been observed at high spatial resolution ($\lesssim$ 1$^{\prime \prime}$) and near a frequency of 5~GHz (1 to 15~GHz) to be included in our analysis.  Archival data observed at frequencies other than 5~GHz were scaled to 5~GHz using high-resolution measurements of the radio spectral index from the literature, if available.  In the absence of radio spectral index information, we assume a flat synchrotron spectral index of $\alpha = -0.1$, where $S \sim \nu^{\alpha}$.  A list of the \atlas\ ETGs with archival nuclear radio data is provided in Table~\ref{tab:archival}.  

\subsection{Properties of the Sample}
The distribution of our sample of ETGs with either new or archival nuclear radio continuum measurements in terms of the main parameters probed by the \atlas\ survey is provided in Figure~\ref{fig:samp_props}.  This figure shows that our sample galaxies span a representative swath of stellar masses (traced by $K$-band magnitude) relative to the full \atlas\ sample.  The fraction of SRs in our high-resolution radio sample is 16 $\pm$ 4\%, which is similar to the fraction of SRs in the full \atlas\ sample (14 $\pm$ 2\%).  Thus, our sample captures the diversity of ETG kinematic states in a statistically similar sense compared to the full \atlas\ sample.

\section{VLA Data}
\label{sec:obs}

\subsection{Observations}
Our sample of 125 local ETGs was observed with the VLA in the A configuration at $C$ band ($4-8$~GHz) over two projects, 11A-226 and 12B-281, spanning a total of 28.5 hours.  A summary of these projects is provided in Table~\ref{tab:projects}.  The Wideband Interferometric Digital Architecture (WIDAR) correlator was configured using the 8-bit samplers with the maximum total bandwidth available during each project.  11A-226 was a ``pilot project" and was observed as part of the Open Shared Risk Observing (OSRO) program, which offered 256~MHz of total bandwidth from March 2010 until September 2011.  With this bandwidth, we required about 30 minutes on source per galaxy to achieve an rms noise of $\sim$ 15~$\mu$Jy beam$^{-1}$.  Beginning in late September 2011, the maximum available bandwidth for OSRO projects was expanded to 2048~MHz, and we were able to utilize this wider bandwidth for our project 12B-281.  This increase in bandwidth during project 12B-281 allowed us to theoretically reach nearly the same rms noise in just one $\sim$ 5-minute-long snapshot per galaxy.  In both projects, the total bandwidth was divided equally into 128-MHz-wide spectral windows (spws) consisting of 64 channels each.  

We divided each project into independent scheduling blocks (SBs) that were designed for optimal efficiency and flexibility for dynamic VLA scheduling.  We phase-referenced each galaxy to a nearby calibrator within 10 degrees and chose calibrators with expected amplitude closure errors of no more than 10\% to ensure robust calibration solutions.  In addition, the positional accuracy of most of our phase calibrators was $<0.002^{\prime \prime}$.  In order to set the amplitude scale to an accuracy of 3\% as well as calibrate the bandpass and instrumental delays, we observed the most conveniently-located standard flux calibrator (3C286, 3C48, 3C147, or 3C138) once per SB \citep{perley+13}.

\subsection{Calibration and Imaging}
\label{sec:cal_im}
Each SB was individually flagged, calibrated, and imaged using the Common Astronomy Software Applications (CASA) package\footnote{http://casa.nrao.edu}  (versions 4.0.0$-$4.1.0).  For each SB, we consulted the observing log and manually inspected the flux calibrator using the CASA PLOTMS tool in order to identify and flag any obviously bad data.  If characteristic Gibbs ringing due to exceptionally bright radio frequency interference (RFI) was apparent in the visibility data, we Hanning-smoothed the data using the CASA task HANNINGSMOTH.  The 12B-281 data obtained during the period of January 7-14, 2013 coincided with the move from the VLA A configuration to the D configuration.  These ``move-time" data were generally well-behaved after antenna baseline position corrections were applied.  For some of the move-time data sets, we had to exclude the data from the shortest baselines in order to mitigate the effects of increased RFI as well as hardware issues on the antennas that had been recently moved into the D configuration.  

The data were calibrated using the CASA VLA calibration pipeline version 1.2.0\footnote{https://science.nrao.edu/facilities/vla/data-processing/pipeline}.  This pipeline performs all standard calibrations, runs an automated RFI flagging algorithm (RFLAG) on the calibrated data, and calculates data weights using the STATWT task.  We used the series of diagnostic plots provided by the pipeline to determine the quality of the calibration solutions and identify bad data for further flagging.  In a few instances (usually involving poor observing conditions or issues with hardware) in which the pipeline did not produce calibrated data of sufficiently high quality, we carefully flagged and calibrated the data by hand using standard procedures.  

We formed and deconvolved images of the Stokes {\it I } emission using the CASA CLEAN task.   Each galaxy was imaged over an extent of at least 5$^{\prime}$ with a cell size of 0.075$^{\prime \prime}$.  In some cases, larger images exceeding the extent of the $C$-band primary beam ($\approx$ 9$^{\prime}$) were generated in order to properly clean the sidelobes of distant confusing sources significantly affecting the noise level at the phase center of the image.  We utilized the Cotton-Schwab algorithm \citep{schwab+83} in the Multi Frequency Synthesis (MFS) mode \citep{conway+90}.  In order to accurately model the frequency dependence of the emission in our wide-bandwidth data, we set the parameter nterms\footnote{When nterms $>$ 1 in the CASA CLEAN task, the MFS algorithm models the frequency-dependent sky brightness as a linear combination of Gaussian-like functions whose amplitudes follow a Taylor-polynomial in frequency.} = 2 \citep{rau+11}.  We chose Briggs weighting \citep{briggs+95} with a robustness parameter of 0.5 to obtain the best compromise among sensitivity, spatial resolution, and sidelobe suppression for our observations.  We utilized the $w$-projection algorithm \citep{cornwell+05} by setting the parameters gridmode = `widefield' and wprojplanes = 128 in order to correct for the effects of non-coplanar baselines at the VLA.  Images with bright emission showing signs of calibration artifacts were carefully re-cleaned with a mask and self-calibrated, if necessary.  For the sources with the largest spatial extents and most complex geometries, the final images were generated using the Multiscale algorithm\footnote{The Multiscale CLEAN algorithm models the sky brightness by the summation of components of emission having different size scales.} \citep{cornwell+08}.  

The full-width at half maximum (FWHM) of the major axis of the synthesized beam in these maps is typically $\theta_{\mathrm{FWHM}}$ $\approx$ 0.5$\arcsec$, though for a few observations that took place at very low elevations or during the move from the A to D configuration, the spatial resolution is significantly lower (in the most extreme case, $\theta_{\mathrm{FWHM}}$ $\approx$ 1.4$\arcsec$).  
Radio continuum maps with contours for the 53 detections from our new 5~GHz observations are shown in Figure~\ref{fig:radio_images}.  Values for the relative contour levels and rms noise in each image associated with a nuclear radio detection are provided in Table~\ref{tab:contours}.

\begin{figure}
\includegraphics[clip=true, trim=0.1cm 0cm 1.5cm 2cm, scale=0.4]{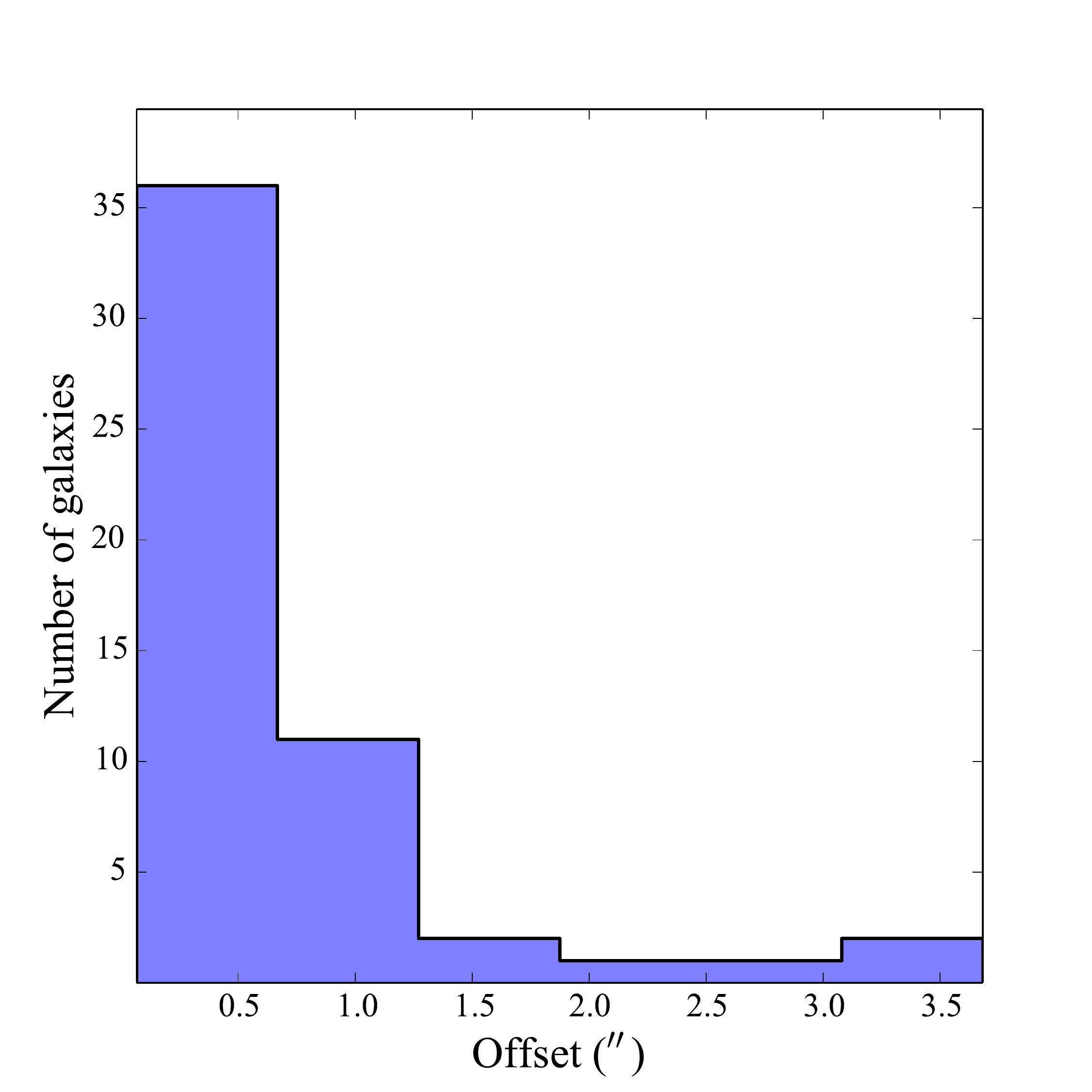}
\caption{Histogram showing the distribution of offsets in arcseconds between the official \atlas\ optical positions \citepalias{cappellari+11a} and the radio positions of the sources detected in our new 5~GHz observations (Table~\ref{tab:gauss}).}
\label{fig:offsets}
\end{figure}

\subsection{Image Analysis}
\label{sec:imanalysis}
For detections, we required a peak flux density $S_{\mathrm{peak}} >$ 5$\sigma$, where $\sigma$ is the rms noise.  Upper limits were set to $S_{\mathrm{peak}} <$ 5$\sigma$.  We required that the radio detections lie within 4$^{\prime \prime}$ of the optical position of the host galaxy from ground-based (SDSS of INT) measurements.  This was to account for the inherent uncertainty in nuclear positions from ground-based optical observations, particularly in galaxy nuclei harboring dust.  Figure~\ref{fig:offsets} shows that the radio-optical positional offsets are generally small among the sources meeting the criteria described in this section.  For the two galaxies with the largest offsets between their optical and radio positions in Figure~\ref{fig:offsets} (IC0676 and NGC5475), the presence of nuclear dust may have hampered the accuracy of optical position measurements.  Summaries of the image parameters and source properties are provided in Tables~\ref{tab:radio_parms}, \ref{tab:gauss}, \ref{tab:multi_flux}, and \ref{tab:multi_spatial}.

For each radio source with a relatively symmetric, Gaussian-like morphology, we determined the source parameters (peak flux density, integrated flux density, deconvolved major and minor axes, and deconvolved position angle) by fitting a single two-dimensional elliptical Gaussian model using the JMFIT task in the 31DEC15 release\footnote{The 31DEC15 version of {\tt AIPS} was the active development version of the software at the time of writing.  We chose this version over previously released versions to take advantage of new improvements to the JMFIT task.} of the Astronomical Image Processing System ({\tt AIPS}).  The flux density errors listed in Tables~\ref{tab:radio_parms} and \ref{tab:multi_flux} were calculated as the sum of the error reported by JMFIT and the standard 3\% VLA calibration error \citep{perley+13}, added in quadrature.  Sources were classified as ``resolved" or ``unresolved" based on the output of the JMFIT task\footnote{For a source to be classified as resolved, JMFIT formally requires successful deconvolution of the major axis.  In other words, if only an upper limit is given for the deconvolved major axis, a source will be considered unresolved.  JMFIT also requires the total integrated flux density to be at least 1$\sigma$ above the peak flux density for resolved sources.}.

For radio sources with more complex morphologies, we manually determined the spatial parameters using the CASA Viewer.  The peak and integrated flux densities were measured using the task IMSTAT.  Flux density errors for these measurements were calculated as $\sqrt{ (N \times \sigma)^2 + (0.03 \times S_{\mathrm{int}})^2 }$, where $N$ is the number of synthesized beams\footnote{If all the pixels within a single synthesized beam area were perfectly independent, one would expect the flux density uncertainty of an extended source to depend not on $N$, but $N^{1/2}$.  However, the pixels in interferometric images are partially correlated.  To account for the increase in the uncertainty in the flux density due to pixel correlation, we conservatively assume a linear scaling of $N$ here.  As a result, we note that the flux density uncertainties of the extended sources in our sample may actually be slightly overestimated.}
 subtended by the source and $S_{\mathrm{int}}$ is the integrated flux density.

\section{Results}
\label{sec:results}
\subsection{Detection Rate}
\label{sec:det_rate}
The detection fractions in projects 11A-226 and 12B-281 are 9/17 and 44/104, respectively\footnote{Although in project 12B-281 we observed 108 \atlas\ ETGs with the VLA, we were not able to successfully mitigate the effects of poor observing conditions and hardware issues for 4 galaxies (NGC6798, PGC056772, PGC058114, and PGC061468) and do not include these galaxies in our subsequent statistical analyses.  Thus, our new VLA observations effectively include a total of 121 galaxies.}.  This corresponds to a detection rate for the new 5~GHz VLA observations of 53/121 (44 $\pm$ 5\%).  
Including the archival data listed in Table~\ref{tab:archival}, the detection rate of nuclear radio emission in the \atlas\ ETGs\footnote{Although 29 galaxies are listed in Table~\ref{tab:archival}, two were already included in our new observations but not detected (NGC4697 and NGC4477).  Thus, the total number of \atlas\ galaxies with high-resolution 5~GHz data is 148.} is 76/148 (51 $\pm$ 4\%).  

\begin{figure*}
\includegraphics[clip=true, trim=0cm 0cm 1.5cm 2cm, scale=0.4]{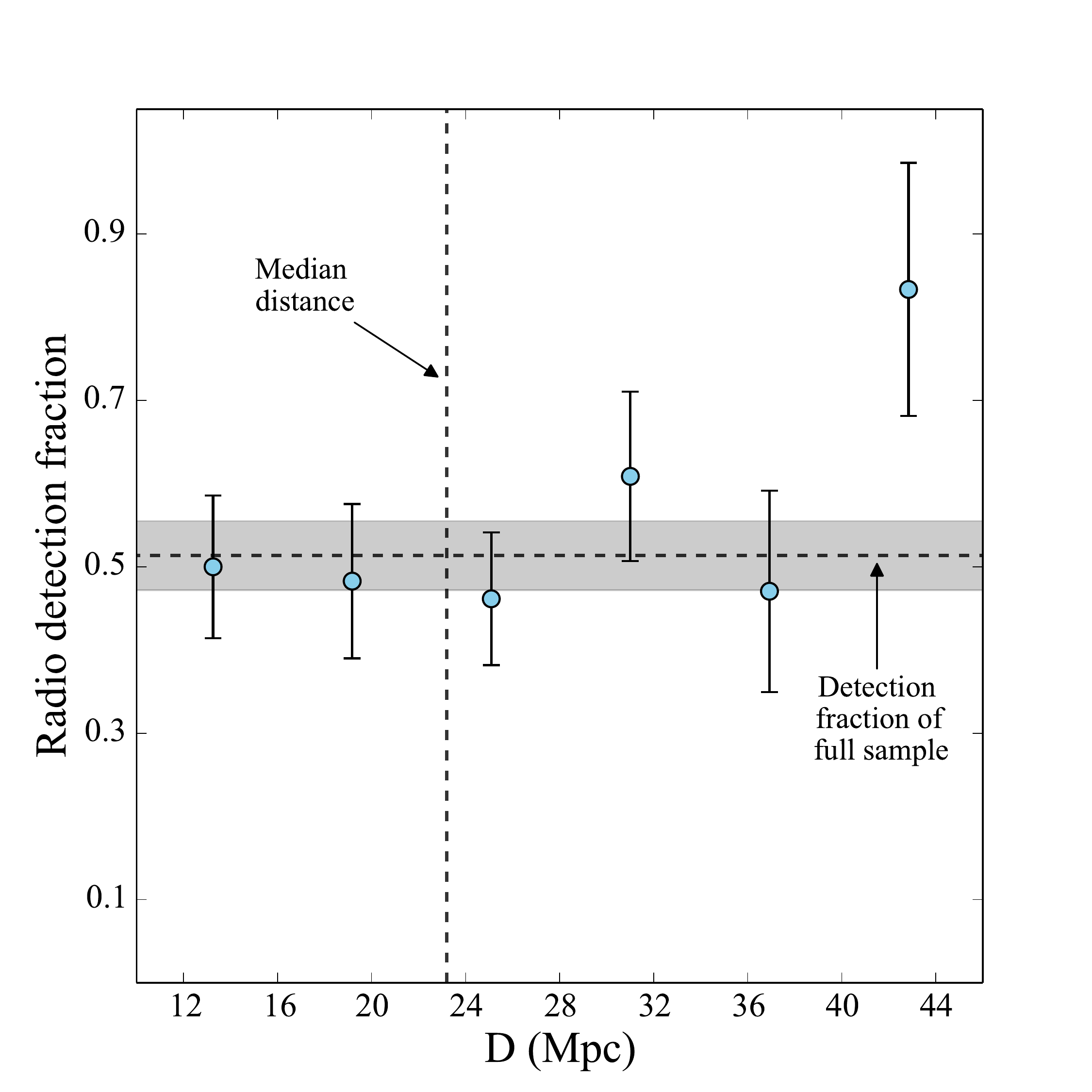}
\includegraphics[clip=true, trim=0cm 0cm 1.5cm 2cm, scale=0.4]{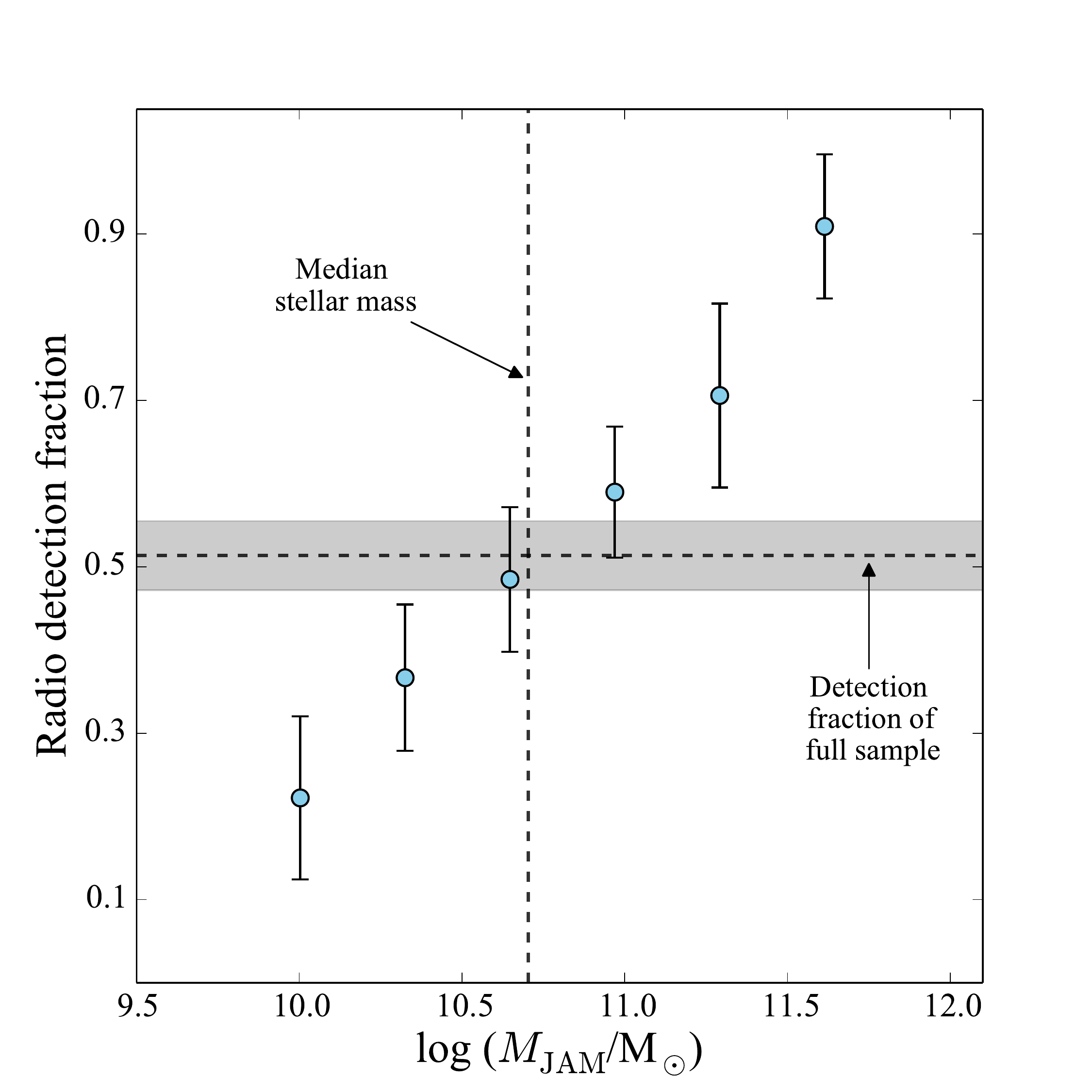}
\caption{{\bf (Left:)} Nuclear radio detection fraction as a function of distance.  The binomial uncertainty ($\sigma_{n_i} = \sqrt{n_i (1-n_i /N)}$, where $n_i$ is the number of ETGs in bin $i$ and $N$ is the total number of ETGs in the sample) for the radio detections in each distance bin is shown by the error bars.  The vertical dashed line denotes the median distance of the full sample.  The dashed horizontal line shows the radio detection fraction for the full sample.  The shaded gray region represents the uncertainty in the detection fraction for the full sample.  {\bf (Right:)}  Same as the left panel, except here the nuclear radio detection fraction is shown as a function of dynamical stellar mass, $M_{\mathrm{JAM}}$ \citepalias{cappellari+13a}.}
\label{fig:biases}
\end{figure*}

A well-known complication in the interpretation of detection rate statistics for flux-limited surveys is that parameters such as galaxy distance and mass tend to bias detection statistics.  We investigate the significance of this effect in Figure~\ref{fig:biases}.  
The left panel of Figure~\ref{fig:biases} shows the radio detection fraction as a function of distance.  The radio detection fraction was calculated by dividing the number of galaxies with nuclear radio detections in each distance bin of width $\approx 6$~Mpc by the total number of galaxies in each bin.  Only the furthest distance bin shows a significantly higher radio detection rate compared to the radio detection rate for the full sample.  Thus, unlike many radio surveys covering larger volumes (e.g., \citealt{mauch+07}), the radio detection rate in our sample is not significantly biased by distance.  

The right panel of Figure~\ref{fig:biases} shows the radio detection fraction as a function of $M_{\mathrm{JAM}}$, which provides our best estimate of the total stellar mass \citepalias{cappellari+13a}.  Mass bins are of width $\log(M_{\mathrm{JAM}}/\mathrm{M}_{\odot}) \sim$ 0.32.
This figure shows a wide range of radio detection fractions, with a radio detection fraction as low as 20\% in the lowest-mass bin and as high as 90\% in the highest-mass bin.  This indicates that stellar mass may bias our radio detection statistics.  
A similar stellar mass bias is seen in larger surveys probing higher-luminosity radio source populations (e.g., \citealt{best+05, mauch+07}).  The dependence of the radio-detected fraction on stellar mass reported in these large surveys is strong and has a mathematical form of $f_{\mathrm{radio-loud}} \propto M_{\mathrm{*}}^{2.0-2.5}$, where ``radio-loud" sources are typically defined as those with $L_{\mathrm{radio}} > 10^{23}$ W~Hz$^{-1}$ at 1.4~GHz.

Direct comparison between the stellar mass bias in our survey and the surveys presented in \citet{best+05} and \citet{mauch+07} is not possible since our sample probes much lower-luminosity radio sources in the range 18.04 $<$ $\log(L_{5 \, \mathrm{GHz}}/$W~Hz$^{-1}$) $<$ 23.04.  After scaling our luminosities to 1.4~GHz, only a few of the massive Virgo galaxies, such as M87, would even qualify as radio-loud in our sample.  Thus, we derive the power-law dependence of the radio detection fraction on stellar mass shown in Figure~\ref{fig:biases} using a simple linear regression and find $f_{\mathrm{radio-detection}} \propto M_{\mathrm{JAM}}^{0.35}$.  This indicates that a dependence of the radio detection fraction on stellar mass in our sample does exist, but the effect is substantially less severe compared to classical, large radio surveys limited to powerful radio sources.  Our high sensitivity is likely the dominant factor in effectively reducing the level of stellar mass bias in our radio detection statistics.  
We further discuss the stellar mass bias in our sample and account for it in the SR/FR statistical tests of Section~\ref{sec:radio_kin}.

We also considered using a fixed luminosity threshold to quantify our detection statistics.  An optimal threshold for our sample that minimizes the number of statistically ``ambiguous" cases (i.e., those with upper limits above the detection threshold) is $\log(L_{5\, \mathrm{GHz}}/$W~Hz$^{-1}$) $\sim 19.2$.  A total of 60/144 (42 $\pm$ 4\%) \atlas\ galaxies have nuclear radio luminosities exceeding this luminosity threshold.  Thus, the detection rate at a fixed-luminosity threshold is similar to the detection rate based on the flux threshold discussed previously in this section.  Most importantly, the number of galaxies that change classification under these two detection standards is small, so the choice of detection standard has minimal effect on the statistical tests.

\subsection{Morphology}
The 53 sources detected in our new 5~GHz observations are typically compact on spatial scales of $\lesssim$ 25 to 110 pc for the nearest ($D = $ 10.3~Mpc) and farthest ($D = $ 45.8~Mpc) ETGs, respectively.  Twenty-two (42 $\pm$ 7\%) of the new 5~GHz detections are classified as resolved by JMFIT (see Table~\ref{tab:gauss}).  
Of the formally resolved sources, 10/22 clearly display multiple components or extended morphologies.  Among these 10 galaxies with well-resolved nuclear radio emission, 3 (NGC3665, NGC4036, and NGC5475) exhibit 5~GHz continuum emission that is clearly distributed among two or three distinct components.  An additional 4/10 of the galaxies with well-resolved 5~GHz emission (NGC1222, NGC1266, NGC2768, and UGC06176) display multiple components or regions with less distinct geometries.  The flux and spatial parameters of the multi-component radio sources in the 7 galaxies discussed here are summarized in Tables~\ref{tab:multi_flux} and \ref{tab:multi_spatial}.  

The remaining 3/10 galaxies with well-resolved nuclear radio emission (IC0676, NGC4684, and UGC05408) are characterized by non-axisymmetric, extended 5~GHz emission.  NGC4111 may also harbor extended emission, however it was observed during a portion of the move from the A to D configuration affected by hardware malfunctions on several VLA antennas.  Thus, it is difficult to determine whether the extension of the radio emission in the nucleus of NGC4111 is real or the result of residual artifacts that have persisted even after careful processing and self calibration.  

For a more detailed discussion of the nuclear radio morphologies of the ETGs in our sample in the context of the origin of the nuclear radio continuum emission, we refer readers to Section \ref{sec:radio_origin}.

\subsection{Comparison to Previous Studies}
\label{sec:previous_studies}

\subsubsection{NVSS}
The NRAO VLA Sky Survey (NVSS; \citealt{condon+98}) provides 1.4~GHz images of the northern sky with a spatial resolution of $\theta_{\mathrm{FWHM}} \sim 45^{\prime \prime}$.  Unfortunately, the low resolution, large positional uncertainty, and relatively shallow depth (rms noise $\sim$ 0.45 mJy beam$^{-1}$) make it a blunt tool for studying the weak, often compact radio emission associated with nearby ETGs.  Of the 260 ETGs in the full \atlas\ sample, 54 are detected in the NVSS catalog within a search radius of 10$^{\prime \prime}$.  Within our 5~GHz, A-configuration VLA sample of 125 \atlas\ ETGs, 35 galaxies are detected in NVSS.  The spatial resolution of the NVSS images of the \atlas\ ETGs corresponds to physical scales of about 2 to 10 kpc.  Except in the presence of relatively powerful radio jets on these spatial scales (e.g., NGC4486), the radio continuum emission on kiloparsec-scales seen in the NVSS observations of the ETGs in our sample is unresolved and may originate from a mixture of AGN emission, low-level SF, and background confusing sources.  

\subsubsection{FIRST}
\label{first}
The Faint Images of the Radio Sky at Twenty Centimeters (FIRST; \citealt{becker+95}) survey provides an order of magnitude higher spatial resolution ($\theta_{\mathrm{FWHM}} \sim 5^{\prime \prime}$), improved positional accuracy, and increased sensitivity (rms noise $\sim$ 0.15 mJy beam$^{-1}$) compared to NVSS.  Fifty-five of the 260 \atlas\ ETGs are detected in FIRST within a search radius of 3$^{\prime \prime}$, 33 of which are also included in our high-resolution 5~GHz VLA sample.  Although the spatial resolution of the FIRST survey is a vast improvement over NVSS, it corresponds to physical scales of about 250~pc to 1~kpc for the \atlas\ ETGs.  A few notable ETGs in our sample have extended radio jets that are well-resolved in the FIRST images (e.g., NGC3665), but the majority of the FIRST detections are characterized by unresolved morphologies.  Since molecular gas disks in ETGs typically have extents similar to the spatial scales probed by FIRST \citepalias{alatalo+13, davis+13}, the FIRST data are well-suited for radio continuum studies of SF (Nyland et al., in preparation), but not necessarily for studies of the nuclear radio activity in nearby ETGs.  

\subsubsection{Wrobel \& Heeschen Survey}
\label{sec:wrobel}
The radio continuum survey at 5~GHz and at a spatial resolution of $\approx$ 5$^{\prime \prime}$ presented by \citet{wrobel+91b} offers an interesting point of comparison to our high-resolution radio study.  This volume-limited, optically-selected sample includes 198 ETGs in the northern hemisphere.  At their 5$\sigma$ detection threshold of 0.5 mJy, \citet{wrobel+91b} reported a detection fraction of 52/198 (26 $\pm$ 3\%) ETGs, 7/52 of which display extended radio continuum emission.  The 5~GHz detection rate in our subarcsecond-scale resolution study (42 $\pm$ 7\%) is significantly higher than that of \citet{wrobel+91b}.  This is not surprising given that our new study is sensitive to radio emission with luminosities an order of magnitude fainter and as low as $\log(L_{\mathrm{\, 5 \, GHz}}$/W~Hz$^{-1}$) = 18.04.

The \atlas\ survey and the ETG sample of \citet{wrobel+91b} share similar selection criteria, and there is significant overlap between the two samples.  In Table~\ref{tab:radio_parms}, we list the lower-resolution 5~GHz flux densities reported by \citet{wrobel+91b}.  Of the 144 ETGs common to both the \citet{wrobel+91b} and \atlas\ samples, 88 also have high-resolution 5~GHz data.  A total of 30/88 ETGs are detected in both samples.  Two galaxies, NGC3032 and NGC4026, are detected in the \citet{wrobel+91b} observations but not in our higher resolution study.  The high-resolution 5~GHz upper limits of these galaxies are factors of 40 and 7, respectively, less than the integrated flux densities reported by \citet{wrobel+91b}.  Thus, NGC3032 and NGC4026 may have significant contributions from SF on $\sim$kiloparsec scales in the radio and only weak or even nonexistent nuclear activity.  There are 9 ETGs (NGC2778, NGC3377, NGC3608, NGC3941, NGC4429, NGC4494, NGC4621, NGC5485, and NGC5631) among the 88 common to both samples that have 5~GHz upper limits in the \citet{wrobel+91b} survey but are detected in our higher-resolution, more sensitive observations.  These galaxies have compact nuclear radio morphologies, making them likely hosts of weakly accreting active nuclei.  

The 5~GHz flux densities of most (24/30) of the ETGs detected in both the \citet{wrobel+91b} study and the new higher-resolution observations presented in this work are within a factor of $\approx$ 2 of each other, suggesting that the majority of the central radio emission in these ETGs is truly localized in their circumnuclear regions.  
A few (5/30) ETGs show larger differences between their arcsecond- and subarcsecond-scale 5~GHz emission.  NGC3619 is over 3 times fainter in our new VLA A-configuration observations, while NGC3648, NGC4435, NGC4710, and NGC6014 are 15, 10, 20, and 9 times weaker, respectively, in our high-resolution data compared to the measurements of \citet{wrobel+91b}.  Our high-resolution 5~GHz image of NGC3648 shows that this galaxy has a nearby, background confusing source to the southwest that may be responsible for the excess radio emission measured in \citet{wrobel+91b}.  NGC3619 and NGC4710 may genuinely have excess radio emission in the \citet{wrobel+91b} study compared to our new high-resolution data.  The radio emission on scales of a few kiloparsecs in these ETGs is similar in extent to their molecular gas disks \citepalias{davis+13}, and is therefore likely to be related to current SF.  Finally, we consider NGC3945.  This ETG is the only one detected in both samples that has an apparent {\it deficit} of radio emission at the lower resolution of the \citet{wrobel+91b} study compared to our VLA A-configuration data (by a factor of 4).  This may be a sign of radio variability, a common feature of LLAGNs over timescales of months and years \citep{ho+08}, over the more than twenty-year gap between the \citet{wrobel+91b} study and our new observations at high resolution.

\section{Origin of the Radio Emission}
\label{sec:radio_origin}

Despite the fact that radio continuum emission is known to be a common occurrence in ETGs (e.g., \citealt{wrobel+91b, brown+11}), its origin is not always clear.  Nuclear radio sources are believed to be produced by AGNs via synchrotron emission along or at the base of jets \citep{nagar+05, balmaverde+06}.  However, if circumnuclear SF is present, the cosmic rays of a recent population of supernovae may also interact with local magnetic fields to produce synchrotron emission at centimeter wavelengths \citep{condon+92}.  Although ETGs were once believed to be devoid of cold gas and recent SF, we now know that this is not necessarily the case (\citealt{knapp+96, welch+03, combes+07, crocker+11}; \citetalias{young+11, serra+12}).  Thus, determining with certainty whether a compact radio source in an ETG originates from low-level circumnuclear SF or from a LLGN is a challenging task, especially in studies of sources with weak radio luminosities.  

High-resolution imaging on sub-kiloparsec scales greatly reduces contamination from radio emission arising from star-forming disks in ETGs that are typically extended on scales of a few kiloparsecs or larger (e.g., \citetalias{davis+13}).  However, even at the high spatial resolutions in the radio study presented here, the detection of nuclear continuum emission does not necessarily guarantee that an LLAGN is present.  In this section, we discuss constraints on the origin of the compact radio sources in our sample using diagnostics at radio, optical, and X-ray wavelengths.

\subsection{Radio Diagnostics}
\subsubsection{Radio Morphology}
\label{morph}
As shown in Figure~\ref{fig:radio_images}, the morphologies of the nuclear 5~GHz sources in our sample are generally compact.  Although the origin of the unresolved radio sources in our sample galaxies is most likely LLAGN emission, supporting evidence is required to robustly exclude alternative possibilities (e.g., SF).
However, if the radio emission exhibits a characteristic AGN-like morphology such as extended jets/lobes or multiple components arranged in a core+jet structure, AGN identification is relatively straightforward (e.g., \citealt{wrobel+91b}).  

The nuclear radio sources in 5/53 (9 $\pm$ 4\%) of the ETGs detected at 5~GHz in our new observations have resolved radio lobes/jets (NGC1266) and core+jet structures (NGC2768, NGC3665, NGC4036, NGC5475).  Considering the \atlas\ ETGs with archival high-resolution data (see Table~\ref{tab:archival}), there are an additional 6 galaxies with extended radio morphologies (NGC4278, NGC4472, NGC4486, NGC5322, NGC5838, and NGC5846) resembling AGN jets/lobes.  Seven objects with milliarcsecond-scale spatial resolution data also have parsec- or sub-parsec-scale jets (NGC4261, NGC4278, NGC4374, NGC4486, NGC4552, NGC5353, and NGC5846; \citealt{nagar+05}).
Thus, 16/148 (11 $\pm$ 3\%) \atlas\ ETGs have resolved radio morphologies in high-resolution observations that are consistent with an AGN origin.

Since radio outflows/jets may exist over a variety of spatial scales, it is also useful to consider the morphology at lower spatial resolution ($\theta_{\mathrm{FWHM}}$ $> 1^{\prime \prime}$).  There are 12/148 ETGs in our sample that display extended jet/lobe-like radio structures on scales of kiloparsecs.  These galaxies are NGC3665 \citep{parma+86}; NGC3998 (Frank et al., in preparation); NGC4261 and NGC4374 \citep{cavagnolo+10}; NGC4472 \citep{biller+04}; NGC4486 (e.g., \citealt{owen+00}); NGC4636 and NGC4649 \citep{dunn+10}; NGC4552 \citep{machacek+06}; NGC5322 \citep{feretti+84}; NGC5813 \citep{randall+11}; and NGC5846 \citep{giacintucci+11}.  Thus, considering both the high- and low-resolution data available, a total of 19/148 \atlas\ ETGs have morphological evidence for the presence of a radio LLAGN on some scale.  The incidence of radio jets in the \atlas\ sample is further discussed in Section~\ref{radio_jets}.

Our sample also includes 6 ETGs (IC0676, NGC1222, NGC4111, NGC4684, UGC05408, and UGC06176) with significantly extended nuclear radio emission displaying more complex morphologies that could be the result of an AGN, SF, or a mixture of these two processes.

\subsubsection{Brightness Temperature}
Sufficiently high (e.g., milliarcsecond-scale) spatial resolution radio continuum images provide a reliable means of identifying AGNs based on their brightness temperatures\footnote{Brightness temperature is defined as $T_{\rm b} \equiv (S/\Omega_{\rm beam})\frac{c^2}{2k\nu^2}$, where $\nu$ is the observing frequency, $S$ is the integrated flux density, and $\Omega_{\rm beam}$ is the beam solid angle.  The constants $c$ and $k$ are the speed of light and the Boltzmann constant, respectively.} \citep{condon+92}.  Since compact nuclear starbursts are limited to \hbox{$T_{\rm b}~\lesssim~10^5$}~K \citep{condon+92}, the detection of compact radio emission with $T_{\rm b}~>~10^5$}~K is generally considered strong evidence for accretion onto a black hole.  

For our 5~GHz \atlas\ data, the spatial resolution of $\approx 0.5^{\prime \prime}$ does not generally correspond to a sufficient brightness temperature sensitivity to unambiguously identify radio continuum emission associated with SMBH accretion.  For instance, our 5$\sigma$ detection threshold of 75 $\mu$Jy beam$^{-1}$ corresponds to $T_{\mathrm{b}} \approx 15$~K, far too low to allow us to distinguish between a SF and LLAGN origin for the weak radio emission.   In fact, the peak flux density required to achieve a brightness temperature sensitivity of 10$^5$ K given our spatial resolution and observing frequency is $\approx$ 500 mJy~beam$^{-1}$.  Thus, only one galaxy in our sample, NGC4486, has a nuclear radio source with a peak flux density greater than this limit\footnote{Based on the 5~GHz VLA peak flux density of $\sim$ 2.88 Jy beam$^{-1}$ reported in the literature for NGC4486 from radio observations with a resolution of $\theta_{\mathrm{FWHM}}$ $\approx 0.5^{\prime \prime}$ \citep{nagar+01}, its brightness temperature is $\log(T_{\mathrm{b}}/$K) $\gtrsim$~5.78.}.

However, higher-resolution (milliarcsecond-scale) data from the literature are available for some of our sample galaxies.  There are 17 \atlas\ ETGs (NGC0524, NGC2768, NGC3226, NGC3998, NGC4143, NGC4168, NGC4203, NGC4261, NGC4278, NGC4374, NGC4472, NGC4486, NGC4552, NGC5322, NGC5353, NGC5846, and NGC5866) with VLBA measurements reported in \citet{nagar+05}.  One additional galaxy from our sample, NGC1266, was detected in the VLBA study by \citet{nyland+13}.  All 18 of the \atlas\ galaxies imaged with the VLBA are characterized by nuclear radio sources with either compact morphologies or evidence for parsec-scale jets and have brightness temperatures in the range 6.8 $\lesssim \log(T_{\mathrm{b}}/$K) $\lesssim$~10.2.  This provides strong evidence that the radio emission detected in these galaxies indeed originates from nuclear activity.  

\subsubsection{Radio Variability}
Radio variability is also a common feature of LLAGNs.  As discussed in Section~\ref{sec:previous_studies}, the flux density of NGC3945 appears to have increased by a factor of $\approx$ 4 between the observations reported in \citet{wrobel+91b} and the new observations reported here.  The compact, variable radio emission in the nucleus of NGC3945 is therefore highly likely to be associated with a LLAGN.  NGC3998 is also known to exhibit radio variability \citep{kharb+12}.  It is possible that the nuclear radio emission in additional galaxies in our sample is also variable.  However, a statistical radio variability analysis of the inhomogeneous and often sparsely-sampled radio observations of the ETGs in the \atlas\ sample is beyond the scope of this paper.

\subsubsection{Radio-FIR Ratio}
\label{q}
The radio-far-infrared (radio-FIR) relation (e.g., \citealt{helou+85, condon+92, yun+01}), which extends over at least three orders of magnitude in ``normal'' star-forming galaxies, is another useful method of identifying radio AGN emission.  In this relation, the FIR emission is believed to arise from thermal infrared dust emission generated by young stars, while the nonthermal radio synchrotron emission is produced by supernova remnants.  The logarithmic radio-FIR ratio\footnote{The logarithmic radio-FIR ratio is defined as $q \equiv \log \left( \frac{\mathrm{FIR}}{3.75 \times 10^{12} \, \mathrm{W} \, \mathrm{m}^{-2}} \right) - \log \left(\frac{S_{1.4 \, \mathrm{GHz}}}{\mathrm{W} \, \mathrm{m}^{-2} \, \mathrm{Hz}^{-1}} \right)$, where $\mathrm{FIR} \equiv 1.26 \times 10^{-14} \, (2.58 S_{60 \, \mu m} + S_{100 \, \mu m})$ W~m$^{-2}$ and $S_{60 \, \mu m}$ and $S_{100 \, \mu m}$ are the Infrared Astronomical Satellite (IRAS) 60 and 100~$\mu$m band flux densities in Jy, respectively.  $S_{1.4 \, \mathrm{GHz}}$ is the 1.4~GHz flux density measured in Jy.} (q-value) can therefore be used to assess whether the level of radio emission is consistent with SF.  Galaxies with q-values indicating an excess of radio emission are likely AGN hosts, while those with an excess of FIR emission have less certain origins, and may signify dusty starbursts, dust-enshrouded AGNs, or systems with intrinsically faint radio emission.  One of the most widely-cited studies, \citet{yun+01}, reports an average q-value of 2.34 and defines radio-excess galaxies as those with $q$ $<$ 1.64.
 
However, more recent studies have demonstrated that the radio-FIR relation can only distinguish radio emission arising from LLAGNs from that produced by SF in relatively powerful, radio-loud AGNs \citep{obric+06, mauch+07, moric+10}.  In fact, \citet{obric+06} conclude that LLAGNs in the Seyfert and low-ionization nuclear emission-line region (LINER; \citealt{heckman+80}) classes often have q-values within the scatter of normal star-forming galaxies despite strong evidence from other tracers that their radio emission is indeed associated with nuclear activity.  More quantitatively, \citet{moric+10} report  an average q-value of the LLAGN host galaxies in their sample of $q$ = 2.27, with Seyfert galaxies and LINERs corresponding to average q-values of 2.14 and 2.29, respectively.  Therefore, we emphasize that in the context of LLAGNs, the radio-FIR relation cannot unambiguously identify the presence of an active nucleus. 
  
With this caveat in mind, we assess the q-values of the 94 \atlas\ ETGs in our sample with available 1.4~GHz data from NVSS, FIRST, or new 5$^{\prime \prime}$-resolution VLA observations (Nyland et al., in preparation), as well as FIR observations from IRAS available in the literature.  We find that 9 ETGs in our sample are in the radio-excess category with $q$ $<$ 1.64 (NGC3665, NGC3998, NGC4261, NGC4278, NGC4374, NGC4486, NGC4552, NGC5322, and NGC5353).  In these galaxies, the majority of the radio emission is highly likely to be associated with an AGN.  The remaining \atlas\ ETGs for which q-values can be calculated\footnote{A few of the \atlas\ ETGs have q-values in the FIR-excess category ($q$ $>$ 3.00; \citealt{yun+01}). The origin of this FIR excess -- or radio deficiency -- in these galaxies is unclear, but may be due to intrinsically weak radio continuum emission as a result of enhanced cosmic ray electron losses or weak magnetic fields in ETGs.} have $q > 1.64$ consistent with either low-level SF or weak LLAGN emission.

\begin{table}
\begin{minipage}{9cm}
\caption{4-6~GHz In-band Spectral Indices}
\label{tab:spix_cband}
\begin{tabular}{lccc}
\hline
\hline
Galaxy      & $\alpha$ & $\sigma_{\alpha}$  \\
\hline 
NGC0524       & 0.44    & 0.25 \\
NGC0936       & -0.37   & 0.28 \\
NGC1266       &  -0.57   &  0.29 \\
NGC2768       & 0.05    & 0.32 \\
NGC2824       & -0.75   &  0.25 \\
NGC2974       &  -0.03   & 0.25 \\
NGC3607       & -0.46   & 0.50 \\
NGC3945       &  -0.01  &  0.31 \\
NGC4036       & -0.91   & 0.40 \\
NGC4526       & -0.02   & 0.28 \\
NGC4546       & 0.04    & 0.28 \\
NGC5198       & -0.20   &  0.32 \\
NGC5475       & -0.91    & 0.33 \\
PGC029321   &  -0.58   & 0.26 \\
\hline
\hline
\end{tabular}
 
\end{minipage} 
\end{table}

\subsubsection{Radio Spectral Indices}
For sufficiently bright sources, the wide bandwidth of the VLA  is an appealing tool for the determination of in-band radio spectral indices.  These radio spectral index measurements may then be used to help constrain the physical origin of the radio emission.  Historically, radio core emission associated with a ``powerful" AGN is expected to have a flat (i.e., $\alpha \gtrsim -0.5$) radio spectrum due to synchrotron self absorption, while synchrotron emission produced by star-forming regions is generally associated with a steep (i.e., $\alpha \lesssim -0.5$) radio spectrum \citep{condon+92}.  However, we note that some studies have demonstrated that the radio spectra of bona-fide LLAGNs can have flat, steep, and even inverted radio spectral indices \citep{ho+01, ho+08}.  

Radio spectral index measurements are generally considered ``useful" if their uncertainties are less than the standard deviation of the distribution of spectral indices measured in large flux-limited samples of $\sim 0.1$ (e.g., \citealt{condon+84}).  
Of the 53 ETGs with detections of nuclear radio emission, 44 were observed over a relatively wide bandwidth of 2~GHz.  For these galaxies, a signal-to-noise ratio (SNR) of $\approx 60$ is required to achieve an in-band spectral index uncertainty of $\lesssim$ 0.1\footnote{Our estimate of the approximate signal-to-noise ratio required to reach this spectral index uncertainty is based on Equation~32 in \citet{condon+15}.}  The observations of the remaining 9 detected ETGs were obtained during the EVLA commissioning period, and have bandwidths of only 256~MHz.  Given this comparatively narrow bandwidth, a SNR of $>$ 500 would be required for reliable spectral index determination.  

Only 14 of the 53 detected galaxies in our high-resolution $C$-band observations meet the bandwidth-dependent SNR requirement for useful determination of the in-band spectral index.  Although our standard CASA imaging products included two-dimensional spectral index maps, spectral index values from these maps tend to be biased steep due to the variation in the synthesized beam solid angle with frequency\footnote{The synthesized beam solid angle in the VLA A-configuration changes by nearly a factor of 4 from 4 to 6 GHz.} as well as ``edge effects" caused by the sharp boundary at the default mask edge used by CASA for calculating the spectral index values.  To avoid these issues, we imaged each galaxy at the center of each baseband for the 14 high-SNR detections with wide-bandwidth data and smoothed the resulting images to a common spatial resolution.  We then used the IMFIT task in CASA to determine the integrated flux density and errors at 4.5 and 5.5 GHz.  The calculated in-band spectral indices are summarized in Table~\ref{tab:spix_cband}.  They span the range $-0.91 < \alpha < 0.44$ with a median value of $\alpha \approx -0.29$.  

Many of the compact radio sources in our sample for which in-band spectral index measurements may be calculated tend to have flat spectral indices consistent with an AGN origin.  For those with steep spectral indices, an AGN origin is still plausible, particularly if other lines of evidence support the presence of an AGN (e.g., the morphological AGN evidence visible in the NGC1266, NGC4036, and NGC5475 maps; Figure~\ref{fig:radio_images}).  However, nuclear SF cannot be completely ruled-out in NGC2824 and PGC029321.  The inverted radio spectrum of NGC0524 indicates optically-thick emission likely due to either synchrotron self-absorption or perhaps free-free absorption, and confirms measurements made in previous studies (e.g., \citealt{filho+04}).  We speculate that the inverted radio spectrum in the nucleus of NGC0524 could be evidence that this galaxy houses a young, recently-ignited radio source, perhaps similar in nature to the class of ``Compact steep spectrum'' or ``Gigahertz-peaked spectrum" sources \citep{odea+98}.

\subsubsection{Limitations}
As discussed throughout this section, the compact nuclear radio sources in our sample of ETGs can conceivably originate from LLAGNs or residual amounts circumnuclear SF.  For the radio sources in our sample that are unresolved on scales of 25-100~pc, lack brightness temperature constraints from the literature, and show no evidence of significant variability, it is difficult to distinguish between the AGN and SF origins definitively.  However, given that the typical spatial extent of the molecular gas in the \atlas\ sample is $\sim$ 1~kpc (\citetalias{davis+13}; \citealt{davis+14}, hereafter \citetalias{davis+14}), we find it unlikely that SF is the primary driver of the radio emission in the ETGs in our sample on spatial scales an order of magnitude or more smaller.  Although we favor a LLAGN origin for the majority of the 5~GHz sources present in the nuclei of our sample galaxies, we explore additional tracers (nebular emission line diagnostics and X-ray properties) in the remainder of this section to help further justify this interpretation.

\subsection{Optical Emission Line Diagnostics}
\label{sec:emission_line_diagnostics}
All 260 of the \atlas\ galaxies have nebular emission-line measurements from observations with the {\tt SAURON} integral-field spectrograph \citepalias{cappellari+11a}. However, this instrument was designed to measure the stellar kinematics and stellar population properties over a relatively large field of view ($30\arcsec \times
40\arcsec$, \citealt{bacon+01}), at the sacrifice of bandwidth (4830 $-$ 5330 \AA).  As a consequence, only two of the strong emission lines
required for traditional diagnostic diagrams (e.g., \citealt{veilleux+87, kewley+06}) are accessible in the {\tt SAURON} data, namely [{\sc O$\,$iii}]$\lambda\lambda4959,5007$ and H$\beta$. One additional weak doublet, [{\sc N$\,$i}]$\lambda\lambda5197,5200$, is detected in a small subset of the \atlas\ sample. When present, the [{\sc N$\,$i}]/H$\beta$ vs. [{\sc O$\,$iii}]/H$\beta$ diagnostic diagram is comparable to the traditional [{\sc O$\,$i}]/H$\alpha$ vs. [{\sc O$\,$iii}]/H$\beta$ diagnostics and can be used to robustly identify the source of ionization of the gas \citep{sarzi+10}.  

Despite the lack of strong emission lines necessary for gauging the hardness of the ionizing radiation field (e.g., [{\sc O$\,$i}], [{\sc S$\,$ii}], [{\sc N$\,$ii}]), we can use
literature data and the central value of the [{\sc O$\,$iii}]/H$\beta$ ratio in concert with ancillary information to help constrain the primary ionization mechanism in the nuclei of the \atlas\ ETGs. 
In the section that follows, we provide a brief explanation of our emission line classification scheme, and utilize these optical emission line classifications to help determine the origin of the compact radio continuum sources in our high-resolution 5~GHz VLA sample.  Details on the extraction of the ionized gas properties from the {\tt SAURON} data\footnote{http://purl.org/atlas3d} are provided in \citetalias{cappellari+11a}.  

\begin{table}
\begin{minipage}{9cm}
\caption{Emission Line Classes}
\label{tab:contingency_optical}
\begin{tabular}{lccc}
\hline
\hline
Classification      & Radio Det. & Radio UL  & Total  \\
\hline 
      All emission classes    &        70          &          31        &     101      \\
      \hspace{0.5cm} Seyfert                 	&          3          &            4         &       7       \\
      \hspace{0.5cm} LINER-AGN         &        31          &            5         &     36       \\
      \hspace{0.5cm} LINER                  	&       19           &         12         &      31       \\
      \hspace{0.5cm} Transition         &       10          &            5         &      15       \\
      \hspace{0.5cm} H{\tt II}     		&          7          &            5         &      12       \\
      \hspace{0.5cm} Passive    		&          6          &          41         &      47       \\
\hline
\hline
\end{tabular}
 
\end{minipage} 
\end{table}

\subsubsection{Emission Line Classification Scheme}
\label{classification}
For several \atlas\ galaxies there exist central spectroscopic data from the SDSS and the Palomar survey  \citep{ho+97} that we can use to robustly classify the central nebular activity of our ETGs, adopting in particular the dividing lines drawn by \citet{schawinski+07} in the [{\sc N$\,$ii}]/H$\alpha$ vs. [{\sc O$\,$iii}]/H$\beta$ diagnostic diagram.  Such a classification based on the SDSS and Palomar data is our preferred method of identifying the dominant source of the ionized emission in this study, followed by the classification based on the central [{\sc N$\,$i}]/H$\beta$ vs. [{\sc O$\,$iii}]/H$\beta$ diagnostic diagram if [{\sc N$\,$i}] is detected. Failing this, we rely on the central value measured within a 3\arcsec-wide aperture of the [{\sc O$\,$iii}]/H$\beta$ ratio from the {\tt SAURON} data.

\begin{figure*}
\includegraphics[clip=true, trim=2.5cm 1cm 3.75cm 2cm, scale=0.5]{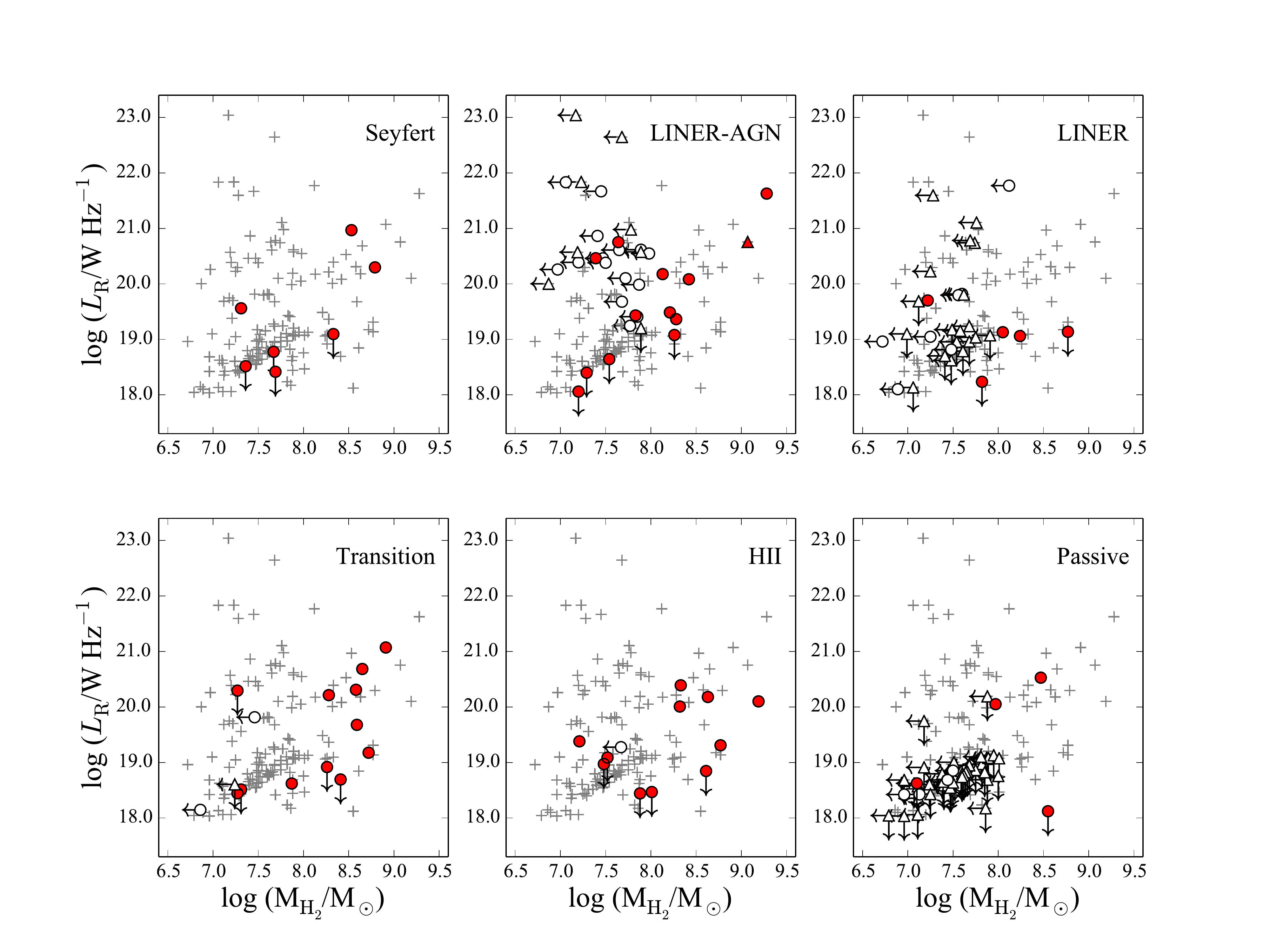}
\caption{5~GHz luminosity vs. H$_2$ mass for each optical emission line classification.  Symbols filled in red represent CO detections; unfilled symbols are CO upper limits \citepalias{young+11}.  Circles represent fast rotators while upward-pointing triangles represent slow rotators \citepalias{emsellem+11}.  Radio luminosity upper limits are denoted by the downward pointing arrows.  Leftward-pointing arrows represent H$_2$ mass upper limits.  The light gray crosses shown in each panel represent the distribution of our full 5~GHz sample.}
\label{fig:emission_radio_H2_sub}
\end{figure*}

For very high ([{\sc O$\,$iii}]/H$\beta > 8.0$) and low ([{\sc O$\,$iii}]/H$\beta < 0.5$) values of this ratio, the ionized gas is most likely excited by a Seyfert nucleus and young O- or B-stars, respectively. For galaxies with more intermediate values of the [{\sc O$\,$iii}]/H$\beta$ ratio, the situation is more complicated. High values of
this ratio can also be observed in star-forming regions if the gas metallicity is low.  AGNs can also lead to modest values of the [{\sc O$\,$iii}]/H$\beta$ ratio, in particular when considering LINERs and composite or transition objects, the latter of which likely contain a mix of SF and an AGN.  

For LINERs, the situation is further complicated since a number of mechanisms besides AGN-driven emission, such as shocks or photoionization from old UV-bright stars, can power this kind of nebular emission (for a review, see \citealt{ho+08}). To help with the nuclear classification at intermediate [{\sc O$\,$iii}]/H$\beta$ ratios, we adapted to the use of our central galactic measurements the so-called ``mass excitation diagram'' of \citet{juneau+11}, where [{\sc O$\,$iii}]/H$\beta$ is juxtaposed to the stellar velocity dispersion, $\sigma_{*}$. The use of this diagram is particularly useful to separate at large [{\sc O$\,$iii}]/H$\beta$ ratios Seyfert and LINERs from SF systems. The latter always tend to have smaller values for the velocity dispersion in the intermediate [{\sc O$\,$iii}]/H$\beta$ ratio regime, which in turn relates to the need for a low gas metallicity that can only be maintained in less massive nuclei. 

Using the Palomar survey database of central emission-line \citep{ho+97} and stellar velocity dispersion measurements \citep{ho+09}, we calibrate the mass excitation diagram for our purposes, finding that a value of $\sigma_{*}=70$ km~s$^{-1}$ constitutes a good threshold for separating SF from Seyfert, LINER and composite activity. The latter classes are then distinguished from one another in the [{\sc O$\,$iii}]/H$\beta$ vs. $\sigma_{*}$ diagnostic diagram for values of $\sigma_{*}  >$ 70 km~s$^{-1}$. Transition objects lie in the region between $0.5 <$ [{\sc O$\,$iii}]/H$\beta < 1$, LINERs between [{\sc O$\,$iii}]/H$\beta > 1$ and the $\log$([{\sc O$\,$iii}]/H$\beta$) = 1.6 $\times$ 10$^{-3}$ $\sigma$ + 0.33 line, and Seyfert nuclei above it.  Finally, to separate among LINERs those that are powered by true AGNs from those dominated by old UV-bright stars, we select AGN-LINERs by requiring equivalent width values for [{\sc O$\,$iii}] above 0.8 \AA, similar to \citet{cid_fernandes+11} who used H$\alpha$.  Galaxies with [{\sc O$\,$iii}] and/or H$\beta$ emission below our detection thresholds are regarded as passive nuclei. 

\subsubsection{The Nuclear Radio Emission Connection}
\label{sec:nuc_rad_conn}
The fraction of ETGs in each optical emission line class for the VLA 5~GHz detections and non-detections is shown in Table~\ref{tab:contingency_optical}.  This table clearly demonstrates that ETGs in the Seyfert, LINER-AGN, LINER, Transition, and H{\tt II} classes are more likely to harbor a nuclear radio source compared to ETGs in the passive emission line group, suggesting a common physical origin for the central ionized gas and radio emission in many of these sources.  ETGs in the LINER-AGN class are especially likely to be detected in our high-resolution observations, consistent with previous studies of nearby LLAGNs (e.g., \citealt{ho+99}).  Among the radio-detected ETGs in our sample with nuclear emission line classifications suggestive of LLAGN emission, the central 5~GHz emission is highly likely to originate from SMBH accretion.

As shown in Table~\ref{tab:contingency_optical}, our sample contains only 7 Seyfert nuclei, 3 of which harbor nuclear radio emission.  Although there are not enough ETGs in our sample classified as Seyferts to perform a robust statistical analysis, it is interesting to consider the lack of nuclear radio emission in 4/7 of the Seyferts.  We speculate that the 4 Seyferts with only upper limits to any nuclear 5~GHz emission may have higher SMBH accretion rates, which would be consistent with their tendency as a class to be less radio loud compared to LINERs \citep{ho+02, ho+08}.  Another possibility is that the Seyferts not detected in our high-resolution radio data are dominated by radio emission on substantially larger scales that has been resolved-out (e.g., \citealt{gallimore+06}).  It is also possible that the criteria we used to identify Seyfert nuclei in our ETG sample are unreliable or insufficient in these systems.  If this is indeed the case, the high ionization nebular emission in the ETGs in the Seyfert class lacking nuclear radio emission may arise from stellar processes rather than SMBH accretion (e.g., \citealt{martins+12}).  The situation is similar for ETGs in the LINER-AGN or LINER classes that lack 5~GHz detections.  In these objects, ultraviolet radiation from post asymptotic giant branch (pAGB) stars may be responsible for the ionized gas emission rather than nuclear activity (e.g., \citealt{sarzi+10, cid_fernandes+11}).  

In Figure~\ref{fig:emission_radio_H2_sub}, we show the radio luminosity as a function of the H$_2$ mass for each optical emission line class separately.  In the remainder of this section, we describe the radio properties of the Seyfet, LINER-AGN, LINER, transition, H{\tt II}, and passive classes and discuss implications for the origin of the radio emission (AGNs or SF activity).
\\

\noindent {\bf \emph{Seyfert Nuclei:}}
All of the ETGs classified as Seyferts are FRs and are rich in molecular gas.  
This is consistent with previous studies concluding that Seyfert nuclei generally reside in gas-rich host galaxies (e.g., \citealt{maiolino+97, hicks+09}).  When detected at 5~GHz (NGC5273, NGC7465, and PGC029321), the Seyferts have radio luminosities of  $L_{\rm R} \lesssim 10^{21}$ W Hz$^{-1}$.  The nuclear radio detection rate of the Seyfert nuclei is 3/7.  Although the small number of ETGs in this  class prevents a robust statistical analysis, we note that other studies of nearby, lower-luminosity Seyferts hosted by ETGs have reported higher radio continuum detection rates (e.g., $\sim$ 100\%; \citealt{nagar+99}, $\sim$ 87\%; \citealt{ho+01}).  This disparity could be due to the selection criteria used to define the samples in \citet{nagar+99} and \cite{ho+01}, which covered larger survey volumes and were constrained by brighter magnitude thresholds compared to the \atlas\ survey.  Differences in the optical emission-line classification methods between studies may contribute to the disparity in the radio detection fraction of Seyferts as well. 

The 4 Seyferts with no detectable nuclear radio continuum emission are NGC3156, NGC3182, NGC3599, and NGC4324.
NGC3182 is detected in lower spatial resolution data at 1.4~GHz (Nyland et al., in preparation) and has a resolved, clumpy, ring-like morphology with an extent of $\approx$ 13$^{\prime \prime}$ ($\approx$ 5~kpc) that traces the molecular gas distribution seen in interferometric CO maps \citepalias{alatalo+13}.  NGC3599 also contains 1.4~GHz continuum emission on larger scales based on WSRT data that is unresolved at a spatial resolution of $\theta_{\mathrm{FWHM}} \approx 35^{\prime \prime}$.  Thus, both of these galaxies almost certainly harbor SF on larger spatial scales that is likely responsible for some (if not all) of the lower-resolution 1.4~GHz radio emission.  However, since the larger-scale radio emission in these galaxies does not display any classic morphological signatures of AGN emission, whether they truly harbor bona fide Seyfert nuclei associated with LLAGNs remains unclear.   

NGC3156 and NGC4324 lack evidence for radio emission even in sensitive, lower-resolution 1.4~GHz observations (Nyland et al., in preparation), indicating that the high-resolution non-detections at 5~GHz in these ETGs are likely not caused by insensitivity to emission on larger spatial scales.  
Given the lack of radio evidence for the presence of an AGN in these galaxies, we conclude that the most likely explanation for the origin of the Seyfert-like emission line properties in NGC3156, NGC3182, NGC3599, and NGC4324 is stellar processes.  For example, photoionization driven by supernova remnants and/or planetary nebulae has been known to mimic the high-ionization nebular emission characteristic of Seyfert nuclei (e.g., \citealt{martins+12}).  The poststarburst features in the optical spectrum of NGC3156 support this scenario \citep{caldwell+96, sarzi+10}.  Furthermore, \citet{sarzi+10} conclude that the Seyfert classification in the nucleus of NGC3156 is likely the result of stellar processes associated with a SF event that began around 100~Myr ago and has now produced a substantial population of stellar remnants capable of generating high ionization emission lines.  
\\

\noindent {\bf \emph{LINER-AGN and LINER Nuclei:}}
The 36 LINER-AGN nuclei and 31 LINERs included in our sample have higher detection rates of nuclear 5~GHz emission (86 $\pm$ 6\% and 61 $\pm$ 9\%, respectively) compared to the Seyferts, but this is only statistically significant for the LINER-AGN nuclei (3$\sigma$).  The most powerful radio sources in our sample are hosted by the LINER-AGN class, though many of the LINERs of less certain origin (i.e., those with [{\sc O$\,$iii}] equivalent widths less than 0.8\AA) also have strong radio emission.  
In the majority of cases, the nuclear radio emission in the ETGs in the LINER-AGN and LINER classes is likely to be associated with genuine AGNs.  However, we caution that the LINER-AGN class does contain two peculiar galaxies, NGC1222 and NGC1266.  NGC1222 is a highly disturbed system harboring a strong starburst that is also in the process of undergoing a complex three-way merger \citep{beck+07}.  The LINER-AGN nuclear emission line classification for NGC1222 is rather uncertain.  In fact, \citet{riffel+13} conclude that NGC1222 actually has a star-forming nucleus based on additional near-infrared diagnostic criteria.  As for NGC1266, while this galaxy does host a radio-emitting AGN \citep{nyland+13}, the low-ionization emission responsible for its LINER-AGN classification is spatially extended in nature and likely driven by shocks rather than AGN photoionization \citep{davis+12, pellegrini+13}.
\\

\noindent {\bf \emph{Transition Nuclei:}}
For the ETGs with Transition nuclei that also harbor molecular gas, nuclear radio emission may be related to compact circumnuclear SF, a LLAGN, or a mixture of these mechanisms.  Here, we emphasize that circumnuclear SF and LLAGNs are not at all mutually exclusive phenomena.  Both simulations (e.g., \citealt{hopkins+10}) and observations \citep{shi+07, watabe+08, esquej+14} suggest that these processes commonly co-exist and may even share evolutionary connections (for a review, see \citealt{heckman+14}).  

The Transition nuclei have a 5~GHz detection rate of 10/15 (67 $\pm$ 12\%).  
All of the ETGs with nuclear radio detections are FRs, and all but two (NGC3301 and NGC4697) contain molecular gas.  As discussed in Section~\ref{morph}, the morphology of the central radio source in the Transition nucleus host NGC3665 strongly supports the presence of an AGN in this object.  NGC4435 and NGC4710, on the other hand, both have a deficit of nuclear 5~GHz emission compared to their radio emission on scales of kiloparsecs of an order of magnitude or more.  These galaxies likely harbor a combination of an AGN in their nuclei and SF on larger scales capable of powering the radio emission visible at high and low spatial resolution, respectively.  NGC4526 is known to harbor low-level SF concentrated in a ring of molecular gas located a few hundred parsecs from its center \citep{sarzi+10, young+08, crocker+11, utomo+15}.  The radio source detected in the nucleus of NGC4526 may originate from weak SMBH accretion, although a circumnuclear SF origin cannot be ruled out.  NGC4697 contains a hard nuclear X-ray source \citep{pellegrini+10}, which indicates the the compact, central radio source in this galaxy is most likely powered by SMBH accretion.  We note that while the classification of UGC06176 as a transition nucleus is based on our preferred diagnostic from SDSS, the [{\sc N$\,$i}]/H$\beta$ vs. [{\sc O$\,$iii}]/H$\beta$ diagnostic diagram suggests this galaxy actually contains an H{\tt II} nucleus.  This suggests that SF may contribute substantially to the production of the central radio emission in UGC06176.  

The 5 ETGs in the Transition nucleus class lacking nuclear radio emission are all FRs with molecular gas, except for NGC4550, which is classified as a SR and lacks evidence for the presence of CO emission.  NGC3245 contains a nuclear hard X-ray source, leading \citet{filho+04} to conclude that AGN photoionization is indeed significant in this galaxy.
In NGC3032 and NGC3245, SF is likely responsible for the extended radio emission on larger scales seen in the FIRST, NVSS, and \citet{wrobel+91b} data.  SF domination in NGC3032 is further supported by evidence for a young stellar population and recent SF \citep{young+08, sarzi+10, shapiro+10, mcdermid+15}.
Thus, nuclear activity produced by SMBH accretion in NGC3032, as well as other transition nuclei lacking central radio sources, is likely extremely weak or essentially absent.
\\

\noindent {\bf \emph{H{\tt II} Nuclei:}}
H{\tt II} nuclei with both molecular gas and radio continuum emission are excellent candidates for ETGs hosting nuclear SF.  There are 12 galaxies with optical emission line ratios consistent with the H{\tt II} nucleus classification.  These galaxies are classified kinematically as FRs, generally contain molecular gas (IC0560 is the only exception), host young central stellar populations \citepalias{mcdermid+15}, and also commonly harbor compact radio sources in their nuclei (58 $\pm$ 14\%).  
The 5 galaxies lacking nuclear radio source detections may harbor diffuse radio emission with disk-like morphologies that is distributed over larger spatial scales and associated with recent SF.  This is likely the case for IC1024 and NGC4684, which are both detected at lower spatial resolution in FIRST.
\\

\noindent {\bf \emph{Passive Nuclei:}}
An additional 47 galaxies included in our high-resolution radio study have passive nuclei.  Of these, 6 (13 $\pm$ 5\%) have nuclear 5~GHz detections. 
Two of the radio-detected ETGs in the Passive nucleus group, NGC5866 and NGC0524, are both listed as transition nuclei in \citet{ho+97}, although those classifications are uncertain.  In addition, there are two galaxies (NGC4283 and NGC4753) in this group that contain molecular gas but lack central radio emission.  For these galaxies, follow-up studies are needed to determine if the molecular gas is forming stars or fueling weak LLAGNs.  NGC2698, NGC2880, NGC3377, and NGC4621 host weak radio nuclei but show no signs of molecular gas, ionized gas, or young stellar populations \citepalias{young+11, mcdermid+15}, suggesting they likely contain LLAGNs accreting at very low levels with optical emission line signatures that are simply below the detection threshold of the {\tt SAURON} data.
\\

\noindent {\bf \emph{Summary:}} We find that the high incidence of nuclear radio sources associated with ETGs hosting LINERs and AGN-LINERs supports a scenario in which the 5 GHz emission in these objects is produced by low-level SMBH accretion.  When present, the origin of the radio emission located in the galactic centers of Seyfert hosts is likely SMBH accretion.  This is also supported by the detection of strong, hard, 4-195 keV X-ray sources in these objects (see Section~\ref{sec:X-rays}).  The high-ionization nebular emission present in the Seyfert nuclei lacking central radio sources is likely the result of SF or other stellar processes.  A few \atlas\ galaxies in the Transition nucleus class may harbor nuclear radio emission produced by an AGN (most notably NGC3665).  However, nuclear SF generally appears to be the best explanation for the origin of the radio emission in Transition nuclei.  The central radio emission detected in a subset of the ETGs in the Passive nucleus class is likely associated with SMBHs accreting at extremely low levels.

\begin{figure*}
\includegraphics[clip=true, trim=0cm 0.1cm 1.5cm 1cm, scale=0.40]{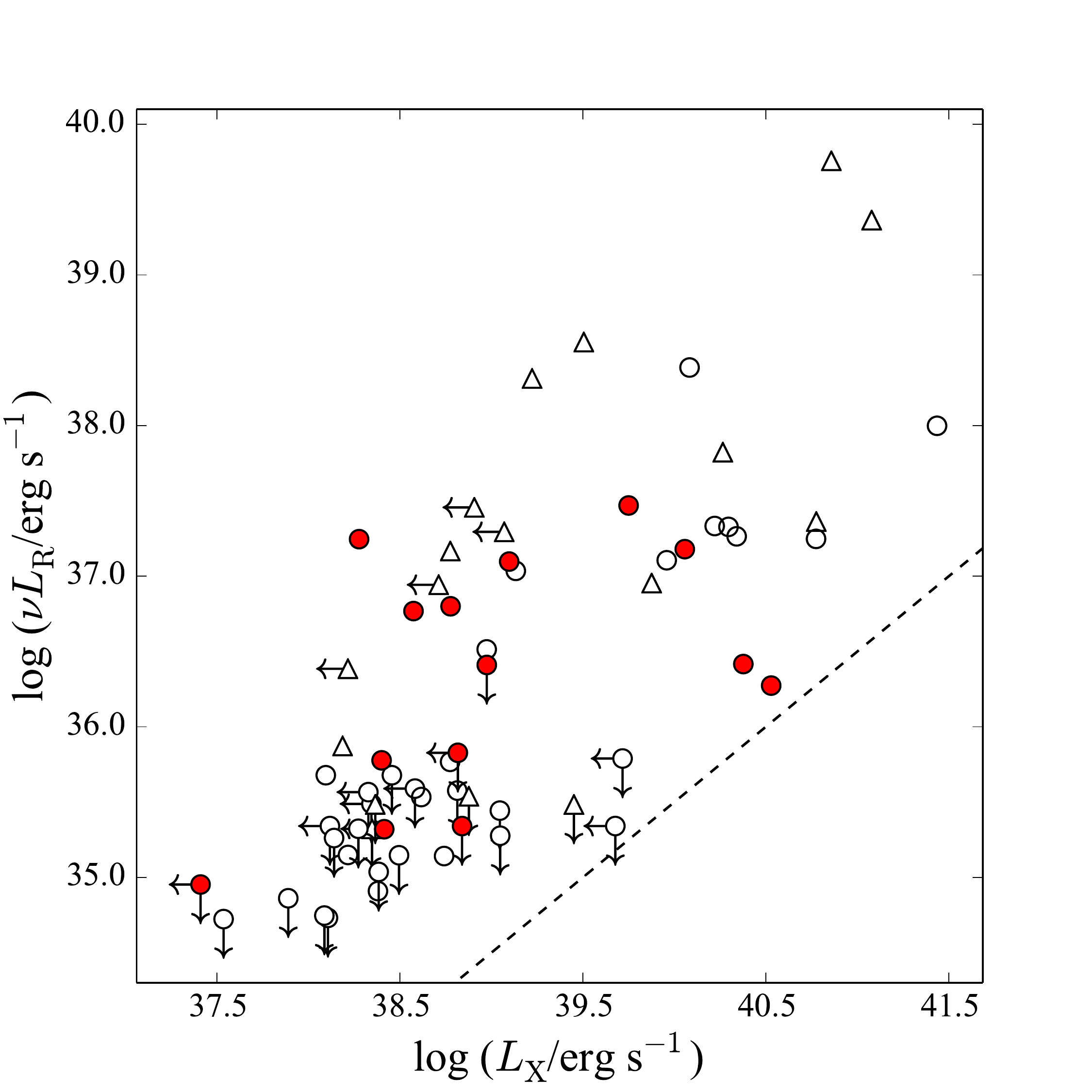}
\includegraphics[clip=true, trim=0cm 0.1cm 1.5cm 1cm, scale=0.40]{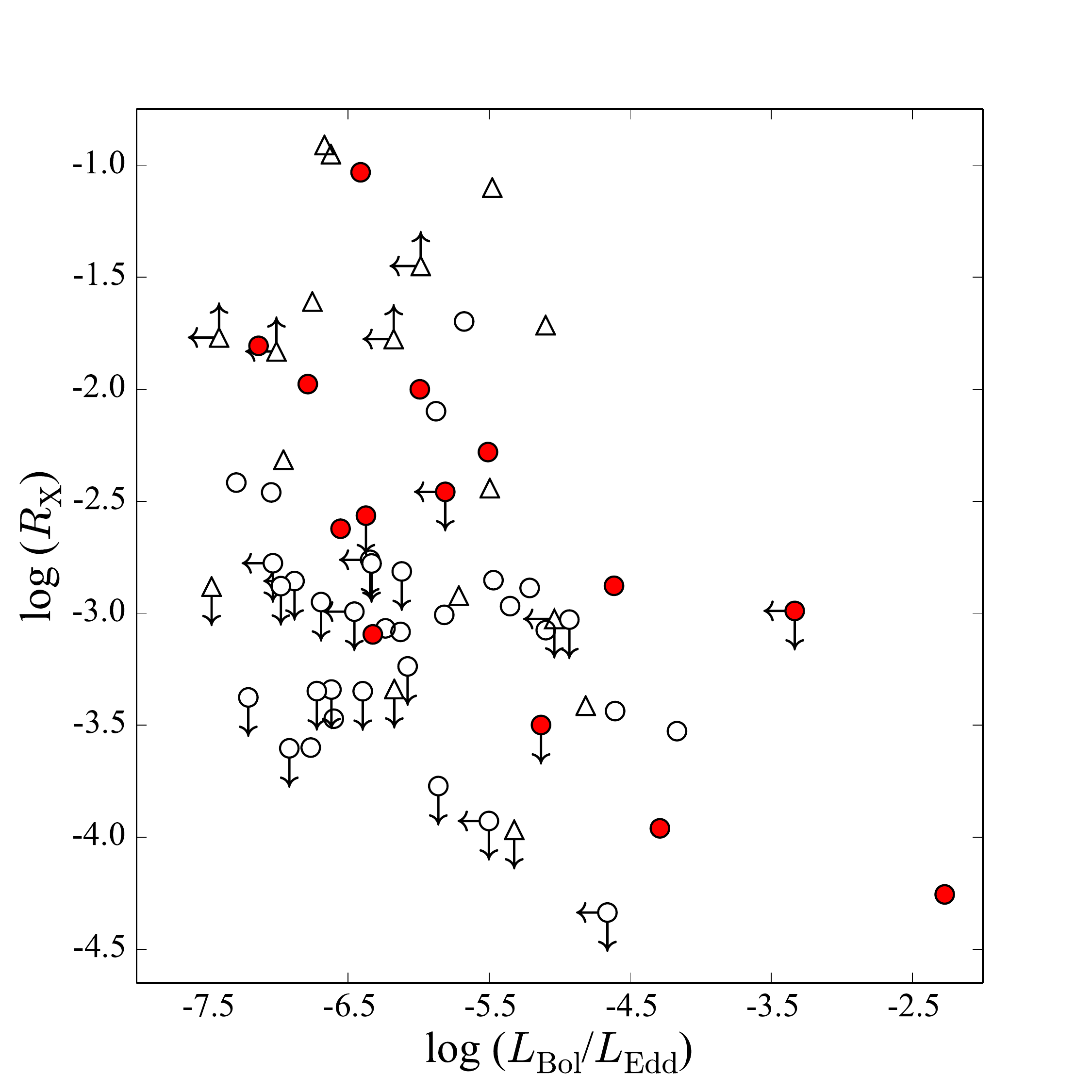}
\caption{{\bf (Left:)} 5~GHz radio luminosity as a function of nuclear X-ray luminosity for the 64 \atlas\ ETGs with both high-resolution radio and X-ray data available.  The dashed black line represents the radio-X-ray ratio conventionally adopted as the division between radio-loud and radio-quiet emission of $\log (R_{\mathrm{X}}) = -4.5$ \citep{terashima+03} for LLAGNs.  All of the \atlas\ ETGs with both nuclear radio and X-ray detections shown in this figure lie above the black dashed line given current observational limits and are therefore formally radio loud.  {\bf (Right:)} Eddington ratio vs. radio-X-ray ratio.  In both panels, symbols filled in red represent CO detections; unfilled symbols are CO upper limits \citepalias{young+11}.  Circles represent fast rotators while upward-pointing triangles represent slow rotators \citepalias{emsellem+11}.  Downward- and leftward-pointing arrows denote upper limits.  Upward-pointing arrows represent lower limits.}
\label{fig:radio_xray}
\end{figure*}

\subsection{X-ray Diagnostics}
\label{sec:X-rays}
The presence of a compact, hard (2-10 keV) X-ray source in a high-resolution X-ray image of a galaxy nucleus provides another excellent means of identifying even LLAGN emission \citep{ho+08}.  The publicly available Swift-BAT\footnote{The BAT refers to the Burst Alert Telescope onboard the Swift gamma-ray observatory.} all-sky hard X-ray survey \citep{baumgartner+13} is sensitive to emission in the 4-195 keV band, although it lacks sensitivity to faint sources.  Nevertheless, three gas-rich galaxies in our sample (NGC3998, NGC5273, and NGC7465) are listed as detections in the Swift-BAT catalog and classified as AGNs.  NGC5273 and NGC7465 are optically classified as Seyferts, while NGC3998 is classified as a LINER-AGN.  Given the presence of hard X-ray emission along with a compact radio core in each of the Swift-BAT detections, the nuclear emission in these ETGs almost certainly originates from an AGN.

Data from an X-ray observatory that is sensitive to much fainter X-ray emission is necessary to further assess the high-energy properties of the ETGs in our sample.  The $Chandra$ observatory offers the best sensitivity and highest spatial resolution X-ray data available.  With a spatial resolution higher than 1$^{\prime \prime}$, $Chandra$ data are well-matched to the properties of our 5~GHz VLA images.  Although no large-scale survey has been conducted with $Chandra$, nuclear X-ray luminosities for many of the ETGs in this study are available in the literature.  \citet{pellegrini+10} reports on a large survey of the nuclear X-ray properties of 112 ETGs, 63 of which overlap with the \atlas\ survey.  Nuclear X-ray measurements for 14 \citep{miller+12} and 3 \citep{kharb+12} additional \atlas\ ETGs are also available.  This yields 80 \atlas\ ETGs with high-spatial-resolution, nuclear, hard X-ray luminosities in the literature.  

Of these 80 galaxies, 64 also have sub-arcsecond resolution radio continuum measurements.  
A total of 34/64 (53 $\pm$ 6\%) \atlas\ ETGs have detections in both the high-resolution 5~GHz and X-ray data.  The detection of both a hard, nuclear X-ray source and central radio emission supports the presence of a LLAGN in these galaxies.  A more rigorous assessment of the astrometric alignment between the nuclear radio and hard X-ray sources would more robustly verify the presence of a LLAGN in these sources, but is beyond the scope of this paper.

There are 9/64 (14 $\pm$ 4\%) galaxies with only upper limits to both their nuclear radio and X-ray emission. 
These ETG nuclei may be dominated by low-level SF or they may be genuinely passive systems. 
A few ETGs (4/64; 6 $\pm$ 3\%) have radio detections, but only upper limits to the presence of a nuclear, hard X-ray point source (NGC4168, NGC4472, NGC4636, and NGC5198).  These are all massive SRs lacking molecular gas and are among the most radio-loud ETGs in our sample.  The nuclear radio sources in these \atlas\ galaxies are likely associated with extremely weak LLAGNs.  Although deeper X-ray observations might reveal fainter emission, the existing limits in these galaxies suggest that their nuclear X-ray emission is likely intrinsically weak due to their inefficient, highly sub-Eddington accretion rates \citep{ho+08}. 
Finally, there are 17/64 (27 $\pm$ 6\%) galaxies with X-ray detections but only radio upper limits. 
The situation is more complicated with this group. One possibility is their SMBHs are accreting in a more radiatively efficient regime in which radio emission is less dominant (e.g., \citealt{ho+02}).  These could also be ``false" X-ray identifications of LLAGNs in which the hard, nuclear X-ray emission is actually due to contamination from stellar-mass X-ray binaries rather than SMBH accretion (e.g., \citealt{mcalpine+11}).  In the future, it would be useful to obtain deeper radio and X-ray observations of this group of ETGs in particular in order to place tighter constraints on the nature of their nuclear emission.

\subsubsection{The Radio-X-Ray Ratio}
\label{sec:radio_xray_ratio}
In the left panel of Figure~\ref{fig:radio_xray}, we have plotted the $Chandra$ nuclear X-ray luminosity vs. the 5~GHz radio luminosity for the 64 galaxies with measurements available in both regimes.  There is an approximate trend of increasing radio luminosity with increasing X-ray luminosity.  We evaluate this relationship in a statistical sense by utilizing the generalized Kendall's $\tau$ correlation test.  Details regarding this statistical test are provided in Appendix~\ref{sec:stats} and Table~\ref{tab:stats}.

Formally, we find $\tau$ = 0.420 $\pm$ 0.043 and $p = 9.5 \times 10^{-7}$ for the relationship between radio luminosity and X-ray luminosity.  Thus, the relation is statistically significant, in agreement with the results of previous studies of low-power AGNs (e.g., \citealt{balmaverde+06, panessa+07}).  In addition, all of the galaxies in our sample with central radio and X-ray detections are formally radio-loud according to the convention for LLAGNs defined by a radio-X-ray ratio\footnote{$R_{\mathrm{X}} = \nu L_{\mathrm{R}}/L_{\mathrm{X}}$, where $\nu$ is the radio frequency in Hz, $L_{\mathrm{R}}$ is the 5~GHz radio luminosity in erg s$^{-1}$ Hz$^{-1}$, and $L_{\mathrm{X}}$ is the hard (2-10 keV) X-ray luminosity in erg s$^{-1}$ \citep{terashima+03}.} of $\log (R_{\mathrm{X}}) > -4.5$ \citep{terashima+03}.  This high fraction of radio-loud ETGs is in agreement with previous studies of the local population of LLAGNs \citep{ho+02, panessa+07}.

Although we do find a relationship between the sub-arcsecond nuclear radio and X-ray luminosities, we emphasize that the X-ray measurements utilized in this analysis are inhomogeneous in nature since they were obtained from multiple sources in the literature.  A future study in which the archival $Chandra$ data are re-analyzed in a consistent way and combined with new $Chandra$ observations of galaxies without existing high-resolution X-ray data (or those with only shallow observations) would improve this analysis.  Such a future study would also allow us to astrometrically match radio and X-ray sources, ensure that the nuclear X-ray emission is not contaminated by diffuse X-ray halo emission \citepalias{sarzi+13}, and improve statistics.

\subsubsection{Inefficient Massive Black Hole Accretion}
\label{sec:inefficient_black_hole_accretion}
In the right panel of Figure~\ref{fig:radio_xray}, we show the radio-X-ray ratio as a function of the Eddington ratio\footnote{The Eddington ratio is defined as $L_{\mathrm{Bol}}/L_{\mathrm{Edd}}$.  $L_{\mathrm{Bol}}$ is the bolometric luminosity defined in this work in the standard manner for LLAGNs as $L_{\mathrm{Bol}} = 16 L_{\mathrm{X}}$, where $L_{\mathrm{X}}$ is the X-ray luminosity in the $2-10$ keV band \citep{ho+08}.  $L_{\mathrm{Edd}}$ is the Eddington luminosity, $ L_{\mathrm{Edd}} \approx 1.26 \times 10^{38} \, \, M_{\mathrm{BH}}/\mathrm{M}_{\odot} \quad \mathrm{erg~s}^{-1}$ \citep{ho+08} where $M_{\mathrm{BH}}$ is the black hole mass.  Our estimates for $M_{\mathrm{BH}}$ are calculated from the $M_{\mathrm{BH}}-\sigma_{*}$ relation \citep{mcconnell+13} or dynamical measurements \citep{kormendy+13}, when available.}, a proxy for black hole accretion rate (e.g., \citealt{kauffmann+09}).   
If the radio sources in our sample are indeed generally associated with LLAGNs, we would expect to find a rough inverse relationship between the radio-loudness and the Eddington ratio in this figure.  This would be in agreement with previous studies of LLAGNs \citep{ho+02, panessa+07, sikora+07, rafter+09} and theoretical predictions of inefficient accretion flows in LLAGNs \citep{narayan+98}.  By eye, there does appear to be a rough inverse correlation between these parameters.  However, it is difficult to establish the statistical significance of this relationship due to the prevalence of upper and lower limits in the radio-X-ray ratio.  The censored Kendall's $\tau$ test gives $\tau$ = $-0.052$ $\pm$ 0.036, $p$ = 5.9 $\times$ 10$^{-1}$, which indicates the relationship in the right panel of Figure~\ref{fig:radio_xray} is not statistically significant.  Nevertheless, the absence of sources in the upper right portion of the figure is consistent with expectations if many of the ETGs in our sample harbor LLAGNs.  It also illustrates the fact that the massive black holes in the nuclei of our sample ETGs are accreting material well below their Eddington limits, in the regime of inefficient black hole accretion \citep{ho+08}.  The topic of SMBH accretion is discussed further in Section~\ref{sec:discussion}.

\section{Nuclear Radio Emission and Galaxy Properties}
\label{sec:galaxy_properties}

\subsection{Radio Luminosity and Host Kinematics}
\label{sec:radio_kin}
\subsubsection{Specific Angular Momentum}
The radio continuum detection rate for the FRs and SRs included in our study is $47 \pm 4$\% and $74 \pm 9$\%, respectively.  This difference in detection rate has a statistical significance of 2.4$\sigma$ and suggests that SRs more commonly host compact radio emission than FRs.  
However, as discussed in Section~\ref{sec:det_rate}, the radio detection rate in our sample is mildly biased by stellar mass (i.e., more massive galaxies are more likely to host a nuclear radio source).  Since the SRs are generally characterized by high galaxy masses, the difference in radio detection rate between FRs and SRs could be due to the underlying dependence of the radio detection fraction on the stellar mass.  

To quantify the effect of mass bias on the FR and SR nuclear radio detection rates, we perform a bootstrap resampling simulation to construct a sample of FRs matched in stellar mass to the SRs.  Although our full sample includes 23 SRs, we discard 6 SRs from our matched sample analysis because their high stellar masses have no counterparts among the FRs.  The radio detection rate of the remaining 17 SRs is 65 $\pm$ 12\%.  For each of the 17 SRs in our resampling simulation, we randomly select one FR with $\log(M_{\mathrm{JAM}}/$M$_{\odot}$) within 0.1 dex\footnote{The tolerance value of 0.1 dex corresponds to the scatter around fundamental plane relations.  For further details, see \citetalias{cappellari+13a}.} of the SR stellar mass.  After performing 10$^4$ iterations, we find a FR detection rate in our matched sample of 59 $\pm$ 10\% (where the detection rate is the median value and the uncertainty is the standard deviation of our simulated distribution of FR detection rates).  Thus, our simulation indicates that for samples matched in stellar mass, no statistically significant difference in nuclear radio detection rate is present.

\begin{figure}
\includegraphics[clip=true, trim=0cm 0cm 1.5cm 2cm, scale=0.4]{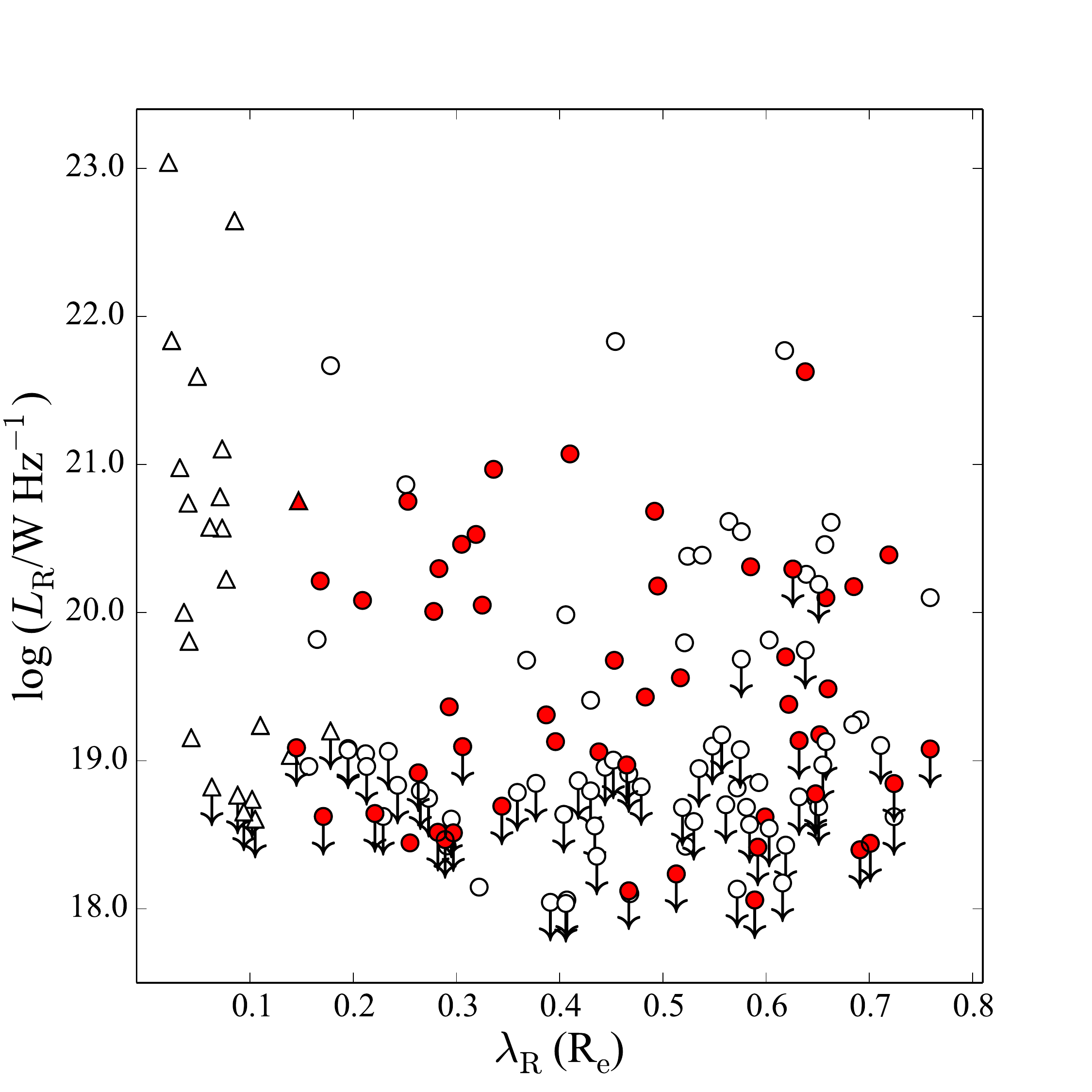}
\caption{5~GHz radio luminosity vs. specific angular momentum ($\lambda_{\mathrm{R}}$; \citetalias{emsellem+11}).  $\lambda_{\mathrm{R}}$ is measured within one effective radius ($R_{\mathrm{e}}$).  Symbols filled in red represent CO detections \citepalias{young+11}; unfilled symbols represent CO non-detections.  Circles represent fast rotators while upward-pointing triangles represent slow rotators \citepalias{emsellem+11}.  Downward-pointing arrows represent 5~GHz upper limits.}
\label{fig:compare_kormendy}
\end{figure}

Figure~\ref{fig:compare_kormendy} shows the nuclear 5~GHz radio luminosity as a function of $\lambda_{\mathrm{R}}$ measured at one effective radius\footnote{We define the effective radius as the circular aperture containing half of the optical light of the galaxy.  Effective radius measurements for all of the \atlas\ ETGs are reported in \citetalias{cappellari+11a}.} ($R_{\mathrm{e}}$; \citetalias{emsellem+11}).  This figure shows that SRs host the most powerful radio continuum sources in our sample, consistent with previous studies (e.g., \citealt{bender+89, kormendy+09, kormendy+13}) that used the photometric isophote shape parameter ($a_4$; \citealt{bender+88}), which is less robust to inclination and projection effects compared to $\lambda_{\mathrm{R}}$, to study the relationship between radio luminosity and galaxy dynamical state.  However, following our discussion of the radio detection rate in FRs and SRs previously in this section, we caution that the relationship between kinematic class and radio luminosity may in fact be driven by an underlying dependence of the radio luminosity on stellar mass.

We also note that our study only includes nuclear radio sources with $\log(L_{R}/$W~Hz$^{-1}$) $\lesssim$ 23.0.  For comparison, Figure~36 in \cite{kormendy+13}, which shows the radio luminosity as a function of the $a_4$ parameter for ellipticals with ``boxy" and ``disky" optical isophotes\footnote{The so-called ``boxiness'' parameter describes the deviation of galaxy isophote shapes from that of an ellipse.  For a review, see \citet{kormendy+96}.  Compared to the kinematic classes defined by the SAURON and \atlas\ surveys, SRs typically have boxy isophote shapes and FRs typically have disky isophote shapes (although about 20\% of the FRs in the \atlas\ survey have boxy isophotes; \citetalias{emsellem+11}).},
includes radio luminosities as high as $\log(L_{R}/$W~Hz$^{-1}$) $\sim$ 26.0 (three orders of magnitude higher than in our sample of \atlas\ ETGs).  

\subsubsection{Kinematic Misalignment Angle}
\label{sec:kin_misalign_ang}
The stellar kinematic misalignment angle $\psi_{\mathrm{kin-phot}}$, defined as the angle between the photometric and stellar kinematic position angles, is a useful parameter for identifying ETGs that may have experienced a recent interaction or merger \citepalias{krajnovic+11}.  For the \atlas\ ETGs, photometric position angles were measured from ground-based $r$-band SDSS and INT images out to $2.5-3.0$ $R_{\mathrm{e}}$.  Stellar kinematic position angles were determined from the {\tt SAURON} two-dimensional stellar velocity maps out to 1.0 $R_{\mathrm{e}}$.  Small values of $\psi_{\mathrm{kin-phot}}$ correspond to aligned, axisymmetric ETGs that likely formed through a series of minor mergers or more passive, secular evolution.  ETGs with large values of $\psi_{\mathrm{kin-phot}}$ ($> 15^{\circ}$), on the other hand, are considered ``misaligned" systems.  These ETGs may exhibit triaxiality in their stellar velocity fields and sometimes harbor stellar bars.  In the grand scheme of galaxy evolution, misaligned ETGs may represent the end-products of more violent interaction histories (e.g., major mergers).

In Figure~\ref{fig:radio_PSI}, we show the radio luminosity as a function of $\psi_{\mathrm{kin-phot}}$.  This figure indicates that only a small fraction of the ETGs in our sample have values of $\psi_{\mathrm{kin-phot}}$ consistent with the misaligned category (13 $\pm$ 3\%).  For our relatively small sample, the difference between the radio detection fractions for aligned and misaligned systems (49 $\pm$ 4\% and 68 $\pm$ 11\%, respectively) is not significant.  In addition, there is no significant relationship between radio luminosity and $\psi_{\mathrm{kin-phot}}$.  However, it is interesting to note that the ETGs with the highest radio luminosities show large kinematic misalignment angles.  This could be driven simply by the fact that the most radio-luminous ETGs in our sample are massive SRs.  Alternatively, this could be interpreted as an indication that the formation mechanisms responsible for producing large values of $\psi_{\mathrm{kin-phot}}$, such as major mergers \citepalias{krajnovic+11}, are substantial drivers of powerful radio emission in nearby ETGs.  

\subsection{Radio Luminosity and Gas}
\label{gas}
In this section we investigate the relationship between the radio luminosity and kinematic misalignment angle between the stars and gas based on a combination of molecular, neutral, and ionized gas position angle measurements.  We also study the relationship between the cold gas mass in the molecular phase and the ionized gas (traced by the O[{\tt III}] luminosity and equivalent width).  We have information on the neutral cold gas phase as well for the 95 galaxies in our study that were also observed in H{\tt I} \citepalias{serra+14}.  Although we investigated the relationship between radio luminosity and H{\tt I} mass, we do not include a plot here since it was extremely difficult to interpret due to the prevalence of upper limits on both parameters and the vast difference in spatial scale between these observations. 

\subsubsection{Gas Misalignment Angle}
\label{misalignment_angle}
The misalignment angle between the angular momenta of the gas and stars in a galaxy provides information on the origin of the gas.  For galaxies in which the stellar and gaseous kinematic position angles are aligned, internal processes, such as stellar mass loss, are likely responsible for the bulk of the interstellar medium (ISM).  Misalignments between the gas and stars, on the other hand, may signify that the gas has either been accreted externally or has been disrupted by a recent major merger \citep{morganti+06}.  \citet{davis+11}, hereafter \citetalias{davis+11}, showed that the molecular, ionized, and atomic gas are always aligned with one another in the \atlas\ sample, even when there is a misalignment between the gas and the stars.  Thus, comparison between the radio luminosity and misalignment angle between the gas and stars provides an additional means of studying the impact of the gas accretion history on the nuclear radio emission in nearby ETGs.  
  
\begin{figure}
\includegraphics[clip=true, trim=0.1cm 0cm 1.5cm 2cm, scale=0.4]{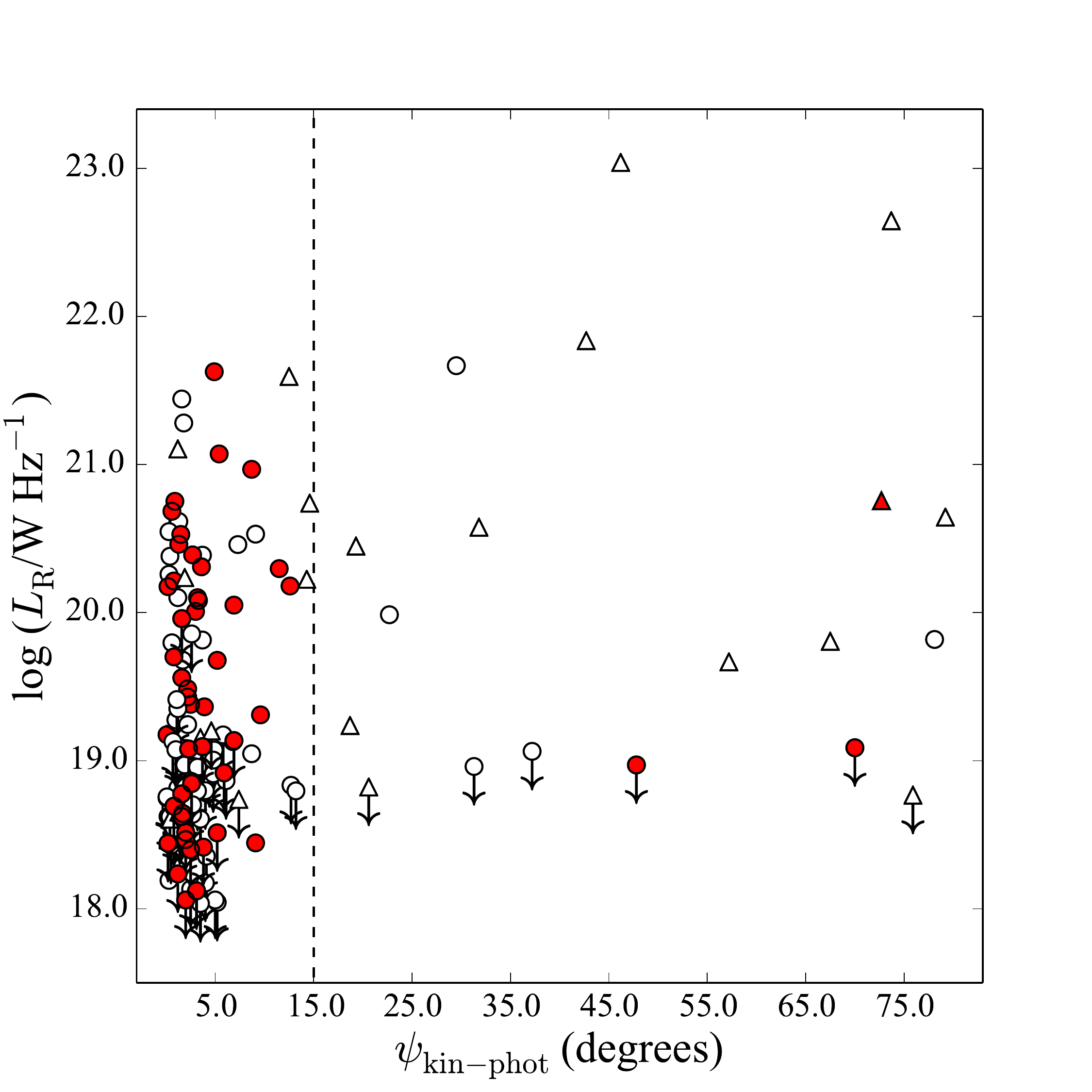}
\caption{5~GHz radio luminosity as a function of the apparent average stellar kinematic misalignment angle, $\psi_{\mathrm{kin-phot}}$ \citepalias{krajnovic+11}.  Symbols filled in red represent CO detections \citepalias{young+11}; unfilled symbols represent CO non-detections.  Circles represent fast rotators while upward-pointing triangles represent slow rotators \citepalias{emsellem+11}.  Downward-pointing arrows represent 5~GHz upper limits.  The vertical dotted line marks $\psi_{\mathrm{kin-phot}}$ = 15$^{\circ}$, above which our sample ETGs are considered to be misaligned \citepalias{krajnovic+11}.}
\label{fig:radio_PSI}
\end{figure}

\begin{figure}
\includegraphics[clip=true, trim=0.1cm 0cm 1.5cm 2cm, scale=0.4]{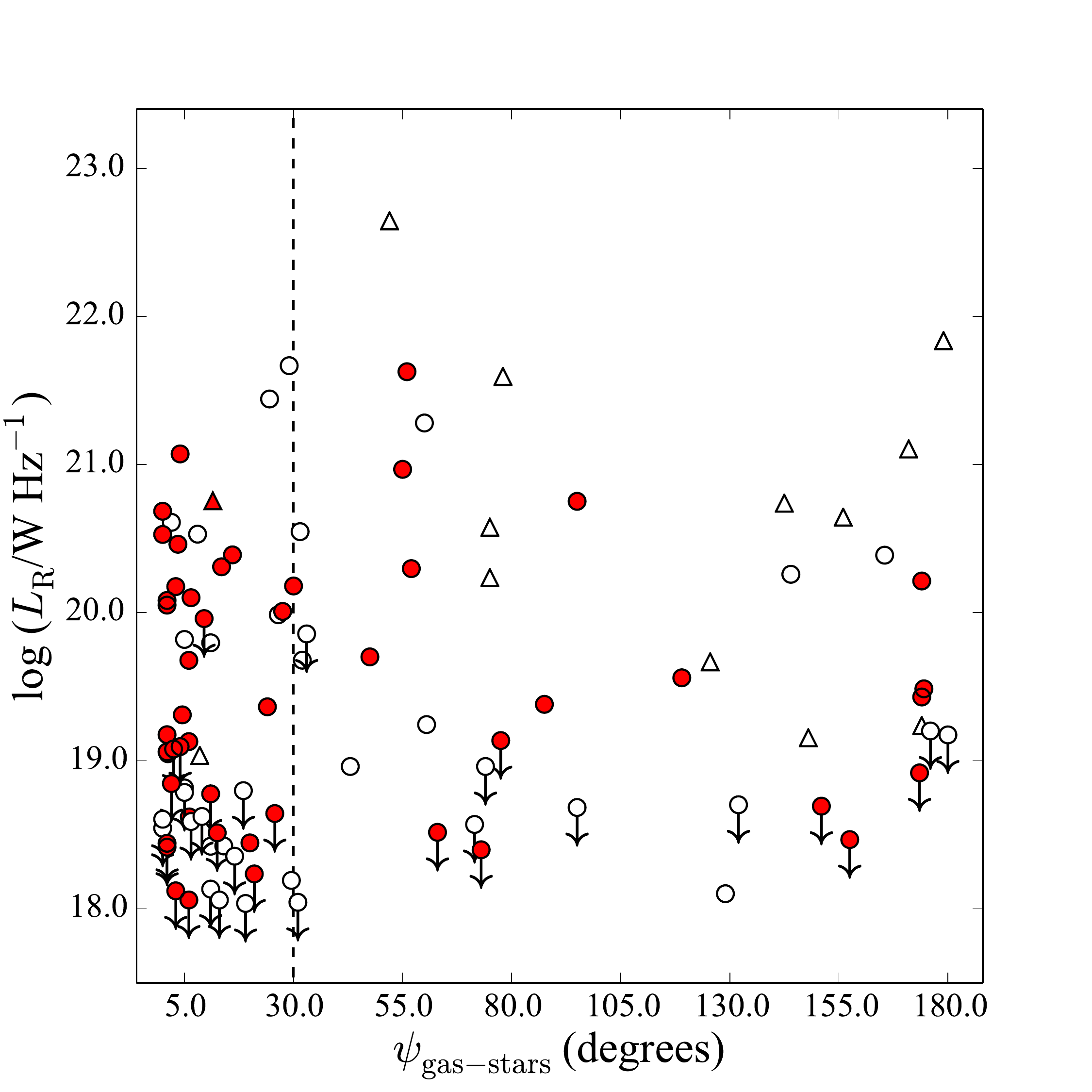}
\caption{5~GHz radio luminosity as a function of the misalignment angle between the kinematic position angles of the molecular gas (or if no molecular gas data are available, the ionized or atomic gas) and the stars, $\psi_{\mathrm{gas-stars}}$ \citepalias{davis+11, serra+14}.  Symbols filled in red represent CO detections \citepalias{young+11}; unfilled symbols represent CO non-detections.  Circles represent fast rotators while upward-pointing triangles represent slow rotators \citepalias{emsellem+11}.  Downward-pointing arrows represent 5~GHz upper limits.  The vertical dotted line marks $\psi_{\mathrm{gas-stars}}$ = 30$^{\circ}$, above which the gas is considered to be misaligned \citepalias{davis+11, serra+14, young+14}.}
\label{fig:radio_PSI_gas_stars}
\end{figure}

In Figure~\ref{fig:radio_PSI_gas_stars}, we show the radio luminosity as function of the difference between the kinematic position angles of the gas and the stars, $\psi_{\mathrm{gas-stars}}$ \citepalias{davis+11, serra+14}.  Of the 136 \atlas\ ETGs with measurements of $\psi_{\mathrm{gas-stars}}$ available, this figure includes the 99 ETGs with both $\psi_{\mathrm{gas-stars}}$ and nuclear 5~GHz data.  The vertical dashed line in Figure~\ref{fig:radio_PSI_gas_stars} separates  ETGs with gas that is kinematically aligned ($\psi_{\mathrm{gas-stars}} < 30^{\circ}$) from those with kinematically misaligned gas ($\psi_{\mathrm{gas-stars}} > 30^{\circ}$) relative to the stars (\citetalias{davis+11, serra+14}; \citealt{young+14}, hereafter \citetalias{young+14}).  
The fraction of ETGs with misaligned gas in Figure~\ref{fig:radio_PSI_gas_stars} is 43 $\pm$ 5\%, indicating that external accretion of gas is likely a common means of gas supply among our sample galaxies \citepalias{davis+11}.  Compact radio emission is detected at similar rates in ETGs with both misaligned (67 $\pm$ 7\%) and aligned (61 $\pm$ 7\%) gas.  Thus, our comparison between the nuclear radio luminosity and the gas kinematic misalignment angle does not provide evidence for a strong link between gas-rich mergers and the presence of nuclear radio activity.  However, we emphasize that the fraction of ETGs with misaligned gas is likely a lower limit since some ETGs show signs of external gas accretion (e.g., in H{\tt I} studies; \citealt{morganti+06, oosterloo+10}) despite the fact that their molecular and ionized gas are both aligned with the stellar body \citepalias{davis+11}.
Additionally, we note that the ETGs with the highest radio luminosities tend to also show large gas kinematic misalignment angles in Figure~\ref{fig:radio_PSI_gas_stars}.  
This is similar to the behavior in Figure~\ref{fig:radio_PSI} that is discussed in Section~\ref{sec:kin_misalign_ang} regarding the relationship between the radio luminosity and the stellar kinematic misalignment angle.

The tendency for the most luminous radio sources in our sample to have large gas misalignment angles is consistent with a scenario in which substantial, recent gas accretion from an external source helped ignite the nuclear radio activity.  Alternatively, we speculate that the radio jets/outflows from these more luminous radio LLAGNs may actually contribute to the observed gas misalignments through the injection of turbulence/generation of shocks.  Further investigation is necessary to distinguish among these possibilities.
 
\begin{figure}
\includegraphics[clip=true, trim=0.1cm 0cm 1.5cm 2cm, scale=0.4]{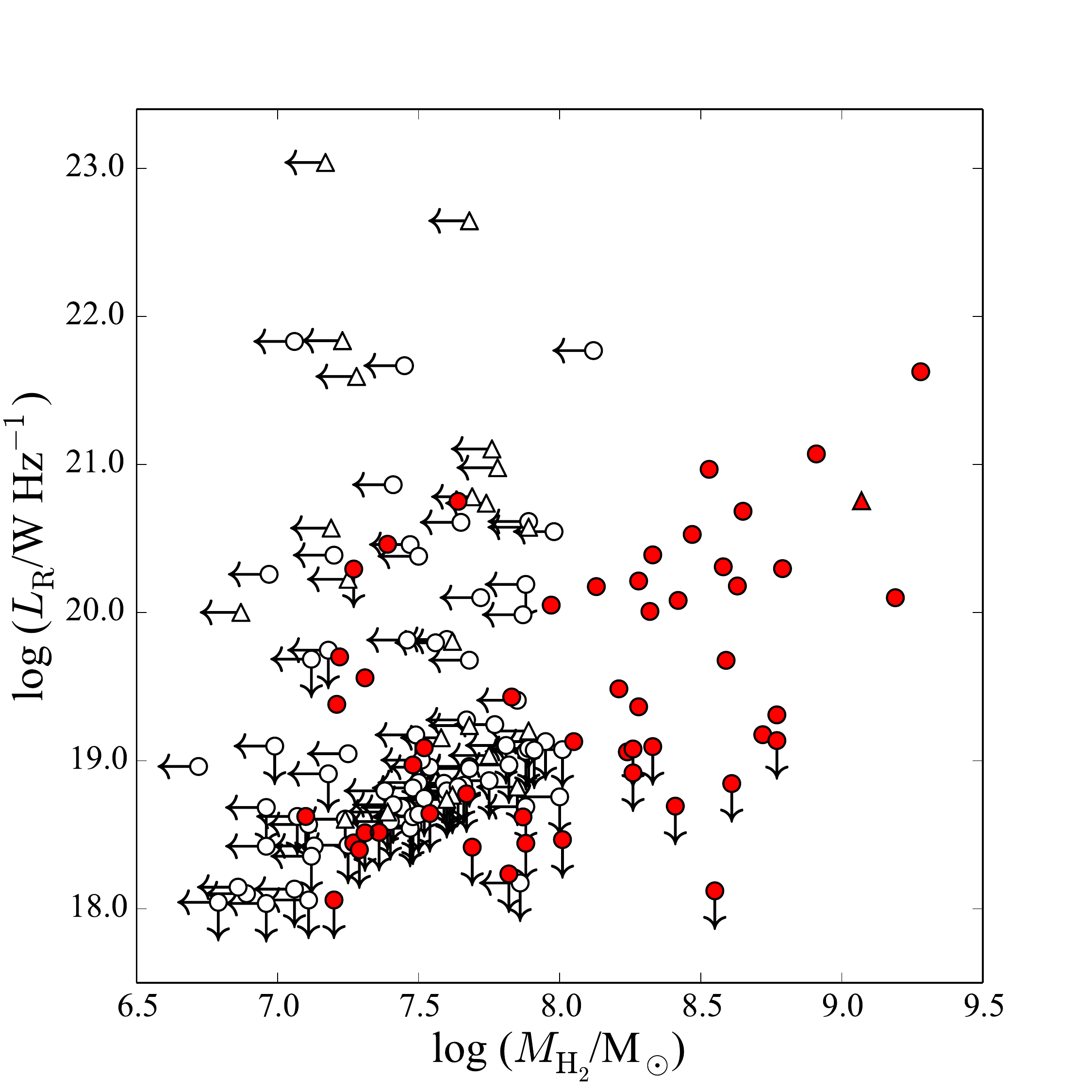}
\caption{5~GHz radio luminosity as a function of H$_2$ mass.  Symbols filled in red represent CO detections \citepalias{young+11}.  Unfilled symbols are CO non-detections (upper limits).  Circles represent fast rotators while upward-pointing triangles represent slow rotators \citepalias{emsellem+11}.  Downward-pointing arrows represent 5~GHz upper limits and leftward-pointing arrows denote H$_2$ mass upper limits.}
\label{fig:radio_H2}
\end{figure}

\subsubsection{Molecular Gas}
\label{radio_molec_gas}
Radio-emitting AGNs draw fuel from cooled halo gas from the intergalactic medium, material expelled during stellar mass loss, a fresh supply of cold gas from a merger or interaction, or residual reservoirs leftover from their assembly (e.g., \citetalias{davis+11}, and references therein).  It is thus natural to examine the relationship between the radio emission and the mass of the molecular gas to determine if the cold gas plays a significant role in AGN fueling.  In Figure~\ref{fig:radio_H2} we show the high-resolution 5~GHz radio luminosity vs. the molecular gas mass for all of the ETGs in our sample.  The H$_2$ masses are from the single-dish measurements reported in \citet{young+11}, resulting in a large discrepancy between the radio and CO spatial resolutions.  

Nevertheless, Figure~\ref{fig:radio_H2} exhibits some interesting features.  ETGs lacking detections of molecular gas in \citetalias{young+11}, which include the majority of the SRs, have the highest radio luminosities and form an almost vertical track in the figure.  Not surprisingly, radio luminosity and molecular gas mass are not significantly correlated among the SRs or CO non-detections in our sample.  The gas-rich, CO-detected ETGs, on the other hand, are generally FRs and populate a somewhat different portion of Figure~\ref{fig:radio_H2}, showing an approximate trend of increasing radio luminosity with increasing molecular gas mass.  This behaviour is similar to that in Figure~\ref{fig:emission_radio_H2_sub} for the LINER-AGN class of objects.  The correlation between these quantities for the FRs in Figure~\ref{fig:radio_H2} is weak but statistically significant (the Kendall's Tau correlation coefficient is $\tau = 0.246 \pm 0.016$ and the probability of the null hypothesis that no correlation exists is $p = 4.8 \times 10^{-5}$).  

Despite the large difference in spatial scales probed by the radio and CO emission in Figure~\ref{fig:radio_H2}, we consider possible origins for the relationship between the cold gas and 5~GHz emission among the FRs/CO detections.  Assuming that the nuclear radio sources are indeed associated with AGNs, the observed relationship could be evidence that, in rejuvenated gas-rich FRs, the accretion of cold gas plays a substantial role in AGN fueling.  This would be in contrast to AGNs located massive SRs, which lack evidence for the presence of cold gas, and are likely fueled by stellar mass loss or the cooling of the hot halos surrounding their host galaxies \citep{hopkins+06, ciotti+07}.  For a more detailed discussion of this possibility in the context of AGN fueling, see Section~\ref{fuel}.

The second possibility is that, despite high-resolution observations, we are in fact observing a SF-related effect.  Correlations between radio continuum emission and CO emission have been reported in the literature (e.g., \citealt{murgia+02, murgia+05}) for studies of spiral galaxies, and are likely driven by the same fundamental processes responsible for the well-established radio-infrared relation \citep{helou+85, condon+92}.  In this scenario, a substantial fraction of the compact radio emission would be produced by regions of compact SF.  Finally, a third potential origin for the trend in Figure~\ref{fig:radio_H2} is a deeper underlying connection between SF and AGNs in our sample galaxies.  Such a SF-AGN connection might arise as a result of the mutual growth of the stellar mass and central SMBH mass with increasing molecular gas reservoir (e.g., \citealt{heckman+14}).  As other authors have suggested in the literature, factors such as negative \citep{fabian+12} and/or positive \citep{gaibler+12, silk+13} AGN feedback would likely contribute to such a connection between SF and nuclear activity.

\begin{figure*}
\includegraphics[clip=true, trim=0cm 0cm 1cm 2cm, scale=0.4]{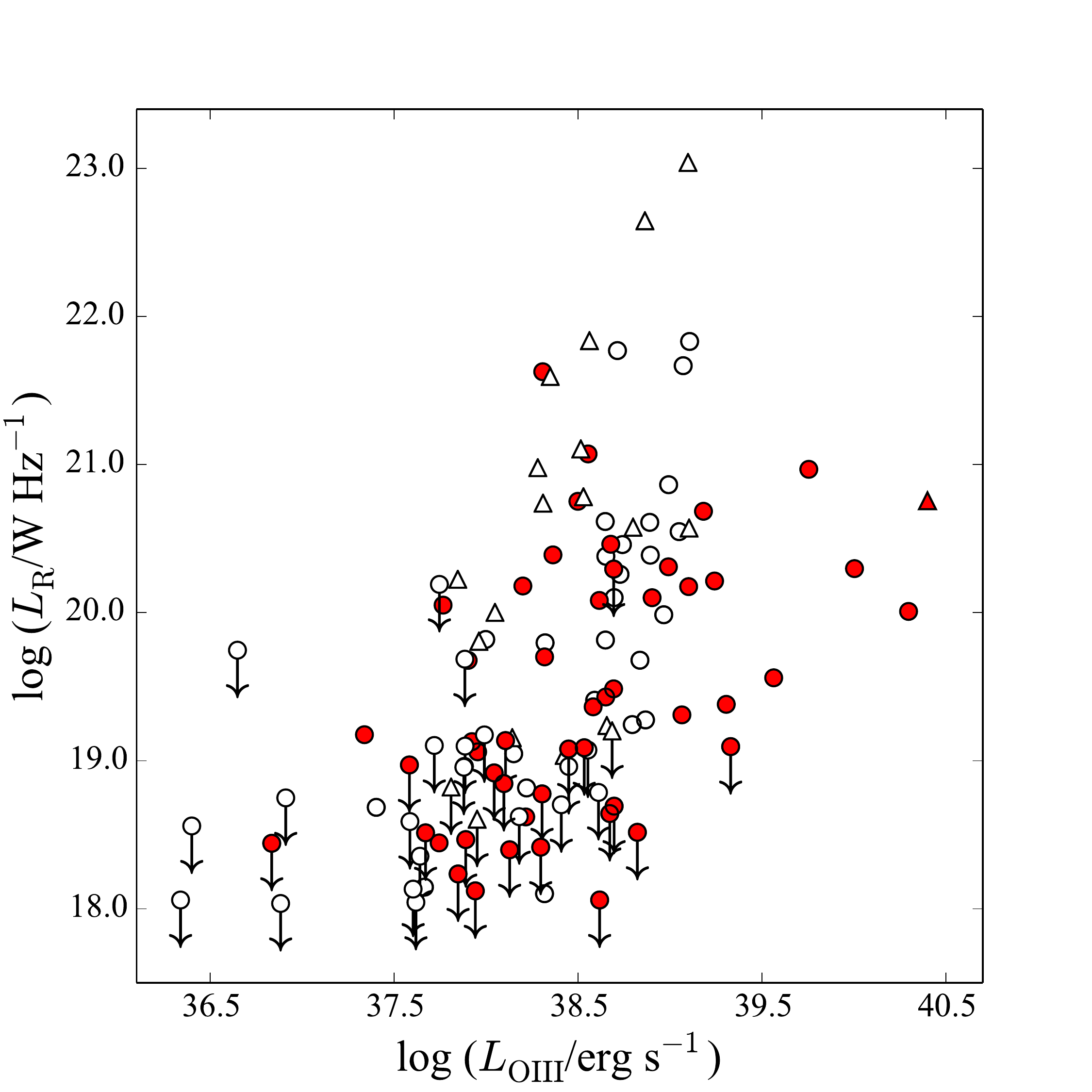}
\includegraphics[clip=true, trim=0cm 0cm 2cm 2cm, scale=0.4]{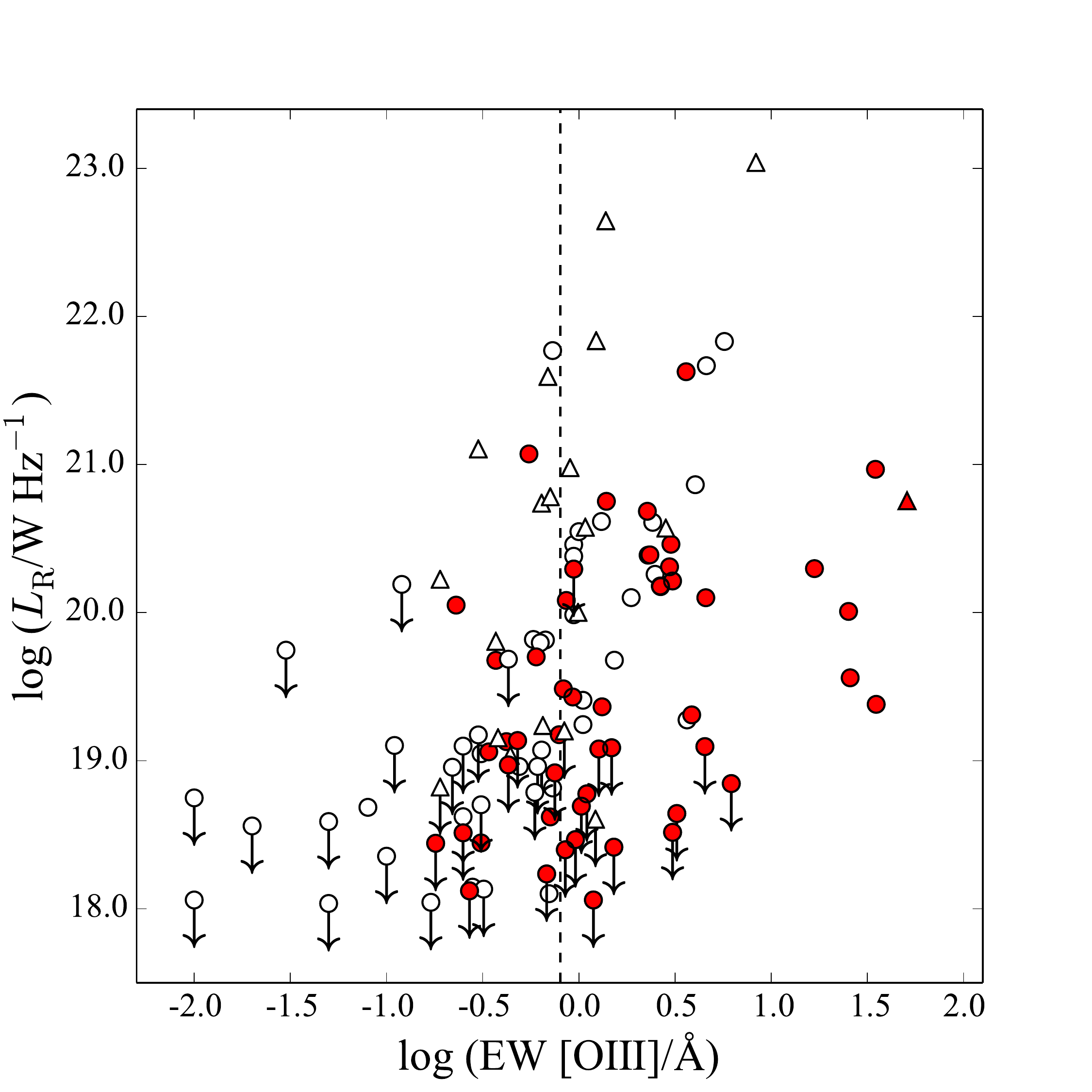}
\caption{{\bf (Left:)} 5~GHz radio luminosity as a function of the nuclear (within a radius of $1.5^{\prime \prime}$) [O{\tt III}] $\lambda \lambda$5007 luminosity (see Table~\ref{tab:high_energy}).  Symbols filled in red represent CO detections \citepalias{young+11}.  Unfilled symbols are CO non-detections (upper limits).  Circles represent fast rotators while upward-pointing triangles represent slow rotators \citepalias{emsellem+11}.  Downward-pointing arrows represent 5~GHz upper limits.   {\bf (Right:)} Same as the left panel, but for the 5~GHz radio luminosity as a function of the nuclear [O{\tt III}] equivalent width.  The vertical, dashed, black line in the right panel indicates an EW [O{\tt III}] of 0.8 \AA, the boundary we adopted (see Section~\ref{classification}) to separate ETGs identified as LINERs that are more likely to be powered by UV-bright stars (EW [O{\tt III}] $<$ 0.8 \AA) from those that are good candidates for the presence of a true AGN (EW [O{\tt III}] $>$ 0.8 \AA).}
\label{fig:radio_opt}
\end{figure*}

\subsubsection{Ionized Gas}
\label{radio_ionized_gas}
The relationships between the radio luminosity and luminosities/line-widths of various optical emission lines have been explored extensively in the literature for samples spanning a wide range of properties \citep{meurs+84, baum+89, sadler+89}.  The most popular tracers have historically been [O{\tt III}] and H$\beta$, which are believed to arise via photoionization from re-processed optical continuum associated with the accretion disk of the AGN \citep{ho+01b}.  As suggested by \citet{ho+01b}, a statistically-significant relationship between radio and optical emission line luminosity (or emission line equivalent width) may provide clues regarding the interplay between AGN fueling and the subsequent production of radio continuum emission.  Previously, a number of authors have concluded that relationships between radio luminosity and optical emission lines are significant (e.g., \citealt{meurs+84, baum+89, tadhunter+98, ho+01b, nagar+05, balmaverde+06, kauffmann+08, park+13}), while others have reported the absence of any statistically significant relationships between these quantities (e.g., \citealt{sadler+89, best+05}).  These discrepant results are likely due to differences in SMBH accretion regime (efficient or inefficient; for a review see \citealt{heckman+14}) and spatial resolution probed by each survey.  Differences in the relative contributions by the various ionization mechanisms (SF, AGNs, shocks, and evolved stars) likely also contribute to the scatter in the relationship between radio luminosity and ionized gas luminosity/equivalent width, as well as discrepancies between studies utilizing samples with different properties and selection criteria.

In Figure~\ref{fig:radio_opt}, we illustrate the high-resolution 5~GHz radio luminosity as a function of the nuclear (within a radius of 1.5$^{\prime \prime}$) [O{\tt III}] luminosity and the [O{\tt III}] equivalent width (EW[O{\tt III}] ) from the \atlas\ observations with the {\tt SAURON} spectrograph (see Table~\ref{tab:high_energy}).  Although not shown here, the behaviour of the radio luminosity as a function of H$\beta$ luminosity and EW is very similar to that of the [O{\tt III}] luminosity and EW.  Figure~\ref{fig:radio_opt} displays a trend of increasing radio luminosity with increasing [O{\tt III}] luminosity\footnote{The relationship between radio luminosity and [O{\tt III}] luminosity may be influenced by the correlation between the distance squared used to calculate each luminosity variable and the luminosity itself.  The relationship between radio luminosity and EW[O{\tt III}]  should be less affected by this issue since the EW does not directly depend on the distance.}, as well as EW[O{\tt III}].

The $\tau$ statistic and $p$-values from the censored Kendall's $\tau$ test for EW[O{\tt III}] vs.\ nuclear radio luminosity are $\tau$ = 0.379 $\pm$ 0.015 and $p = 5.5 \times 10^{-8}$ for the FRs, and $\tau$ = 0.259 $\pm$ 0.017 and $p = 1.1 \times 10^{-1}$ for the SRs.  Thus, the correlation is statistically significant for the FRs, but not the SRs.  This could be a manifestation of the greater availability of material for massive black hole accretion in FRs.  Since the relatively small number of SRs included in the sample could cause relationships such as these to appear less significant, studies that include larger numbers of SRs will be necessary to further investigate this possibility.

\begin{figure*}
\includegraphics[clip=true, trim=0cm 0.25cm 1cm 2cm, scale=0.4]{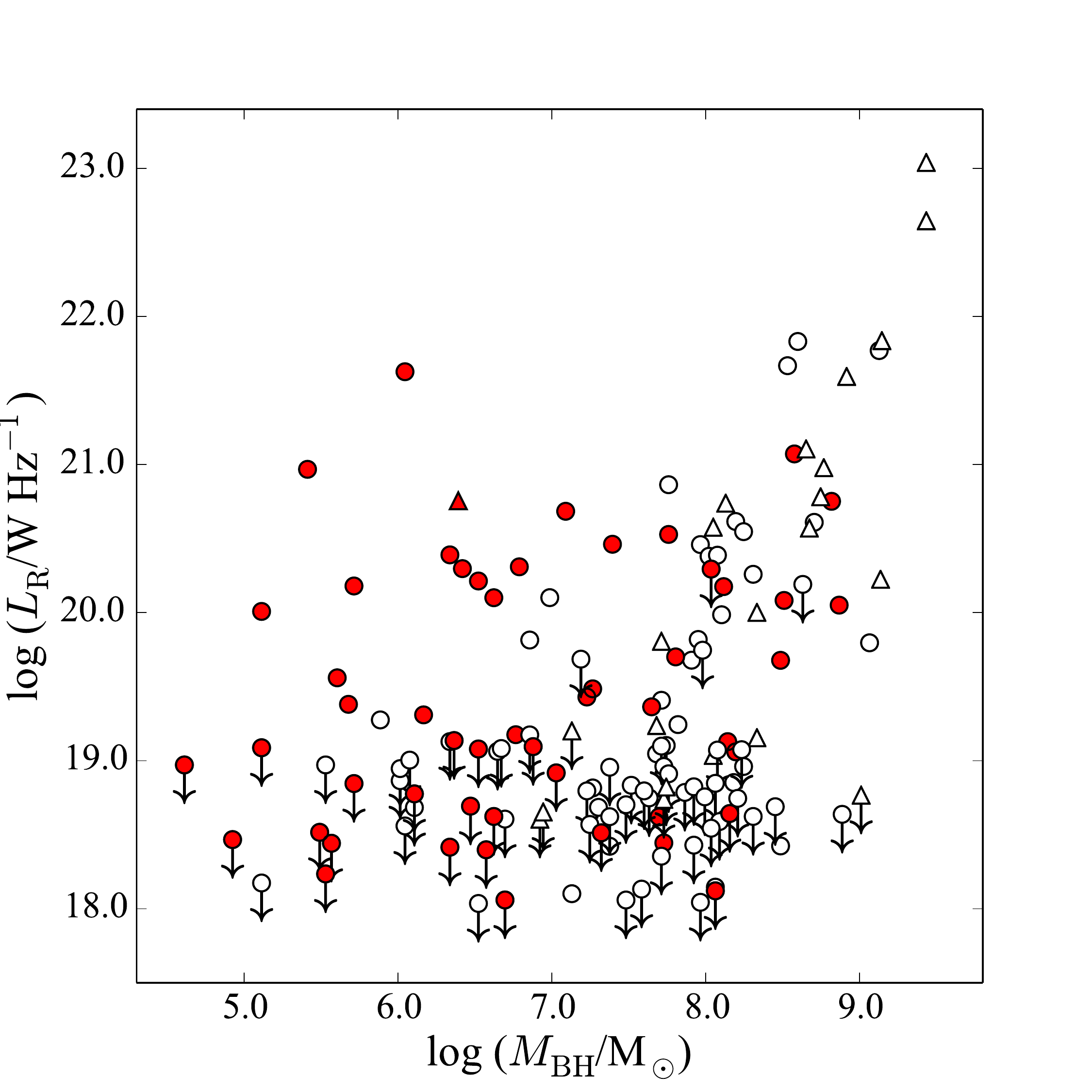}
\includegraphics[clip=true, trim=0cm 0.25cm 2cm 2cm, scale=0.4]{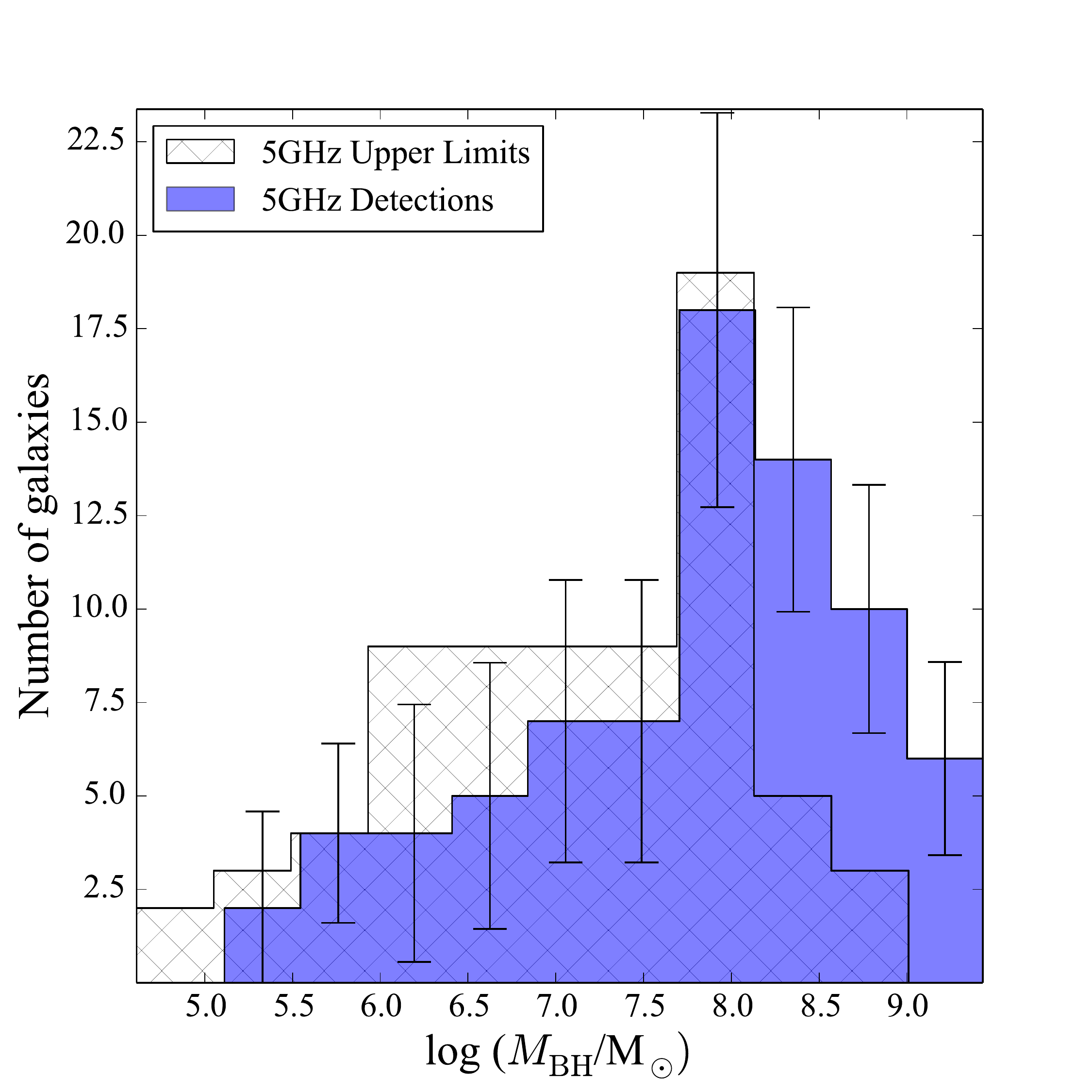}
\caption{
{\bf (Left:)} 5~GHz radio luminosity vs. SMBH mass.  SMBH masses are based on dynamical measurements when available.  If no dynamical SMBH mass was available, the mass was estimated using the $M_{\mathrm{BH}}-\sigma_{*}$ relation \citep{mcconnell+13}.  The stellar velocity dispersions ($\sigma_{*}$) within 1 $R_{\mathrm{e}}$ used to estimate the black hole masses are from {\tt SAURON} measurements \citepalias{cappellari+13a}.  Symbols filled in red represent CO detections \citepalias{young+11}; unfilled symbols represent CO non-detections.  Circles represent fast rotators while upward-pointing triangles represent slow rotators \citepalias{emsellem+11}.  Downward-pointing arrows represent 5~GHz upper limits.
{\bf (Right:)} Radio detection fraction as a function of SMBH mass for the \atlas\ ETGs with high-resolution 5~GHz data.  The blue-filled histogram represents the radio continuum detections and the black-hatched histogram represents the radio upper limits.  Error bars denote the binomial uncertainties.  At black hole masses above and below $\log (M_{\mathrm{BH}}$/M$_{\odot}$) $\approx$ 8.0, ETGs are more and less likely to harbor a compact radio source in their centers, respectively.}
\label{fig:radio_BH}
\end{figure*}


\subsection{Radio Luminosity and Black Hole Mass}
\label{sec:bh}
The relationship between radio luminosity and SMBH mass has been the subject of much debate in the literature, with some studies finding evidence for the presence of a correlation \citep{franceschini+98, jarvis+02, nagar+05} and others concluding that the two parameters are unrelated \citep{ho+02, balmaverde+06, park+13}.
In the left panel of Figure~\ref{fig:radio_BH} we have plotted the 5~GHz radio luminosity vs. SMBH mass.  Whenever possible, black hole masses are drawn from dynamical black hole mass estimates available in the literature (\citealt{kormendy+13}, and references therein).  For the ETGs with no dynamical SMBH mass measurement available in the literature, we used the $M_{\mathrm{BH}}-\sigma_{*}$ relation\footnote{\citet{mcconnell+13} report a relationship of the form $\log(M_{\mathrm{BH}}$/M$_{\odot}$) = $\alpha + \beta \log(\sigma_{*} / 200 \, \mathrm{km} \, \mathrm{s^{-1}})$, where $\sigma_{*}$ is the velocity dispersion (measured within one $R_{\mathrm{e}}$), $\alpha = 8.32 \pm 0.05$, and $\beta = 5.64 \pm 0.32$.} 
from \citet{mcconnell+13} to estimate the mass.  
Nuclear radio luminosity and SMBH mass are correlated for the full sample and also for FRs and SRs separately, however the correlation is strongest and most significant among the SRs ($\tau$ = 0.528 $\pm$ 0.020, $p = 4.2 \times 10^{-4}$).
In addition, the left panel of Figure~\ref{fig:radio_BH} shows that the most powerful radio sources in our sample reside in the ETGs with the most massive black holes (e.g., NGC4486, NGC4261, and NGC4374).

The histogram shown in the right panel of Figure~\ref{fig:radio_BH} further illustrates the fact that ETGs with massive central black holes ($\log(M_{\mathrm{BH}}/$M$_{\odot}$) $\gtrsim$ 8.0) are more likely to harbor central radio sources compared to ETGs with less massive black holes ($\log(M_{\mathrm{BH}}/$M$_{\odot}$) $\lesssim$ 8.0).  A two-sample Kolmogorov-Smirnov (KS) test \citep{peacock+83} on the SMBH mass distributions of the radio detections and non-detections yields a probability of $p = 4.0 \times 10^{-3}$ that the two samples come from the same parent sample.  Thus, the ETGs in our sample with larger black hole masses are more likely to harbor nuclear radio emission, consistent with previous studies (e.g., \citealt{balmaverde+06}).  

\subsection{Radio Luminosity and Stellar Mass}
\label{stellar_mass}
Given the established relationships linking black hole mass to a variety of stellar bulge properties including bulge velocity dispersion, luminosity, and mass (e.g., \citealt{mcconnell+13}), we expect the relationship between radio luminosity and stellar mass to be similar to that between radio luminosity and black hole mass.  Indeed, the fact that the most powerful radio galaxies have high stellar masses is well-established (e.g., \citealt{condon+02, mauch+07}).  However, the dependence of radio luminosity on stellar mass for weaker radio sources is less clear.  

In many previous studies \citep{sadler+89, wrobel+91b, capetti+09,  brown+11}, the optical magnitude was used as a proxy for stellar mass and a constant mass-to-light ratio (M/L; \citealt{bell+03}) was assumed for all galaxies.  For the \atlas\ sample, the abundant stellar kinematic data have provided more accurate estimates of the stellar masses based on dynamical models \citepalias{cappellari+13a}.  As discussed in \citetalias{cappellari+13a}, this dynamical mass, denoted $M_{\mathrm{JAM}}$, represents an accurate estimate of the total galaxy stellar mass, $M_{\mathrm{*}}$.  In fact, it accounts for both variations in the stellar M/L due to the age and metallicity, as well as to systematic variations in the IMF \citep{cappellari+12}.

\begin{figure}
\includegraphics[clip=true, trim=0cm 0.25cm 2cm 1cm, scale=0.4]{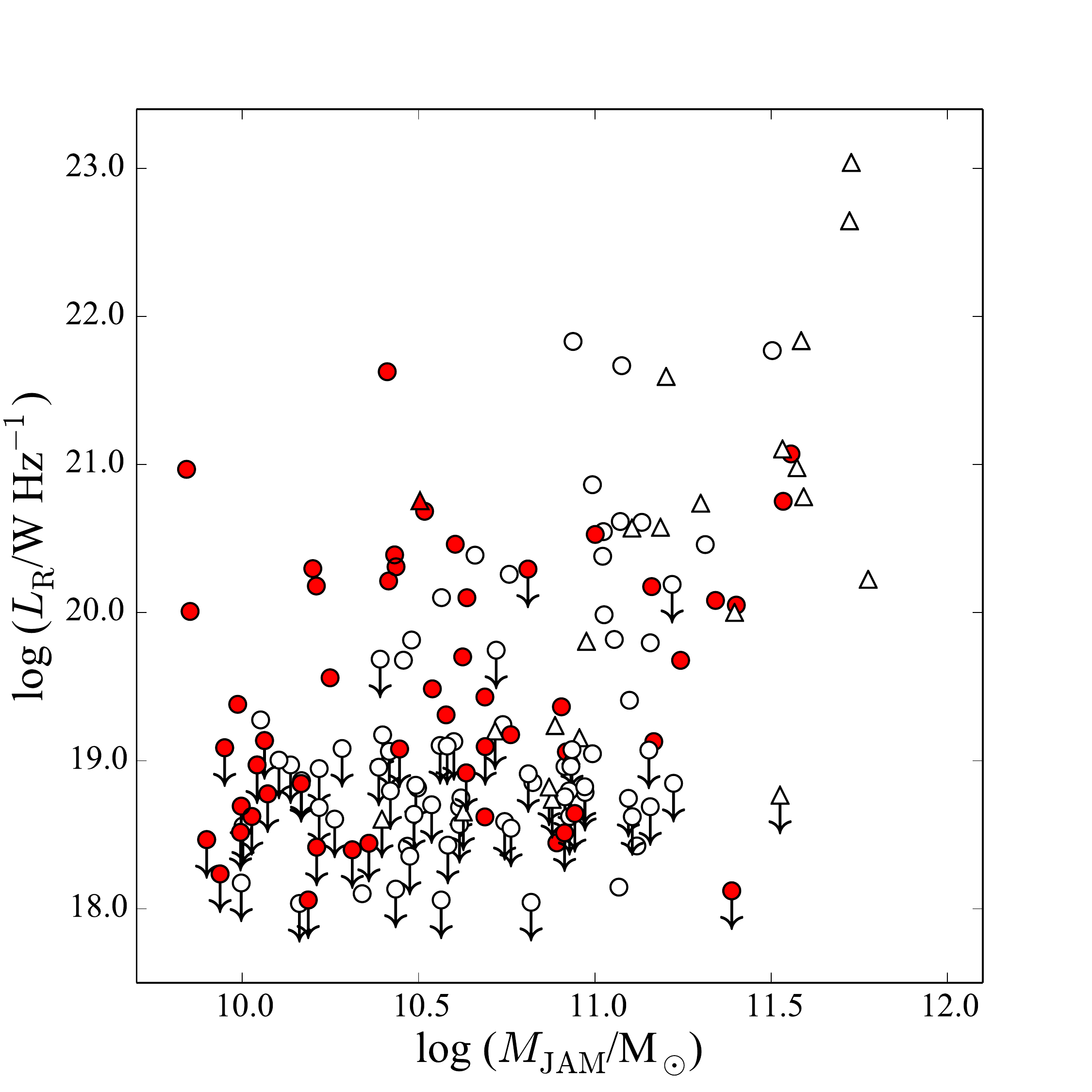}
\caption{5~GHz radio luminosity as a function of the dynamically-determined stellar mass, $M_{\mathrm{JAM}}$ \citepalias{cappellari+13a}.  Symbols filled in red represent CO detections \citepalias{young+11}; unfilled symbols represent CO non-detections.  Circles represent fast rotators while upward-pointing triangles represent slow rotators \citepalias{emsellem+11}.  Downward-pointing arrows represent 5~GHz upper limits.}
\label{fig:radio_MS}
\end{figure}

A plot of the radio luminosity as a function of the dynamical stellar mass is shown in Figure~\ref{fig:radio_MS}.  Although the relationship between stellar mass and radio luminosity is not statistically significant over our full sample, there are significant correlations when the FR and SR kinematic classes are considered separately.  The Kendall's $\tau$ statistic for the correlation between radio luminosity and dynamical stellar mass for the SRs is $\tau$ = 0.565 $\pm$ 0.022, and the probability of the null hypothesis that there is no correlation is $p = 1.6 \times 10^{-4}$.  For the FRs, $\tau$ = 0.239 $\pm$ 0.019 and the probability of no correlation is $p = 7.8 \times 10^{-4}$.  Thus, radio luminosity and stellar mass are slightly less strongly correlated among the FRs, which comprise the majority of our sample.  
However, we caution that any differences in the relationship between stellar mass and nuclear radio luminosity for SRs and FRs could be a loose consequence of an underlying relationship with environment (see Section~\ref{sec:density}) since the most massive galaxies in the \atlas\ sample reside in the Virgo cluster and are also SRs.  

\subsection{Radio Continuum and Optical Properties}
\label{morph_features}
\subsubsection{Morphological Features}
\begin{figure*}
\includegraphics[clip=true, trim=3cm 2cm 2cm 0cm, scale=0.58]{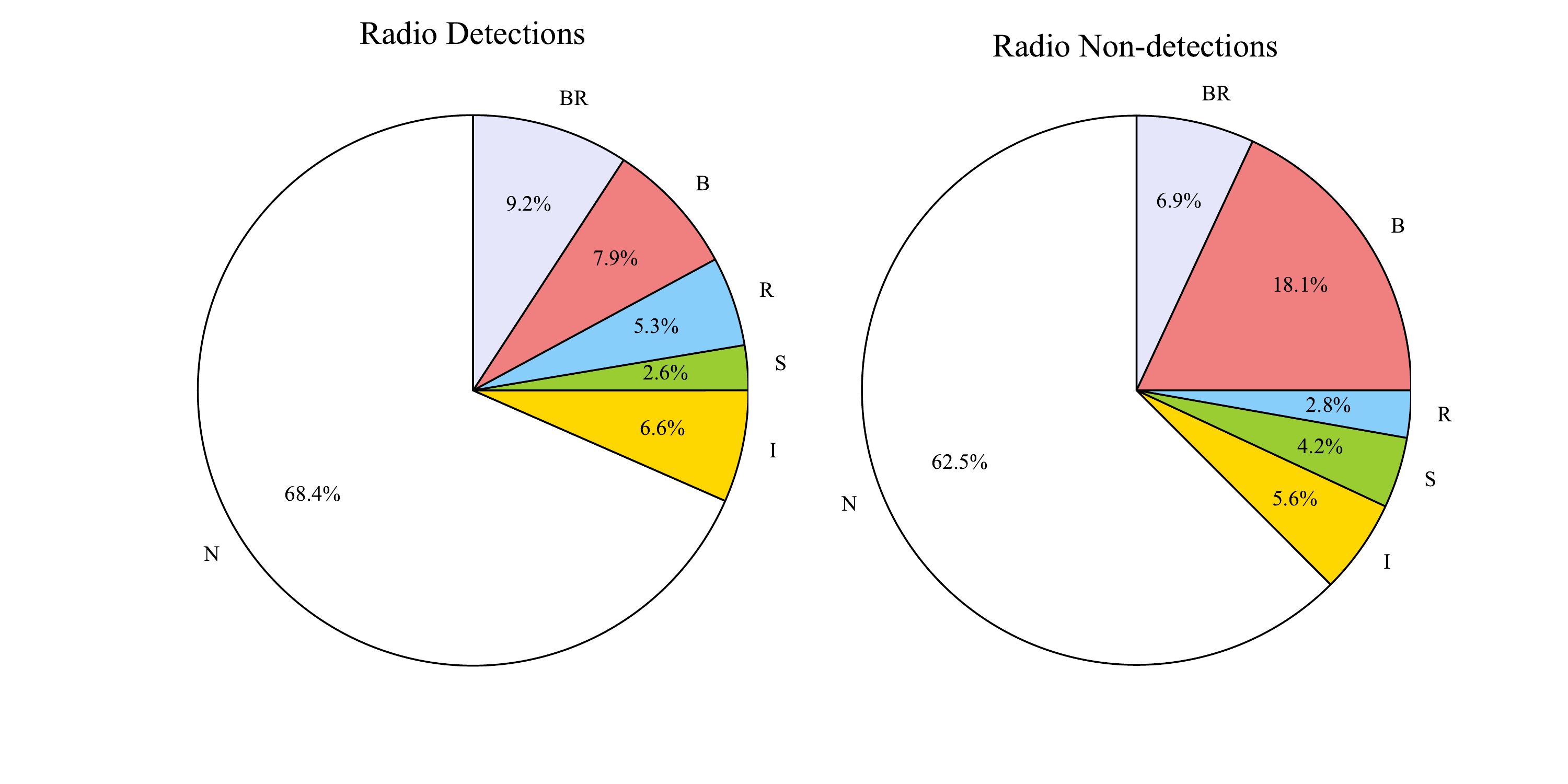}
\caption{{\bf (Left:)} Proportion of various morphological features for the galaxies with nuclear radio detections.  The morphological features are based on SDSS and INT optical images and are tabulated and described in detail in \citetalias{krajnovic+11}.  N = no feature, BR = bar + ring, R = ring, B = bar, S = shell, and I = interaction feature.  {\bf (Right:)}  Same as the left panel, but for the galaxies with only upper limits to the presence of nuclear 5~GHz emission.}
\label{fig:radio_morph}
\end{figure*}

Fossil clues regarding the mechanisms responsible for AGN triggering can be surmised from information on the optical morphological features in galaxies combined with tracers of the radiative signatures of massive black hole accretion, such as compact radio continuum emission.  In Figure~\ref{fig:radio_morph}, we illustrate the proportions of 5~GHz radio detections and non-detections that have various morphological features based on SDSS and INT optical images\footnote{A detailed description of the morphological classes is provided in \citetalias{krajnovic+11}.}  This figure clearly shows that the majority of both the radio detections and radio non detections have no discernible morphological features in the ground-based optical images.  Table~\ref{tab:contingency_morph} provides this information in a more quantitative manner.  However, we note that due to the sensitivity and spatial resolution limitations of the SDSS and INT observations, the number of \atlas\ ETGs reported to be devoid of these features is an overestimate.  On-going efforts to obtain extremely deep, multi-band optical images of the full \atlas\ sample using the MegaCam instrument at the Canada-France-Hawaii Telescope are in progress, and a preliminary report of these observations of 92 of the \atlas\ ETGs is provided in \citetalias{duc+15}.  

We find that the incidence of interaction features, shells, rings, and bars + rings are similar for galaxies both with and without compact radio continuum emission, suggesting that major mergers do not play a significant role in initiating radio activity in the majority of nearby ETG nuclei.  Table~\ref{tab:contingency_morph} hints at the possibility that ETGs with only upper limits to the presence of a radio core are more likely to harbor a bar than ETGs with detections of 5 GHz emission in our study.  However, a simple test of the $\chi^{2}$ statistic for the 2 $\times$ 2 contingency table for the detection rate of high-resolution radio emission in ETGs with and without bars reveals no statistically significant difference.  Thus, our data indicate that bars do not have a significant impact on the level of nuclear radio emission and, by extension, do not substantially influence AGN fueling.  This is consistent with previous studies of the relationship between stellar bars and emission-line AGNs \citep{ho+97b}.

\begin{table}
\caption{Summary of Morphological Features}
\label{tab:contingency_morph}
\begin{tabular}{lccc}
\hline
\hline
Feature      & Radio Det. & Radio UL  & Total  \\
\hline 
      No feature              &          52          &            45             &       97       \\
      All features                 &          24          &             27            &        51      \\
       \hspace{0.5cm} Bar                  &            6          &            13             &       19       \\
       \hspace{0.5cm} Ring                &            4          &             2              &         6       \\
       \hspace{0.5cm} Bar + Ring     &            7          &             5              &       12       \\
       \hspace{0.5cm} Shell               &            2          &             3              &         5        \\
       \hspace{0.5cm} Interaction     &            5          &             4              &         9        \\
\hline
\hline
\end{tabular}
\end{table}

\begin{table}
\begin{minipage}{10.5cm}
\caption{Summary of Dust Features}
\label{tab:contingency_dust}
\begin{tabular}{lccc}
\hline
\hline
Feature      & Radio Det. & Radio UL  & Total  \\
\hline 
      No feature              		&        48          &         58           &      106      \\
      All features                   		&        45          &         24           &        69      \\
       \hspace{0.5cm} Blue Nucleus                &           8          &           6           &        14       \\
       \hspace{0.5cm} Ring                		&           5          &           3           &          8       \\
       \hspace{0.5cm} Blue Nucleus + Ring   &           5          &           3           &          8        \\
       \hspace{0.5cm} Dusty Disk              	&         14          &           6          &         20       \\
       \hspace{0.5cm} Filament     			&         13          &           6          &         19       \\
\hline
\hline
\end{tabular}
 
\end{minipage} 
\end{table}

\begin{figure*}
\includegraphics[clip=true, trim=3cm 2cm 2cm 0cm, scale=0.58]{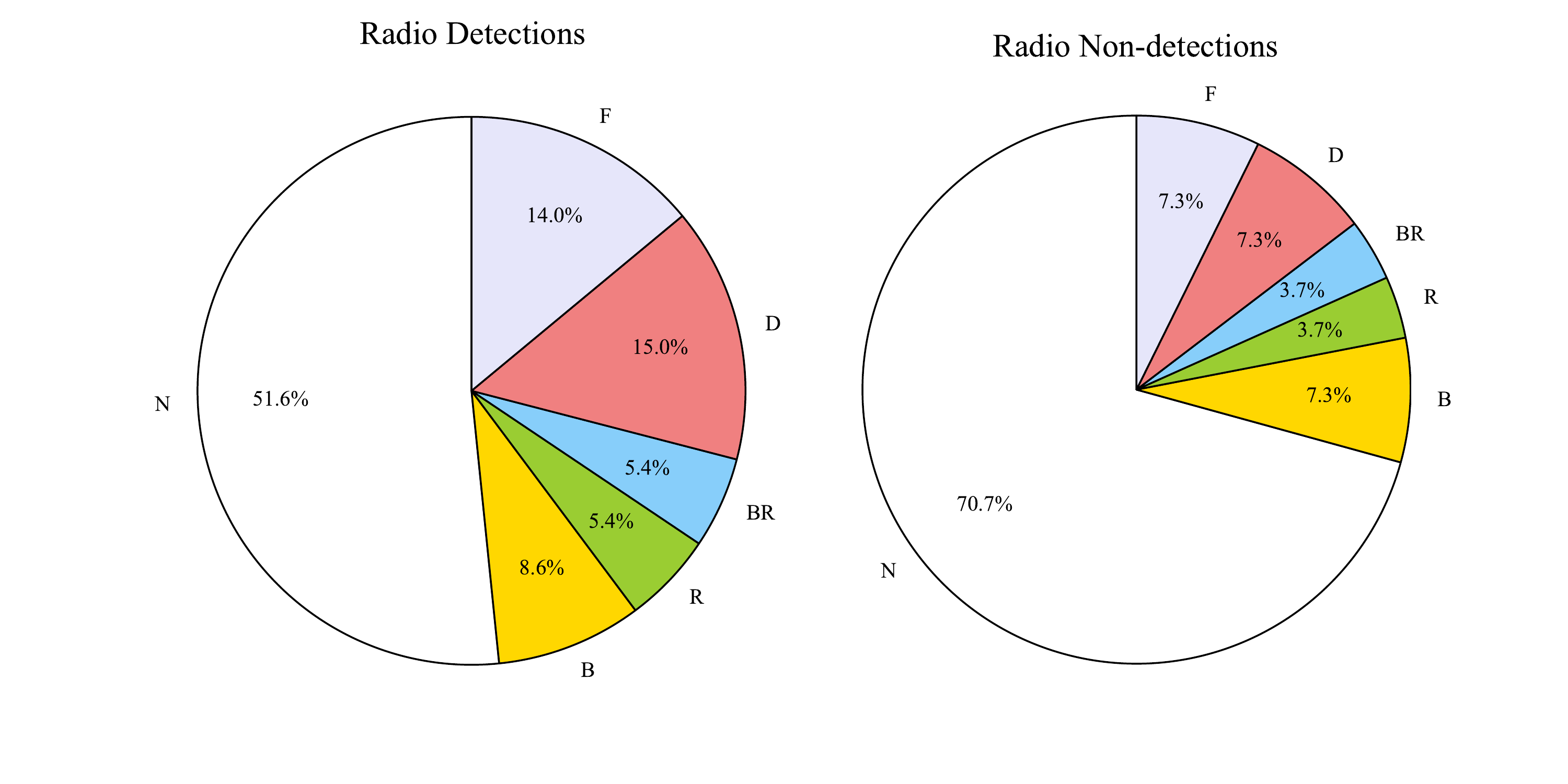}
\caption{{\bf (Left:)} Proportion of various optical dust features for the galaxies with nuclear radio detections.  The dust features are based on SDSS and INT optical images and are described in detail in \citetalias{krajnovic+11}.  N = no feature, BR = blue nucleus + ring, R = ring, B = blue nucleus, D = dusty disk, and F = filament.  {\bf (Right:)}  Same as the left panel, but for the galaxies with only upper limits to the presence of nuclear 5~GHz emission.}
\label{fig:radio_dust}
\end{figure*}

\subsubsection{Dust}
\label{dust}
In Figure~\ref{fig:radio_dust} and Table~\ref{tab:contingency_dust}, we assess the dust features of our sample galaxies using ground-based SDSS and INT images.  Our goal is to study the association of dust features with the presence/absence of compact, nuclear 5~GHz emission.  None of the individual dust categories (blue nucleus+ring, ring, blue nucleus, dusty disk, and filament\footnote{Note that we include dust lanes (which are quite common in ETGs; see \citealt{kaviraj+12}) in our use of the word ``filament" here.}) is preferentially associated with the presence or absence of nuclear radio emission at a statistically significant level.  However, the values in Table~\ref{tab:contingency_dust} suggest that ETGs with nuclear radio continuum emission may be more likely to exhibit a dust feature from any category compared to their counterparts with only 5~GHz upper limits.  This is consistent with the results of previous studies of the relationships between dust and AGN properties \citep{vandokkum+95, tran+01, krajnovic+02, kaviraj+12, shabala+12, martini+13}. 
The tendency for ETGs with dust to preferentially harbor nuclear radio emission is marginally significant with a confidence interval of 2.35$\sigma$.  This may be an indication that the underlying processes responsible for the deposition of gas and the generation of dust disks and filaments are connected to the production of radio emission associated with LLAGNs.

We also consider the origin of the dust.  As numerous studies have noted (e.g., \citetalias{davis+11}), 
ETGs may be rejuvenated with dust from either internal processes (e.g., stellar mass loss from evolved stars) or external processes (mergers and/or cold mode accretion from the ambient intergalactic medium).  Recent studies have found that a large fraction of dust-rich ETGs have acquired their gas from minor mergers and interactions 
\citep{martel+04, davis+13, kaviraj+12, shabala+12}.
Since ETGs with radio continuum emission also commonly contain dust, this is consistent with a scenario in which late-time minor mergers play an important role in triggering nuclear radio emission associated with LLAGNs.  A more detailed study incorporating the results of extremely deep imaging observations that are in currently in progress \citepalias{duc+15} is necessary to further investigate the relationship between the nuclear radio emission and the dust properties.

\subsection{Radio Luminosity and Environment}
\label{sec:density}
\begin{figure}
\includegraphics[clip=true, trim=0cm 0.1cm 2cm 2cm, scale=0.4]{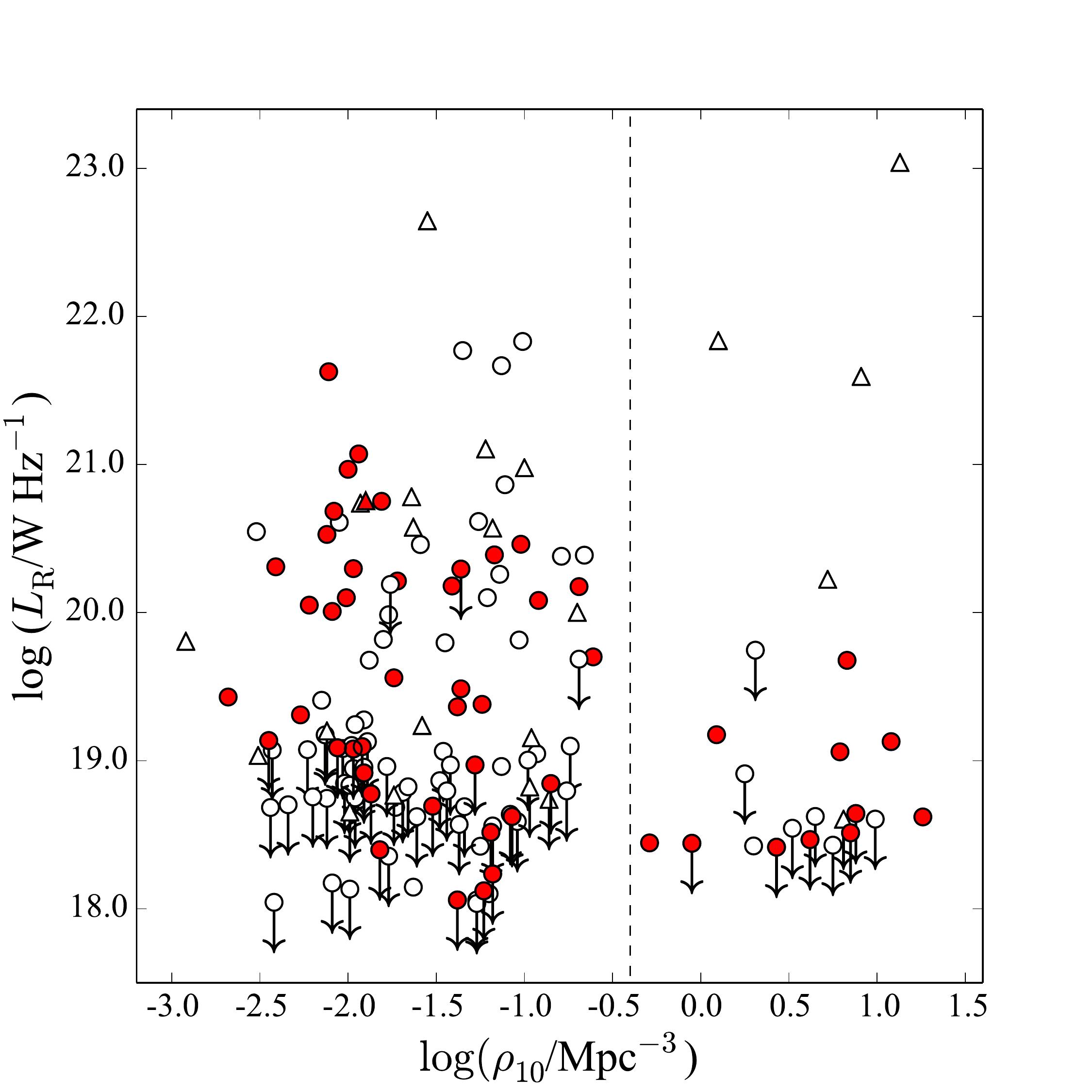}
\caption{5~GHz radio luminosity as a function of the local galaxy volume density \citepalias{cappellari+11b}.  Symbols filled in red represent the CO detections \citepalias{young+11}.  Unfilled symbols are CO non-detections (upper limits).  Circles represent fast rotators while upward-pointing triangles represent slow rotators \citepalias{emsellem+11}.  Downward-pointing arrows represent 5~GHz upper limits.  The volume density is based on a sphere centered around each galaxy and containing the 10 nearest neighbors with $M_{\mathrm{K}} < -21$.  The vertical dotted black line at $\log (\rho_{10}$/Mpc$^{-3}$) = $-0.4$ separates Virgo and non-Virgo cluster members to the right and left, respectively.}
\label{fig:radio_density}
\end{figure}

The connection between radio luminosity and environment, though relevant in terms of exploring the importance of galaxy interactions in triggering nuclear activity, remains a subject of active debate.  Confounding factors known to correlate with galaxy environment, such as galaxy mass and morphology, complicate the interpretation of any association (or lack thereof) between radio luminosity and local galaxy density.  Nevertheless, in Figure~\ref{fig:radio_density} we explore the relationship between the radio luminosity and the mean volume galaxy number density for our sample of \atlas\ ETGs.  The volume density parameter, $\rho_{10}$, is tabulated and defined in \citetalias{cappellari+11b}.  Briefly, $\rho_{10}$ is calculated by measuring the volume of a sphere centered around each galaxy that contains the 10 nearest neighbors with $M_{\mathrm{K}} < -21$.  The two distinct clumps visible to the left and right of $\log(\rho_{10}$/Mpc$^{-3}$) = $-0.4$ in Figure~\ref{fig:radio_density} correspond to non-members and members of the Virgo Cluster, respectively.  This figure does not support any association between local galaxy volume density and radio luminosity in our sample.

We performed the censored Kendall's $\tau$ test to check for correlations between nuclear radio luminosity and local galaxy density parametrized by $\rho_{10}$ for our full sample, the FRs, and the SRs.  No statistically-significant correlations were found.  In addition, we find no significant difference between the radio detection fractions of the ETGs in the Virgo cluster (48 $\pm$ 14\%) and those in lower-density environments (52 $\pm$ 6\%).  We performed the same statistical tests on the dependence of nuclear radio luminosity on galaxy environment using the local galaxy surface density parameters\footnote{As described in detail in \citetalias{cappellari+11b}, $\Sigma_{10}$ is the mean surface density of galaxies within a cylinder of height h = 600 km~s$^{-1}$ centered on each galaxy and containing the 10 nearest neighbors.  $\Sigma_{3}$ is defined in the same way but for the 3 nearest neighbors.} $\Sigma_{10}$ and $\Sigma_{3}$ as defined in \citetalias{cappellari+11b}.  No relationships with a significance greater than 2$\sigma$ were found.

The similarity between the 5~GHz detection fractions of ETGs in high and low-density environments, as well as the lack of a trend between radio luminosity and $\rho_{10}$ in our sample, suggests that environment does not significantly affect the production of nuclear radio emission.  This is consistent with the results of some previous studies (e.g., \citealt{ledlow+95}), although more recent studies utilizing larger sample sizes \citep{hickox+09, vanvelzen+12, sabater+13} have reported statistically significant associations between the clustering of galaxies and radio source properties.  Of course, comparison to these studies in the literature is complicated by the fact that our radio study was performed at much higher spatial resolution.  We also note that the average radio luminosity, as well as the range of radio luminosities, represented in our sample is much smaller than in the literature studies we have cited.  Finally, our small sample size, and poor representation of cluster ETGs, may also contribute to the lack of any relationship between radio luminosity and local galaxy volume density.  

\section{Discussion}
\label{sec:discussion}
In Section~\ref{sec:radio_origin}, we described a number of LLAGN/SF diagnostics available for our sample of ETGs.  Although in some cases the nuclear 5~GHz sources may be associated with compact circumnuclear SF, the majority likely originate from LLAGNs.  This is supported by the compactness of the radio emission (typically unresolved on scales of $\lesssim$ 25$-$110 pc), the association between the emission-line classifications and the presence of a nuclear radio source, and correlations between the nuclear radio luminosity and various galaxy parameters such as ionized gas luminosity and X-ray luminosity.  

The physical origin of the majority of the compact radio sources is most likely synchrotron emission from the base of a radio jet \citep{nagar+05}.  As other authors have suggested, the compact nature of the radio emission associated with LLAGNs lacking large-scale, collimated radio structures might be caused by intermittent SMBH accretion, a recent onset of SMBH accretion, lower bulk jet velocities (i.e., non-relativistic jet propagation), low SMBH spin, entrainment with a dense ISM, or perhaps a combination of these mechanisms \citep{ulvestad+99, sikora+09, chai+12, baldi+15}.

In the remainder of this section, we discuss the implications of our nuclear radio emission study of the \atlas\ sample in the broader context of SMBH growth, AGN fueling, AGN triggering, and AGN feedback under the assumption that the radio sources in our sample are indeed predominantly associated with LLAGNs.

\subsection{Supermassive Black Hole Growth}
The detection rate of compact, nuclear radio emission for the \atlas\ galaxies in this study is relatively high compared to detection rates in previous surveys, yet only about half of the ETGs in our sample were detected at 5~GHz.  In fact, many of the ETGs in our sample lack evidence for active SMBH accretion down to radio luminosities as low as 10$^{18}$ W~Hz$^{-1}$.  Although this is a clear indication that SMBH accretion rates in our nearby sample of ETGs are indeed quite low ($L_{\mathrm{bol}}$/$L_{\mathrm{Edd}}$ $\sim$ 5 $\times$ 10$^{-3}$ to 3 $\times$ 10$^{-8}$; see Figure~\ref{fig:radio_xray}), the detection fraction would likely increase if deeper radio observations become available.  For instance, to be able to detect 5~GHz continuum emission in an \atlas\ ETG with similar radio properties to those of the weak Galactic center source Sgr A$^{*}$, which has a SMBH mass similar to the lower-mass ETGs in our sample, significantly deeper radio data would be required.  Sgr A$^{*}$ has a radio luminosity of $L_{5\mathrm{GHz}}$ $\sim 4.6 \times 10^{15}$ W Hz$^{-1}$ at its distance of $\sim$ 8~kpc \citep{melia+01, falcke+04}.  Thus, to detect 5~GHz emission associated with a Sgr A$^{*}$ analog in even the nearest \atlas\ ETG, an rms noise of less than 60 nJy beam$^{-1}$ would be necessary for a 5$\sigma$ detection.  This is about 250 times lower than the $\sim$ 15 $\mu$Jy beam$^{-1}$ rms noise achieved in the highest-quality 5~GHz observations presented here.  Thus, we conclude that the low radio luminosities and Eddington ratios that characterize the nearby ETGs in our sample are an indication that their central SMBHs are typically not experiencing significant growth in the current epoch. 

\subsection{AGN Fueling}
\label{fuel}
As discussed previously, AGNs may accrete material from a number of sources including cooling hot gas from X-ray halos, winds from evolved stars, and the deposition of cold gas from external sources (\citealt{heckman+14}, and references therein).
However, determining the dominant SMBH accretion fuel source in LLAGNs is challenging, especially in samples of galaxies dominated by radio luminosities below $\log (L_{5\mathrm{GHz}}/$W~Hz$^{-1}$) $\sim$ 23.0.
The weak associations in our sample between the radio emission and dust/cold gas properties in sections \ref{radio_molec_gas} and \ref{dust} provide some important clues on this topic.  For instance, the statistical relationship between the radio luminosity and the molecular gas mass could be an indication that the accretion of cold gas is a significant fueling mode among nearby LLAGNs hosted by ETGs able to sustain reservoirs of cold gas.

Another possibility is that, in spite of our high-resolution observations, we are in fact observing a SF-related effect.  Correlations between radio continuum emission and CO emission have been reported in the literature (e.g., \citealt{murgia+02, murgia+05}) for studies of spiral galaxies, and are likely driven by the same fundamental processes responsible for the well-established radio-FIR relation (see Section~\ref{q}).  In this scenario, either a substantial fraction of the nuclear radio emission in our sample is driven by SF-related activities, or the trend is the result of the deeper underlying connections between SF and AGNs that have been suggested in the literature (e.g., \citealt{heckman+14}).
An example of how the relationship in Figure~\ref{fig:radio_H2} might result indirectly from a connection between SF and AGNs is as follows.  After a galaxy is rejuvenated with cold gas, the gas is first converted into a population of intermediate-mass (i.e., 1-8 M$_{\odot}$) stars in the bulge of the galaxy.  After a delay on the order of tens to hundreds of Myrs, these stars would enter the AGB phase and begin to produce winds that could ultimately serve as a fuel source for the SMBH \citep{schawinski+07, davies+07, wild+10, hopkins+12}.  
Unfortunately, given the limitations of the data currently available, it is not possible to robustly determine the dominant fueling mechanism in the \atlas\ ETGs containing reservoirs of molecular gas as well as evidence for an LLAGN.  Future studies will require higher-resolution molecular gas observations with an instrument such as the Atacama Large Millimeter/Submillimeter Array (ALMA), as well as higher-resolution measurements of the stellar population from two-dimensional optical spectroscopy, to directly map the location and properties of the cold gas in the vicinity of nearby LLAGNs.

Among the \atlas\ galaxies in our sample lacking evidence for the presence of cold gas or dust, their LLAGNs may be fueled by hot gas through advection dominated accretion flows or Bondi accretion (\citealt{yuan+14}, and references therein).  In these systems, SMBH accretion from material that has cooled out of an X-ray halo may be a less likely SMBH fuel source than direct Bondi accretion of hot gas from pAGB-star stellar winds since any substantial gas cooling might lead to SF and the presence of cold gas.  This is tentatively supported by the dissipationless and merger-driven formation history of SRs advocated by \citetalias{khochfar+11} and \citetalias{naab+14}.  However, we caution that confirmation of this statement will require future studies, and may be complicated by the fact that very little cold gas is actually required to ignite weak LLAGNs.

\subsection{Triggering Nuclear Radio Activity}
The triggering of radio AGNs has long been a subject of much debate.  
The popular consensus in the literature is that the most luminous AGNs in the radio are triggered by major mergers/strong tidal interactions (e.g., \citealt{hopkins+06, hopkins+08, ramos-almeida+11, tadhunter+11}).  Although this is true for powerful AGNs or those residing at high redshifts, \citet{heckman+14} concluded in their review of AGNs in the nearby universe that such violent interactions are not necessary for the triggering of weak AGNs.  Thus, based on these previous studies alone, we would expect that major mergers/strong tidal interactions are not the primary drivers of the AGN emission in the \atlas\ sample.

A number of lines of evidence presented in this study support the view that major mergers are not necessary to ignite the nuclear engines in nearby ETGs.  While some of the SR ETGs in our sample with relatively high radio luminosities ($L_{5\mathrm{GHz}}$ $>$ 10$^{21}$ W~Hz$^{-1}$) have tentative fossil evidence for significant mergers/interactions during their formation histories (e.g., large kinematic misalignment angles; see Section~\ref{sec:kin_misalign_ang} of this paper and \citetalias{duc+15}), the majority of the ETGs in our study lack such features and tend to have more modest radio luminosities ($L_{5\mathrm{GHz}}$ $<$ 10$^{21}$ W~Hz$^{-1}$).  These ETGs likely developed their current characteristics via ``gentler" mechanisms such as secular processes (e.g., gas transport to the center of the galaxy by dynamical structures such as bars; \citealt{garcia-burillo+05}) or minor mergers.  

In Section~\ref{dust}, we examined the incidence of dust in our sample of nearby ETGs and found that ETGs with nuclear 5~GHz emission are statistically more likely to harbor dust than ETGs lacking central radio sources.  Since previous studies have found that dust in ETGs likely has an external origin associated with minor mergers or, in more massive ETGs, the deposition of cooled hot halo gas (e.g., \citealt{martini+13}), we suggest that these processes may commonly trigger the most recent bouts of nuclear radio activity in nearby ETGs.  This is consistent with the conclusions of recent studies of the triggering of AGNs with higher radio luminosities \citep{ellison+15}.

\subsection{AGN-driven Feedback}
\label{AGN_Fback}
Evidence is mounting that AGN feedback may be responsible for the observed scaling relations between black hole and host galaxy properties, the regulation of cooling flows in clusters, galaxy-scale outflows, the suppression of SF, and the build-up of the red sequence of ETGs \citep{heckman+14}.  AGN feedback is likely carried out through radiative winds from energetic quasars and mechanical energy from radio jets in lower-accretion rate AGNs (e.g., \citealt{ciotti+10}).  Powerful radio-mode AGN feedback may be capable of directly expelling gas from the host galaxy, thereby reducing the amount of future SF in the system (e.g., \citealt{morganti+13}).  Radio jets may also disrupt future SF through the injection of turbulent energy into the ambient ISM \citep{alatalo+15, guillard+15}.  Since the nuclear radio and X-ray properties of the nearby ETGs in our sample are consistent with inefficient SMBH accretion (Section~\ref{sec:X-rays}), we would therefore expect any significant AGN feedback in these galaxies to be mechanical in nature and related to the presence of radio outflows/jets. 

\subsubsection{Incidence of Radio Jets/Lobes}
\label{radio_jets}
As discussed in Section~\ref{morph}, the fraction of ETGs in our sample with radio jets is 19/148 ($<$ 13 $\pm$ 3\%).  If no other galaxies in the \atlas\ sample currently lacking radio observations of sufficient depth/appropriate angular scale harbor central radio sources with jet/lobe morphologies, the fraction of \atlas\ ETGs with AGN-like radio structures may be as low as 19/260 (7 $\pm$ 2\%).  This implies that the duty cycle of this type of radio activity is likely quite low among nearby ETGs, consistent with previous studies over comparable ranges in stellar mass, SMBH mass, and radio luminosity (e.g., Figure~4 in \citealt{shabala+08}).  The majority of the ETGs in our sample with radio jets are massive (10.4 $<$ $\log(M_{\mathrm{jam}}$/M$_{\odot})$ $<$ 11.8; see Section~\ref{stellar_mass} for a description of the $M_{\mathrm{jam}}$ parameter), reside in relatively dense cluster or group environments (NGC1266, NGC3665, and NGC4636 are exceptions), and lack cold gas (NGC1266, NGC3665, and NGC3998 are exceptions).  Any putative energetic feedback from the radio jets in these ETGs is likely operating in ``maintenance mode" \citep{fabian+12} and may help keep SFRs low.

\subsubsection{Impact on Star Formation}
Of the \atlas\ ETGs with radio jets/lobes, only NGC3665 and NGC1266 harbor reservoirs of molecular gas \citepalias{young+11, davis+13, alatalo+13}.  Extensive studies of NGC1266 in the literature have identified the presence of a massive molecular outflow \citep{alatalo+11}, and a number of lines of evidence suggest that AGN feedback is indeed responsible for the expulsion of the cold gas \citep{nyland+13} and subsequent suppression of SF \citep{alatalo+14, alatalo+15}.  Although additional observations of NGC3665 are necessary to rule-out the possibility of an interaction between the radio jets and molecular gas in that ETG, NGC1266 is currently the only \atlas\ ETG with evidence of an AGN directly impacting the ISM of its host galaxy.  Thus, it is not clear if NGC1266 represents a common phase of galaxy evolution with a short duty cycle or is instead an evolutionary anomaly.  The constraint on the duty cycle of radio jets/lobes in the \atlas\ sample of roughly 5\% is consistent with the presence of only one ETG with evidence of direct AGN-driven feedback associated with a molecular outflow.  If NGC1266 is truly an anomaly, the implication is that unless special ISM conditions prevail (e.g., dense, compact, centrally-concentrated molecular gas likely driven inward by a minor merger within the last Gyr; \citealt{alatalo+13, davis+12}), substantial AGN feedback capable of directly suppressing SF is largely negligible in the current epoch.  

\subsubsection{Connection to Global Kinematic Classification}
Previous studies of the duty cycles of radio-loud activity in nearby AGNs (e.g., \citealt{shabala+08}) suggest that galaxies with the most massive SMBHs have the highest duty cycles.  Since the SRs in our sample are dominated by their bulge/spheroid components, their total dynamical stellar masses should also roughly scale with their SMBH masses.  Thus, it would be reasonable to expect that the most massive ETGs in our sample, which tend to be classified kinematically as SRs (see Figure~\ref{fig:radio_MS}), should have higher duty cycles of nuclear radio activity.  Indeed, SRs harbor the most powerful radio sources in our sample, and a simplistic assessment reveals a statistically significant difference in the nuclear radio detection fraction of FRs and SRs in Section~\ref{stellar_mass}.  However, as we showed later in Section~\ref{stellar_mass} in our bootstrap resampling simulation, when samples of FRs and SRs matched in stellar mass are considered, the difference in nuclear radio detection rate between the two kinematic classes is no longer significant.  
 
 Thus, we do not find evidence for a dependence of the radio detection fraction, or correspondingly the radio duty cycle, on global kinematic classification.  Future studies spanning a wider range of galaxy masses and a larger proportion of massive, slowly-rotating ETGs will be needed to further address this issue.

\section{Summary and Conclusions}
\label{sec:summary}
We report new 5~GHz VLA observations at a resolution of $\theta_{\mathrm{FWHM}}$ = 0.5$^{\prime \prime}$ of the nuclear radio emission in 121 ETGs selected from the \atlas\ sample.  We also include measurements from the literature.  In total, our study encompasses a representative sample of 148 nearby ETGs.  

We find that nuclear 5~GHz sources are detected in 42 $\pm$ 4\% of our new VLA observations at a spatial resolution corresponding to a linear scale of $\approx 25-110$~pc.  Considering the archival data as well, we find that 76/148 (51 $\pm$ 4\%) of the \atlas\ ETGs included in our sample contain nuclear radio sources.  The range of radio luminosities of the detected sources (18.04 $<$ $\log(L_{5 \, \mathrm{GHz}}/$W~Hz$^{-1}$) $<$ 23.04) and lack of detectable radio emission in almost half of our sample confirms the general consensus that nuclear accretion rates in nearby ETGs are extremely low.  Thus, the SMBHs residing in the local population of ETGs are not experiencing significant growth in the current epoch, consistent with previous studies.  

In terms of their radio morphologies, a few of our sample ETGs are characterized by emission distributed in complex or disk-like structures that may be related to circumnuclear SF rather than SMBH accretion.  However, the vast majority of the nuclear sources detected in our study are compact.  Of the sources with linearly extended 5~GHz morphologies in our high-resolution data or in observations at lower resolution available in the literature, 19 contain classic AGN-like jets/lobes or core+jet structures on some scale.  
  Additional radio continuum information available for some sources, including evidence for radio variability, q-values in the radio-excess regime, flat or inverted in-band nuclear radio spectra, and high brightness temperatures from VLBI observations in the literature, suggests that the compact radio sources detected in our study are typically associated with LLAGNs.

To help further constrain the most likely origin of the nuclear radio sources (AGN vs. SF), we also incorporated optical and X-ray AGN diagnostics in our study.  We found that ETGs with central ionized gas emission are significantly more likely to harbor nuclear radio sources compared to those lacking strong nebular emission lines.  LINERs comprise the most prevalent optical emission line class among our sample galaxies.  The LINERs with EW[O{\tt III}] in excess of 0.8 \AA, which are likely to be associated with a genuine LLAGN as opposed to pAGB stars or shocks, show a particularly strong association with the presence of compact radio emission compared to other classes at the 3$\sigma$ level.  This supports the interpretation that the majority of the nuclear radio sources detected in our study are indeed associated with LLAGNs.  This is further supported by available X-ray data in the literature.  The radio-X-ray ratios for the subset of the ETGs in our sample with both nuclear radio and X-ray measurements available are consistent with radio-loud emission (as defined by \citealt{terashima+03}), similar to typical nearby LLAGNs \citep{ho+08}.  Since radio loudness is known to scale inversely with Eddington ratio, this indicates that the SMBH accretion in our sample of local ETGs is dominated by radiatively inefficient mechanisms.

We use the multiwavelength data available for the \atlas\ galaxies to investigate the relationships between their nuclear radio activity and various galaxy properties as a function of the cold gas content and kinematic state of the host galaxy.  Our main conclusions from these analyses are as follows:\\

\begin{enumerate}
\renewcommand{\theenumi}{(\arabic{enumi})}

\item For the first time, we studied the relationship between the nuclear radio luminosity and the specific stellar angular momentum, $\lambda_{\mathrm{R}}$, measured from the two-dimensional integral-field spectroscopic data (\citealt{emsellem+07}; \citetalias{emsellem+11}).  Although SRs contain the most powerful radio sources in our sample, we do not find a significant trend between radio luminosity and $\lambda_{\mathrm{R}}$.  A higher proportion of SRs are detected in our nuclear radio observations compared to FRs, however the radio detection rates of these two kinematic classes are not statistically different when samples matched in stellar mass are considered.

\item Radio luminosity and stellar mass are correlated for both the FRs and the SRs.  The correlation is slightly stronger among the SRs, in line with previous studies of massive ETGs.  As expected based on the mutual growth of  stellar bulge mass and SMBHs, the relationships between radio luminosity and SMBH mass follow the same trends observed with stellar mass.  These trends may be due to the different dominant late-time galaxy assembly mechanisms (dry minor mergers vs.\ gas-rich major/minor mergers) and/or differences in typical environment (clusters vs. the field) for SRs and FRs, respectively.

\item Although no significant relationship exists between the radio luminosity and the kinematic misalignment angle, the most powerful radio sources have significant misalignments.  A similar analysis of the radio luminosity as a function of the angle between the kinematic axes of the stellar body and the gas leads to the same conclusion.  This supports a scenario in which massive SRs with strong radio emission are built-up through major mergers.

\item Nuclear radio continuum emission often coexists with molecular gas.  When radio luminosity is compared to molecular gas mass, two populations are evident.  Massive SRs with no evidence for the presence of molecular gas populate the left portion of Figure~\ref{fig:radio_H2}.  These ETGs show no statistical relationship between molecular gas mass and radio luminosity.  The second population consists of gas-rich ETGs in the FR kinematic class.  These ETGs display a statistically-significant trend of increasing radio luminosity with increasing molecular gas mass.  This behaviour may be an extension of the well-known radio-infrared relation thought to stem from SF.  Alternatively, if we assume the nuclear radio emission indeed originates from LLAGNs, the correlation between molecular gas mass and radio luminosity among the FRs could be an indication that cold gas accretion plays an important role in fueling their central engines.

\item ETGs with compact radio emission are statistically more likely to contain dust compared to those with only radio upper limits, though no dependence on the type of dust feature is observed.  Building on the results of previous studies of the origin of dust in ETGs (e.g., \citealt{martini+13}), we suggest that late-time minor mergers/the deposition of cooled halo gas likely play important roles in triggering LLAGNs in the nearby ETG population.  

\item We do not observe a significant trend between radio luminosity and local galaxy density.  However, the most powerful radio sources are found in the densest environment included in the \atlas\ survey (the Virgo cluster).  This could be a result of the known effects of galaxy mass and the morphology-density relation on correlations between any galaxy property and environment.  We also emphasize that our sample encompasses a relatively small proportion of ETGs in dense environments.  A future study that includes a larger sample of ETGs with two-dimensional kinematic and cold gas measurements is necessary to better establish the role of environment in influencing nuclear radio activity.

\end{enumerate}

Although this study sets the stage for connecting host ETG properties and formation histories to the properties of their nuclear radio emission, additional work is needed.  Future studies should focus on imaging the molecular gas at higher spatial resolution and sensitivity with ALMA to study the importance of cold gas in LLAGN fueling, investigating the dust mass and distribution at high-resolution to explore LLAGN triggering and gas transport, obtaining sensitive optical emission line observations across a wider range of wavelengths to facilitate more robust emission line ratio diagnostics of the nuclear activity, and performing a robust X-ray analysis in a self-consistent manner to better characterize SMBH accretion properties.  Expansion of the \atlas\ sample to include a wider range of stellar masses, galaxy environments, and more equal representation of fast and slow rotators is also necessary to improve statistics.  A high-resolution radio study of the MASSIVE sample of 116 ETGs within 108~Mpc and with masses greater than 10$^{11.5}$~M$_{\odot}$ \citep{ma+14}, combined with the \atlas\ radio study presented here, would span the necessary range of properties for a more robust statistical analysis in the future.

\section*{Acknowledgments}
We thank the referee for carefully reading our paper and for providing us with constructive comments that have strengthened it substantially.  We also thank Rene Breton for helpful discussions on accurately calculating statistical parameters and for generously providing us with a Python script for calculating the censored Kendall's $\tau$ statistic.  The National Radio Astronomy Observatory is a facility of the National Science Foundation operated under cooperative agreement by Associated Universities, Inc.  This research was funded in part by National Science Foundation grant 1109803. 
The research leading to these results has received funding from the European Research Council under the European Union's Seventh Framework Programme (FP/2007-2013) / ERC Advanced Grant RADIOLIFE-320745.
MC acknowledges support from a Royal Society University Research Fellowship. This work was supported by the rolling grants `Astrophysics at Oxford' PP/E001114/1 and ST/H002456/1 and visitor grants PPA/V/S/2002/00553, PP/E001564/1 and ST/H504862/1 from the UK Research Councils. RLD acknowledges travel and computer grants from Christ Church, Oxford and support from the Royal Society in the form of a Wolfson Merit Award 502011.K502/jd. TN acknowledges support from the DFG Cluster of Excellence Origin and Structure of the Universe. MS acknowledges support from a STFC Advanced Fellowship ST/F009186/1. TAD acknowledges the support provided by an ESO fellowship. The research leading to these results has received funding from the European Community's Seventh Framework Programme (/FP7/2007-2013/) under grant agreement No 229517. The authors acknowledge financial support from ESO. SK acknowledges support from the Royal Society Joint Projects Grant JP0869822. NS acknowledges support of Australian Research Council grant DP110103509.  AW acknowledges support of a Leverhulme Trust Early Career Fellowship.

{\it Facilities:} NRAO
\\

\footnotesize{
\bibliographystyle{mnras}
\bibliography{llagn_v31}
}

\appendix


\section{Data Tables}

\begin{table*}
\begin{minipage}{18cm}
\caption{VLA 5~GHz Sample and Flux Density Measurements
\label{tab:radio_parms}}
\begin{tabular*}{18cm}{lcccccccccc}
\hline
\hline
Galaxy & D & Virgo & F/S & $\log(M_{\mathrm{JAM}}$) & $\log(M_{\mathrm{H2}}$) & $S_\mathrm{W91}$ & rms  & $S_\mathrm{peak}$ & $S_\mathrm{int}$ & $\log(L)$\\
  & (Mpc) & & & (M$_{\odot}$) & (M$_{\odot}$) & (mJy) & ($\mu$Jy b$^{-1}$) & (mJy b$^{-1}$) & (mJy) & (W Hz$^{-1}$)\\
(1) & (2) & (3) & (4) & (5) & (6) & (7) & (8) & (9) & (10) & (11)\\
\hline
IC0560                     &  27.2 &  0 &  F &    10.05 &  $<$  7.67 &          \nodata                 &       22 &             0.16   $\pm$   0.02 &  \nodata              &            19.15 \\ 
$^{*}$IC0676               &  24.6 &  0 &  F &    10.21 &       8.63 &          \nodata                 &       16 &             0.44   $\pm$   0.02 &     2.14 $\pm$   0.72 &            20.19 \\ 
IC0719                     &  29.4 &  0 &  F &    10.63 &       8.26 &              3.6   $\pm$     0.2 &       16 &    $<$      0.08                &  \nodata              &    $<$     18.92 \\ 
IC1024                     &  24.2 &  0 &  F &    10.17 &       8.61 &    $<$       0.5                 &       20 &    $<$      0.10                &  \nodata              &    $<$     18.85 \\ 
NGC0474                    &  30.9 &  0 &  F &    10.93 &  $<$  7.68 &          \nodata                 &       16 &    $<$      0.08                &  \nodata              &    $<$     18.96 \\ 
NGC0502                    &  35.9 &  0 &  F &    10.42 &  $<$  7.88 &             11.3   $\pm$       1 &       15 &    $<$      0.07                &  \nodata              &    $<$     19.06 \\ 
NGC0509                    &  32.3 &  0 &  F &    10.04 &       7.48 &          \nodata                 &       15 &    $<$      0.07                &  \nodata              &    $<$     18.97 \\ 
NGC0516                    &  34.7 &  0 &  F &    10.14 &  $<$  7.82 &          \nodata                 &       13 &    $<$      0.07                &  \nodata              &    $<$     18.97 \\ 
NGC0524                    &  23.3 &  0 &  F &    11.40 &       7.97 &    $<$       0.5                 &       13 &             1.69   $\pm$   0.01 &     1.73 $\pm$   0.06 &            20.05 \\ 
NGC0525                    &  30.7 &  0 &  F &    10.17 &  $<$  7.75 &    $<$       0.5                 &       13 &    $<$      0.07                &  \nodata              &    $<$     18.87 \\ 
NGC0661                    &  30.6 &  0 &  S &    10.93 &  $<$  7.75 &          \nodata                 &       14 &             0.08   $\pm$   0.02 &  \nodata              &            18.95 \\ 
NGC0680                    &  37.5 &  0 &  F &    11.03 &  $<$  7.87 &          \nodata                 &       15 &             0.59   $\pm$   0.02 &  \nodata              &            20.00 \\ 
NGC0770                    &  36.7 &  0 &  F &    10.28 &  $<$  7.89 &              1.4   $\pm$     0.1 &       15 &    $<$      0.07                &  \nodata              &    $<$     19.08 \\ 
NGC0821                    &  23.4 &  0 &  F &    11.09 &  $<$  7.52 &          \nodata                 &       17 &    $<$      0.09                &  \nodata              &    $<$     18.75 \\ 
NGC0936                    &  22.4 &  0 &  F &    11.31 &  $<$  7.47 &          \nodata                 &       20 &             4.64   $\pm$   0.02 &     4.80 $\pm$   0.15 &            20.46 \\ 
NGC1023                    &  11.1 &  0 &  F &    10.82 &  $<$  6.79 &              0.6   $\pm$     0.1 &       15 &    $<$      0.07                &  \nodata              &    $<$     18.04 \\ 
NGC1121                    &  35.3 &  0 &  F &    10.56 &  $<$  7.81 &          \nodata                 &       17 &    $<$      0.09                &  \nodata              &    $<$     19.10 \\ 
$^{*}$NGC1222$^{\dagger}$  &  33.3 &  0 &  S &    10.50 &       9.07 &    $<$       0.5                 &       20 &             0.47   $\pm$   0.02 &     4.29 $\pm$   1.18 &            20.76 \\ 
NGC1248                    &  30.4 &  0 &  F &    10.22 &  $<$  7.68 &          \nodata                 &       16 &    $<$      0.08                &  \nodata              &    $<$     18.95 \\ 
$^{*}$NGC1266$^{\dagger}$  &  29.9 &  0 &  F &    10.41 &       9.28 &          \nodata                 &       18 &            12.24   $\pm$   0.37 &    39.55 $\pm$   7.09 &            21.63 \\ 
NGC1289                    &  38.4 &  0 &  S &    10.72 &  $<$  7.89 &          \nodata                 &       18 &    $<$      0.09                &  \nodata              &    $<$     19.20 \\ 
NGC1665                    &  37.5 &  0 &  F &    10.60 &  $<$  7.95 &          \nodata                 &       16 &    $<$      0.08                &  \nodata              &    $<$     19.13 \\ 
NGC2549                    &  12.3 &  0 &  F &    10.44 &  $<$  7.06 &          \nodata                 &       15 &    $<$      0.07                &  \nodata              &    $<$     18.13 \\ 
NGC2685                    &  16.7 &  0 &  F &    10.31 &       7.29 &          \nodata                 &       15 &    $<$      0.07                &  \nodata              &    $<$     18.40 \\ 
NGC2695                    &  31.5 &  0 &  F &    10.94 &  $<$  8.01 &          \nodata                 &       20 &    $<$      0.10                &  \nodata              &    $<$     19.07 \\ 
NGC2698                    &  27.1 &  0 &  F &    10.82 &  $<$  7.50 &          \nodata                 &       20 &             0.12   $\pm$   0.02 &  \nodata              &            19.02 \\ 
NGC2699                    &  26.2 &  0 &  F &    10.39 &  $<$  7.54 &          \nodata                 &       22 &    $<$      0.11                &  \nodata              &    $<$     18.96 \\ 
NGC2764                    &  39.6 &  0 &  F &    10.64 &       9.19 &          \nodata                 &       14 &             0.27   $\pm$   0.01 &     0.66 $\pm$   0.05 &            20.09 \\ 
NGC2768$^{\dagger}$        &  21.8 &  0 &  F &    11.53 &       7.64 &          \nodata                 &       25 &             9.94   $\pm$   0.02 &    10.62 $\pm$   0.32 &            20.78 \\ 
NGC2778                    &  22.3 &  0 &  F &    10.50 &  $<$  7.48 &          \nodata                 &       14 &             0.11   $\pm$   0.01 &  \nodata              &            18.82 \\ 
NGC2824                    &  40.7 &  0 &  F &    10.52 &       8.65 &          \nodata                 &       14 &             2.25   $\pm$   0.01 &     2.44 $\pm$   0.08 &            20.68 \\ 
NGC2852                    &  28.5 &  0 &  F &    10.46 &  $<$  7.68 &          \nodata                 &       16 &             0.53   $\pm$   0.02 &  \nodata              &            19.71 \\ 
NGC2859                    &  27.0 &  0 &  F &    10.97 &  $<$  7.61 &          \nodata                 &       14 &    $<$      0.07                &  \nodata              &    $<$     18.79 \\ 
NGC2880                    &  21.3 &  0 &  F &    10.62 &  $<$  7.44 &          \nodata                 &       15 &             0.09   $\pm$   0.02 &  \nodata              &            18.69 \\ 
NGC2950                    &  14.5 &  0 &  F &    10.47 &  $<$  7.12 &          \nodata                 &       18 &    $<$      0.09                &  \nodata              &    $<$     18.35 \\ 
NGC2962                    &  34.0 &  0 &  F &    11.10 &  $<$  7.85 &          \nodata                 &       20 &             0.21   $\pm$   0.02 &  \nodata              &            19.46 \\ 
NGC2974                    &  20.9 &  0 &  F &    11.13 &  $<$  7.65 &          \nodata                 &       25 &             7.63   $\pm$   0.02 &  \nodata              &            20.60 \\ 
NGC3032                    &  21.4 &  0 &  F &    10.00 &       8.41 &          \nodata                 &       18 &    $<$      0.09                &  \nodata              &    $<$     18.69 \\ 
NGC3073                    &  32.8 &  0 &  F &     9.95 &       7.52 &    $<$       0.6                 &       19 &    $<$      0.10                &  \nodata              &    $<$     19.09 \\ 
NGC3156                    &  21.8 &  0 &  F &    10.07 &       7.67 &          \nodata                 &       21 &    $<$      0.10                &  \nodata              &    $<$     18.78 \\ 
NGC3182                    &  34.0 &  0 &  F &    10.69 &       8.33 &          \nodata                 &       18 &    $<$      0.09                &  \nodata              &    $<$     19.10 \\ 
NGC3193                    &  33.1 &  0 &  F &    11.15 &  $<$  7.91 &    $<$       0.5                 &       18 &    $<$      0.09                &  \nodata              &    $<$     19.07 \\ 
NGC3301                    &  22.8 &  0 &  F &    10.48 &  $<$  7.46 &          \nodata                 &       18 &             0.64   $\pm$   0.02 &     1.04 $\pm$   0.05 &            19.81 \\ 
NGC3377                    &  10.9 &  0 &  F &    10.47 &  $<$  6.96 &          \nodata                 &       15 &             0.19   $\pm$   0.01 &  \nodata              &            18.43 \\ 
NGC3379                    &  10.3 &  0 &  F &    10.91 &  $<$  6.72 &          \nodata                 &       15 &             0.71   $\pm$   0.01 &  \nodata              &            18.95 \\ 
NGC3384                    &  11.3 &  0 &  F &    10.56 &  $<$  7.11 &          \nodata                 &       15 &    $<$      0.07                &  \nodata              &    $<$     18.06 \\ 
NGC3412                    &  11.0 &  0 &  F &    10.16 &  $<$  6.96 &              3.7   $\pm$     0.7 &       15 &    $<$      0.07                &  \nodata              &    $<$     18.04 \\ 
NGC3489                    &  11.7 &  0 &  F &    10.19 &       7.20 &    $<$       0.5                 &       14 &    $<$      0.07                &  \nodata              &    $<$     18.06 \\ 
NGC3599                    &  19.8 &  0 &  F &     9.99 &       7.36 &          \nodata                 &       14 &    $<$      0.07                &  \nodata              &    $<$     18.52 \\ 
NGC3605                    &  20.1 &  0 &  F &    10.00 &  $<$  7.48 &    $<$       0.5                 &       15 &    $<$      0.07                &  \nodata              &    $<$     18.56 \\ 
NGC3607                    &  22.2 &  0 &  F &    11.34 &       8.42 &          \nodata                 &       16 &             1.91   $\pm$   0.02 &     2.05 $\pm$   0.07 &            20.08 \\ 
NGC3608                    &  22.3 &  0 &  S &    10.96 &  $<$  7.58 &    $<$       0.5                 &       14 &             0.26   $\pm$   0.01 &  \nodata              &            19.19 \\ 
NGC3610                    &  20.8 &  0 &  F &    10.74 &  $<$  7.40 &              3.6   $\pm$     0.2 &       15 &    $<$      0.07                &  \nodata              &    $<$     18.59 \\ 
NGC3619                    &  26.8 &  0 &  F &    10.90 &       8.28 &          \nodata                 &       15 &             0.32   $\pm$   0.02 &  \nodata              &            19.44 \\ 
NGC3626                    &  19.5 &  0 &  F &    10.54 &       8.21 &              3.3   $\pm$     0.2 &       17 &             0.43   $\pm$   0.02 &     0.66 $\pm$   0.05 &            19.48 \\ 
NGC3630                    &  25.0 &  0 &  F &    10.62 &  $<$  7.60 &    $<$       0.5                 &       15 &    $<$      0.07                &  \nodata              &    $<$     18.75 \\ 
NGC3640                    &  26.3 &  0 &  F &    11.22 &  $<$  7.59 &          \nodata                 &       17 &    $<$      0.09                &  \nodata              &    $<$     18.85 \\ 
NGC3641                    &  25.9 &  0 &  F &    10.49 &  $<$  7.66 &    $<$       0.5                 &       17 &    $<$      0.09                &  \nodata              &    $<$     18.83 \\ 
NGC3648                    &  31.9 &  0 &  F &    10.74 &  $<$  7.77 &              0.7   $\pm$     0.1 &       16 &             0.10   $\pm$   0.02 &  \nodata              &            19.09 \\ 
\end{tabular*}
\end{minipage}
\end{table*}

\begin{table*}
\begin{minipage}{18cm}
\contcaption{} 
\begin{tabular*}{18cm}{lcccccccccc}
\hline
\hline
Galaxy & D & Virgo & F/S & $\log(M_{\mathrm{JAM}}$) & $\log(M_{\mathrm{H2}}$) & $S_\mathrm{W91}$ & rms  & $S_\mathrm{peak}$ & $S_\mathrm{int}$ & $\log(L)$\\
  & (Mpc) & & & (M$_{\odot}$) & (M$_{\odot}$) & (mJy) & ($\mu$Jy b$^{-1}$) & (mJy b$^{-1}$) & (mJy) & (W Hz$^{-1}$)\\
(1) & (2) & (3) & (4) & (5) & (6) & (7) & (8) & (9) & (10) & (11)\\
\hline
NGC3665$^{\dagger}$        &  33.1 &  0 &  F &    11.56 &       8.91 &    $<$       0.5                 &       20 &             7.59   $\pm$   0.02 &     9.00 $\pm$   0.05 &            21.07 \\ 
NGC3941                    &  11.9 &  0 &  F &    10.34 &  $<$  6.89 &          \nodata                 &       12 &             0.09   $\pm$   0.01 &  \nodata              &            18.18 \\ 
NGC3945                    &  23.2 &  0 &  F &    11.02 &  $<$  7.50 &    $<$       0.5                 &       15 &             2.81   $\pm$   0.02 &  \nodata              &            20.26 \\ 
NGC4036$^{\dagger}$        &  24.6 &  0 &  F &    11.16 &       8.13 &              5.0   $\pm$     0.2 &       17 &             1.68   $\pm$   0.02 &     3.73 $\pm$   0.05 &            20.43 \\ 
NGC4111                    &  14.6 &  0 &  F &    10.62 &       7.22 &    $<$       0.5                 &       25 &             0.15   $\pm$   0.01 &     2.09 $\pm$   0.19 &            19.73 \\ 
NGC4119                    &  16.5 &  1 &  F &    10.36 &       7.88 &    $<$       0.5                 &       17 &    $<$      0.09                &  \nodata              &    $<$     18.44 \\ 
NGC4150                    &  13.4 &  0 &  F &     9.94 &       7.82 &    $<$       0.5                 &       16 &    $<$      0.08                &  \nodata              &    $<$     18.24 \\ 
NGC4251                    &  19.1 &  0 &  F &    10.62 &  $<$  7.11 &          \nodata                 &       17 &    $<$      0.09                &  \nodata              &    $<$     18.57 \\ 
NGC4283                    &  15.3 &  0 &  F &    10.03 &       7.10 &          \nodata                 &       30 &    $<$      0.15                &  \nodata              &    $<$     18.62 \\ 
NGC4324                    &  16.5 &  1 &  F &    10.21 &       7.69 &          \nodata                 &       16 &    $<$      0.08                &  \nodata              &    $<$     18.42 \\ 
NGC4365                    &  23.3 &  0 &  S &    11.53 &  $<$  7.62 &    $<$       0.5                 &       18 &    $<$      0.09                &  \nodata              &    $<$     18.77 \\ 
NGC4429                    &  16.5 &  1 &  F &    11.17 &       8.05 &    $<$       0.5                 &       16 &             0.40   $\pm$   0.02 &  \nodata              &            19.11 \\ 
NGC4435                    &  16.7 &  1 &  F &    10.69 &       7.87 &    $<$       0.5                 &       20 &             0.15   $\pm$   0.02 &  \nodata              &            18.70 \\ 
NGC4459                    &  16.1 &  1 &  F &    10.92 &       8.24 &              2.6   $\pm$     0.1 &       15 &             0.29   $\pm$   0.02 &     0.37 $\pm$   0.03 &            19.06 \\ 
NGC4473                    &  15.3 &  1 &  F &    10.93 &  $<$  7.07 &    $<$       0.5                 &       30 &    $<$      0.15                &  \nodata              &    $<$     18.62 \\ 
NGC4477                    &  16.5 &  1 &  F &    10.94 &       7.54 &    $<$       0.5                 &       27 &    $<$      0.14                &  \nodata              &    $<$     18.64 \\ 
NGC4494                    &  16.6 &  0 &  F &    10.99 &  $<$  7.25 &    $<$       0.5                 &       16 &             0.29   $\pm$   0.02 &  \nodata              &            18.98 \\ 
NGC4526                    &  16.4 &  1 &  F &    11.24 &       8.59 &              0.9   $\pm$     0.1 &       18 &             1.50   $\pm$   0.02 &  \nodata              &            19.68 \\ 
NGC4546                    &  13.7 &  0 &  F &    10.76 &  $<$  6.97 &              1.4   $\pm$     0.1 &       25 &             7.92   $\pm$   0.02 &     8.08 $\pm$   0.24 &            20.26 \\ 
NGC4550                    &  15.5 &  1 &  S &    10.40 &  $<$  7.24 &    $<$       0.5                 &       28 &    $<$      0.14                &  \nodata              &    $<$     18.60 \\ 
NGC4551                    &  16.1 &  1 &  F &    10.26 &  $<$  7.24 &    $<$       0.5                 &       26 &    $<$      0.13                &  \nodata              &    $<$     18.61 \\ 
NGC4564                    &  15.8 &  1 &  F &    10.58 &  $<$  7.25 &    $<$       0.5                 &       18 &    $<$      0.09                &  \nodata              &    $<$     18.43 \\ 
NGC4570                    &  17.1 &  1 &  F &    10.76 &  $<$  7.47 &              2.1   $\pm$     0.1 &       20 &    $<$      0.10                &  \nodata              &    $<$     18.54 \\ 
NGC4596                    &  16.5 &  1 &  F &    10.91 &       7.31 &    $<$       0.5                 &       20 &    $<$      0.10                &  \nodata              &    $<$     18.51 \\ 
NGC4643                    &  16.5 &  1 &  F &    10.89 &       7.27 &          \nodata                 &       18 &             0.12   $\pm$   0.02 &  \nodata              &            18.59 \\ 
$^{*}$NGC4684              &  13.1 &  0 &  F &     9.99 &       7.21 &    $<$       0.5                 &       18 &             0.57   $\pm$   0.02 &     1.47 $\pm$   0.54 &            19.48 \\ 
NGC4694                    &  16.5 &  1 &  F &     9.90 &       8.01 &              0.8   $\pm$     0.1 &       18 &    $<$      0.09                &  \nodata              &    $<$     18.47 \\ 
NGC4697                    &  11.4 &  0 &  F &    11.07 &  $<$  6.86 &    $<$       0.5                 &       15 &    $<$      0.07                &  \nodata              &    $<$     18.07 \\ 
NGC4710                    &  16.5 &  1 &  F &    10.76 &       8.72 &          \nodata                 &       26 &             0.34   $\pm$   0.03 &     0.54 $\pm$   0.06 &            19.25 \\ 
NGC4753                    &  22.9 &  0 &  F &    11.39 &       8.55 &          \nodata                 &       17 &    $<$      0.09                &  \nodata              &    $<$     18.73 \\ 
NGC4754                    &  16.1 &  1 &  F &    10.81 &  $<$  7.18 &    $<$       0.5                 &       26 &    $<$      0.13                &  \nodata              &    $<$     18.61 \\ 
NGC4762                    &  22.6 &  0 &  F &    11.11 &  $<$  7.48 &              1.0   $\pm$     0.1 &       27 &    $<$      0.14                &  \nodata              &    $<$     18.92 \\ 
NGC5173                    &  38.4 &  0 &  F &    10.42 &       8.28 &          \nodata                 &       17 &             0.94   $\pm$   0.02 &  \nodata              &            20.22 \\ 
NGC5198                    &  39.6 &  0 &  S &    11.19 &  $<$  7.89 &              1.4   $\pm$     0.1 &       17 &             2.03   $\pm$   0.02 &  \nodata              &            20.58 \\ 
NGC5273                    &  16.1 &  0 &  F &    10.25 &       7.31 &              2.9   $\pm$     0.2 &       17 &             0.58   $\pm$   0.02 &     1.19 $\pm$   0.06 &            19.57 \\ 
NGC5308                    &  31.5 &  0 &  F &    11.16 &  $<$  7.88 &              0.6   $\pm$     0.1 &       16 &    $<$      0.08                &  \nodata              &    $<$     18.98 \\ 
NGC5379                    &  30.0 &  0 &  F &    10.43 &       8.33 &              2.3   $\pm$     0.1 &       16 &             0.12   $\pm$   0.01 &     0.23 $\pm$   0.04 &            19.39 \\ 
NGC5475$^{\dagger}$        &  28.6 &  0 &  F &    10.57 &  $<$  7.72 &          \nodata                 &       15 &             1.24   $\pm$   0.01 &     2.28 $\pm$   0.07 &            20.35 \\ 
NGC5485                    &  25.2 &  0 &  F &    11.05 &  $<$  7.60 &              6.7   $\pm$     0.3 &       17 &             0.88   $\pm$   0.01 &  \nodata              &            19.83 \\ 
NGC5574                    &  23.2 &  0 &  F &    10.10 &  $<$  7.51 &    $<$       0.5                 &       17 &    $<$      0.09                &  \nodata              &    $<$     18.74 \\ 
NGC5576                    &  24.8 &  0 &  S &    10.88 &  $<$  7.60 &              4.5   $\pm$     0.2 &       17 &    $<$      0.09                &  \nodata              &    $<$     18.80 \\ 
NGC5631                    &  27.0 &  0 &  S &    10.89 &  $<$  7.68 &    $<$       0.5                 &       17 &             0.17   $\pm$   0.01 &  \nodata              &            19.17 \\ 
NGC5638                    &  25.6 &  0 &  F &    10.93 &  $<$  7.60 &    $<$       0.5                 &       17 &    $<$      0.09                &  \nodata              &    $<$     18.82 \\ 
NGC5687                    &  27.2 &  0 &  F &    10.97 &  $<$  7.64 &             12.5   $\pm$     0.4 &       17 &    $<$      0.09                &  \nodata              &    $<$     18.88 \\ 
NGC5831                    &  26.4 &  0 &  S &    10.87 &  $<$  7.85 &    $<$       0.5                 &       15 &    $<$      0.07                &  \nodata              &    $<$     18.80 \\ 
NGC5839                    &  22.0 &  0 &  F &    10.42 &  $<$  7.38 &              1.9   $\pm$     0.1 &       15 &    $<$      0.07                &  \nodata              &    $<$     18.64 \\ 
NGC5845                    &  25.2 &  0 &  F &    10.49 &  $<$  7.50 &          \nodata                 &       15 &    $<$      0.07                &  \nodata              &    $<$     18.76 \\ 
NGC6014                    &  35.8 &  0 &  F &    10.58 &       8.77 &    $<$       0.5                 &       15 &             0.09   $\pm$   0.02 &  \nodata              &            19.14 \\ 
NGC6547                    &  40.8 &  0 &  F &    10.91 &  $<$  8.00 &    $<$       0.5                 &       15 &    $<$      0.07                &  \nodata              &    $<$     19.17 \\ 
NGC6548                    &  22.4 &  0 &  F &    10.87 &  $<$  7.58 &          \nodata                 &  \nodata &          \nodata                &  \nodata              &          \nodata \\ 
NGC6703                    &  25.9 &  0 &  S &    10.98 &  $<$  7.62 &          \nodata                 &       15 &             0.75   $\pm$   0.01 &  \nodata              &            19.78 \\ 
NGC6798                    &  37.5 &  0 &  F &    10.69 &       7.83 &    $<$       0.5                 &       15 &             0.11   $\pm$   0.01 &     0.16 $\pm$   0.03 &            19.43 \\ 
NGC7280                    &  23.7 &  0 &  F &    10.40 &  $<$  7.49 &    $<$      10.0                 &       15 &    $<$      0.07                &  \nodata              &    $<$     18.70 \\ 
NGC7332                    &  22.4 &  0 &  F &    10.54 &  $<$  7.41 &    $<$       0.5                 &       15 &    $<$      0.07                &  \nodata              &    $<$     18.65 \\ 
NGC7454                    &  23.2 &  0 &  S &    10.63 &  $<$  7.39 &          \nodata                 &       15 &    $<$      0.07                &  \nodata              &    $<$     18.68 \\ 
NGC7457                    &  12.9 &  0 &  F &    10.22 &  $<$  6.96 &    $<$       0.5                 &       15 &    $<$      0.07                &  \nodata              &    $<$     18.17 \\ 
NGC7465                    &  29.3 &  0 &  F &    10.20 &       8.79 &          \nodata                 &       17 &             1.06   $\pm$   0.02 &     1.93 $\pm$   0.07 &            20.30 \\ 
NGC7693                    &  35.4 &  0 &  F &    10.00 &  $<$  7.86 &    $<$       0.5                 &       15 &    $<$      0.07                &  \nodata              &    $<$     19.05 \\ 
PGC016060                  &  37.8 &  0 &  F &    10.45 &       8.26 &    $<$       2.0                 &       16 &    $<$      0.08                &  \nodata              &    $<$     19.14 \\ 
PGC029321                  &  40.9 &  0 &  F &     9.84 &       8.53 &    $<$       0.5                 &       13 &             4.48   $\pm$   0.01 &     4.64 $\pm$   0.14 &            20.97 \\ 
PGC056772                  &  39.5 &  0 &  F &    10.24 &       8.19 &    $<$       0.5                 &  \nodata &          \nodata                &  \nodata              &          \nodata \\ 
PGC058114                  &  23.8 &  0 &  F &  \nodata &       8.60 &    $<$       0.5                 &  \nodata &          \nodata                &  \nodata              &          \nodata \\ 
\end{tabular*}
\end{minipage} 
\end{table*}

\begin{table*}
\begin{minipage}{18cm}
\contcaption{} 
\begin{tabular*}{18cm}{lcccccccccc}
\hline
\hline
Galaxy & D & Virgo & F/S & $\log(M_{\mathrm{JAM}}$) & $\log(M_{\mathrm{H2}}$) & $S_\mathrm{W91}$ & rms  & $S_\mathrm{peak}$ & $S_\mathrm{int}$ & $\log(L)$\\
  & (Mpc) & & & (M$_{\odot}$) & (M$_{\odot}$) & (mJy) & ($\mu$Jy b$^{-1}$) & (mJy b$^{-1}$) & (mJy) & (W Hz$^{-1}$)\\
(1) & (2) & (3) & (4) & (5) & (6) & (7) & (8) & (9) & (10) & (11)\\
\hline
PGC061468                  &  36.2 &  0 &  F &    10.24 &       8.00 &    $<$       0.5                 &  \nodata &          \nodata                &  \nodata              &          \nodata \\ 
UGC05408                   &  45.8 &  0 &  F &     9.85 &       8.32 &    $<$       1.0                 &       17 &             0.14   $\pm$   0.02 &  \nodata              &            19.55 \\ 
UGC06176$^{\dagger}$       &  40.1 &  0 &  F &    10.44 &       8.58 &    $<$       0.5                 &       14 &             0.20   $\pm$   0.02 &     1.06 $\pm$   0.38 &            20.44 \\ 
UGC09519                   &  27.6 &  0 &  F &    10.06 &       8.77 &    $<$       0.5                 &       16 &    $<$      0.08                &  \nodata              &    $<$     18.86 \\ 
\hline
\hline
\end{tabular*}
 
\medskip
{\bf Notes.} Column 1: galaxy name.  Column 2: \atlas\ distance \citepalias{cappellari+11a}.  Column 3: Virgo membership.  Column 4: kinematic class \citepalias{emsellem+11} of either fast rotator (F) or slow rotator (S).  Column 5: dynamically-modeled stellar mass \citepalias{cappellari+13a}.  Column 6: molecular hydrogen mass \citep{young+11}.  Column 7: integrated flux density from the 5~GHz study of nearby ETGs carried-out by \citet{wrobel+91b} at 5$^{\prime \prime}$ resolution.  Column 8: average rms noise.  This column and all subsequent columns refer to data from our new high-resolution 5~GHz VLA observations.  Column 9:  peak flux density.   Column 10: integrated flux density.  Note that measurements of the integrated flux density are only given for sources that are formally resolved by JMFIT.  Column 11: 5~GHz radio luminosity.  When an integrated flux density is given, $L$ is based on the integrated flux density.  If only a peak flux density is given (either a measurement or an upper limit), then $L$ is based on the peak flux density.

\medskip
$^{*}$ Extended source not well-represented by a single two-dimensional Gaussian model.  The peak and integrated flux densities were calculated by drawing an aperture at the 3 $\times$ $\sigma_{\mathrm{rms}}$ level around the source in the CASA Viewer and then using the IMSTAT task to determine the flux parameters.  
 
\medskip
$^{\dagger}$ Multi-component source.  The integrated flux density refers to the sum of all components.  See Table~\ref{tab:multi_flux} for information on the properties of individual components.
\end{minipage} 
\end{table*}

\begin{sidewaystable*}
\centering
\begin{minipage}{20.5cm}
\caption{5~GHz Spatial Parameters of Detections 
\label{tab:gauss}}
\begin{tabular*}{20.5cm}{lcccccccc}
\hline
\hline
Galaxy & Morph. & R.A. & DEC.  & Beam & B.P.A. & $\theta_{M} \times \theta_{m}$ & P.A. & $M \times m$ \\
      & & (J2000) & (J2000) & (arcsec) & (deg) & (arcsec) & (deg) & (pc) \\
 (1) & (2) & (3) & (4) & (5) & (6) & (7) & (8) & (9) \\
 \hline
IC0560                   & U  & 09:45:53.436 &  -00:16:5.90 & 0.43 $\times$ 0.28 &  -46.12 &  $<$  0.39                                        &        \nodata       &  $<$ 51.43                \\ 
$^{*}$IC0676             & R  & 11:12:39.751 &  09:03:23.77 & 0.71 $\times$ 0.34 &  -45.19 &       3.84              $\times$  1.38            &        \nodata       &     457.97  $\times$164.58 \\ 
NGC0524                  & R  & 01:24:47.745 &  09:32:20.13 & 0.70 $\times$ 0.39 &  -55.99 &       0.12  $\pm$  0.03 $\times$  0.00 $\pm$ 0.02 &  68.17     $\pm$ 15.89 &      13.56  $\times$ 0.00 \\ 
NGC0661                  & U  & 01:44:14.597 &  28:42:21.11 & 0.34 $\times$ 0.28 &  -49.82 &  $<$  0.87                                        &        \nodata       &  $<$129.07                \\ 
NGC0680                  & U  & 01:49:47.291 &  21:58:15.12 & 0.90 $\times$ 0.31 &   56.12 &  $<$  0.14                                        &        \nodata       &  $<$ 25.45                \\ 
NGC0936                  & R  & 02:27:37.447 & -01:09:21.61 & 0.81 $\times$ 0.31 &   59.03 &       0.09  $\pm$  0.04 $\times$  0.07 $\pm$ 0.04 &  58.72     $\pm$ 32.88 &       9.77  $\times$ 7.60 \\ 
$^{*}$NGC1222$^{\dagger}$ & R  & 03:08:56.750 & -02:57:18.66 & 0.84 $\times$ 0.31 &   59.26 &       3.47              $\times$  1.33            &        \nodata       &     560.21  $\times$214.72 \\ 
$^{*}$NGC1266$^{\dagger}$ & R  &  03:16:0.747 & -02:25:38.69 & 0.50 $\times$ 0.32 &   60.80 &      10.01              $\times$  6.97            &        \nodata       &     1451.04  $\times$1010.37 \\ 
NGC2698                  & U  & 08:55:36.544 &  -03:11:0.90 & 0.52 $\times$ 0.32 &   60.32 &  $<$  0.21                                        &        \nodata       &  $<$ 27.59                \\ 
NGC2764                  & R  & 09:08:17.519 &  21:26:36.09 & 0.51 $\times$ 0.32 &   60.84 &       0.68  $\pm$  0.09 $\times$  0.36 $\pm$ 0.10 &  56.49     $\pm$ 12.87 &     130.55  $\times$69.11 \\ 
NGC2768$^{\dagger}$      & U  & 09:11:37.407 &  60:02:14.88 & 0.55 $\times$ 0.32 &   67.40 &  $<$  0.06                                        &        \nodata       &  $<$  6.34                \\ 
NGC2778                  & U  & 09:12:24.378 &  35:01:39.13 & 0.61 $\times$ 0.32 &   64.85 &  $<$  0.34                                        &        \nodata       &  $<$ 36.76                \\ 
NGC2824                  & R  &  09:19:2.222 &  26:16:11.97 & 0.62 $\times$ 0.31 &   62.84 &       0.15  $\pm$  0.03 $\times$  0.10 $\pm$ 0.02 & 115.19     $\pm$ 21.13 &      29.60  $\times$19.73 \\ 
NGC2852                  & U  & 09:23:14.603 &  40:09:49.75 & 0.75 $\times$ 0.31 &   61.15 &  $<$  0.13                                        &        \nodata       &  $<$ 17.96                \\ 
NGC2880                  & U  & 09:29:34.546 &  62:29:26.14 & 0.91 $\times$ 0.29 &   47.57 &  $<$  1.18                                        &        \nodata       &  $<$121.85                \\ 
NGC2962                  & U  & 09:40:53.932 &  05:09:57.02 & 0.45 $\times$ 0.35 &   62.02 &  $<$  0.28                                        &        \nodata       &  $<$ 46.15                \\ 
NGC2974                  & U  & 09:42:33.291 & -03:41:57.02 & 0.81 $\times$ 0.28 &   46.56 &  $<$  0.08                                        &        \nodata       &  $<$  8.11                \\ 
NGC3301                  & R  & 10:36:56.040 &  21:52:55.53 & 0.44 $\times$ 0.35 &  -19.63 &       0.46  $\pm$  0.04 $\times$  0.24 $\pm$ 0.03 &  48.66     $\pm$  6.82 &      50.85  $\times$26.53 \\ 
NGC3377                  & U  & 10:47:42.314 &   13:59:9.02 & 0.76 $\times$ 0.28 &   45.04 &  $<$  0.31                                        &        \nodata       &  $<$ 16.38                \\ 
NGC3379                  & U  & 10:47:49.599 &  12:34:53.95 & 0.70 $\times$ 0.31 &   47.42 &  $<$  0.15                                        &        \nodata       &  $<$  7.49                \\ 
NGC3607                  & R  & 11:16:54.668 &   18:03:6.43 & 0.78 $\times$ 0.30 &   45.95 &       0.11  $\pm$  0.03 $\times$  0.09 $\pm$ 0.03 & 115.99     $\pm$ 39.49 &      11.84  $\times$ 9.69 \\ 
NGC3608                  & U  & 11:16:58.957 &  18:08:55.22 & 0.81 $\times$ 0.26 &   42.06 &  $<$  0.16                                        &        \nodata       &  $<$ 17.30                \\ 
NGC3619                  & U  & 11:19:21.583 &  57:45:27.98 & 0.86 $\times$ 0.30 &  -62.58 &  $<$  0.15                                        &        \nodata       &  $<$ 19.49                \\ 
NGC3626                  & R  &  11:20:3.800 &  18:21:24.61 & 1.01 $\times$ 0.30 &  -56.48 &       0.48  $\pm$  0.07 $\times$  0.18 $\pm$ 0.13 & 135.64     $\pm$ 11.57 &      45.38  $\times$17.02 \\ 
NGC3648                  & U  & 11:22:31.504 &  39:52:36.76 & 0.41 $\times$ 0.27 &  -38.78 &  $<$  0.66                                        &        \nodata       &  $<$102.07                \\ 
NGC3665$^{\dagger}$      & U  & 11:24:43.626 &  38:45:46.28 & 0.40 $\times$ 0.27 &  -36.75 &  $<$  0.07                                        &        \nodata       &  $<$ 11.23                \\ 
NGC3941                  & U  & 11:52:55.366 &  36:59:11.06 & 0.41 $\times$ 0.28 &  -37.53 &  $<$  0.40                                        &        \nodata       &  $<$ 23.08                \\ 
NGC3945                  & U  & 11:53:13.607 &  60:40:32.11 & 0.62 $\times$ 0.32 &  -85.20 &  $<$  0.09                                        &        \nodata       &  $<$ 10.12                \\ 
NGC4036$^{\dagger}$      & R  & 12:01:26.475 &  61:53:44.02 & 0.95 $\times$ 0.27 &  -55.43 &       0.19  $\pm$  0.01 $\times$  0.14 $\pm$ 0.02 &  34.08     $\pm$ 15.97 &      22.66  $\times$16.70 \\ 
NGC4111                  & R  &  12:07:3.120 &  43:03:56.35 & 0.58 $\times$ 0.31 &   88.37 &       1.56  $\pm$  0.18 $\times$  1.02 $\pm$ 0.13 &  29.35     $\pm$ 11.16 &     110.42  $\times$72.20 \\ 
NGC4429                  & U  & 12:27:26.503 &  11:06:27.58 & 0.62 $\times$ 0.31 &  -85.83 &  $<$  0.29                                        &        \nodata       &  $<$ 23.20                \\ 
NGC4435                  & U  & 12:27:40.504 &  13:04:44.41 & 0.44 $\times$ 0.35 &   33.25 &  $<$  0.29                                        &        \nodata       &  $<$ 23.48                \\ 
\hline
\hline
 \end{tabular*}
\end{minipage} 
 \end{sidewaystable*}

\begin{sidewaystable*}
\centering
\begin{minipage}{20.5cm}
\contcaption{}
\begin{tabular*}{20.5cm}{lcccccccc}
\hline
\hline
Galaxy & Morph. & R.A. & DEC.  & Beam & B.P.A. & $\theta_{M} \times \theta_{m}$ & P.A. & $M \times m$ \\
      & & (J2000) & (J2000) & (arcsec) & (deg) & (arcsec) & (deg) & (pc) \\
 (1) & (2) & (3) & (4) & (5) & (6) & (7) & (8) & (9) \\
 \hline
NGC4459                  & R  &  12:29:0.033 &  13:58:42.83 & 0.59 $\times$ 0.31 &   89.76 &       0.37  $\pm$  0.06 $\times$  0.00 $\pm$ 0.05 &   8.09     $\pm$ 78.47 &      28.88  $\times$ 0.00 \\ 
NGC4494                  & U  & 12:31:24.037 &  25:46:30.10 & 0.96 $\times$ 0.28 &  -53.27 &  $<$  0.35                                        &        \nodata       &  $<$ 28.17                \\ 
NGC4526                  & U  &  12:34:2.998 &  07:41:57.91 & 1.10 $\times$ 0.27 &  -50.82 &  $<$  0.09                                        &        \nodata       &  $<$  7.16                \\ 
NGC4546                  & R  & 12:35:29.494 & -03:47:35.38 & 0.40 $\times$ 0.28 &  -53.17 &       0.08  $\pm$  0.01 $\times$  0.04 $\pm$ 0.02 &  58.37     $\pm$ 14.74 &       5.31  $\times$ 2.66 \\ 
NGC4643                  & U  & 12:43:20.143 &  01:58:41.71 & 0.46 $\times$ 0.27 &  -42.29 &  $<$  0.29                                        &        \nodata       &  $<$ 23.20                \\ 
$^{*}$NGC4684            & R  & 12:47:17.537 & -02:43:37.57 & 0.44 $\times$ 0.37 &  -82.14 &       2.37              $\times$  0.68            &        \nodata       &     150.52  $\times$43.19 \\ 
NGC4710                  & R  & 12:49:38.833 &  15:09:56.99 & 1.43 $\times$ 0.27 &  -47.95 &       0.34  $\pm$  0.09 $\times$  0.08 $\pm$ 0.11 & 162.00     $\pm$ 69.83 &      27.20  $\times$ 6.40 \\ 
NGC5173                  & U  & 13:28:25.282 &  46:35:29.91 & 0.43 $\times$ 0.28 &  -51.55 &  $<$  0.10                                        &        \nodata       &  $<$ 18.62                \\ 
NGC5198                  & U  & 13:30:11.386 &  46:40:14.67 & 1.28 $\times$ 0.26 &  -47.06 &  $<$  0.09                                        &        \nodata       &  $<$ 17.28                \\ 
NGC5273                  & R  &  13:42:8.380 &  35:39:15.42 & 0.50 $\times$ 0.37 &  -73.14 &       0.55  $\pm$  0.03 $\times$  0.18 $\pm$ 0.06 & 166.76     $\pm$  4.81 &      42.93  $\times$14.05 \\ 
NGC5379                  & R  & 13:55:34.302 &  59:44:33.90 & 0.53 $\times$ 0.35 &   61.71 &       0.45  $\pm$  0.21 $\times$  0.33 $\pm$ 0.29 & 110.91     $\pm$ 43.81 &      65.45  $\times$48.00 \\ 
NGC5475$^{\dagger}$      & U  & 14:05:12.764 &  55:44:29.47 & 0.53 $\times$ 0.34 &   55.94 &  $<$  0.14                                        &        \nodata       &  $<$ 19.41                \\ 
NGC5485                  & U  & 14:07:11.348 &   55:00:6.02 & 0.53 $\times$ 0.34 &   54.58 &  $<$  0.13                                        &        \nodata       &  $<$ 15.88                \\ 
NGC5631                  & U  & 14:26:33.288 &  56:34:57.42 & 0.45 $\times$ 0.31 &   57.36 &  $<$  0.32                                        &        \nodata       &  $<$ 41.89                \\ 
NGC6014                  & U  & 15:55:57.446 &  05:55:54.32 & 0.50 $\times$ 0.35 &   52.85 &  $<$  0.80                                        &        \nodata       &  $<$138.85                \\ 
NGC6703                  & U  & 18:47:18.818 &   45:33:2.28 & 0.59 $\times$ 0.33 &  -44.73 &  $<$  0.17                                        &        \nodata       &  $<$ 21.35                \\ 
NGC6798                  & R  &  19:24:3.164 &  53:37:29.44 & 0.43 $\times$ 0.31 &  -70.40 &       0.52  $\pm$  0.18 $\times$  0.00 $\pm$ 0.11 &  69.46     $\pm$ 20.03 &      94.54  $\times$ 0.00 \\ 
NGC7465                  & R  &  23:02:0.961 &  15:57:53.22 & 0.46 $\times$ 0.31 &  -71.20 &       0.40  $\pm$  0.02 $\times$  0.27 $\pm$ 0.02 &  19.63     $\pm$  6.39 &      56.82  $\times$38.35 \\ 
PGC029321                & R  & 10:05:51.187 &  12:57:40.62 & 0.46 $\times$ 0.31 &  -71.22 &       0.09  $\pm$  0.01 $\times$  0.05 $\pm$ 0.02 & 145.74     $\pm$ 14.05 &      17.85  $\times$ 9.91 \\ 
UGC05408                 & U  & 10:03:51.931 &  59:26:10.69 & 0.45 $\times$ 0.31 &  -70.16 &  $<$  1.35                                        &        \nodata       &  $<$299.76                \\ 
UGC06176$^{\dagger}$     & R  & 11:07:24.675 &  21:39:25.22 & 0.50 $\times$ 0.30 &  -88.62 &       2.43  $\pm$  0.32 $\times$  0.69 $\pm$ 0.15 &  24.40     $\pm$  4.30 &     472.42  $\times$134.14 \\ 
\hline
\hline
\end{tabular*}
 
\medskip
{\bf Notes.}  Column 1: galaxy name.  Column 2: radio morphology based on the output of the JMFIT task in AIPS.  R = resolved and U = unresolved.  Column 3: right ascension of the emission at the peak flux density.  For sources with multiple components denoted by a $\dagger$ symbol, the position listed is that of the component closest to the optical position of the nucleus.  The format is sexagesimal and the epoch is J2000.  Column 4: declination of the central position of the emission, determined in the same manner as the right ascension in Column 3.  Column 5: angular dimensions of the synthesized beam (major $\times$ minor axis).  Column 6: beam position angle, measured anti clockwise from North.  Column 7: angular dimensions of the emission (major $\times$ minor axis).  If JMFIT was only able to deconvolve the major axis of the source, then the minor axis extent is given as 0.00.  The errors are from JMFIT and are only given if the emission was successfully deconvolved in at least one dimension and categorized as resolved.  For non-Gaussian sources, source dimensions were determined using the CASA Viewer and no error is reported.  Column 8: position angle of the emission from JMFIT.  Column 9: linear dimensions of the emission (major $\times$ minor axis) in physical units.

\medskip
$^{*}$ Extended source not well-represented by a two-dimensional Gaussian model.  The source dimensions were measured using the CASA Viewer.  

\medskip
$^{\dagger}$ Multi-component source.  The spatial dimensions refer to the source closest to the center of the galaxy based on ground-based optical measurements.  See Table~\ref{tab:multi_spatial} for information on individual components.  
\end{minipage}
\end{sidewaystable*}


\begin{table*}
\begin{minipage}{15cm}
\caption{5~GHz Image Properties of Sources with Multiple Components}
\label{tab:multi_flux}
\begin{tabular*}{15cm}{ccccccc}
\hline
\hline
 Galaxy & Component & R.A. & DEC. & $S_{\mathrm{peak}}$ & $S_{\mathrm{int}}$ & $\log(L)$ \\
               &   & (J2000) & (J2000) & (mJy beam$^{-1}$) &  (mJy)  &  (W Hz$^{-1}$)  \\
 (1) & (2) & (3) & (4) & (5) & (6) & (7)  \\
\hline 
NGC1222     &   Eastern source & 03:08:56.721 & -02:57:18.63 &   0.44   $\pm$    0.02 &     2.45   $\pm$      0.13       &  20.51 \\ 
                       &   Western source & 03:08:56.818 & -02:57:18.78 &   0.28   $\pm$    0.02 &     2.12   $\pm$      0.15       &  20.45 \\ 
NGC1266     &   Core & 03:16:0.747 & -02:25:38.69 &  11.56   $\pm$    0.02 &    17.91   $\pm$      0.54       &  21.28 \\ 
                       &   Southern lobe & 03:16:0.822 & -02:25:42.09 &   0.71   $\pm$    0.01 &     8.60   $\pm$      0.32       &  20.96 \\ 
                       &   Northern lobe & 03:16:0.828 & -02:25:38.96 &   0.22   $\pm$    0.01 &    19.15   $\pm$      0.74       &  21.31 \\ 
NGC2768     &   Central source & 09:11:37.407 & 60:02:14.88 &   9.93   $\pm$    0.02 &  \nodata                         &  20.75 \\ 
                       &   Southern source & 09:11:37.510 & 60:02:13.86 &   0.53   $\pm$    0.02 &  \nodata                         &  19.48 \\ 
NGC3665     &   Southwestern source & 11:24:43.626 & 38:45:46.28 &   7.59   $\pm$    0.02 &  \nodata                         &  21.00 \\ 
                       &   Northeastern source & 11:24:43.492 & 38:45:47.95 &   0.67   $\pm$    0.02 &     1.39   $\pm$      0.06       &  20.26 \\ 
NGC4036     &   Eastern jet & 12:01:26.475 & 61:53:44.02 &   1.68   $\pm$    0.02 &     2.07   $\pm$      0.07       &  20.18 \\ 
                       &   Core & 12:01:26.747 & 61:53:44.60 &   1.01   $\pm$    0.02 &     1.26   $\pm$      0.05       &  19.96 \\ 
                       &   Western jet & 12:01:27.041 & 61:53:45.26 &   0.25   $\pm$    0.02 &     0.36   $\pm$      0.04       &  19.42 \\ 
NGC5475     &   Southwestern source & 14:05:12.764 & 55:44:29.47 &   1.24   $\pm$    0.01 &  \nodata                         &  20.08 \\ 
                       &   Northeastern source & 14:05:12.415 & 55:44:30.68 &   0.32   $\pm$    0.01 &     0.98   $\pm$      0.06       &  19.98 \\ 
UGC06176   &   Southern source & 11:07:24.666 & 21:39:24.96 &   0.19   $\pm$    0.01 &     0.75   $\pm$      0.07       &  20.16 \\ 
                       &   Northern source & 11:07:24.701 & 21:39:26.12 &   0.11   $\pm$    0.01 &     0.53   $\pm$      0.08       &  20.01 \\ 
\hline
\hline
\end{tabular*}
 
\medskip
{\bf Notes.}  Column 1: galaxy name.  Column 2: radio component name.  The brightest component in each galaxy is listed first.  Column 3: right ascension at the location of the peak flux density.  The format is sexagesimal and the epoch is J2000.  Column 4: declination, determined in the same manner as the right ascension in Column 3.  Column 5: peak flux density.  Column 6: integrated flux density.  Column 7: radio luminosity.  When an integrated flux density is given, $L$ is based on the integrated flux density.  If only a peak flux density is given (either a measurement or an upper limit), then $L$ is based on the peak flux density.

\end{minipage}
\end{table*}

\begin{table*}
\begin{minipage}{15cm}
\caption{5~GHz Spatial Properties of Sources with Multiple Components}
\label{tab:multi_spatial}
\begin{tabular*}{15cm}{ccccccccc}
\hline
\hline
 Galaxy & Component & Morph. & $\theta_{\mathrm{M}} \times \theta_{\mathrm{m}}$ & P.A.  & $M \times m$ \\
               &                       &               & (arcsec)                                                     & (deg) & (pc) \\
 (1) & (2) & (3) & (4) & (5) & (6)  \\
\hline 
NGC1222     & Eastern source  &    R &         1.14  $\pm$  0.06 $\times$  0.56 $\pm$ 0.04 &  91.54     $\pm$    3.49 &         184.04  $\times$   90.41 \\ 
                       & Western source  &    R &         1.37  $\pm$  0.11 $\times$  0.70 $\pm$ 0.07 & 112.60     $\pm$    5.33 &         221.18  $\times$  113.01 \\ 
NGC1266     & Core  &    R &         0.36  $\pm$  0.05 $\times$  0.29 $\pm$ 0.00 &  71.38     $\pm$    1.95 &          52.19  $\times$   42.04 \\ 
                       & Southern lobe  &    R &         2.67  $\pm$  0.07 $\times$  0.88 $\pm$ 0.03 &  22.55     $\pm$    0.97 &         387.04  $\times$  127.56 \\ 
                       & Northern lobe  &    R &         8.55  $\pm$  0.27 $\times$  2.14 $\pm$ 0.07 & 153.61     $\pm$    0.72 &        1239.40  $\times$  310.21 \\ 
NGC2768     & Central source  &    U &    $<$  0.06                                        &        \nodata         &    $<$    6.34                   \\ 
                       & Southern source &    U &    $<$  0.50                                        &        \nodata         &    $<$   52.84                   \\ 
NGC3665     & Southwestern source  &    U &    $<$  0.07                                        &        \nodata         &    $<$   11.23                   \\ 
                       & Northeastern source &    R &         0.73  $\pm$  0.04 $\times$  0.33 $\pm$ 0.03 & 119.12     $\pm$    3.75 &         117.15  $\times$   52.96 \\ 
NGC4036     & Eastern jet  &    R &         0.19  $\pm$  0.01 $\times$  0.14 $\pm$ 0.02 &  34.08     $\pm$   15.97 &          22.66  $\times$   16.70 \\ 
                       & Core  &    R &         0.25  $\pm$  0.03 $\times$  0.02 $\pm$ 0.05 &  63.13     $\pm$    7.89 &          29.82  $\times$    2.39 \\ 
                       & Western jet  &    R &         0.32  $\pm$  0.09 $\times$  0.14 $\pm$ 0.12 &  64.47     $\pm$   29.36 &          38.16  $\times$   16.70 \\ 
NGC5475     & Southwestern source  &    U &    $<$  0.14                                        &        \nodata         &    $<$   19.41                   \\ 
                       & Northeastern source  &    R &         0.81  $\pm$  0.06 $\times$  0.40 $\pm$ 0.04 &  92.81     $\pm$    4.95 &         112.31  $\times$   55.46 \\ 
UGC06176  & Southern source  &    R &         0.84  $\pm$  0.11 $\times$  0.63 $\pm$ 0.13 &  53.90     $\pm$   34.87 &         163.30  $\times$  122.48 \\ 
                       & Northern source  &    R &         1.08  $\pm$  0.22 $\times$  0.61 $\pm$ 0.19 &   6.08     $\pm$   69.58 &         209.96  $\times$  118.59 \\ 
\hline
\hline
\end{tabular*}
 
\medskip
{\bf Notes.} Column 1: galaxy name.  Column 2: radio component name.  Column 3: radio morphology as reported by JMFIT.  R = resolved and U = unresolved.  Column 4: angular dimensions of the emission (major $\times$ minor axis) and errors as determined by JMFIT.  Column 5: position angle and uncertainty.  Angles are measured anti-clockwise from north.  Column 6: linear dimensions of the emission (major $\times$ minor axis).

\end{minipage}
\end{table*}

\begin{table*}
\begin{minipage}{18.5cm}
\caption{Archival High-Resolution Radio Continuum Data}
\label{tab:archival}
\begin{tabular*}{18.5cm}{ccccccccccccc}
\hline
\hline
Galaxy & D & Virgo & F/S & $\log(M_{\mathrm{JAM}}$) & $\log(M_{\mathrm{H2}}$) & $S_\mathrm{W91}$ & Freq. & Res. & $S_\mathrm{5GHz}$ & $\log(L)$ & Ref.\\
  & (Mpc) & & & (M$_{\odot}$) & (M$_{\odot}$) & (mJy) & (GHz) & ($\prime \prime$) & (mJy) & (W Hz$^{-1}$)\\
(1) & (2) & (3) & (4) & (5) & (6) & (7) & (8) & (9) & (10) & (11) & (12)\\
\hline 
\hline
NGC3226                    & 22.9 &   0 &   F & 10.99 &   $<$  7.41 &        3.60 $\pm$  0.20 &      5.0 &     0.15 &               3.88   $\pm$    0.12 &              20.39 &        1 \\ 
NGC3245                    & 20.3 &   0 &   F & 10.81 &        7.27 &        3.30 $\pm$  0.20 &      5.0 &     0.15 &      $<$      0.50                 &      $<$     19.39 &        2 \\ 
NGC3414                    & 24.5 &   0 &   S & 11.11 &   $<$  7.19 &        5.00 $\pm$  0.20 &      5.0 &     0.15 &               1.13   $\pm$    0.03 &              19.91 &        1 \\ 
NGC3998                    & 13.7 &   0 &   F & 10.94 &   $<$  7.06 &     \nodata             &      5.0 &      0.5 &             302.00   $\pm$    9.06 &              21.83 &        3 \\ 
NGC4026                    & 13.2 &   0 &   F & 10.58 &   $<$  6.99 &        1.40 $\pm$  0.10 &      5.0 &      0.5 &      $<$      0.20                 &      $<$     18.62 &        3 \\ 
NGC4143                    & 15.5 &   0 &   F & 10.66 &   $<$  7.20 &        6.70 $\pm$  0.30 &      5.0 &      0.5 &               8.50   $\pm$    0.27 &              20.39 &        4 \\ 
NGC4168                    & 30.9 &   0 &   S & 11.30 &   $<$  7.74 &        4.50 $\pm$  0.20 &      5.0 &      0.5 &               4.80   $\pm$    0.14 &              20.74 &        4 \\ 
NGC4203                    & 14.7 &   0 &   F & 10.60 &        7.39 &       12.50 $\pm$  0.40 &      5.0 &      0.5 &              11.20   $\pm$    0.35 &              20.46 &        4 \\ 
NGC4233                    & 33.9 &   0 &   F & 11.07 &   $<$  7.89 &        1.90 $\pm$  0.10 &      5.0 &      0.5 &               3.00   $\pm$    0.10 &              20.62 &        4 \\ 
NGC4261$^{\dagger}$        & 30.8 &   0 &   S & 11.72 &   $<$  7.68 &     \nodata             &      5.0 &    0.005 &             390.00   $\pm$   19.50 &              22.65 &        5 \\ 
NGC4278                    & 15.6 &   0 &   F & 11.08 &   $<$  7.45 &     \nodata             &      5.0 &      0.5 &             159.80   $\pm$    4.80 &              21.67 &        4 \\ 
NGC4281                    & 24.4 &   0 &   F & 11.22 &   $<$  7.88 &     \nodata             &     15.0 &     0.15 &      $<$      1.15                 &      $<$     19.91 &        5 \\ 
NGC4346                    & 13.9 &   0 &   F & 10.39 &   $<$  7.12 &     \nodata             &     15.0 &     0.15 &      $<$      1.15                 &      $<$     19.42 &        5 \\ 
NGC4350                    & 15.4 &   1 &   F & 10.72 &   $<$  7.18 &     \nodata             &     15.0 &     0.15 &      $<$      1.00                 &      $<$     19.45 &        5 \\ 
NGC4374                    & 18.5 &   1 &   S & 11.59 &   $<$  7.23 &     \nodata             &      5.0 &      0.5 &             167.60   $\pm$    5.03 &              21.84 &        4 \\ 
NGC4472                    & 17.1 &   1 &   S & 11.78 &   $<$  7.25 &     \nodata             &      5.0 &      0.5 &               4.80   $\pm$    0.18 &              20.23 &        4 \\ 
NGC4477$^{\ddagger}$       & 16.5 &   1 &   F & 10.94 &        7.54 &     \nodata             &      5.0 &      0.3 &               0.12   $\pm$    0.04 &              18.59 &        6 \\ 
NGC4486                    & 17.2 &   1 &   S & 11.73 &   $<$  7.17 &     \nodata             &      5.0 &      0.5 &            3097.10   $\pm$   92.91 &              23.04 &        4 \\ 
NGC4552                    & 15.8 &   1 &   S & 11.20 &   $<$  7.28 &     \nodata             &      5.0 &      0.5 &             131.70   $\pm$    3.95 &              21.59 &        4 \\ 
NGC4621                    & 14.9 &   1 &   F & 11.12 &   $<$  7.13 &     \nodata             &      8.5 &      0.3 &               0.10   $\pm$    0.02 &              18.44 &        7 \\ 
NGC4636                    & 14.3 &   0 &   S & 11.40 &   $<$  6.87 &     \nodata             &     15.0 &     0.15 &               2.12   $\pm$    0.12 &              19.71 &        5 \\ 
NGC4697$^{\ddagger}$       & 11.4 &   0 &   F & 11.07 &   $<$  6.86 &     \nodata             &      8.5 &      0.3 &               0.09   $\pm$    0.02 &              18.16 &        7 \\ 
NGC5322                    & 30.3 &   0 &   S & 11.53 &   $<$  7.76 &     \nodata             &      5.0 &      0.5 &              11.60   $\pm$    0.35 &              21.11 &        3 \\ 
NGC5353                    & 35.2 &   0 &   F & 11.50 &   $<$  8.12 &     \nodata             &     15.0 &     0.15 &              17.42   $\pm$    1.19 &              21.41 &        1 \\ 
NGC5813                    & 31.3 &   0 &   S & 11.59 &   $<$  7.69 &        2.10 $\pm$  0.10 &      5.0 &      0.4 &               5.18   $\pm$    0.16 &              20.78 &        5 \\ 
NGC5838                    & 21.8 &   0 &   F & 11.16 &   $<$  7.56 &        2.00 $\pm$  0.10 &      5.0 &      0.1 &               1.10   $\pm$    0.06 &              19.80 &        3 \\ 
NGC5846$^{*}$              & 24.2 &   0 &   S & 11.57 &   $<$  7.78 &        5.30 $\pm$  0.30 &      8.4 &      0.3 &               6.50   $\pm$    0.41 &              20.66 &        2 \\ 
NGC5866                    & 14.9 &   0 &   F & 11.00 &        8.47 &        7.40 $\pm$  0.30 &      5.0 &      0.5 &              12.70   $\pm$    0.39 &              20.53 &        4 \\ 
NGC6278                    & 42.9 &   0 &   F & 11.02 &   $<$  7.98 &        1.20 $\pm$  0.10 &      5.0 &      0.5 &               1.60   $\pm$    0.07 &              20.55 &        3 \\ 
\hline
\hline
\end{tabular*}
 
\medskip
{\bf Notes.} The additional \atlas\ galaxies included in this study with radio continuum observations at high spatial resolution ($\theta_{\mathrm{FWHM}} \lesssim$ 1$^{\prime \prime}$) and near a frequency of 5 GHz in the literature.  Column 1: galaxy name.  Column 2: official \atlas\ distance \citepalias{cappellari+11a}.  Column 3: Virgo membership.  Column 4: kinematic class \citepalias{emsellem+11} of either fast rotator (F) or slow rotator (S).  Column 5: dynamically-modeled stellar mass \citepalias{cappellari+13a}.  Column 6: 
molecular hydrogen mass \citepalias{young+11}.  Column 7: integrated flux density from the 5~GHz study of nearby ETGs by \citet{wrobel+91b} at 5$^{\prime \prime}$ resolution.  Column 8: original observing frequency.  Column 9: approximate synthesized beam major axis.  Column 10: integrated nuclear flux density.  For measurements originally at frequencies other than 5~GHz, the flux densities listed here have been scaled to 5~GHz using the source spectral index reported in the literature.  If no such spectral index information is available, the flux density is scaled to 5~GHz assuming a flat spectral index of $\alpha = -0.1$, where $S \sim \nu^{\alpha}$.  Uncertainties are reported as given in the literature when possible.  If errors were not reported in the literature, we estimate the flux density uncertainty as described in Section~\ref{sec:results}.  Column 11: nuclear radio luminosity.  Column 12: reference.

\medskip
$^{\dagger}$ For NGC4261, a milliarcsecond-scale spatial resolution VLBA radio flux density is reported.  

\medskip
$^{\ddagger}$ Although NGC4697 and NGC4477 were included in our new 5~GHz VLA observations, the sensitivity and quality of the final images were not sufficient to detect them.  We include a literature detection of NGC4697 \citep{wrobel+08} as well as a detection of NGC4477 from project 12B-191 based on our own independent analysis.    

\medskip
$^{*}$For NGC5846, the flux density has been extrapolated from to 5~GHz using a spectral index of $-0.03$.  This spectral index was calculated based on high-resolution measurements at 8.4~GHz and 15~GHz in \citet{filho+04} and \citet{nagar+05}, respectively.

\medskip
{\bf References.} (1) \citealt{filho+06}; (2) \citealt{filho+04}; (3) \citealt{kharb+12}; (4) \citealt{nagar+01}; (5) \citealt{nagar+05}; (6) Project 12B-191; (7)\citealt{wrobel+08}.

\end{minipage}
\end{table*}


\begin{table*}
\begin{minipage}{17cm}
\caption{Additional Galaxy Properties}
\label{tab:high_energy}
\begin{tabular*}{17cm}{cccccccccccccccc}
\hline
\hline
Galaxy & $\log(M_{\mathrm{BH}})$ & Ref. & $\log(L_{\mathrm{X}})$ & Ref. & $\log(L_{\mathrm{Bol}}/L_{\mathrm{Edd}})$ & $\log(R_{\mathrm{X}})$ & $\log(L_{[\textsc{O iii}]})$ & $\log$(EW[\textsc{O$\,$iii}]) & Class. \\
  & (M$_{\odot}$) & & (erg s$^{-1}$) &  & & & (erg s$^{-1}$) & (\AA) \\
(1) & (2) & (3) & (4) & (5) & (6) & (7) & (8) & (9) & (10) \\
\hline 
\hline
IC0560       &       5.88 &   \nodata &         \nodata &    \nodata &                 \nodata &          \nodata &           38.87 &             0.56 &          H \\ 
IC0676       &       5.71 &   \nodata &         \nodata &    \nodata &                 \nodata &          \nodata &           38.20 &             0.42 &          H \\ 
IC0719       &       7.03 &   \nodata &         \nodata &    \nodata &                 \nodata &          \nodata &           38.04 &            -0.12 &          T \\ 
IC1024       &       5.71 &   \nodata &         \nodata &    \nodata &                 \nodata &          \nodata &           38.10 &             0.79 &          H \\ 
NGC0474      &       7.73 &   \nodata &           38.46 &         15 &                   -6.18 &    $<$     -2.78 &           38.45 &            -0.21 &          L \\ 
NGC0502      &       6.65 &   \nodata &         \nodata &    \nodata &                 \nodata &          \nodata &         \nodata &          \nodata &          P \\ 
NGC0509      &       4.61 &   \nodata &         \nodata &    \nodata &                 \nodata &          \nodata &           37.58 &            -0.37 &          H \\ 
NGC0516      &       5.53 &   \nodata &         \nodata &    \nodata &                 \nodata &          \nodata &         \nodata &          \nodata &          P \\ 
NGC0524      &       8.94 &         1 &           38.57 &         15 &                   -7.28 &            -1.80 &           37.77 &            -0.64 &          P \\ 
NGC0525      &       6.01 &   \nodata &         \nodata &    \nodata &                 \nodata &          \nodata &         \nodata &          \nodata &          P \\ 
NGC0661      &       8.05 &   \nodata &         \nodata &    \nodata &                 \nodata &          \nodata &           38.42 &            -0.36 &          L \\ 
NGC0680      &       8.10 &   \nodata &         \nodata &    \nodata &                 \nodata &          \nodata &           38.97 &            -0.03 &      L-AGN \\ 
NGC0770      &       6.67 &   \nodata &         \nodata &    \nodata &                 \nodata &          \nodata &         \nodata &          \nodata &          P \\ 
NGC0821      &       8.22 &         2 &    $<$    38.34 &         15 &      $<$          -6.79 &          \nodata &         \nodata &          \nodata &          P \\ 
NGC0936      &       7.96 &   \nodata &         \nodata &    \nodata &                 \nodata &          \nodata &           38.74 &            -0.03 &      L-AGN \\ 
NGC1023      &       7.62 &         3 &           38.11 &         15 &                   -6.42 &    $<$     -3.35 &           37.62 &            -0.77 &          P \\ 
NGC1121      &       7.74 &   \nodata &         \nodata &    \nodata &                 \nodata &          \nodata &           37.72 &            -0.96 &          P \\ 
NGC1222      &       6.39 &   \nodata &         \nodata &    \nodata &                 \nodata &          \nodata &           40.40 &             1.71 &      L-AGN \\ 
NGC1248      &       6.01 &   \nodata &         \nodata &    \nodata &                 \nodata &          \nodata &         \nodata &          \nodata &          P \\ 
NGC1266      &       6.04 &   \nodata &         \nodata &    \nodata &                 \nodata &          \nodata &           38.31 &             0.56 &      L-AGN \\ 
NGC1289      &       7.13 &   \nodata &         \nodata &    \nodata &                 \nodata &          \nodata &           38.69 &            -0.08 &      L-AGN \\ 
NGC1665      &       6.34 &   \nodata &         \nodata &    \nodata &                 \nodata &          \nodata &         \nodata &          \nodata &          P \\ 
NGC2549      &       7.16 &         1 &         \nodata &    \nodata &                 \nodata &          \nodata &           37.60 &            -0.49 &          L \\ 
NGC2685      &       6.57 &   \nodata &         \nodata &    \nodata &                 \nodata &          \nodata &           38.13 &            -0.07 &      L-AGN \\ 
NGC2695      &       8.23 &   \nodata &         \nodata &    \nodata &                 \nodata &          \nodata &         \nodata &          \nodata &          P \\ 
NGC2698      &       8.18 &   \nodata &         \nodata &    \nodata &                 \nodata &          \nodata &         \nodata &          \nodata &          P \\ 
NGC2699      &       7.38 &   \nodata &         \nodata &    \nodata &                 \nodata &          \nodata &           37.88 &            -0.66 &          L \\ 
NGC2764      &       6.62 &   \nodata &         \nodata &    \nodata &                 \nodata &          \nodata &           38.90 &             0.66 &          H \\ 
NGC2768      &       8.82 &   \nodata &           39.75 &         16 &                   -5.98 &            -2.25 &           38.50 &             0.14 &      L-AGN \\ 
NGC2778      &       7.16 &         2 &           38.62 &         16 &                   -5.45 &            -3.09 &           38.22 &            -0.14 &          L \\ 
NGC2824      &       7.09 &   \nodata &         \nodata &    \nodata &                 \nodata &          \nodata &           39.18 &             0.36 &          T \\ 
NGC2852      &       7.91 &   \nodata &         \nodata &    \nodata &                 \nodata &          \nodata &           38.84 &             0.18 &      L-AGN \\ 
NGC2859      &       7.86 &   \nodata &         \nodata &    \nodata &                 \nodata &          \nodata &           38.61 &            -0.23 &          L \\ 
NGC2880      &       7.30 &   \nodata &         \nodata &    \nodata &                 \nodata &          \nodata &           37.40 &            -1.10 &          P \\ 
NGC2950      &       7.71 &   \nodata &         \nodata &    \nodata &                 \nodata &          \nodata &           37.64 &            -1.00 &          P \\ 
NGC2962      &       7.71 &   \nodata &         \nodata &    \nodata &                 \nodata &          \nodata &           38.59 &             0.02 &      L-AGN \\ 
NGC2974      &       8.70 &   \nodata &           40.30 &         15 &                   -5.31 &            -2.97 &           38.89 &             0.38 &      L-AGN \\ 
NGC3032      &       6.47 &   \nodata &         \nodata &    \nodata &                 \nodata &          \nodata &           38.70 &             0.01 &          T \\ 
NGC3073      &       5.11 &   \nodata &    $<$    38.82 &         16 &      $<$          -3.20 &          \nodata &           38.53 &             0.17 &          H \\ 
NGC3156      &       6.11 &   \nodata &         \nodata &    \nodata &                 \nodata &          \nodata &           38.30 &             0.04 &          S \\ 
NGC3182      &       6.88 &   \nodata &         \nodata &    \nodata &                 \nodata &          \nodata &           39.33 &             0.66 &          S \\ 
NGC3193      &       8.08 &   \nodata &    $<$    39.72 &         15 &      $<$          -5.27 &          \nodata &           38.55 &            -0.19 &          L \\ 
NGC3226      &       7.76 &   \nodata &           40.77 &         15 &                   -3.90 &            -3.19 &           38.99 &             0.61 &      L-AGN \\ 
NGC3245      &       8.38 &         4 &           38.97 &         15 &                   -6.32 &    $<$     -2.23 &           38.69 &            -0.03 &          T \\ 
NGC3301      &       6.86 &   \nodata &         \nodata &    \nodata &                 \nodata &          \nodata &           38.65 &            -0.17 &          T \\ 
NGC3377      &       8.25 &         2 &           38.22 &         15 &                   -6.94 &            -3.07 &         \nodata &          \nodata &          P \\ 
NGC3379      &       8.62 &         5 &           38.10 &         15 &                   -7.43 &            -2.42 &           37.88 &            -0.31 &          L \\ 
NGC3384      &       7.03 &         2 &           38.09 &         15 &                   -5.85 &    $<$     -3.31 &           36.34 &            -2.00 &          P \\ 
NGC3412      &       6.52 &   \nodata &           37.54 &         15 &                   -5.89 &    $<$     -2.79 &           36.88 &            -1.30 &          P \\ 
NGC3414      &       8.67 &   \nodata &           39.88 &         15 &                   -5.70 &            -2.59 &           39.10 &             0.45 &      L-AGN \\ 
NGC3489      &       6.77 &         6 &         \nodata &    \nodata &                 \nodata &          \nodata &           38.62 &             0.08 &      L-AGN \\ 
NGC3599      &       5.49 &   \nodata &         \nodata &    \nodata &                 \nodata &          \nodata &           38.82 &             0.49 &          S \\ 
NGC3605      &       6.04 &   \nodata &         \nodata &    \nodata &                 \nodata &          \nodata &           36.40 &            -1.70 &          P \\ 
NGC3607      &       8.14 &         7 &           38.78 &         15 &                   -6.27 &            -1.98 &           38.62 &            -0.07 &      L-AGN \\ 
NGC3608      &       8.67 &         2 &           38.19 &         15 &                   -7.39 &            -2.32 &           38.14 &            -0.42 &          L \\ 
NGC3610      &       8.09 &   \nodata &           39.05 &         16 &                   -5.95 &    $<$     -3.74 &           37.59 &            -1.30 &          P \\ 
NGC3619      &       7.65 &   \nodata &         \nodata &    \nodata &                 \nodata &          \nodata &           38.58 &             0.12 &      L-AGN \\ 
NGC3626      &       7.26 &   \nodata &         \nodata &    \nodata &                 \nodata &          \nodata &           38.69 &            -0.08 &      L-AGN \\ 
NGC3630      &       7.63 &   \nodata &         \nodata &    \nodata &                 \nodata &          \nodata &           36.91 &            -2.00 &          P \\ 
NGC3640      &       8.06 &   \nodata &    $<$    38.58 &         16 &      $<$          -6.39 &          \nodata &         \nodata &          \nodata &          P \\ 
NGC3641      &       7.52 &   \nodata &           38.81 &         16 &                   -5.62 &    $<$     -3.26 &         \nodata &          \nodata &          P \\ 
NGC3648      &       7.82 &   \nodata &         \nodata &    \nodata &                 \nodata &          \nodata &           38.79 &             0.02 &      L-AGN \\ 
\end{tabular*}
\end{minipage}
\end{table*}

\begin{table*}
\begin{minipage}{17cm}
\contcaption{} 
\begin{tabular*}{17cm}{cccccccccccccccc}
\hline
\hline
Galaxy & $\log(M_{\mathrm{BH}})$ & Ref. & $\log(L_{\mathrm{X}})$ & Ref. & $\log(L_{\mathrm{Bol}}/L_{\mathrm{Edd}})$ & $\log(R_{\mathrm{X}})$ & $\log(L_{[\textsc{O iii}]})$ & $\log$(EW[\textsc{O$\,$iii}]) & Class. \\
  & (M$_{\odot}$) & & (erg s$^{-1}$) &  & & & (erg s$^{-1}$) & (\AA) \\
(1) & (2) & (3) & (4) & (5) & (6) & (7) & (8) & (9) & (10) \\
\hline 
\hline
NGC3665      &       8.58 &   \nodata &         \nodata &    \nodata &                 \nodata &          \nodata &           38.55 &            -0.26 &          T \\ 
NGC3941      &       7.13 &   \nodata &         \nodata &    \nodata &                 \nodata &          \nodata &           38.32 &            -0.15 &          L \\ 
NGC3945      &       6.94 &         7 &           39.13 &         15 &                   -4.72 &            -2.15 &           38.65 &            -0.03 &      L-AGN \\ 
NGC3998      &       8.93 &         8 &           41.44 &         15 &                   -4.40 &            -2.89 &           39.11 &             0.76 &      L-AGN \\ 
NGC4026      &       8.26 &         7 &           38.38 &         15 &                   -6.79 &    $<$     -3.04 &           37.89 &            -0.60 &          L \\ 
NGC4036      &       8.12 &   \nodata &           39.10 &         15 &                   -5.93 &            -1.95 &           39.10 &             0.42 &      L-AGN \\ 
NGC4111      &       7.80 &   \nodata &           40.38 &         15 &                   -4.33 &            -3.94 &           38.32 &            -0.22 &          L \\ 
NGC4119      &       5.57 &   \nodata &         \nodata &    \nodata &                 \nodata &          \nodata &           36.84 &            -0.74 &          H \\ 
NGC4143      &       8.08 &   \nodata &           39.96 &         15 &                   -5.03 &            -2.85 &           38.89 &             0.36 &      L-AGN \\ 
NGC4150      &       5.53 &   \nodata &    $<$    37.41 &         15 &      $<$          -5.03 &          \nodata &           37.85 &            -0.17 &          L \\ 
NGC4168      &       8.13 &   \nodata &    $<$    38.91 &         15 &      $<$          -6.13 &    $>$     -1.45 &           38.31 &            -0.19 &          L \\ 
NGC4203      &       7.39 &   \nodata &           40.06 &         15 &                   -4.24 &            -2.88 &           38.68 &             0.48 &      L-AGN \\ 
NGC4233      &       8.19 &   \nodata &           40.22 &         17 &                   -4.88 &            -2.89 &           38.65 &             0.12 &      L-AGN \\ 
NGC4251      &       7.25 &   \nodata &         \nodata &    \nodata &                 \nodata &          \nodata &         \nodata &          \nodata &          P \\ 
NGC4261      &       8.72 &         9 &           41.08 &         15 &                   -4.55 &            -1.72 &           38.86 &             0.14 &      L-AGN \\ 
NGC4278      &       8.53 &   \nodata &           40.08 &         15 &                   -5.36 &            -1.70 &           39.07 &             0.66 &      L-AGN \\ 
NGC4281      &       8.63 &   \nodata &         \nodata &    \nodata &                 \nodata &          \nodata &           37.75 &            -0.92 &          P \\ 
NGC4283      &       6.62 &   \nodata &           38.84 &         16 &                   -4.69 &    $<$     -3.50 &         \nodata &          \nodata &          P \\ 
NGC4324      &       6.34 &   \nodata &         \nodata &    \nodata &                 \nodata &          \nodata &           38.30 &             0.18 &          S \\ 
NGC4346      &       7.19 &   \nodata &         \nodata &    \nodata &                 \nodata &          \nodata &           37.88 &            -0.37 &          L \\ 
NGC4350      &       7.98 &   \nodata &         \nodata &    \nodata &                 \nodata &          \nodata &           36.65 &            -1.52 &          P \\ 
NGC4365      &       9.01 &   \nodata &           38.37 &         15 &                   -7.55 &    $<$     -2.89 &         \nodata &          \nodata &          P \\ 
NGC4374      &       8.97 &        10 &           39.50 &         15 &                   -6.38 &            -0.95 &           38.56 &             0.09 &      L-AGN \\ 
NGC4429      &       8.14 &   \nodata &         \nodata &    \nodata &                 \nodata &          \nodata &           37.92 &            -0.38 &          L \\ 
NGC4435      &       7.70 &   \nodata &           38.41 &         15 &                   -6.20 &            -3.13 &           38.22 &            -0.15 &          T \\ 
NGC4459      &       7.84 &        11 &           38.40 &         15 &                   -6.35 &            -2.62 &           37.95 &            -0.47 &          L \\ 
NGC4472      &       9.40 &        12 &    $<$    38.71 &         15 &      $<$          -7.60 &    $>$     -1.77 &           37.85 &            -0.72 &          L \\ 
NGC4473      &       7.95 &         2 &    $<$    38.12 &         15 &      $<$          -6.74 &          \nodata &         \nodata &          \nodata &          P \\ 
NGC4477      &       8.16 &   \nodata &         \nodata &    \nodata &                 \nodata &          \nodata &           38.67 &             0.51 &      L-AGN \\ 
NGC4486      &       9.79 &        13 &           40.86 &         15 &                   -5.84 &            -1.10 &           39.10 &             0.92 &          P \\ 
NGC4494      &       7.68 &   \nodata &           38.77 &         15 &                   -5.82 &            -3.00 &           38.15 &            -0.51 &          L \\ 
NGC4526      &       8.65 &        14 &         \nodata &    \nodata &                 \nodata &          \nodata &           37.90 &            -0.43 &          T \\ 
NGC4546      &       8.31 &   \nodata &         \nodata &    \nodata &                 \nodata &          \nodata &           38.73 &             0.40 &      L-AGN \\ 
NGC4550      &       6.92 &   \nodata &    $<$    38.35 &         15 &      $<$          -5.48 &          \nodata &           37.95 &             0.09 &          T \\ 
NGC4551      &       6.69 &   \nodata &         \nodata &    \nodata &                 \nodata &          \nodata &         \nodata &          \nodata &          P \\ 
NGC4552      &       8.92 &   \nodata &           39.22 &         15 &                   -6.60 &            -0.91 &           38.35 &            -0.16 &          L \\ 
NGC4564      &       7.94 &         2 &           38.50 &         15 &                   -6.35 &    $<$     -3.35 &         \nodata &          \nodata &          P \\ 
NGC4570      &       8.03 &   \nodata &           38.14 &         15 &                   -6.80 &    $<$     -2.88 &         \nodata &          \nodata &          P \\ 
NGC4596      &       7.88 &        11 &         \nodata &    \nodata &                 \nodata &          \nodata &           37.67 &            -0.60 &          T \\ 
NGC4621      &       8.49 &   \nodata &           38.74 &         15 &                   -6.66 &            -3.60 &         \nodata &          \nodata &          P \\ 
NGC4636      &       8.33 &   \nodata &    $<$    38.22 &         15 &      $<$          -7.02 &    $>$     -1.50 &           38.05 &            -0.00 &      L-AGN \\ 
NGC4643      &       7.73 &   \nodata &         \nodata &    \nodata &                 \nodata &          \nodata &           37.74 &            -0.51 &          T \\ 
NGC4684      &       5.68 &   \nodata &         \nodata &    \nodata &                 \nodata &          \nodata &           39.31 &             1.55 &          H \\ 
NGC4694      &       4.92 &   \nodata &         \nodata &    \nodata &                 \nodata &          \nodata &           37.89 &            -0.02 &          H \\ 
NGC4697      &       8.31 &         2 &           38.38 &         15 &                   -6.84 &            -3.52 &           37.67 &            -0.55 &          T \\ 
NGC4710      &       6.76 &   \nodata &         \nodata &    \nodata &                 \nodata &          \nodata &           37.34 &            -0.10 &          T \\ 
NGC4753      &       8.06 &   \nodata &         \nodata &    \nodata &                 \nodata &          \nodata &           37.94 &            -0.57 &          P \\ 
NGC4754      &       7.76 &   \nodata &           38.27 &         15 &                   -6.40 &    $<$     -2.95 &         \nodata &          \nodata &          P \\ 
NGC4762      &       7.38 &   \nodata &         \nodata &    \nodata &                 \nodata &          \nodata &           38.18 &            -0.60 &          L \\ 
NGC5173      &       6.52 &   \nodata &         \nodata &    \nodata &                 \nodata &          \nodata &           39.24 &             0.49 &          T \\ 
NGC5198      &       8.05 &   \nodata &    $<$    39.07 &         17 &      $<$          -5.89 &    $>$     -1.77 &           38.80 &             0.03 &      L-AGN \\ 
NGC5273      &       5.60 &   \nodata &           40.53 &         15 &                   -1.98 &            -4.25 &           39.56 &             1.41 &          S \\ 
NGC5308      &       8.45 &   \nodata &         \nodata &    \nodata &                 \nodata &          \nodata &         \nodata &          \nodata &          P \\ 
NGC5322      &       8.65 &   \nodata &           40.26 &         15 &                   -5.30 &            -2.44 &           38.52 &            -0.52 &          L \\ 
NGC5353      &       9.13 &   \nodata &         \nodata &    \nodata &                 \nodata &          \nodata &           38.71 &            -0.14 &          L \\ 
NGC5379      &       6.34 &   \nodata &         \nodata &    \nodata &                 \nodata &          \nodata &           38.36 &             0.37 &          H \\ 
NGC5475      &       6.99 &   \nodata &         \nodata &    \nodata &                 \nodata &          \nodata &           38.70 &             0.27 &      L-AGN \\ 
NGC5485      &       7.95 &   \nodata &         \nodata &    \nodata &                 \nodata &          \nodata &           38.00 &            -0.24 &          L \\ 
NGC5574      &       6.08 &   \nodata &         \nodata &    \nodata &                 \nodata &          \nodata &         \nodata &          \nodata &          P \\ 
NGC5576      &       8.44 &         7 &           38.88 &         16 &                   -6.47 &    $<$     -3.37 &         \nodata &          \nodata &          P \\ 
NGC5631      &       7.68 &   \nodata &         \nodata &    \nodata &                 \nodata &          \nodata &           38.66 &            -0.19 &          L \\ 
NGC5638      &       7.60 &   \nodata &    $<$    38.33 &         16 &      $<$          -6.18 &          \nodata &         \nodata &          \nodata &          P \\ 
NGC5687      &       7.92 &   \nodata &         \nodata &    \nodata &                 \nodata &          \nodata &         \nodata &          \nodata &          P \\ 
\end{tabular*}
\end{minipage}
\end{table*}


\begin{table*}
\begin{minipage}{17cm}
\contcaption{} 
\begin{tabular*}{17cm}{cccccccccccccccc}
\hline
\hline
Galaxy & $\log(M_{\mathrm{BH}})$ & Ref. & $\log(L_{\mathrm{X}})$ & Ref. & $\log(L_{\mathrm{Bol}}/L_{\mathrm{Edd}})$ & $\log(R_{\mathrm{X}})$ & $\log(L_{[\textsc{O iii}]})$ & $\log$(EW[\textsc{O$\,$iii}]) & Class. \\
  & (M$_{\odot}$) & & (erg s$^{-1}$) &  & & & (erg s$^{-1}$) & (\AA) \\
(1) & (2) & (3) & (4) & (5) & (6) & (7) & (8) & (9) & (10) \\
\hline 
\hline
NGC5813      &       8.75 &   \nodata &           38.78 &         15 &                   -6.88 &            -1.28 &           38.53 &            -0.15 &          L \\ 
NGC5831      &       7.74 &   \nodata &           39.45 &         16 &                   -5.20 &    $<$     -3.94 &           37.81 &            -0.72 &          P \\ 
NGC5838      &       9.06 &   \nodata &           38.97 &         15 &                   -7.00 &            -2.46 &           38.32 &            -0.20 &          L \\ 
NGC5839      &       7.23 &   \nodata &         \nodata &    \nodata &                 \nodata &          \nodata &         \nodata &          \nodata &          P \\ 
NGC5845      &       8.69 &         2 &           39.05 &         15 &                   -6.55 &    $<$     -3.58 &         \nodata &          \nodata &          P \\ 
NGC5846      &       8.77 &   \nodata &           40.78 &         15 &                   -4.90 &            -3.08 &           38.28 &            -0.05 &      L-AGN \\ 
NGC5866      &       7.76 &   \nodata &           38.28 &         15 &                   -6.39 &            -1.03 &         \nodata &          \nodata & \nodata \\ 
NGC6014      &       6.17 &   \nodata &         \nodata &    \nodata &                 \nodata &          \nodata &           39.06 &             0.59 &          H \\ 
NGC6278      &       8.25 &   \nodata &           40.34 &         17 &                   -4.82 &            -3.08 &           39.05 &             0.00 &      L-AGN \\ 
NGC6547      &       7.99 &   \nodata &         \nodata &    \nodata &                 \nodata &          \nodata &         \nodata &          \nodata &          P \\ 
NGC6703      &       7.71 &   \nodata &         \nodata &    \nodata &                 \nodata &          \nodata &           37.96 &            -0.43 &          L \\ 
NGC6798      &       7.23 &   \nodata &         \nodata &    \nodata &                 \nodata &          \nodata &           38.65 &            -0.03 &      L-AGN \\ 
NGC7280      &       6.86 &   \nodata &         \nodata &    \nodata &                 \nodata &          \nodata &           37.99 &            -0.52 &          L \\ 
NGC7332      &       7.48 &   \nodata &    $<$    39.68 &         15 &      $<$          -4.71 &          \nodata &           38.41 &            -0.51 &          L \\ 
NGC7454      &       6.94 &   \nodata &         \nodata &    \nodata &                 \nodata &          \nodata &         \nodata &          \nodata &    \nodata \\ 
NGC7457      &       6.95 &         2 &           37.89 &         15 &                   -5.97 &    $<$     -3.00 &         \nodata &          \nodata &    \nodata \\ 
NGC7465      &       6.42 &   \nodata &         \nodata &    \nodata &                 \nodata &          \nodata &           40.00 &             1.22 &          S \\ 
NGC7693      &       5.11 &   \nodata &         \nodata &    \nodata &                 \nodata &          \nodata &         \nodata &          \nodata &          P \\ 
PGC016060    &       6.52 &   \nodata &         \nodata &    \nodata &                 \nodata &          \nodata &           38.45 &             0.10 &      L-AGN \\ 
PGC029321    &       5.41 &   \nodata &         \nodata &    \nodata &                 \nodata &          \nodata &           39.75 &             1.54 &          S \\ 
UGC05408     &       5.11 &   \nodata &         \nodata &    \nodata &                 \nodata &          \nodata &           40.30 &             1.40 &          H \\ 
UGC06176     &       6.79 &   \nodata &         \nodata &    \nodata &                 \nodata &          \nodata &           38.99 &             0.47 &          T \\ 
UGC09519     &       6.36 &   \nodata &         \nodata &    \nodata &                 \nodata &          \nodata &           38.11 &            -0.32 &          L \\ 
\hline
\hline
\end{tabular*}
 
\medskip
{\bf Notes.} Column 1: galaxy name.  Column 2: black hole mass.  Column 3: black hole mass reference (see list of references below) for galaxies with dynamical mass estimates available in the literature.  If no reference is given, the black hole mass is estimated from the $M_{\mathrm{BH}}-\sigma_{*}$ relation \citep{mcconnell+13}.  Column 4: $2-10$ keV X-ray luminosity from the literature.  Column 5: X-ray luminosity reference (see list of references below).  Column 6: Eddington ratio.  Details on the computation of this parameter are provided in Section~\ref{sec:inefficient_black_hole_accretion}.  Column 7: radio-X-ray ratio as defined in \citet{terashima+03}.  Column 8: nuclear [{\sc O$\,$iii}] luminosity.  Column 9: nuclear [{\sc O$\,$iii}] equivalent width.  Column 10: nuclear emission line classification.\\

{\bf References.} 1. \citealt{krajnovic+09};  2. \citealt{schulze+11}; 3. \citealt{bower+01}; 4. \citealt{barth+01}; 5. \citealt{vandenbosch+10}; 6. \citealt{nowak+10}; 7. \citealt{gultekin+09}; 8.  \citealt{walsh+12}; 9. \citealt{ferrarese+96}; 10. \citealt{walsh+10}; 11. \citealt{sarzi+01}; 12. \citealt{rusli+13}; 13. \citealt{gebhardt+11}; 14. \citealt{davis+13}; 15. \citealt{pellegrini+10}; 16. \citealt{miller+12}; 17.  \citealt{kharb+12}.


\end{minipage}
\end{table*}

\clearpage
\section{Statistics}
\label{sec:stats}
Throughout this study, we have utilized the generalized Kendall's $\tau$ correlation test \citep{isobe+86} to help quantify whether a relationship exists between two parameters.  Our implementation of this statistical test has been modified to account for data with both uncertainties and censored values (i.e., upper limits).  We perform a Monte Carlo sample draw with 10,000 iterations from a given pair of parameters (e.g., radio luminosity and molecular gas mass), sampling each data point according to a normal distribution with the location and scale determined by the $X$ and $Y$ values and errors.  If no uncertainty is available for a given parameter, the exact value is used.  We assume the errors are normally distributed and values must be strictly positive.  Censored data points are included as values with zero mean and a standard deviation given by the one-sigma rms noise value.  For calculation of the normal distribution, we use the truncnorm function available in the SciPy\footnote{http://www.scipy.org} (version 0.13.3) Python module.  


In this statistical test, $\tau$ = 1 implies a direct correlation and $\tau$ = $-1$ implies an inverse correlation.  If $\tau$ = 0, then the two parameters are uncorrelated.  Relationships are considered significant if the probability of the null hypothesis that no correlation exists is less than 2$\sigma$.  In other words, we require $p < 0.05$ to ensure that a relationship is significant at a confidence interval of 95\% or higher.  We provide a summary of the results of our implementation of the Kendall $\tau$ correlation test in Table~\ref{tab:stats}.

\begin{table}
\begin{minipage}{9cm}
\caption{Summary of Kendall's $\tau$ Correlation Tests}
\label{tab:stats}
\begin{tabular*}{9cm}{lcccccc}
\hline
\hline
$X$      & $Y$ & $N$ & $\tau$ & $\sigma_{\tau}$ & $P_{\mathrm{null}}$ &\\
(1)      & (2) & (3) & (4) & (5) & (6) &\\
\hline 
All & & & & & & \\
\hline
$\lambda_{\mathrm{R}}$   &       $L_{5\,\mathrm{GHz}}$ &  148 &      -0.108 &       0.0196 &      5.2e-02 & \\ 
$\psi_{\mathrm{kin-phot}}$      &       $L_{5\,\mathrm{GHz}}$ &  148 &       0.101 &       0.0177 &      6.8e-02 & \\ 
$\psi_{\mathrm{gas-stars}}$  &       $L_{5\,\mathrm{GHz}}$ &  148 &       0.084 &       0.0268 &      2.2e-01 & \\ 
$M_{\mathrm{BH}}$      &       $L_{5\,\mathrm{GHz}}$ & 148  &       0.253 &       0.0168 &      5.1e-06 & \\ 
$M_{\mathrm{JAM}}$     &       $L_{5\,\mathrm{GHz}}$ & 148  &       0.311 &       0.0151 &      2.0e-08 & \\ 
$M_{\mathrm{H2}}$      &       $L_{5\,\mathrm{GHz}}$ & 148  &      -0.014 &       0.0524 &      5.2e-01 & \\ 
$\rho_{10}$    &       $L_{5\,\mathrm{GHz}}$ &  148 &      -0.017 &       0.0154 &      7.6e-01 & \\ 
$L_{[\textsc{O iii}]}$      &       $L_{5\,\mathrm{GHz}}$ & 115 &       0.402 &       0.0125 &      2.0e-10 & \\ 
EW[\textsc{O$\,$iii}]     &       $L_{5\,\mathrm{GHz}}$ & 115 &       0.338 &       0.0114 &      8.6e-08 & \\ 
$L_{\mathrm{X}}(2-10\,\mathrm{kev})$       &       $\nu L_{5\,\mathrm{GHz}}$ & 64 &       0.420 &       0.0434 &      9.5e-07 & \\ 
$L_{\mathrm{Edd}}/L_{\mathrm{Bol}}$      &       $R_{\mathrm{x}}$ & 64 &      -0.052 &       0.0358 &      5.9e-01 & \\ 
\hline 
Slow Rotators & & & & & & \\
\hline
$\lambda_{\mathrm{R}}$   &       $L_{5\,\mathrm{GHz}}$ & 23  &      -0.341 &       0.0231 &      2.3e-02 & \\ 
$\psi_{\mathrm{kin-phot}}$      &       $L_{5\,\mathrm{GHz}}$ & 23 &       0.265 &       0.0197 &      7.7e-02 & \\ 
$\psi_{\mathrm{gas-stars}}$  &       $L_{5\,\mathrm{GHz}}$ & 23 &       0.029 &       0.0881 &      7.3e-01 & \\ 
$M_{\mathrm{BH}}$      &       $L_{5\,\mathrm{GHz}}$ & 23 &       0.528 &       0.0195 &      4.2e-04 & \\ 
$M_{\mathrm{JAM}}$     &       $L_{5\,\mathrm{GHz}}$ & 23  &       0.565 &       0.0220 &      1.6e-04 & \\ 
$M_{\mathrm{H2}}$      &       $L_{5\,\mathrm{GHz}}$ & 23  &      -0.043 &       0.0248 &      7.7e-01 & \\ 
$\rho_{10}$    &       $L_{5\,\mathrm{GHz}}$ &  23 &       0.150 &       0.0210 &      3.2e-01 & \\ 
$L_{[\textsc{O iii}]}$      &       $L_{5\,\mathrm{GHz}}$ & 20 &       0.358 &       0.0190 &      2.7e-02 & \\ 
EW[\textsc{O$\,$iii}]     &       $L_{5\,\mathrm{GHz}}$ & 20 &       0.259 &       0.0172 &      1.1e-01 & \\ 
$L_{\mathrm{X}}(2-10\,\mathrm{kev})$       &       $\nu L_{5\,\mathrm{GHz}}$ & 17 &       0.426 &       0.0706 &      1.7e-02 & \\ 
$L_{\mathrm{Edd}}/L_{\mathrm{Bol}}$      &       $R_{\mathrm{x}}$ & 17 &       0.152 &       0.0789 &      4.9e-01 & \\ 
\hline
Fast Rotators & & & & & & \\
\hline
$\lambda_{\mathrm{R}}$   &       $L_{5\,\mathrm{GHz}}$ & 125 &       0.015 &       0.0214 &      7.7e-01 & \\ 
$\psi_{\mathrm{kin-phot}}$      &       $L_{5\,\mathrm{GHz}}$ & 125 &      -0.003 &       0.0215 &      8.1e-01 & \\ 
$\psi_{\mathrm{gas-stars}}$  &       $L_{5\,\mathrm{GHz}}$ & 125 &      -0.005 &       0.0315 &      7.7e-01 & \\ 
$M_{\mathrm{BH}}$      &       $L_{5\,\mathrm{GHz}}$ & 125 &       0.180 &       0.0203 &      2.8e-03 & \\ 
$M_{\mathrm{JAM}}$     &       $L_{5\,\mathrm{GHz}}$ & 125 &       0.239 &       0.0194 &      7.8e-05 & \\ 
$M_{\mathrm{H2}}$      &       $L_{5\,\mathrm{GHz}}$ &  125 &       0.246 &       0.0161 &      4.8e-05 & \\ 
$\rho_{10}$    &       $L_{5\,\mathrm{GHz}}$ & 125 &      -0.049 &       0.0183 &      4.2e-01 & \\ 
$L_{[\textsc{O iii}]}$      &       $L_{5\,\mathrm{GHz}}$ & 95 &       0.439 &       0.0164 &      2.9e-10 & \\ 
EW[\textsc{O$\,$iii}]     &       $L_{5\,\mathrm{GHz}}$ & 95 &       0.379 &       0.0147 &      5.5e-08 & \\ 
$L_{\mathrm{X}}(2-10\,\mathrm{kev})$       &       $\nu L_{5\,\mathrm{GHz}}$ & 47 &       0.448 &       0.0489 &      9.1e-06 & \\ 
$L_{\mathrm{Edd}}/L_{\mathrm{Bol}}$      &       $R_{\mathrm{x}}$ & 47 &       0.030 &       0.0503 &      7.1e-01 & \\ 
\hline
CO Detections & & & & & & \\
\hline
$\psi_{\mathrm{gas-stars}}$  &       $L_{5\,\mathrm{GHz}}$ & 49 &      -0.025 &       0.0466 &      7.2e-01 & \\ 
   $M_{\mathrm{H2}}$     &  $L_{5\,\mathrm{GHz}}$  & 52 &       0.333 &       0.0208 &      5.0e-04 & \\ 
$L_{\mathrm{Edd}}/L_{\mathrm{Bol}}$      &       $R_{\mathrm{x}}$ & 14 &      -0.394 &       0.0750 &      7.5e-02 & \\ 
\hline
CO Upper Limits & & & & & & \\
\hline
$\psi_{\mathrm{gas-stars}}$  &       $L_{5\,\mathrm{GHz}}$ & 50 &       0.187 &       0.0375 &      5.5e-02 & \\ 
    $M_{\mathrm{H2}}$    &  $L_{5\,\mathrm{GHz}}$  & 96 &       0.134 &       0.0223 &      5.3e-02 & \\ 
$L_{\mathrm{Edd}}/L_{\mathrm{Bol}}$      &       $R_{\mathrm{x}}$ & 50 &       0.194 &       0.0534 &      8.2e-02 & \\ 
\hline
\hline
\end{tabular*}

\medskip
{\bf Notes.} Column 1: Variable $X$.  Column 2: Variable $Y$.  Column 3: Number of objects in the subsample.  Column 4: Generalized Kendall's $\tau$ correlation coefficient as described in Section~\ref{sec:stats}.  Column 5: Uncertainty in the generalized Kendall's $\tau$ correlation coefficient given in Column 4.  Column 6: Probability for accepting the null hypothesis that no correlation exists.

\end{minipage} 
\end{table}

\clearpage
\section{Radio Continuum Maps}
For each ETG included in our new 5~GHz VLA observations, we provide a map of the radio continuum emission with contours in Figure~\ref{fig:radio_images}.  The rms noise level and relative contours for each detected ETG are listed in Table~\ref{tab:contours}.

\begin{table*}
\begin{minipage}{9.75cm}
\caption{Relative Contour Levels in the 5~GHz Continuum Maps}
\label{tab:contours}
\begin{tabular*}{9.75cm}{lcr}
\hline
\hline
Galaxy      & rms & Relative Contours \\
                   & ($\mu$Jy beam$^{-1}$) &  \\
\hline 
IC0560     &     22 &                        [-3, 3, 4.5, 6, 6.75]  \\ 
IC0676     &     16 &                       [-3, 3, 6, 12, 20, 25]  \\ 
NGC0524    &     13 &                         [-3, 3, 15, 55, 115]  \\ 
NGC0661    &     14 &                          [-3, 3, 4, 5, 5.75]  \\ 
NGC0680    &     15 &                      [-3, 3, 10, 20, 32, 37]  \\ 
NGC0936    &     20 &                        [-3, 3, 25, 115, 210]  \\ 
NGC1222    &     20 &                       [-3, 3, 6, 10, 16, 23]  \\ 
NGC1266    &     18 &   [-3, 3, 9, 15, 25, 40, 100, 250, 500, 650]  \\ 
NGC2698    &     19 &                           [-3, 3, 4.5, 5.75]  \\ 
NGC2764    &     14 &                       [-3, 3, 6, 10, 16, 19]  \\ 
NGC2768    &     25 &            [-3, 3, 9, 18, 60, 150, 300, 375]  \\ 
NGC2778    &     14 &                           [-3, 3, 4.5, 6, 7]  \\ 
NGC2824    &     14 &                    [-3, 3, 12, 40, 100, 150]  \\ 
NGC2852    &     16 &                           [-3, 3, 9, 22, 32]  \\ 
NGC2880    &     15 &                          [-3, 3, 4, 5, 5.75]  \\ 
NGC2962    &     20 &                            [-3, 3, 5, 8, 10]  \\ 
NGC2974    &     25 &                    [-3, 3, 24, 75, 185, 280]  \\ 
NGC3301    &     18 &                       [-3, 3, 6, 15, 27, 36]  \\ 
NGC3377    &     15 &                           [-3, 3, 6, 10, 12]  \\ 
NGC3379    &     15 &                       [-3, 3, 9, 18, 36, 46]  \\ 
NGC3607    &     16 &                     [-3, 3, 12, 40, 85, 115]  \\ 
NGC3608    &     14 &                       [-3, 3, 6, 10, 15, 18]  \\ 
NGC3619    &     15 &                         [-3, 3, 8, 15, 19.5]  \\ 
NGC3626    &     17 &                           [-3, 3, 9, 18, 26]  \\ 
NGC3648    &     16 &                             [-3, 3, 4, 5, 6]  \\ 
NGC3665    &     20 &             [-3, 3, 8, 20, 32, 85, 200, 350]  \\ 
NGC3941    &     12 &                           [-3, 3, 4.5, 6, 7]  \\ 
NGC3945    &     15 &                    [-3, 3, 12, 36, 100, 175]  \\ 
NGC4036    &     17 &                      [-3, 3, 12, 30, 60, 95]  \\ 
NGC4111    &     25 &                         [-3, 3, 4.5, 6.6, 8]  \\ 
NGC4429    &     16 &                           [-3, 3, 8, 16, 24]  \\ 
NGC4435    &     20 &                           [-3, 3, 4.5, 6, 7]  \\ 
NGC4459    &     15 &                           [-3, 3, 6, 14, 19]  \\ 
NGC4494    &     16 &                           [-3, 3, 7, 14, 18]  \\ 
NGC4526    &     18 &                       [-3, 3, 9, 25, 55, 78]  \\ 
NGC4546    &     25 &                    [-3, 3, 18, 75, 200, 300]  \\ 
NGC4643    &     18 &                                [-3, 3, 5, 6]  \\ 
NGC4684    &     18 &                       [-3, 3, 6, 15, 25, 31]  \\ 
NGC4710    &     26 &                           [-3, 3, 6, 10, 13]  \\ 
NGC5173    &     17 &                       [-3, 3, 9, 20, 40, 53]  \\ 
NGC5198    &     17 &                     [-3, 3, 12, 40, 86, 114]  \\ 
NGC5273    &     17 &                       [-3, 3, 8, 16, 27, 35]  \\ 
NGC5379    &     16 &                         [-3, 3, 4.5, 6, 7.5]  \\ 
NGC5475    &     15 &                   [-3, 3, 9, 16, 23, 55, 80]  \\ 
NGC5485    &     17 &                       [-3, 3, 9, 20, 38, 49]  \\ 
NGC5631    &     17 &                           [-3, 3, 5, 8, 9.5]  \\ 
NGC6014    &     15 &                          [-3, 3, 4, 5, 5.75]  \\ 
NGC6703    &     15 &                       [-3, 3, 8, 20, 35, 48]  \\ 
NGC6798    &     15 &                           [-3, 3, 4.5, 6, 7]  \\ 
NGC7465    &     17 &                       [-3, 3, 6, 15, 35, 60]  \\ 
PGC029321  &     13 &                    [-3, 3, 12, 50, 175, 300]  \\ 
UGC05408   &     17 &                           [-3, 3, 5, 7, 8.5]  \\ 
UGC06176   &     14 &                        [-3, 3, 6, 9, 12, 14]  \\ 
\hline
\hline
\end{tabular*}

\end{minipage} 
\end{table*}


\begin{figure*}
{\label{fig:sub:IC0560}\includegraphics[clip=True, trim=0cm 0cm 0cm 0cm, scale=0.27]{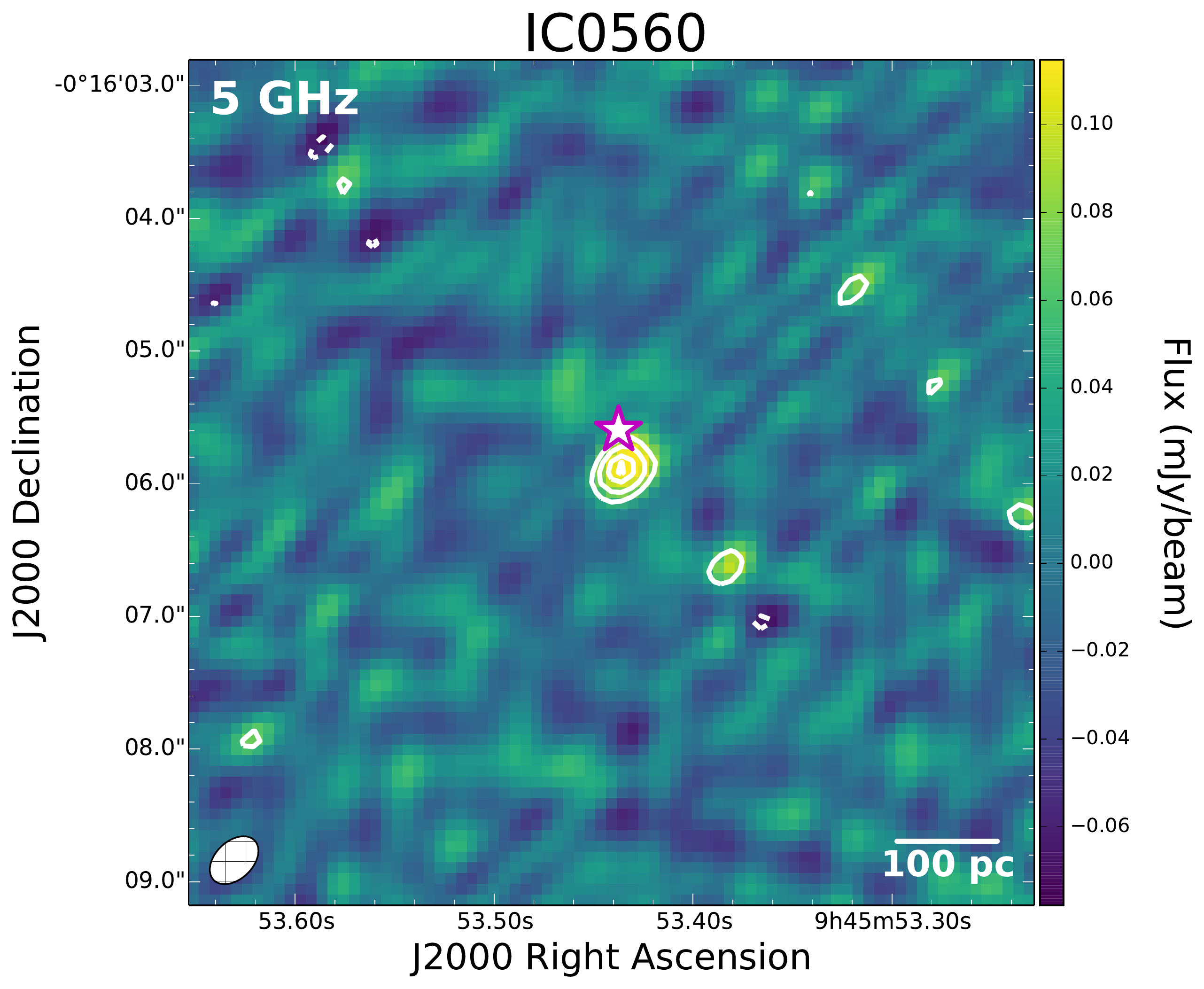}}
{\label{fig:sub:IC0676}\includegraphics[clip=True, trim=0cm 0cm 0cm 0cm, scale=0.27]{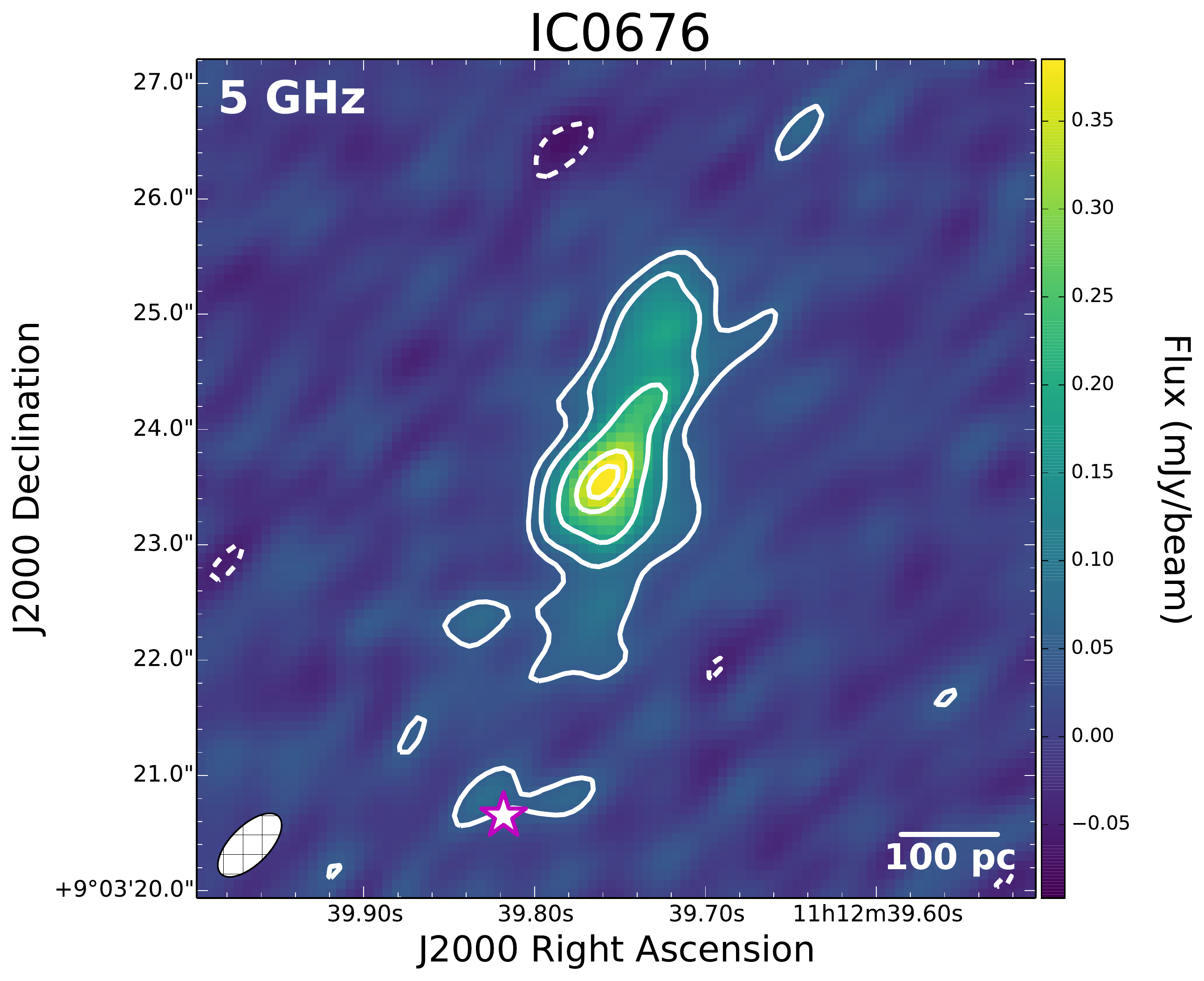}}
{\label{fig:sub:NGC0524}\includegraphics[clip=True, trim=0cm 0cm 0cm 0cm, scale=0.27]{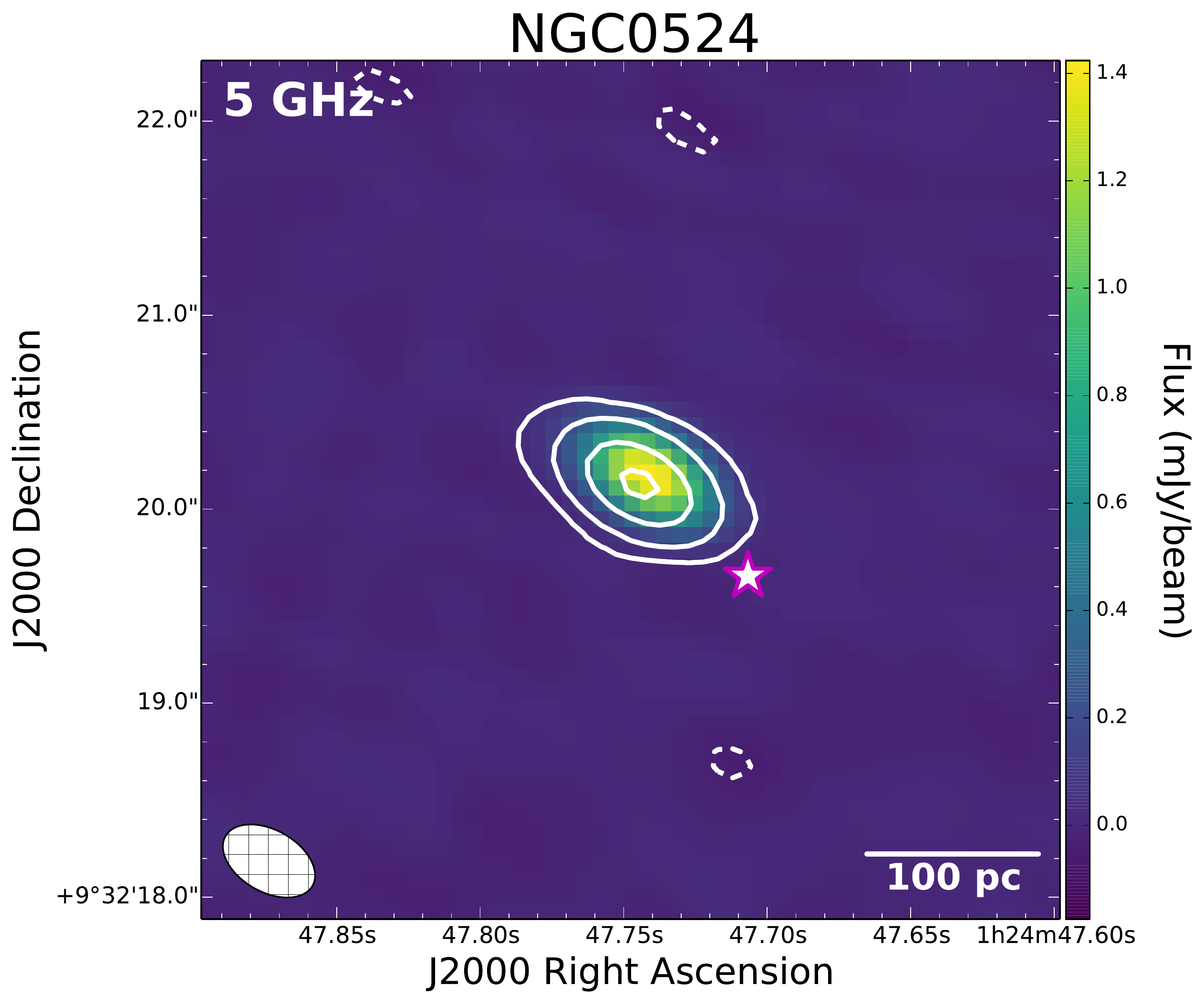}}
{\label{fig:sub:NGC0661}\includegraphics[clip=True, trim=0cm 0cm 0cm 0cm, scale=0.27]{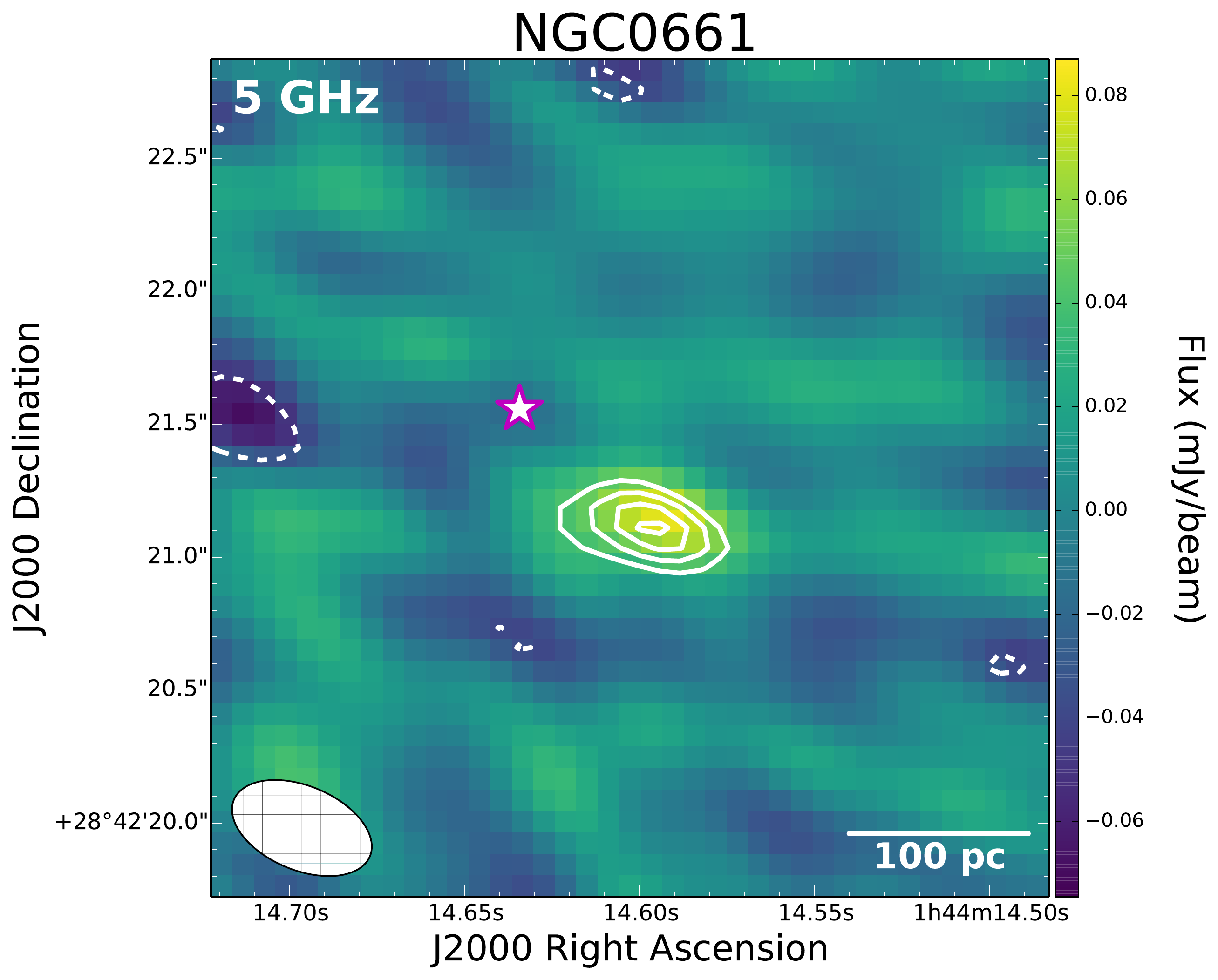}}
{\label{fig:sub:NGC0680}\includegraphics[clip=True, trim=0cm 0cm 0cm 0cm, scale=0.27]{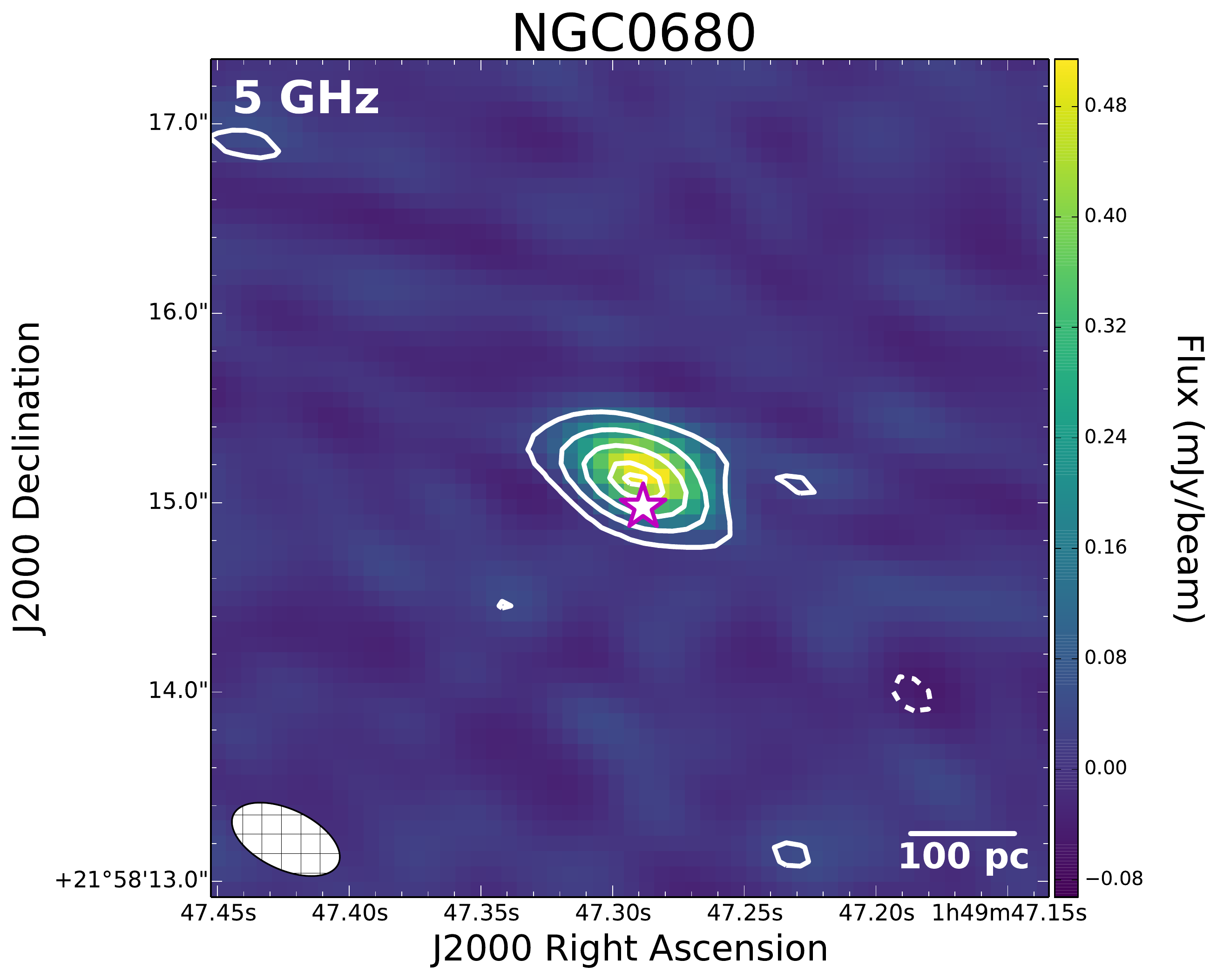}}
{\label{fig:sub:NGC0936}\includegraphics[clip=True, trim=0cm 0cm 0cm 0cm, scale=0.27]{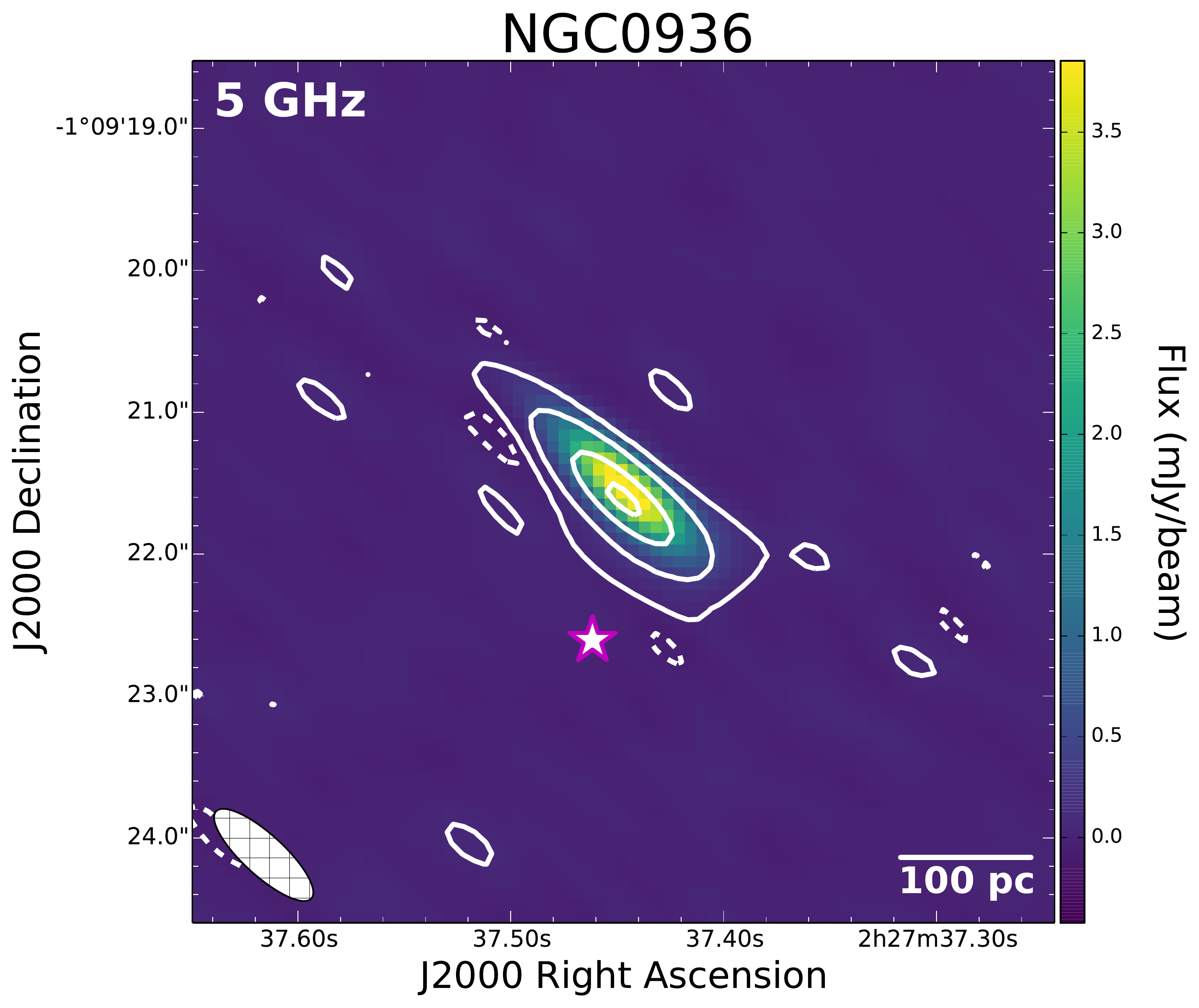}}
\end{figure*}
 
\begin{figure*}
{\label{fig:sub:NGC1222}\includegraphics[clip=True, trim=0cm 0cm 0cm 0cm, scale=0.27]{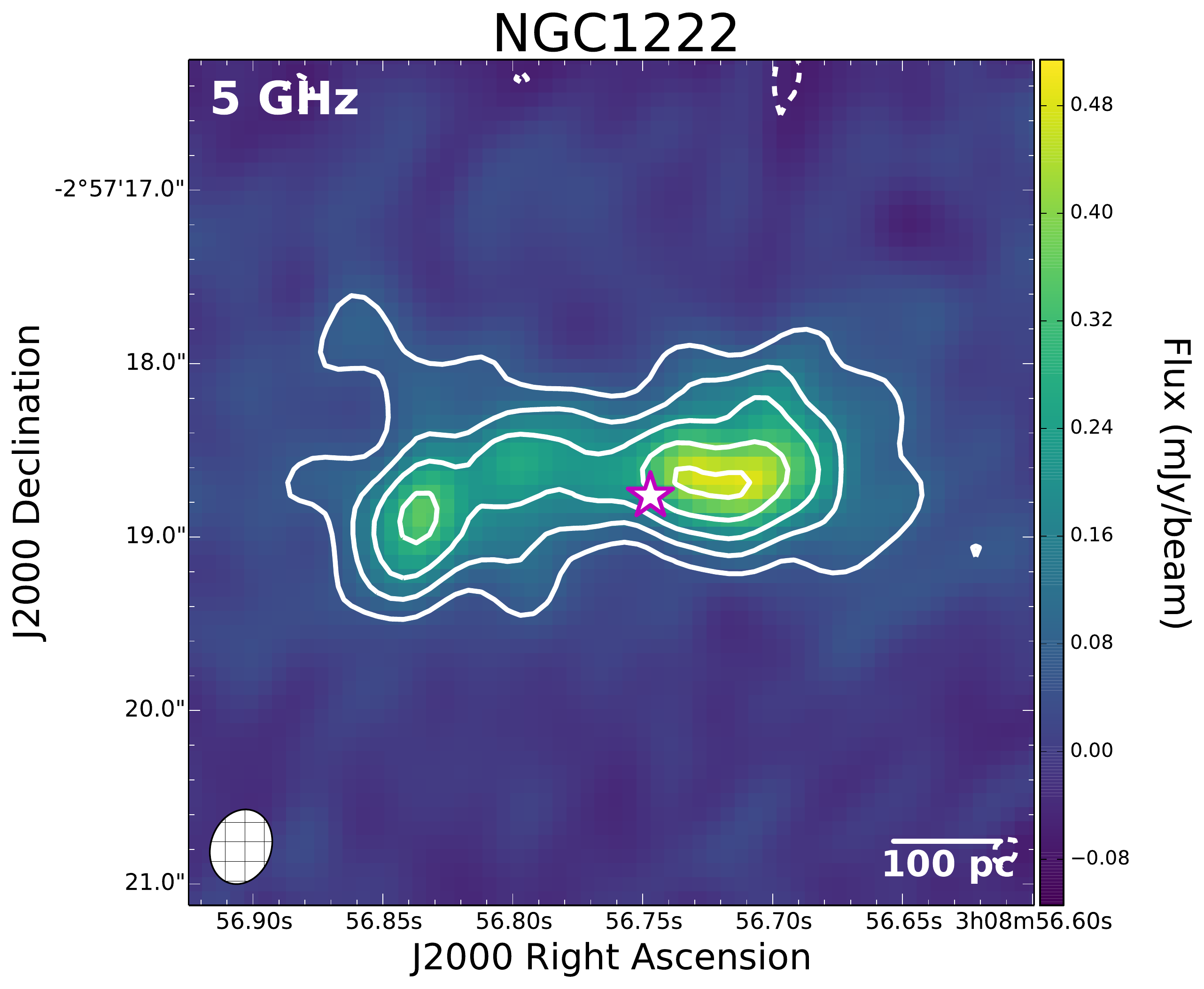}}
{\label{fig:sub:NGC1266}\includegraphics[clip=True, trim=0cm 0cm 0cm 0cm, scale=0.27]{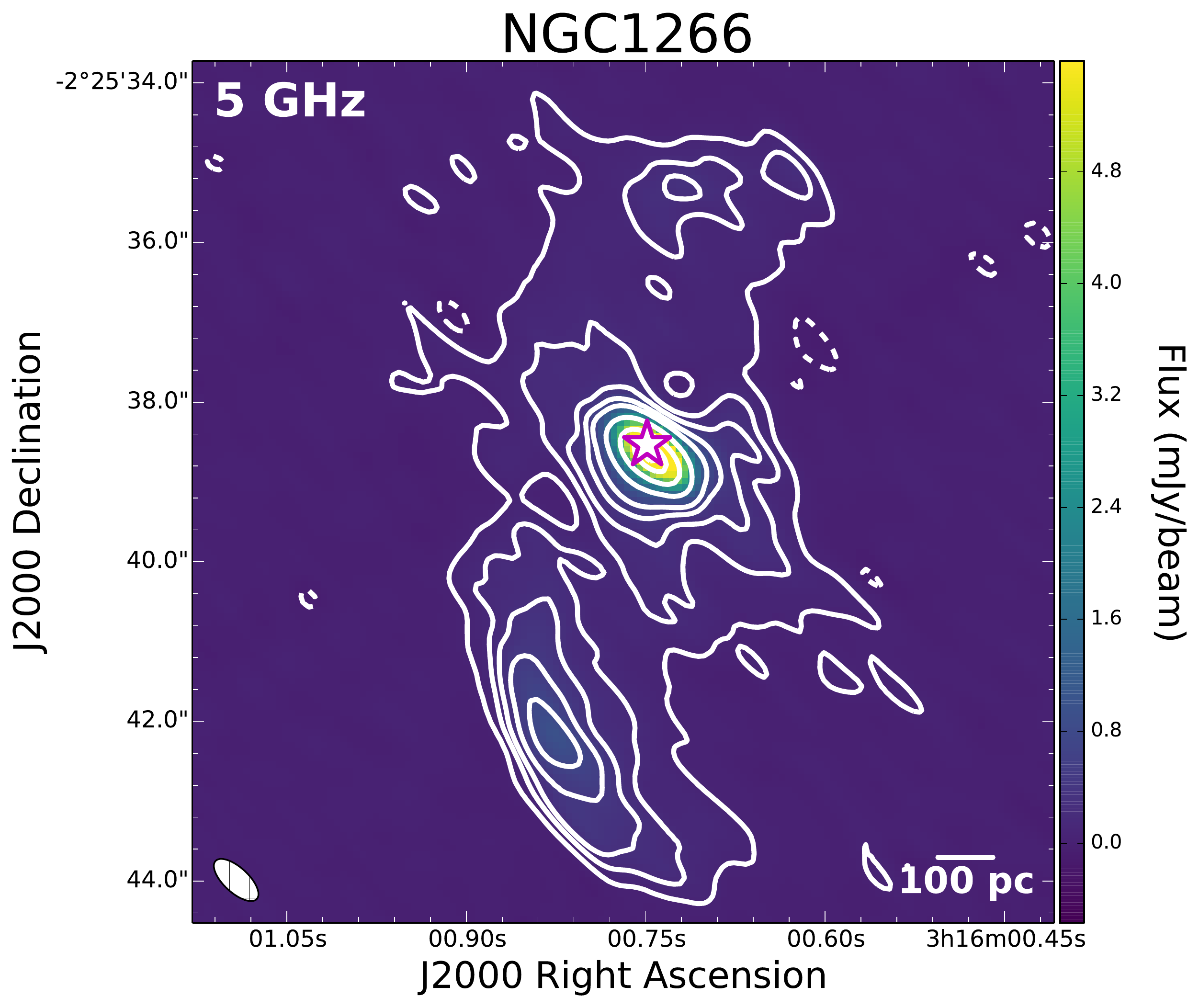}}
{\label{fig:sub:NGC2698}\includegraphics[clip=True, trim=0cm 0cm 0cm 0cm, scale=0.27]{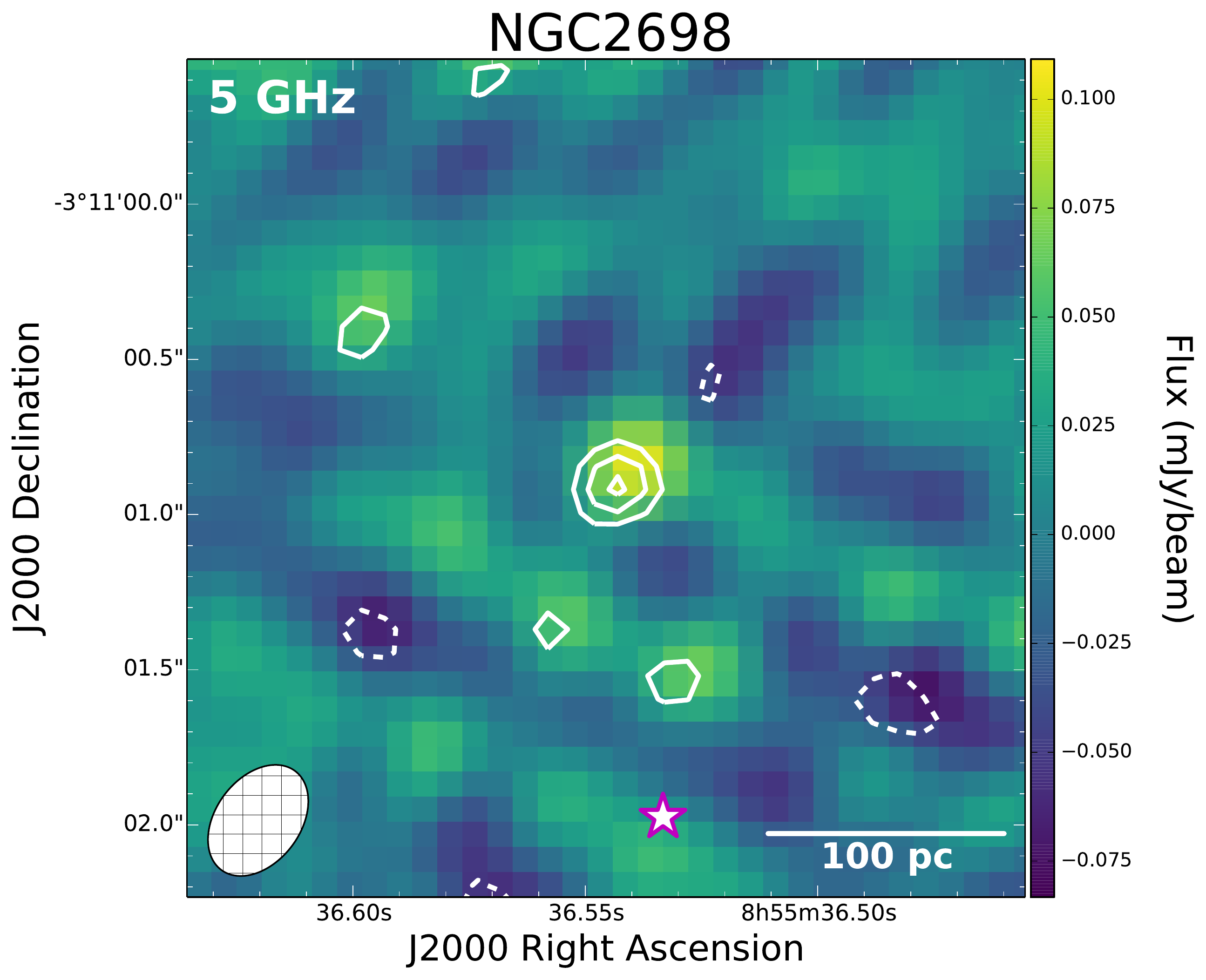}}
{\label{fig:sub:NGC2764}\includegraphics[clip=True, trim=0cm 0cm 0cm 0cm, scale=0.27]{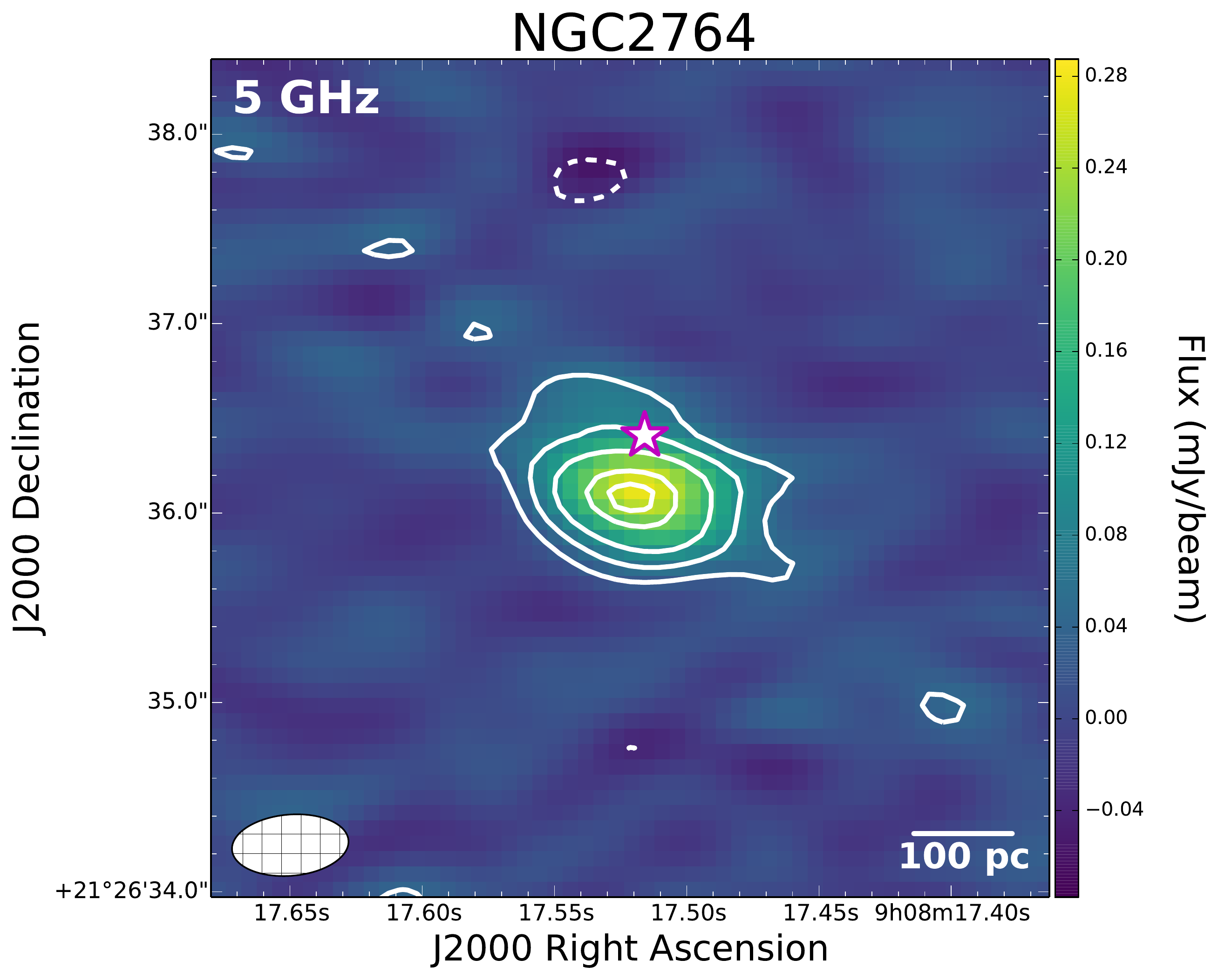}}
{\label{fig:sub:NGC2768}\includegraphics[clip=True, trim=0cm 0cm 0cm 0cm, scale=0.27]{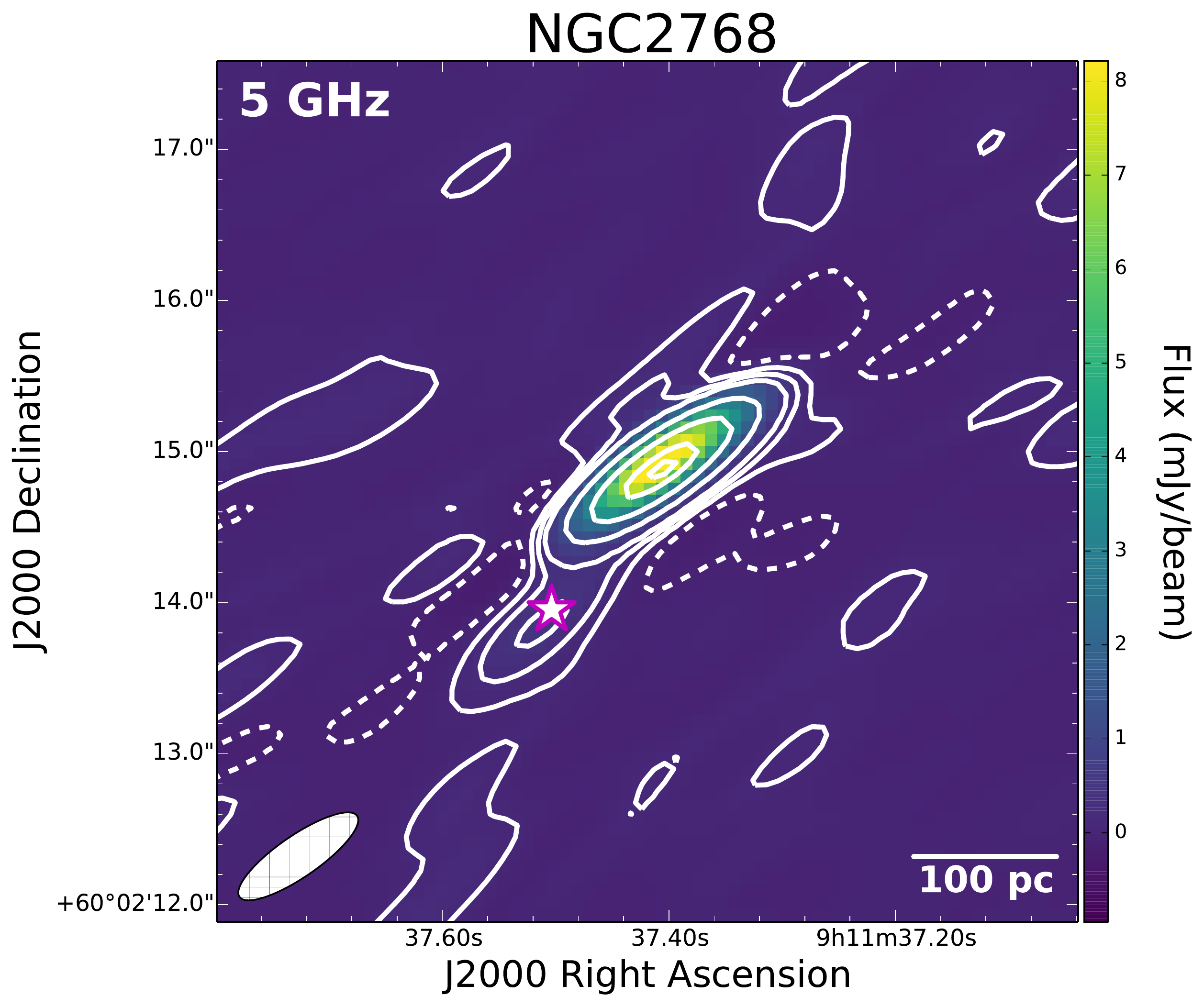}}
{\label{fig:sub:NGC2778}\includegraphics[clip=True, trim=0cm 0cm 0cm 0cm, scale=0.27]{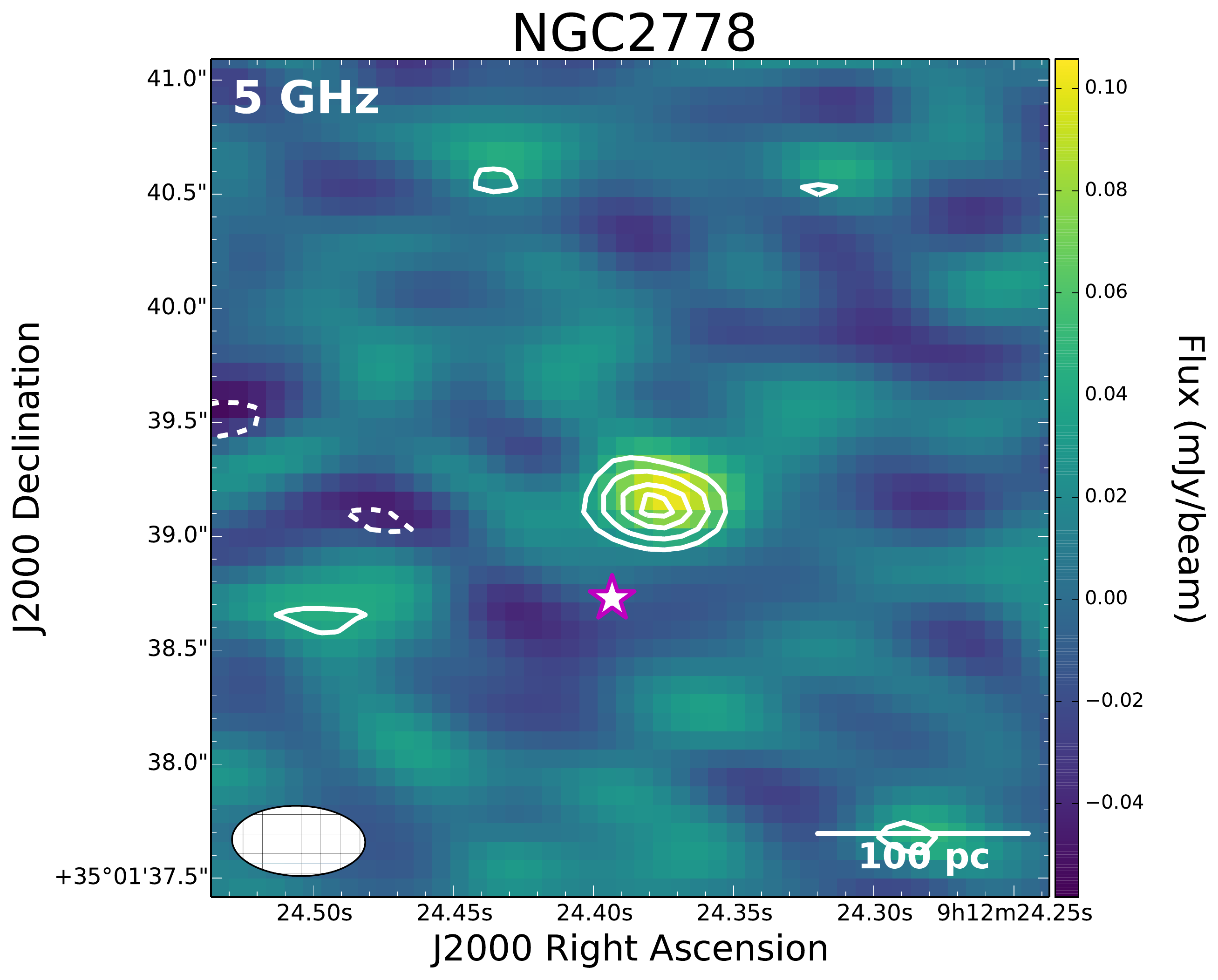}}
\end{figure*}
 
\begin{figure*}
{\label{fig:sub:NGC2824}\includegraphics[clip=True, trim=0cm 0cm 0cm 0cm, scale=0.27]{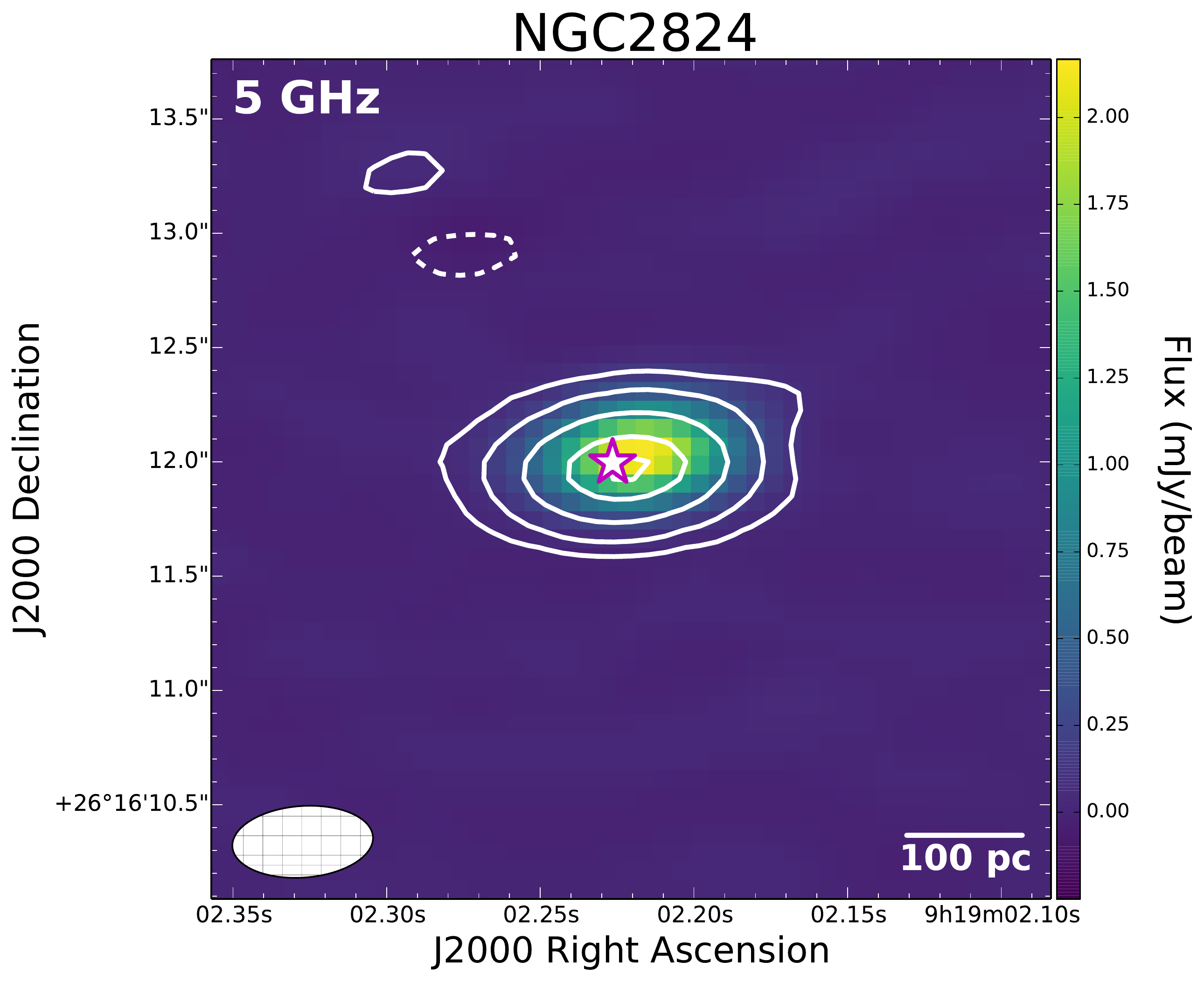}}
{\label{fig:sub:NGC2852}\includegraphics[clip=True, trim=0cm 0cm 0cm 0cm, scale=0.27]{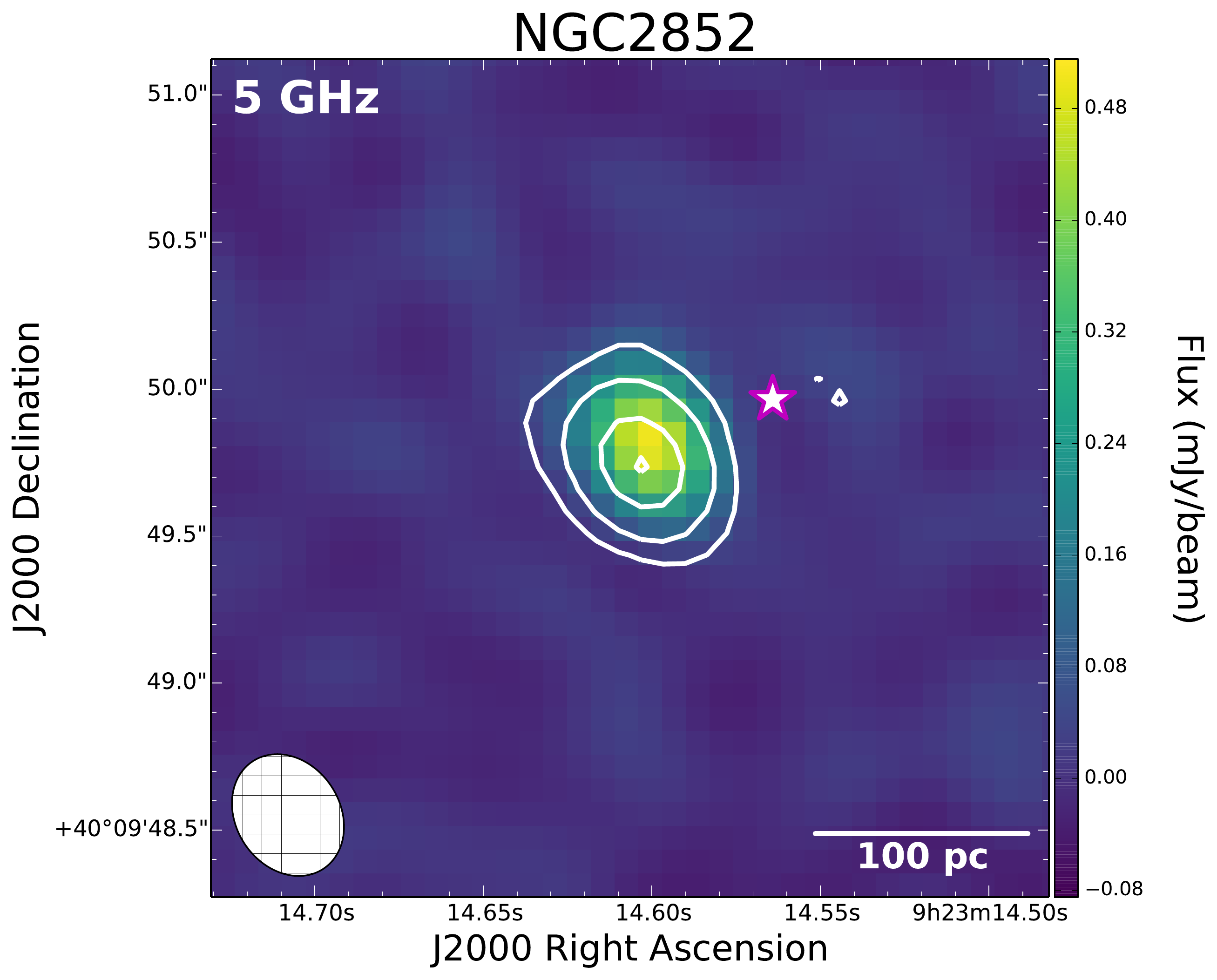}}
{\label{fig:sub:NGC2880}\includegraphics[clip=True, trim=0cm 0cm 0cm 0cm, scale=0.27]{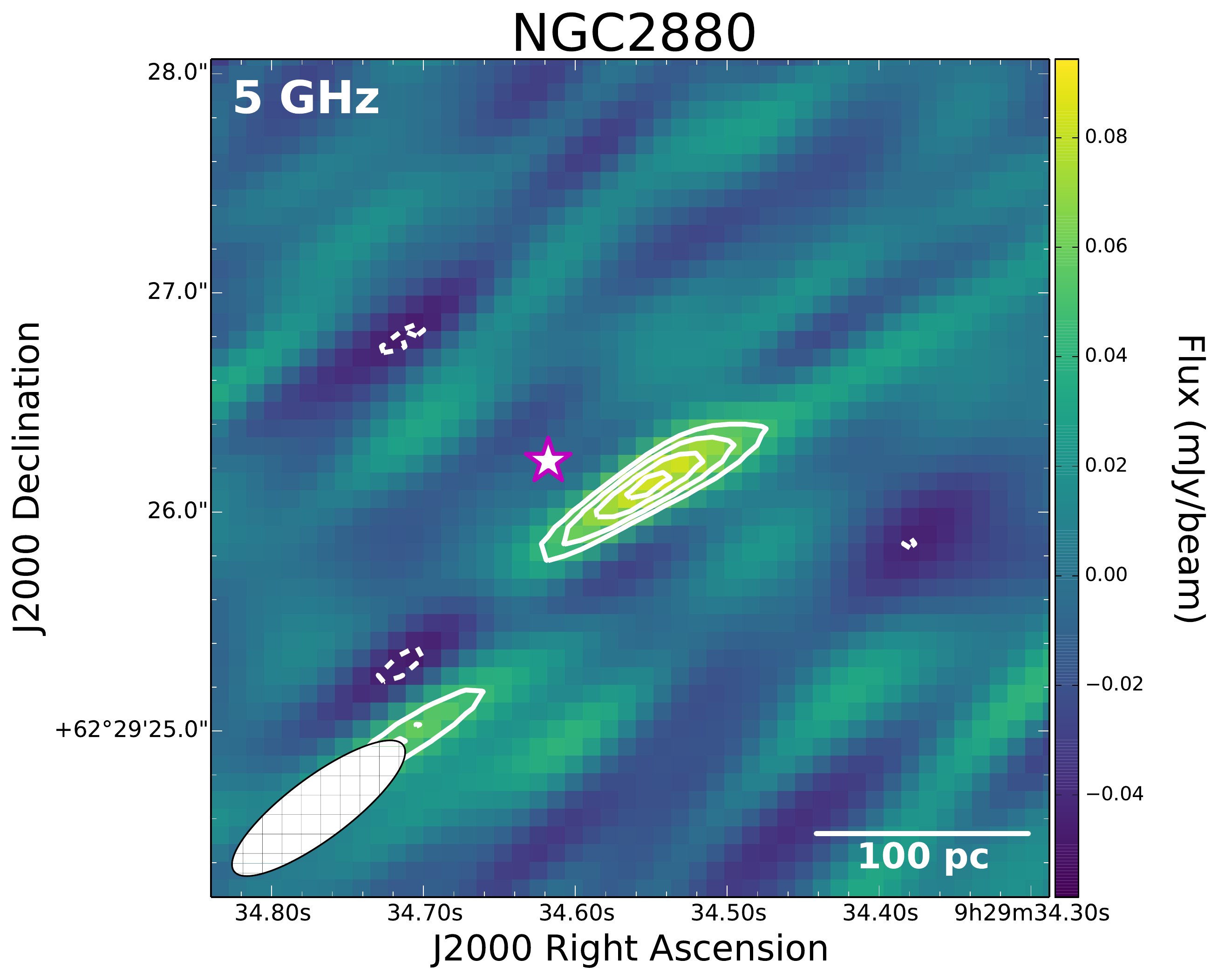}}
{\label{fig:sub:NGC2962}\includegraphics[clip=True, trim=0cm 0cm 0cm 0cm, scale=0.27]{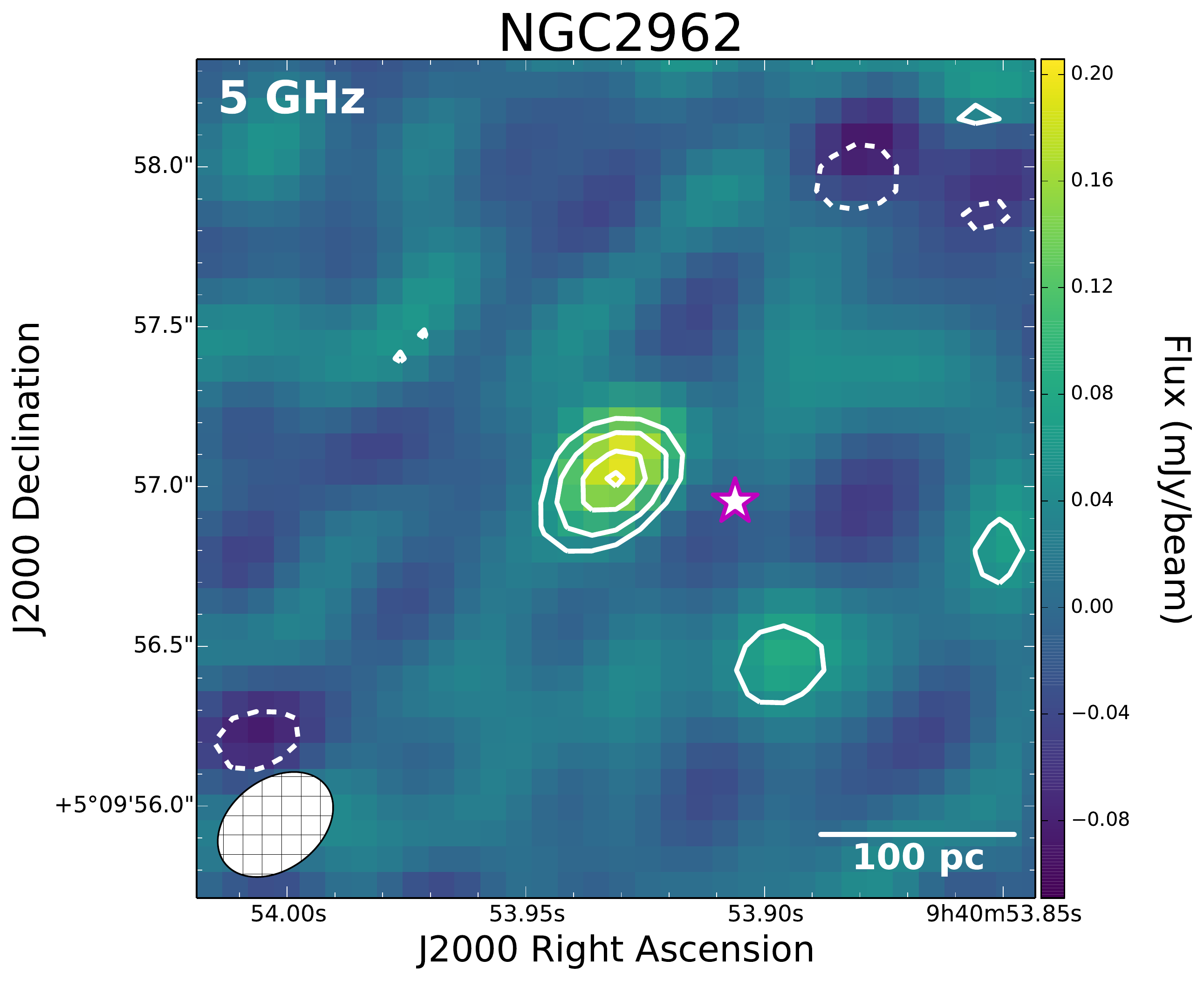}}
{\label{fig:sub:NGC2974}\includegraphics[clip=True, trim=0cm 0cm 0cm 0cm, scale=0.27]{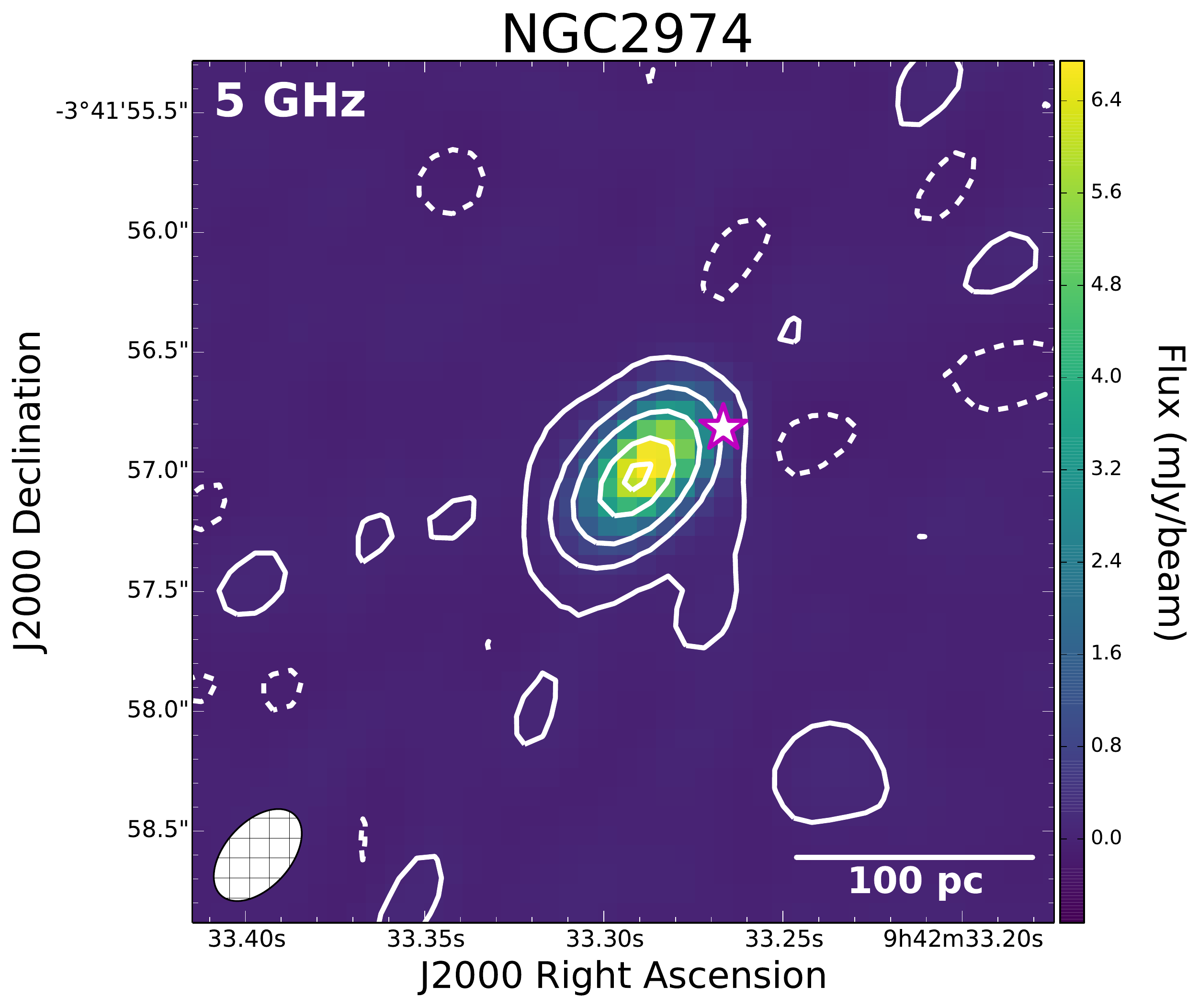}}
{\label{fig:sub:NGC3301}\includegraphics[clip=True, trim=0cm 0cm 0cm 0cm, scale=0.27]{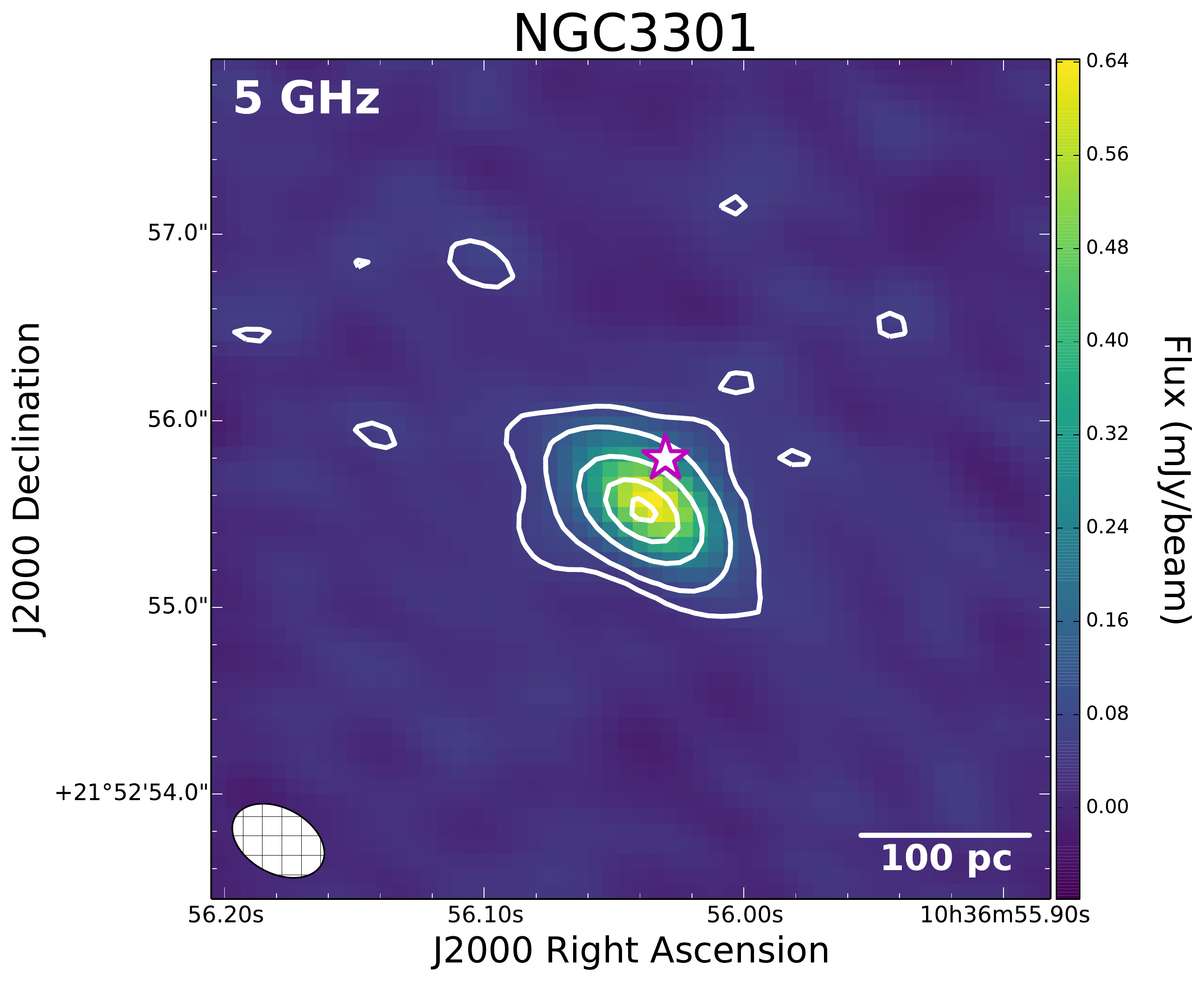}}
\end{figure*}

\begin{figure*}
{\label{fig:sub:NGC3377}\includegraphics[clip=True, trim=0cm 0cm 0cm 0cm, scale=0.27]{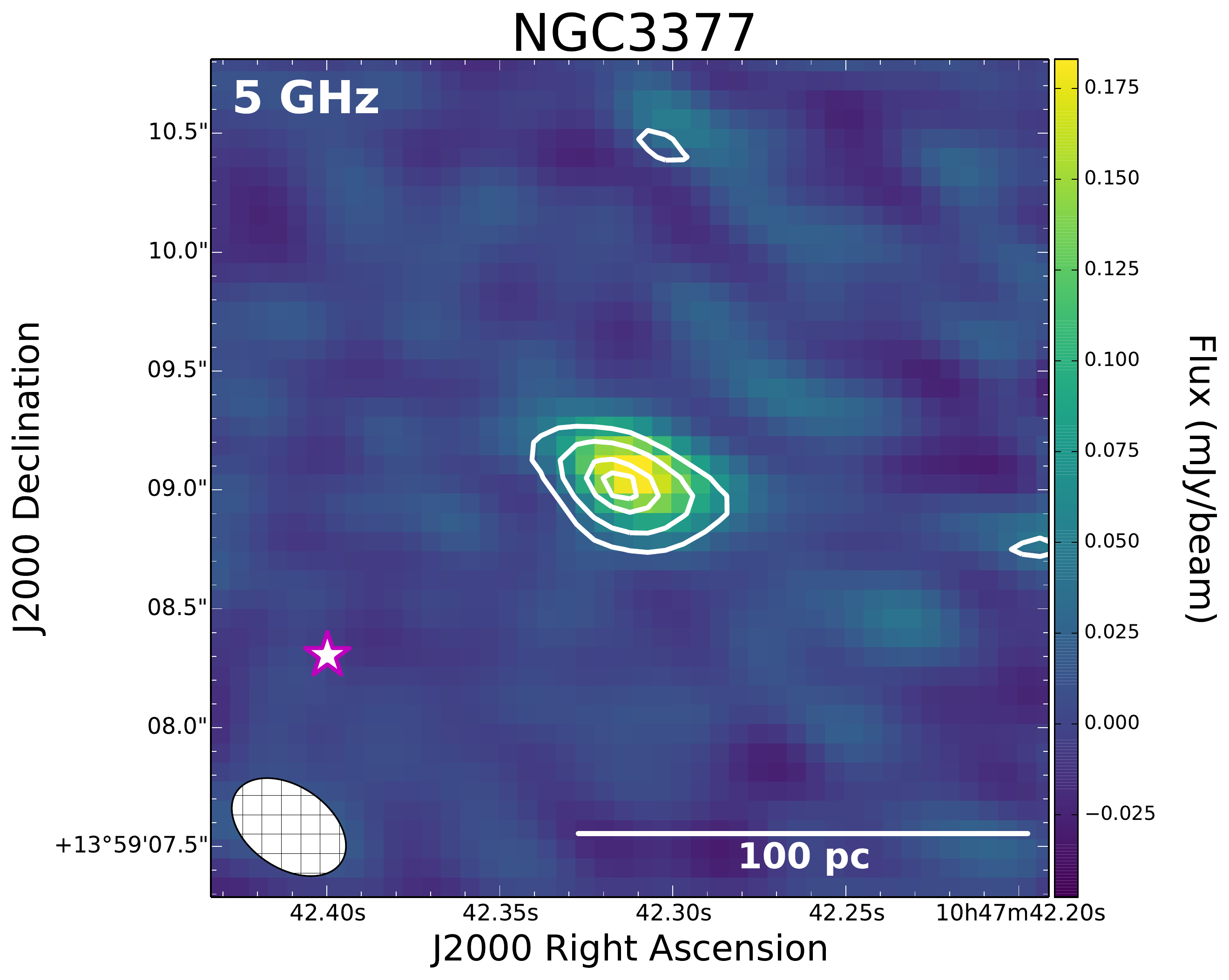}}
{\label{fig:sub:NGC3379}\includegraphics[clip=True, trim=0cm 0cm 0cm 0cm, scale=0.27]{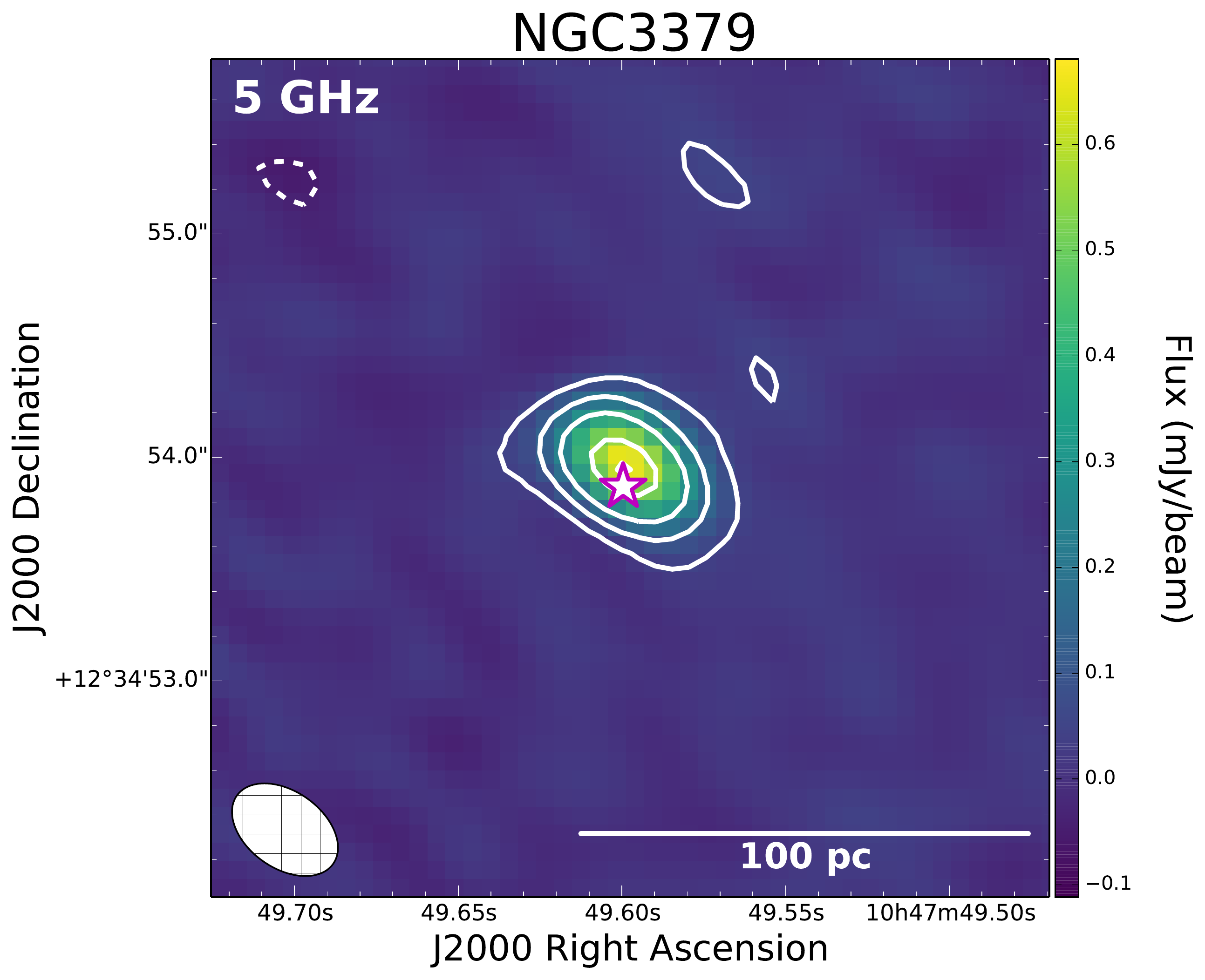}}
{\label{fig:sub:NGC3607}\includegraphics[clip=True, trim=0cm 0cm 0cm 0cm, scale=0.27]{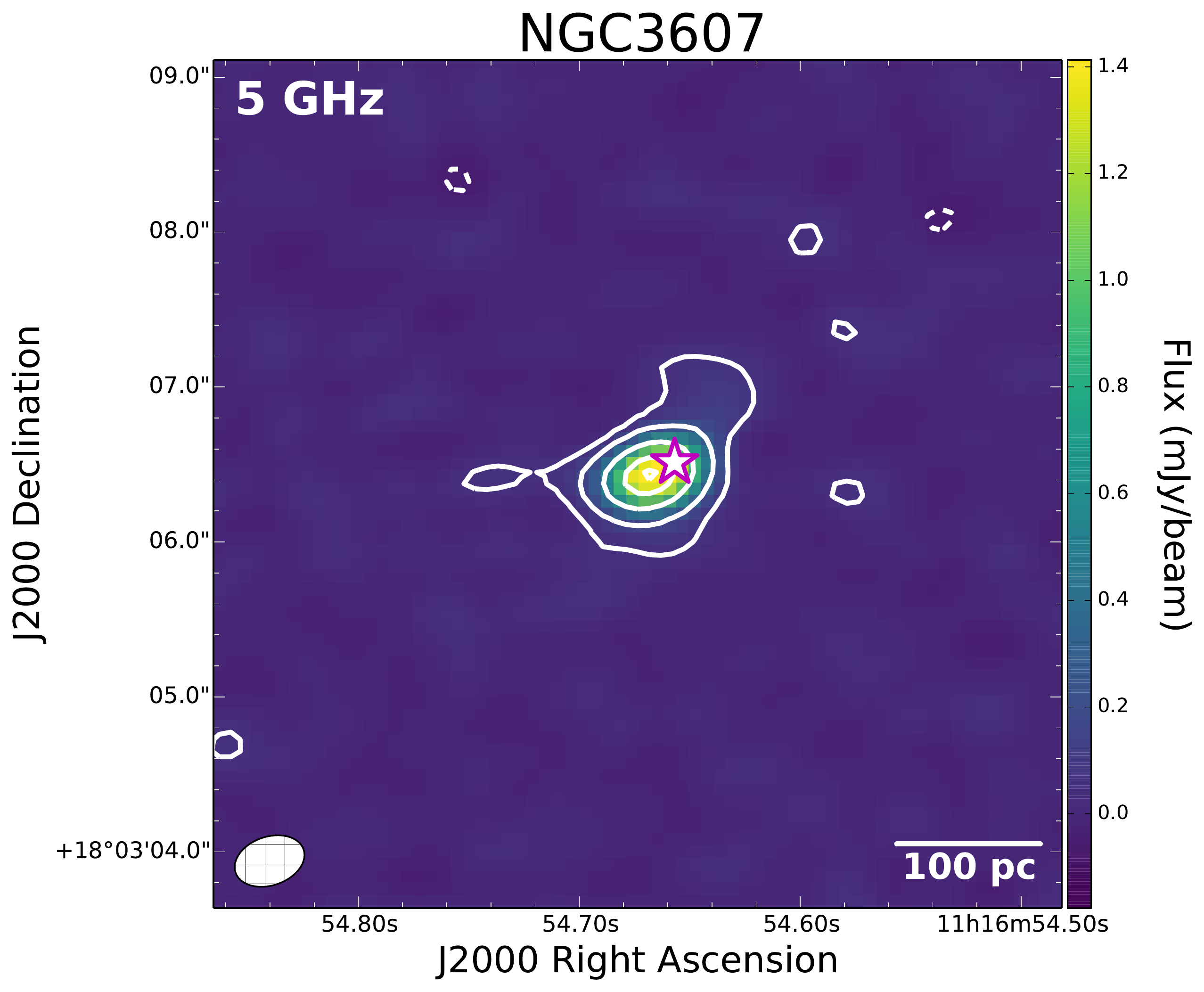}}
{\label{fig:sub:NGC3608}\includegraphics[clip=True, trim=0cm 0cm 0cm 0cm, scale=0.27]{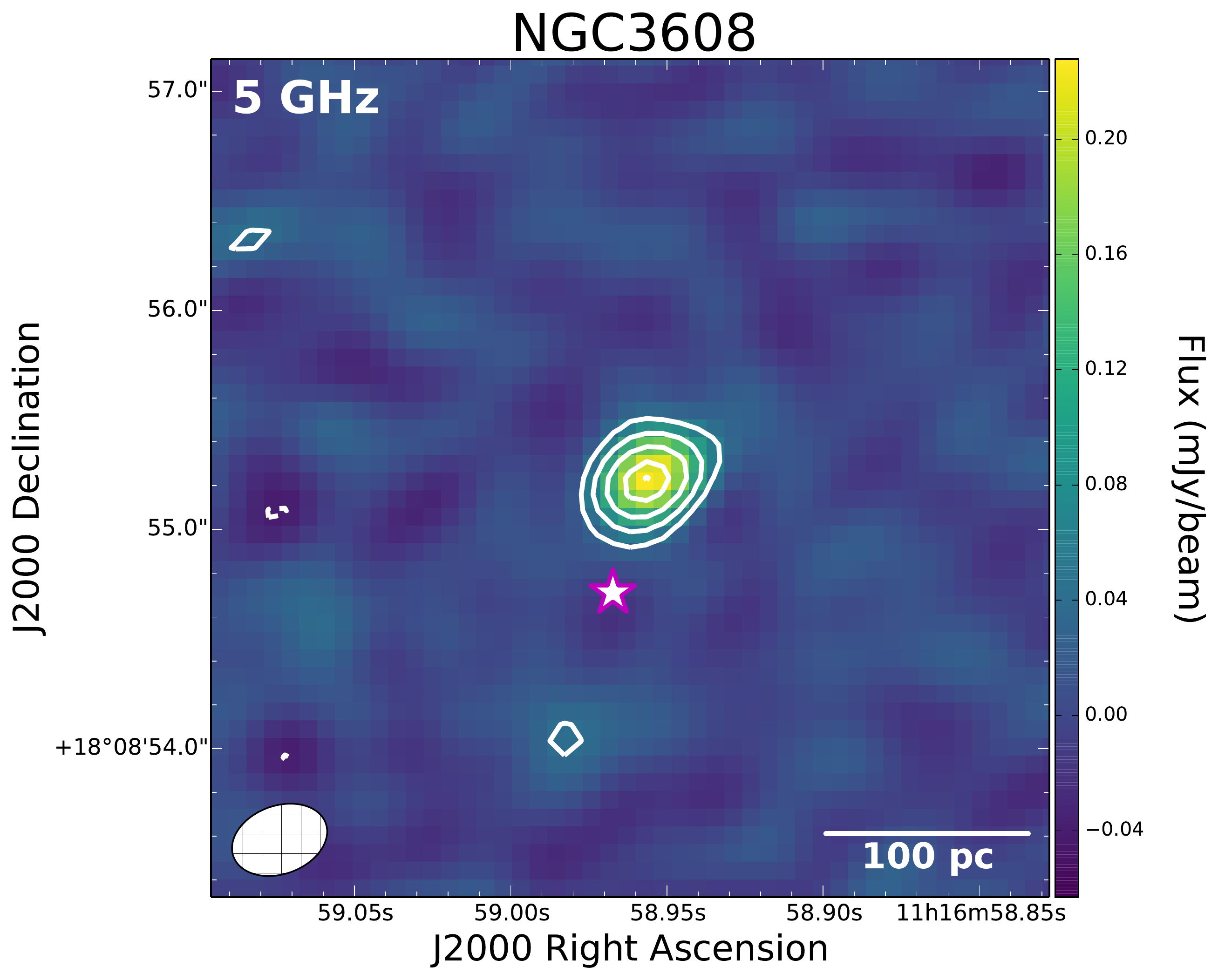}}
{\label{fig:sub:NGC3619}\includegraphics[clip=True, trim=0cm 0cm 0cm 0cm, scale=0.27]{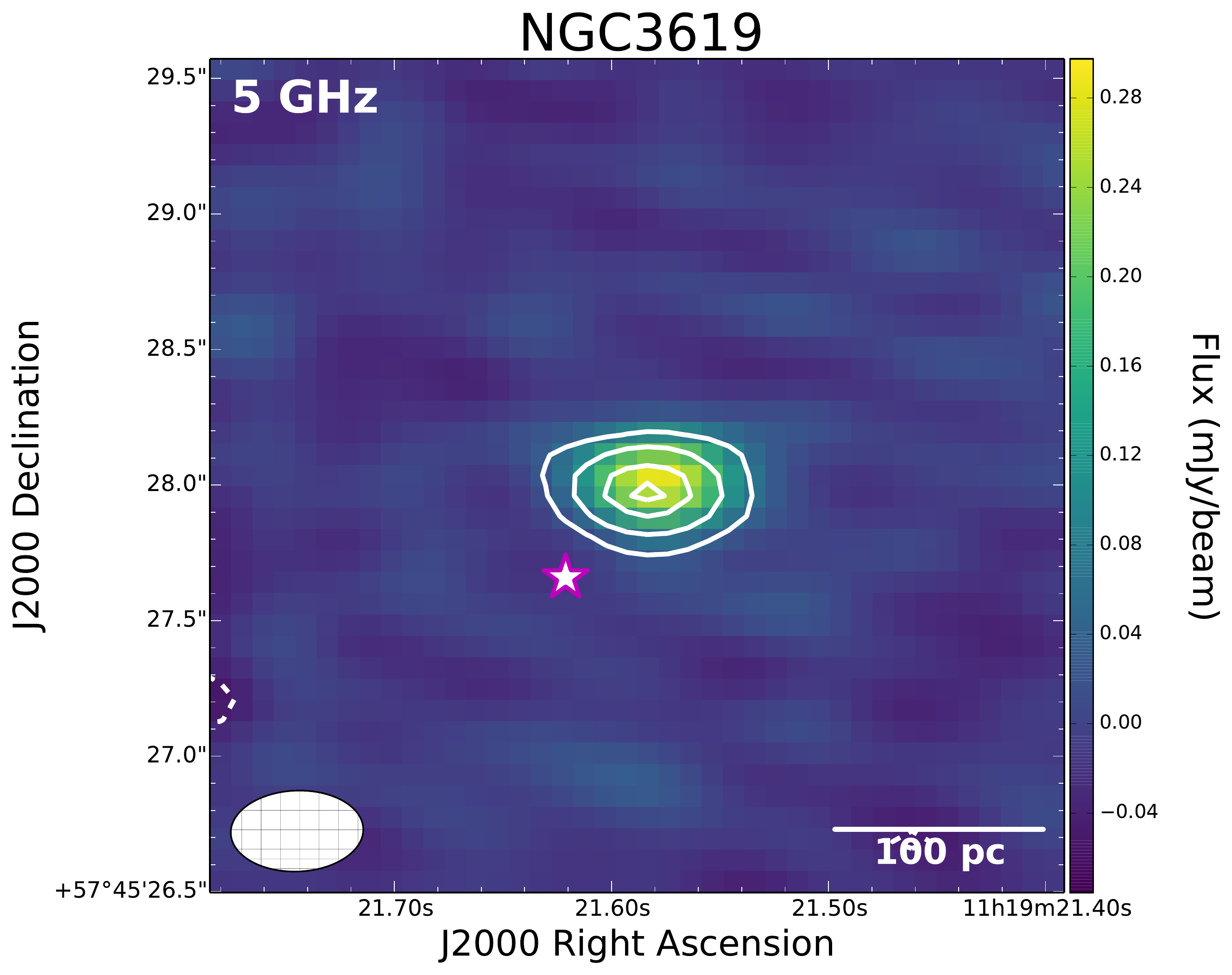}}
{\label{fig:sub:NGC3626}\includegraphics[clip=True, trim=0cm 0cm 0cm 0cm, scale=0.27]{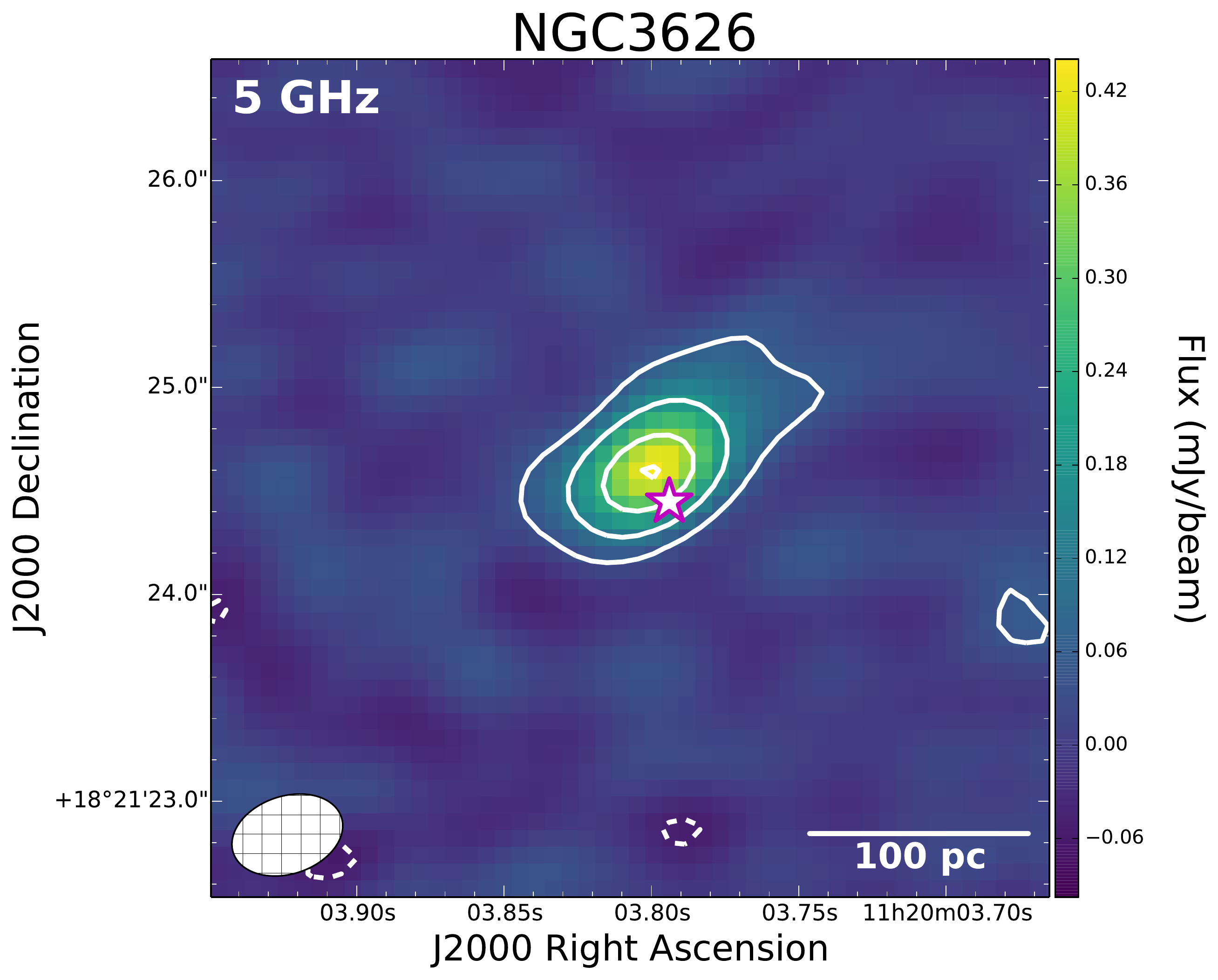}}
\end{figure*}

\begin{figure*}
{\label{fig:sub:NGC3648}\includegraphics[clip=True, trim=0cm 0cm 0cm 0cm, scale=0.27]{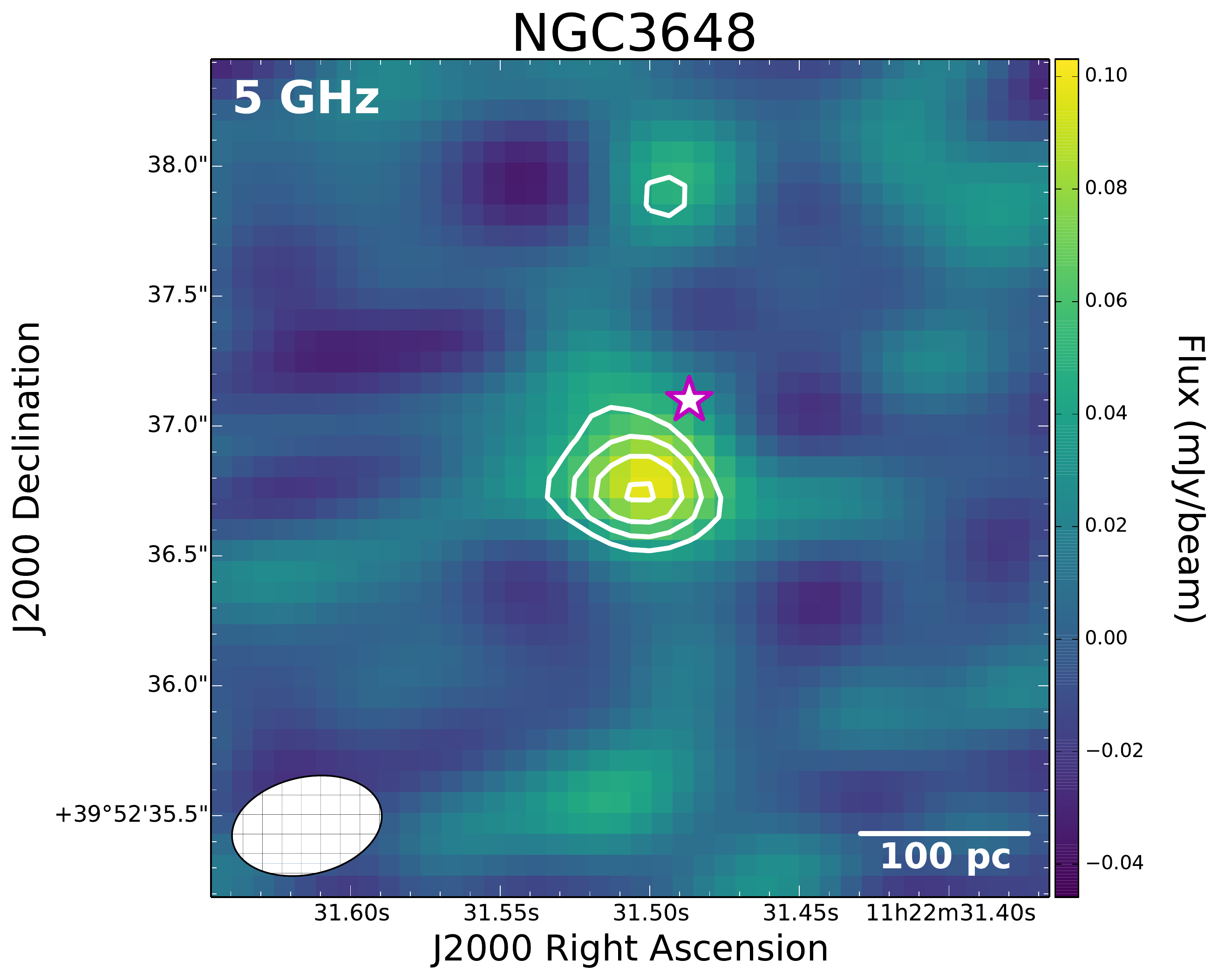}}
{\label{fig:sub:NGC3665}\includegraphics[clip=True, trim=0cm 0cm 0cm 0cm, scale=0.27]{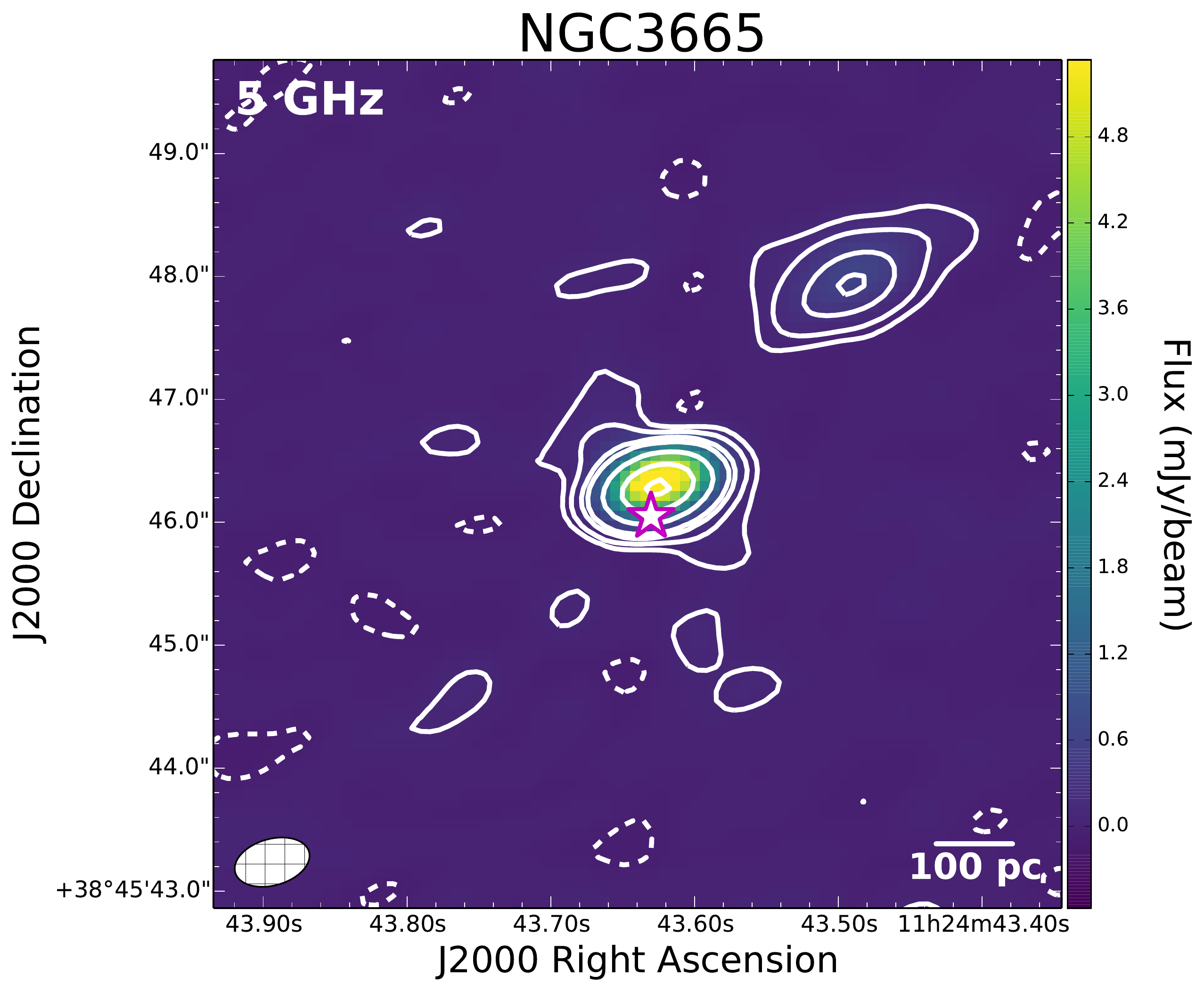}}
{\label{fig:sub:NGC3941}\includegraphics[clip=True, trim=0cm 0cm 0cm 0cm, scale=0.27]{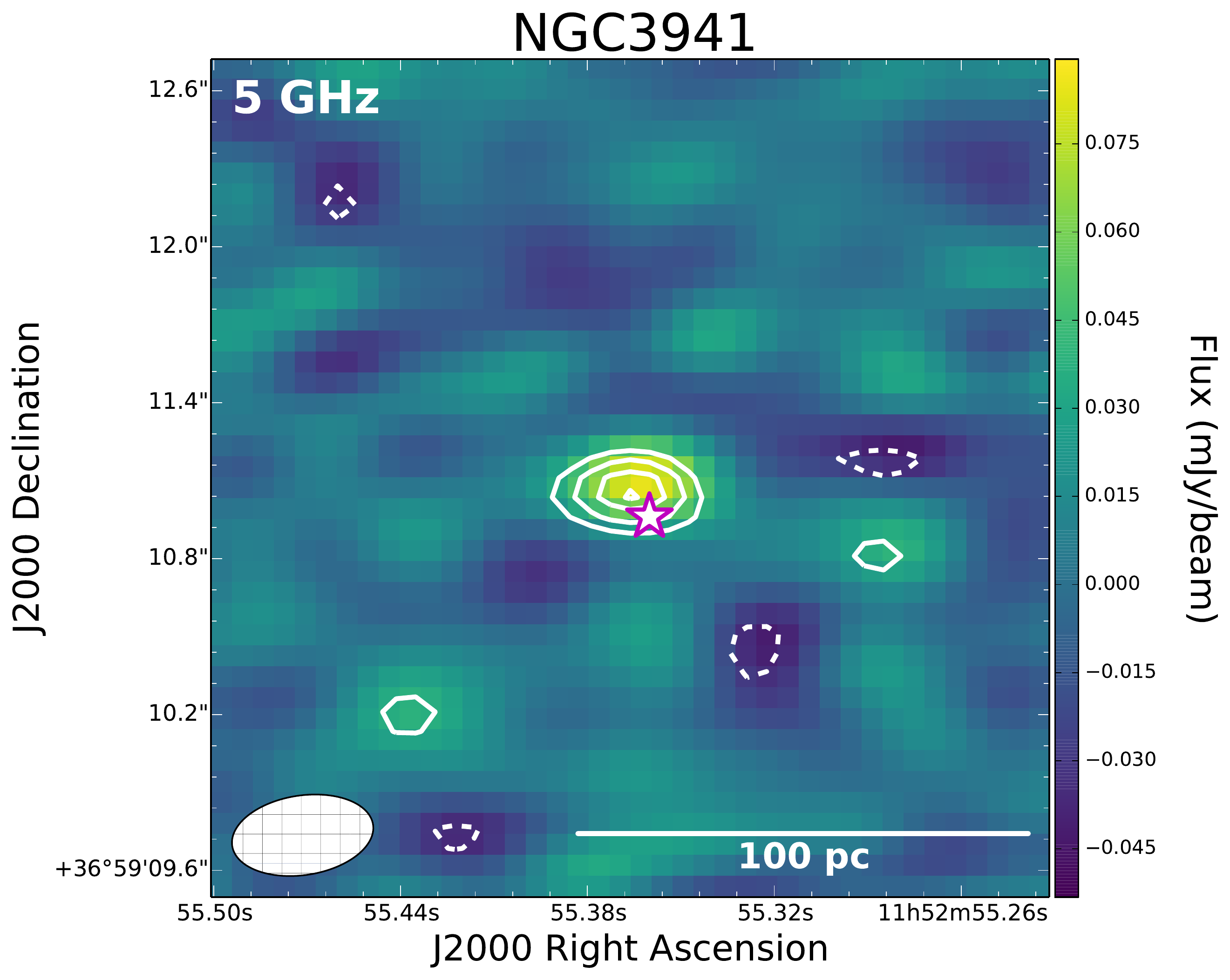}}
{\label{fig:sub:NGC3945}\includegraphics[clip=True, trim=0cm 0cm 0cm 0cm, scale=0.27]{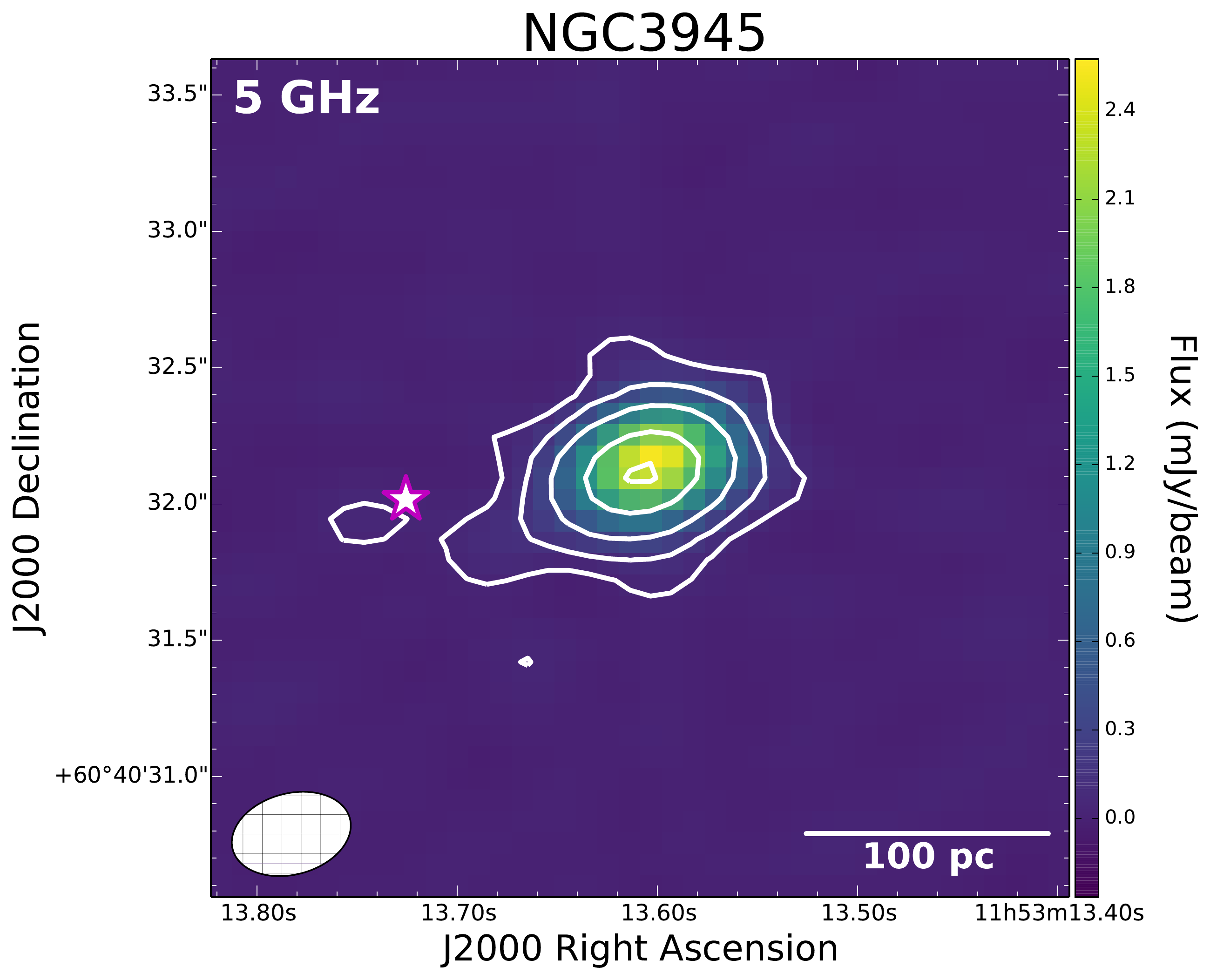}}
{\label{fig:sub:NGC4036}\includegraphics[clip=True, trim=0cm 0cm 0cm 0cm, scale=0.27]{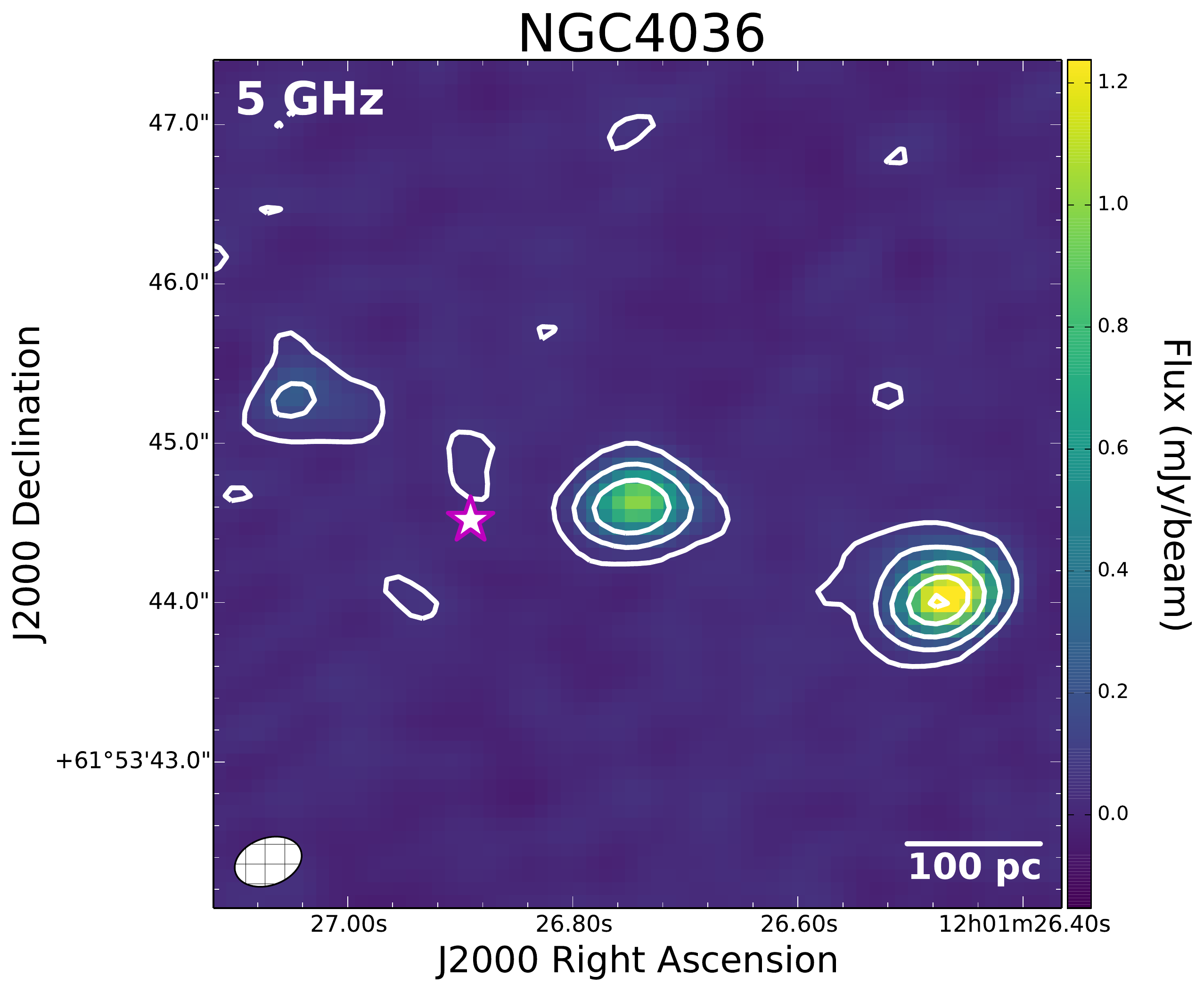}}
{\label{fig:sub:NGC4111}\includegraphics[clip=True, trim=0cm 0cm 0cm 0cm, scale=0.27]{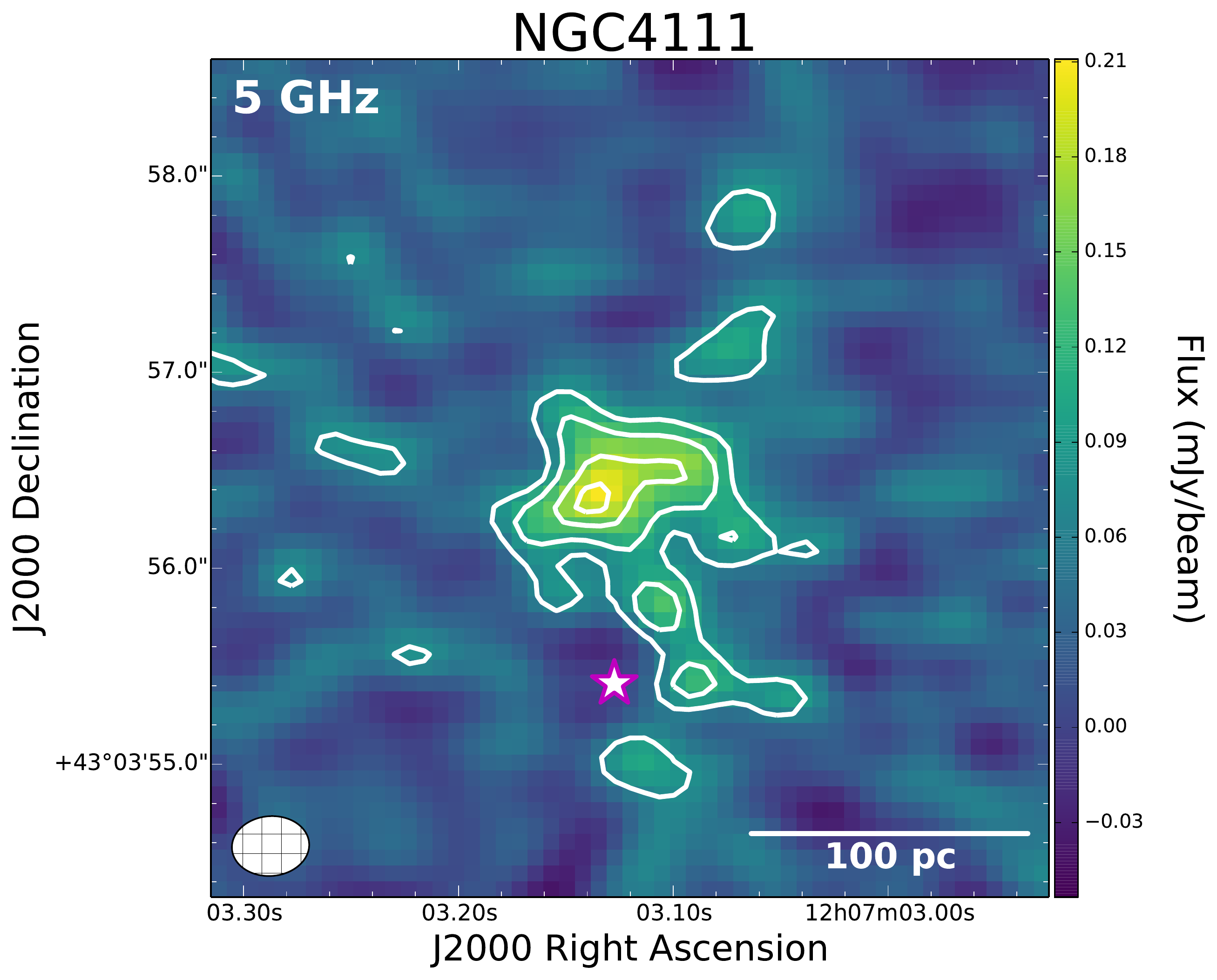}}
\end{figure*}
 
\begin{figure*}
{\label{fig:sub:NGC4429}\includegraphics[clip=True, trim=0cm 0cm 0cm 0cm, scale=0.27]{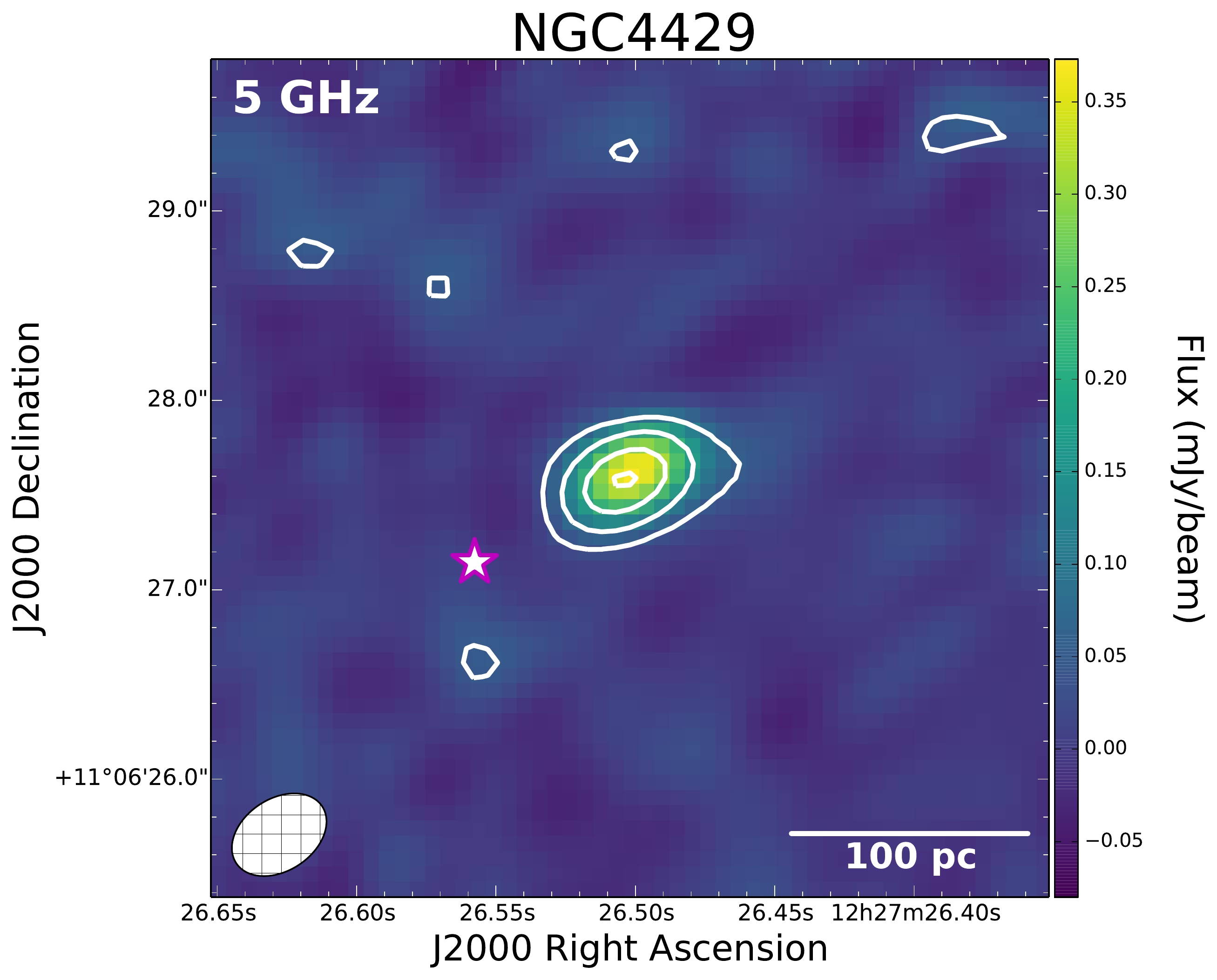}}
{\label{fig:sub:NGC4435}\includegraphics[clip=True, trim=0cm 0cm 0cm 0cm, scale=0.27]{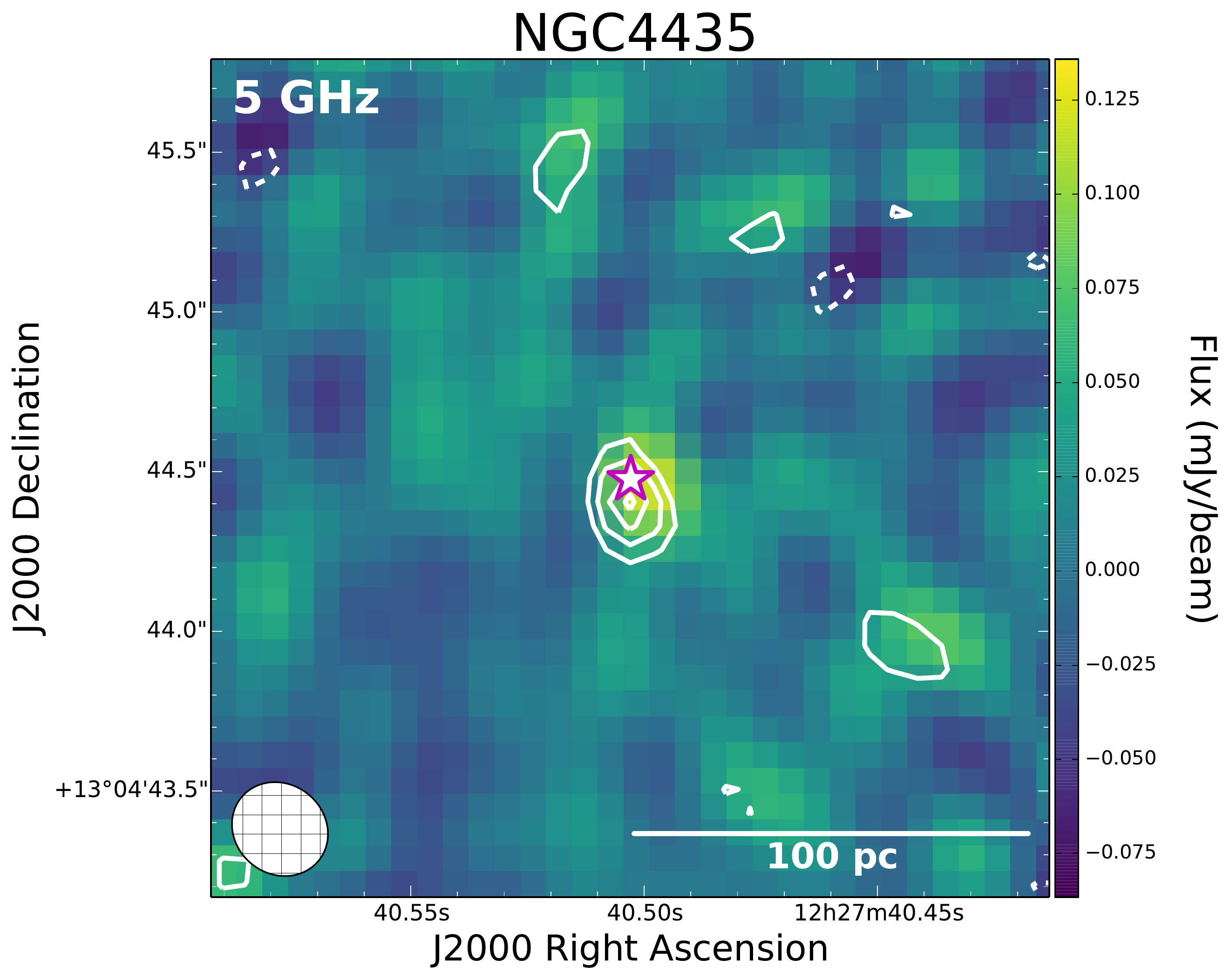}}
{\label{fig:sub:NGC4459}\includegraphics[clip=True, trim=0cm 0cm 0cm 0cm, scale=0.27]{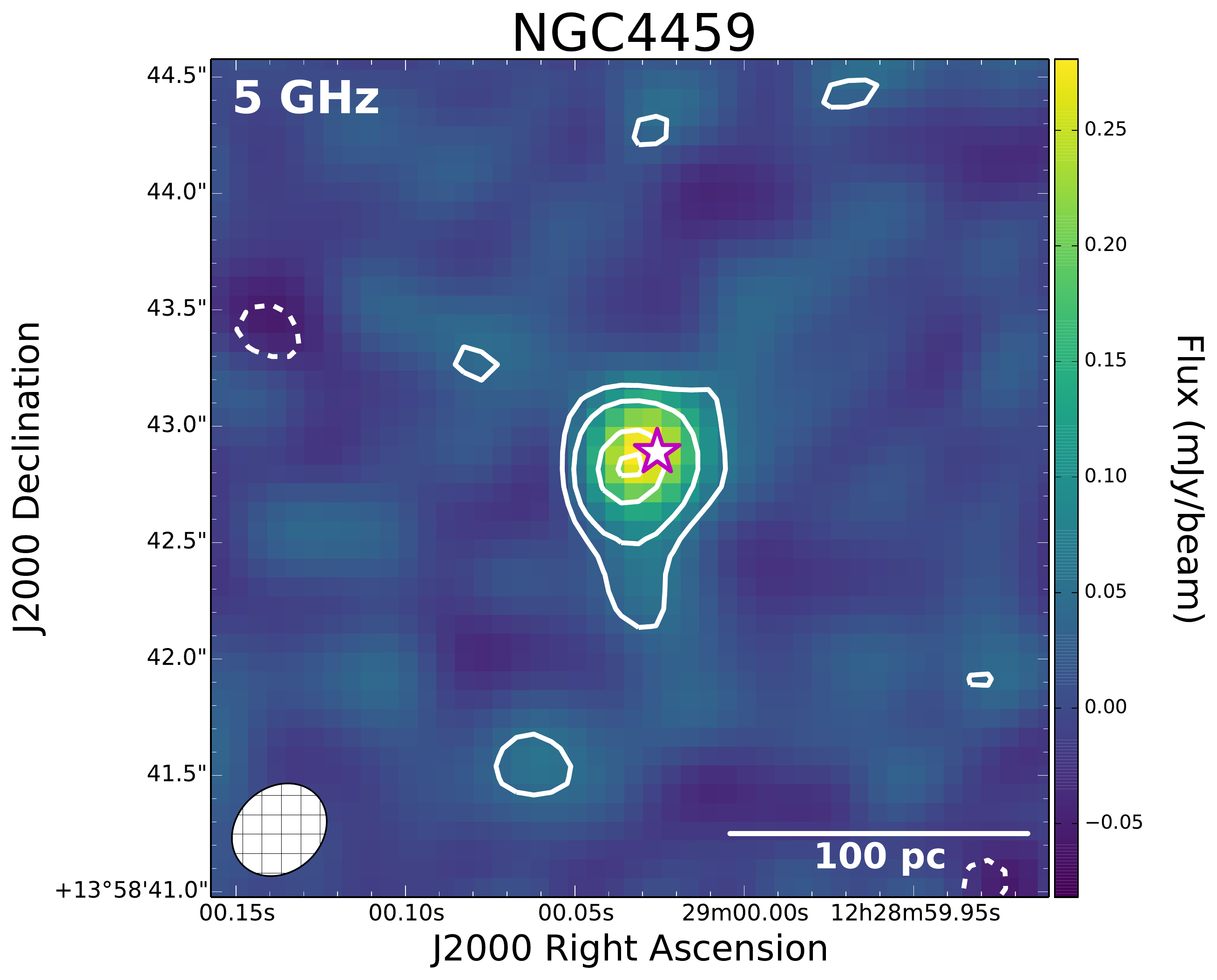}}
{\label{fig:sub:NGC4494}\includegraphics[clip=True, trim=0cm 0cm 0cm 0cm, scale=0.27]{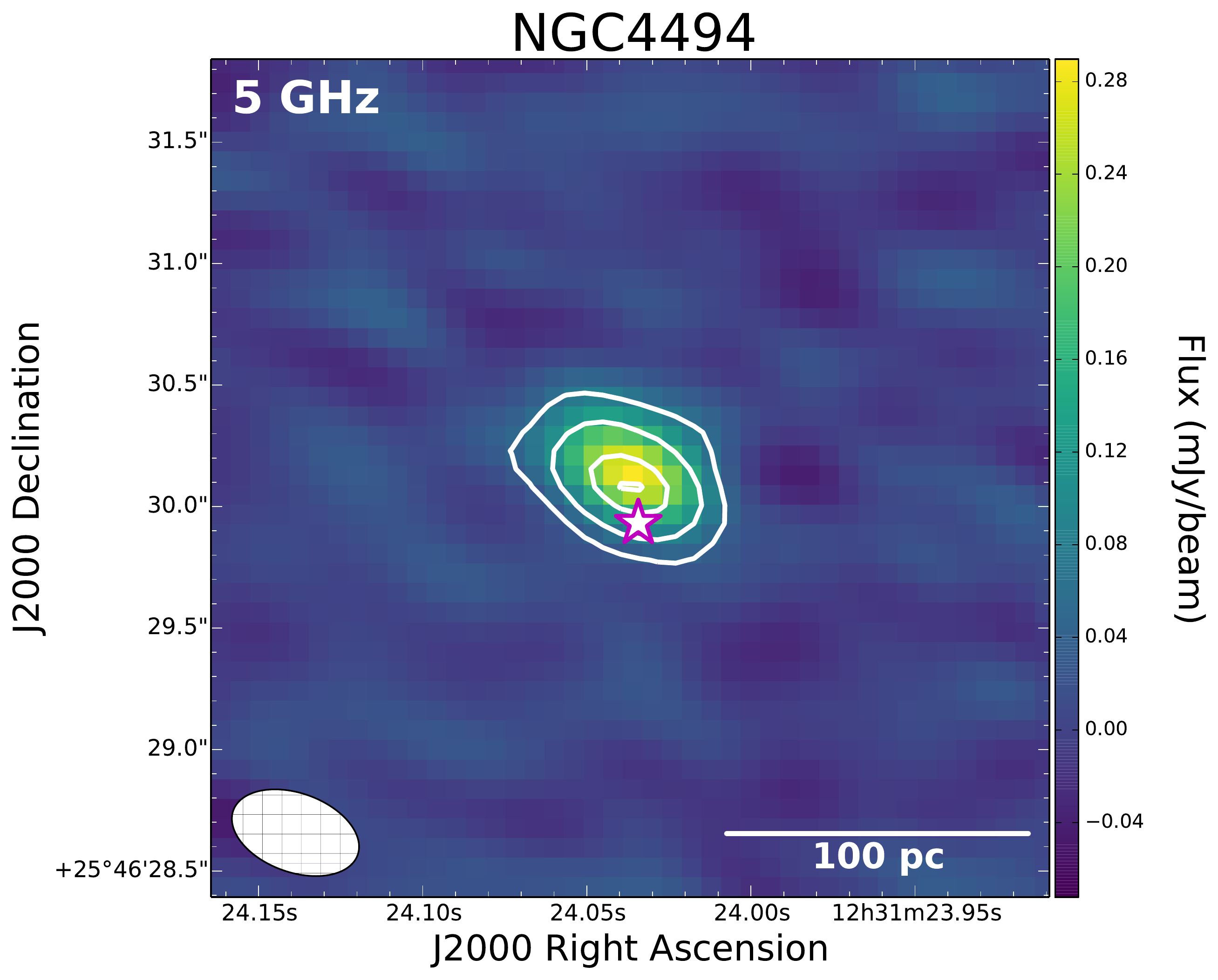}}
{\label{fig:sub:NGC4526}\includegraphics[clip=True, trim=0cm 0cm 0cm 0cm, scale=0.27]{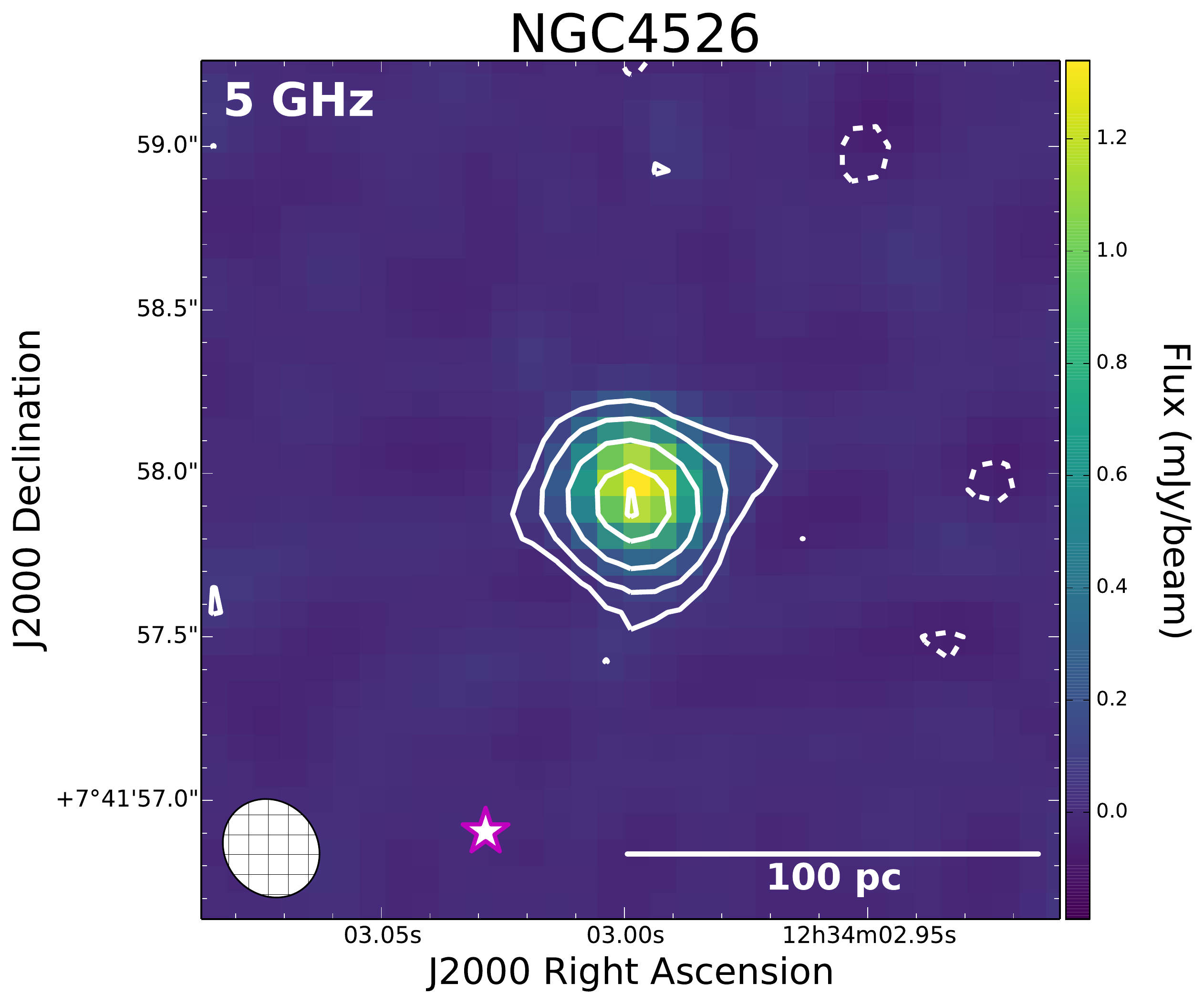}}
{\label{fig:sub:NGC4546}\includegraphics[clip=True, trim=0cm 0cm 0cm 0cm, scale=0.27]{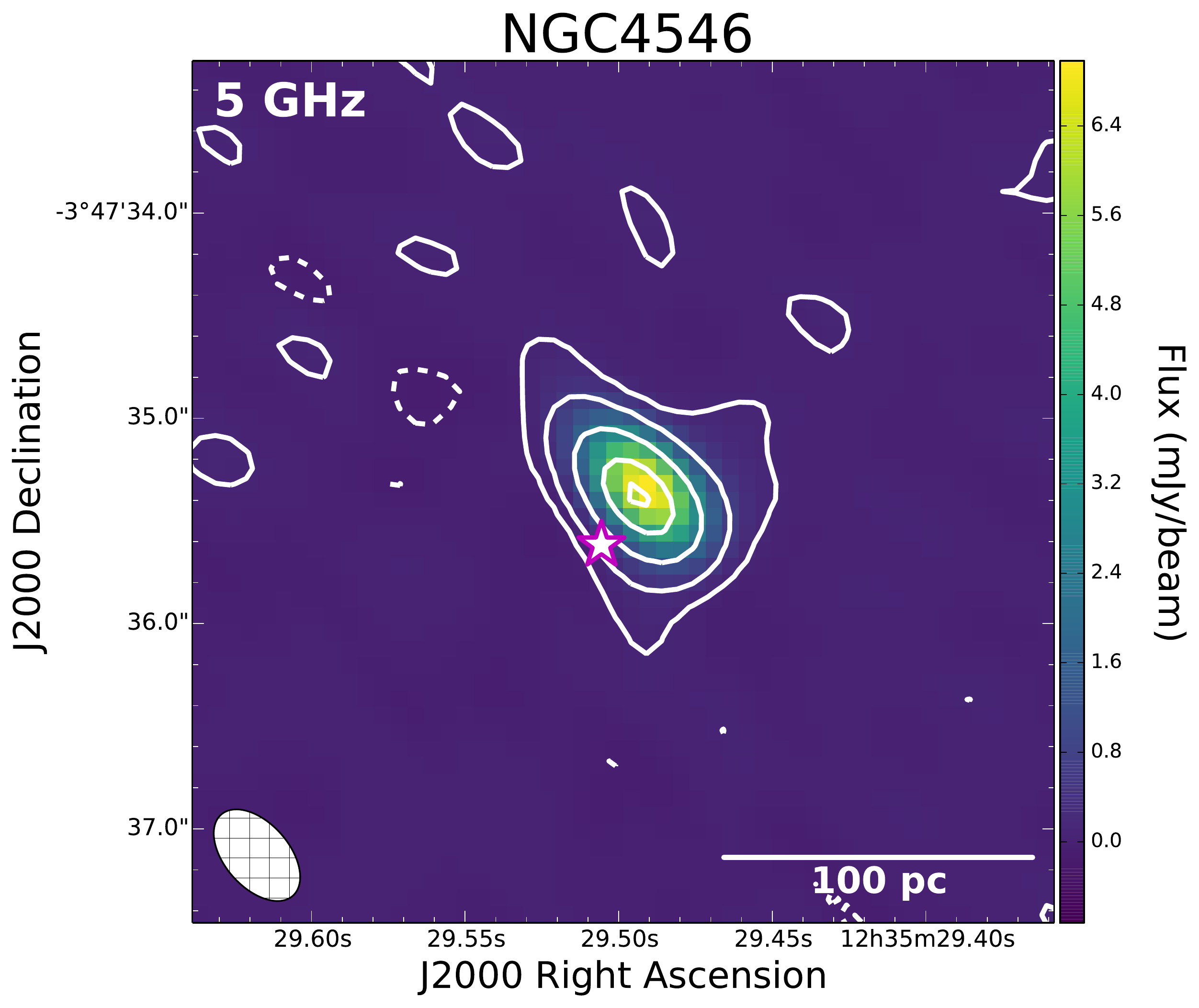}}
\end{figure*}
 
\begin{figure*}
{\label{fig:sub:NGC4643}\includegraphics[clip=True, trim=0cm 0cm 0cm 0cm, scale=0.27]{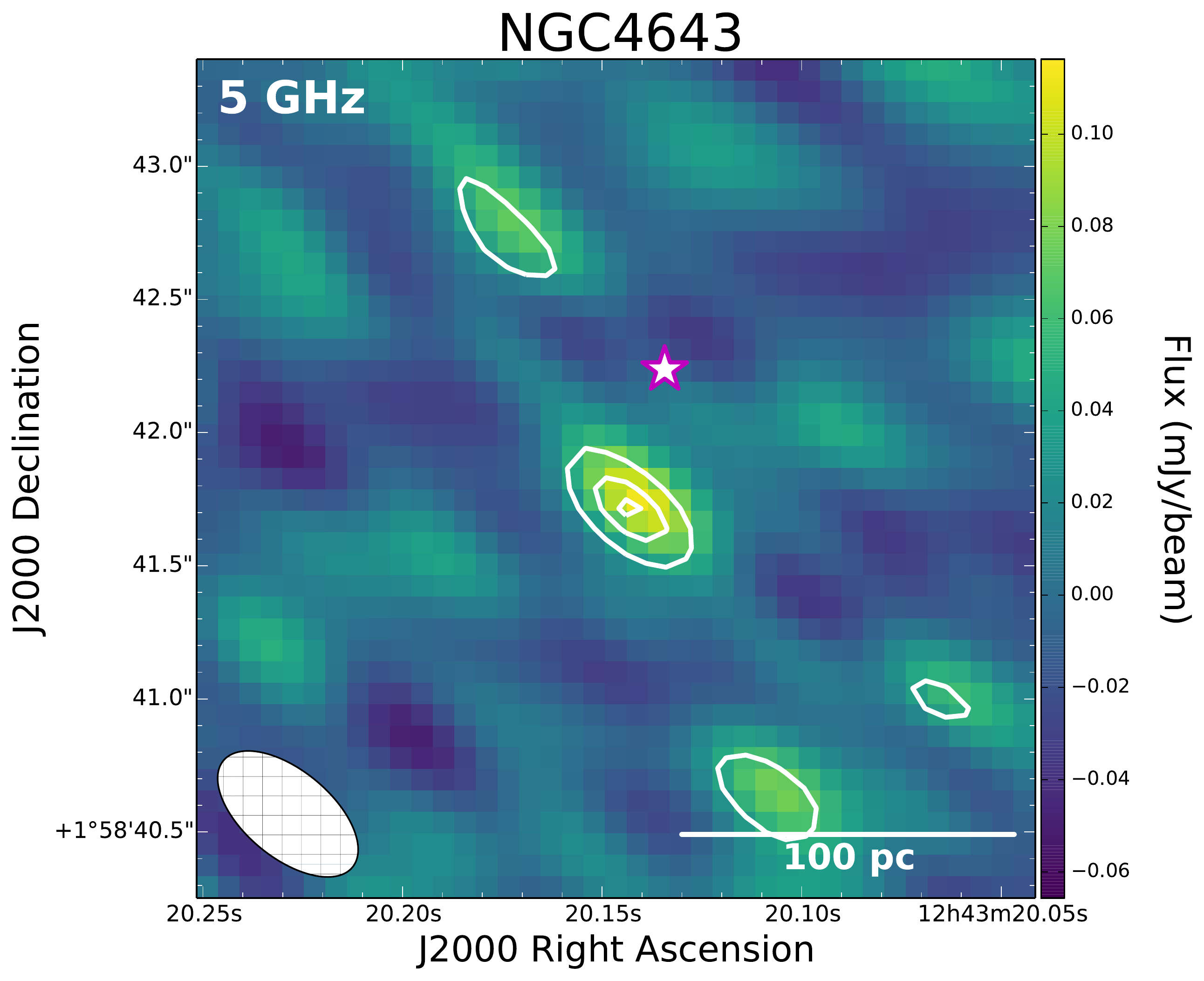}}
{\label{fig:sub:NGC4684}\includegraphics[clip=True, trim=0cm 0cm 0cm 0cm, scale=0.27]{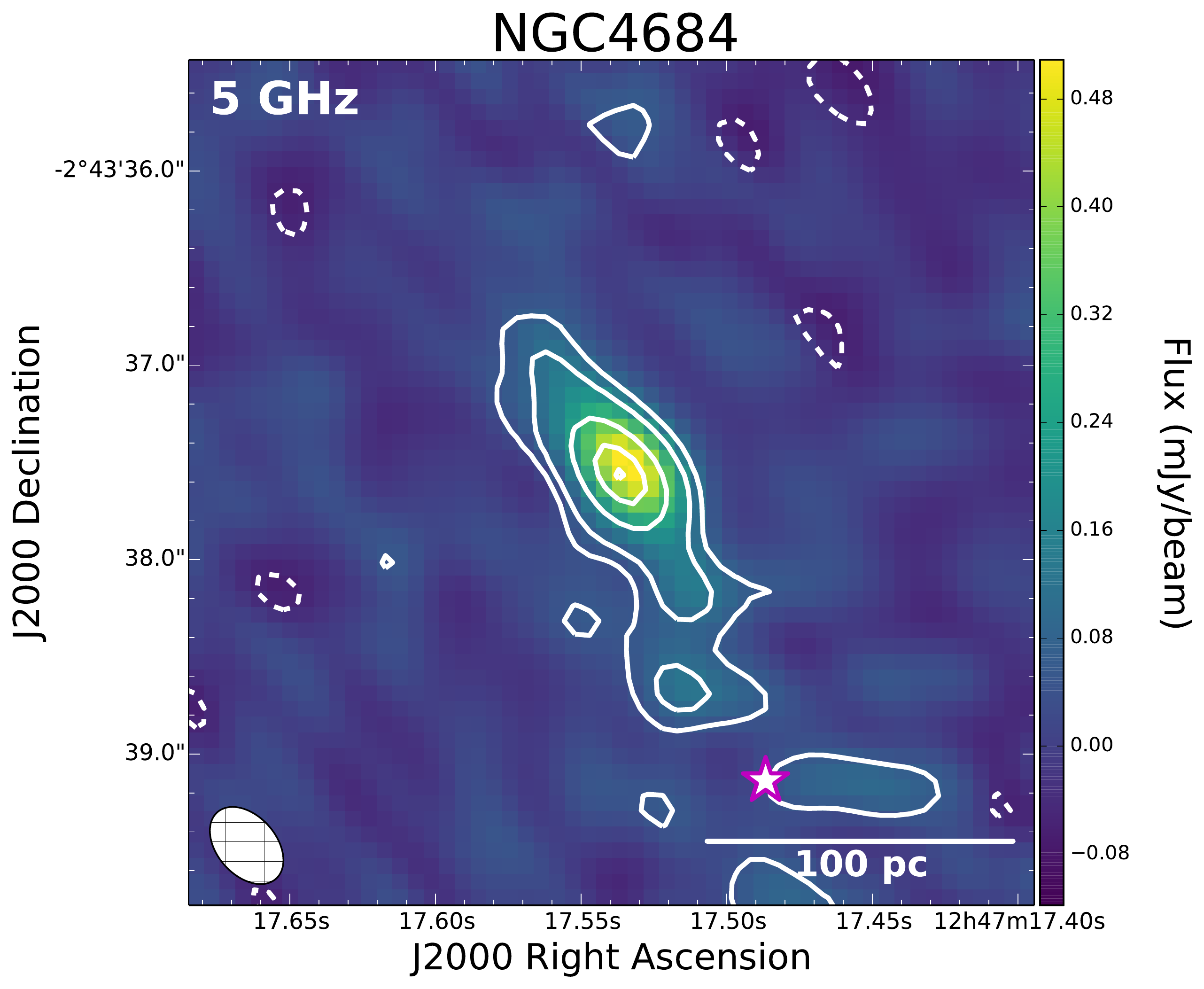}}
{\label{fig:sub:NGC4710}\includegraphics[clip=True, trim=0cm 0cm 0cm 0cm, scale=0.27]{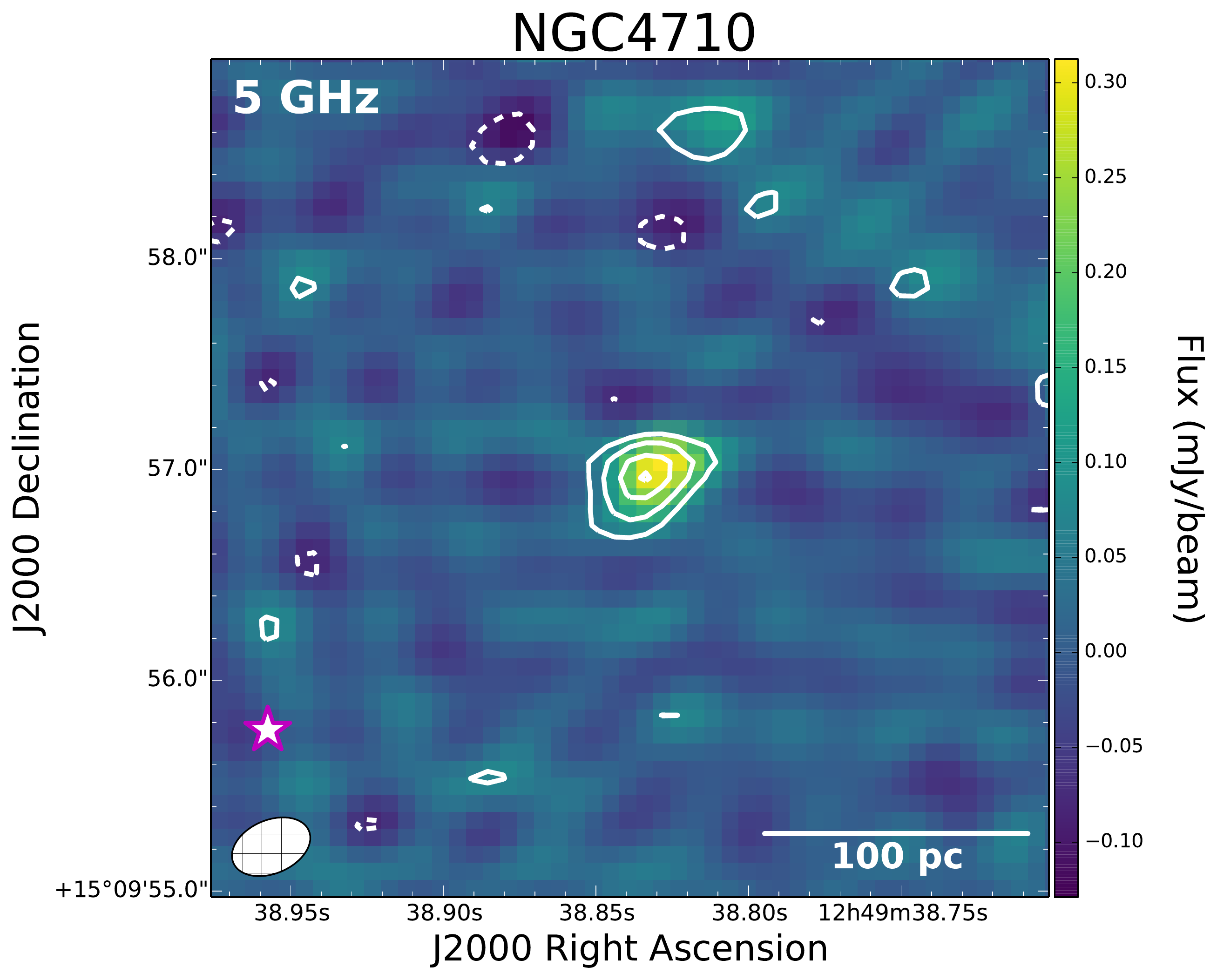}}
{\label{fig:sub:NGC5173}\includegraphics[clip=True, trim=0cm 0cm 0cm 0cm, scale=0.27]{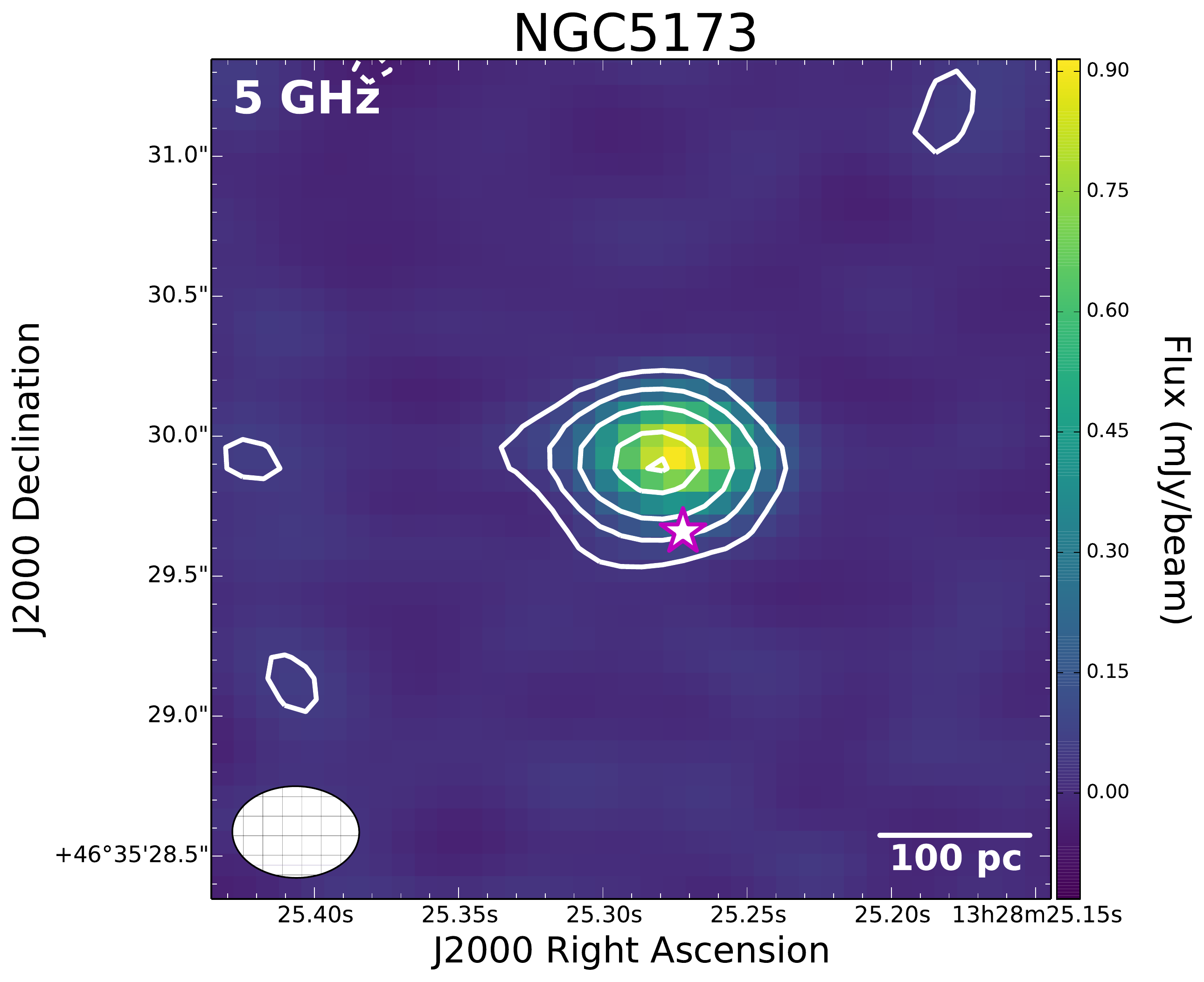}}
{\label{fig:sub:NGC5198}\includegraphics[clip=True, trim=0cm 0cm 0cm 0cm, scale=0.27]{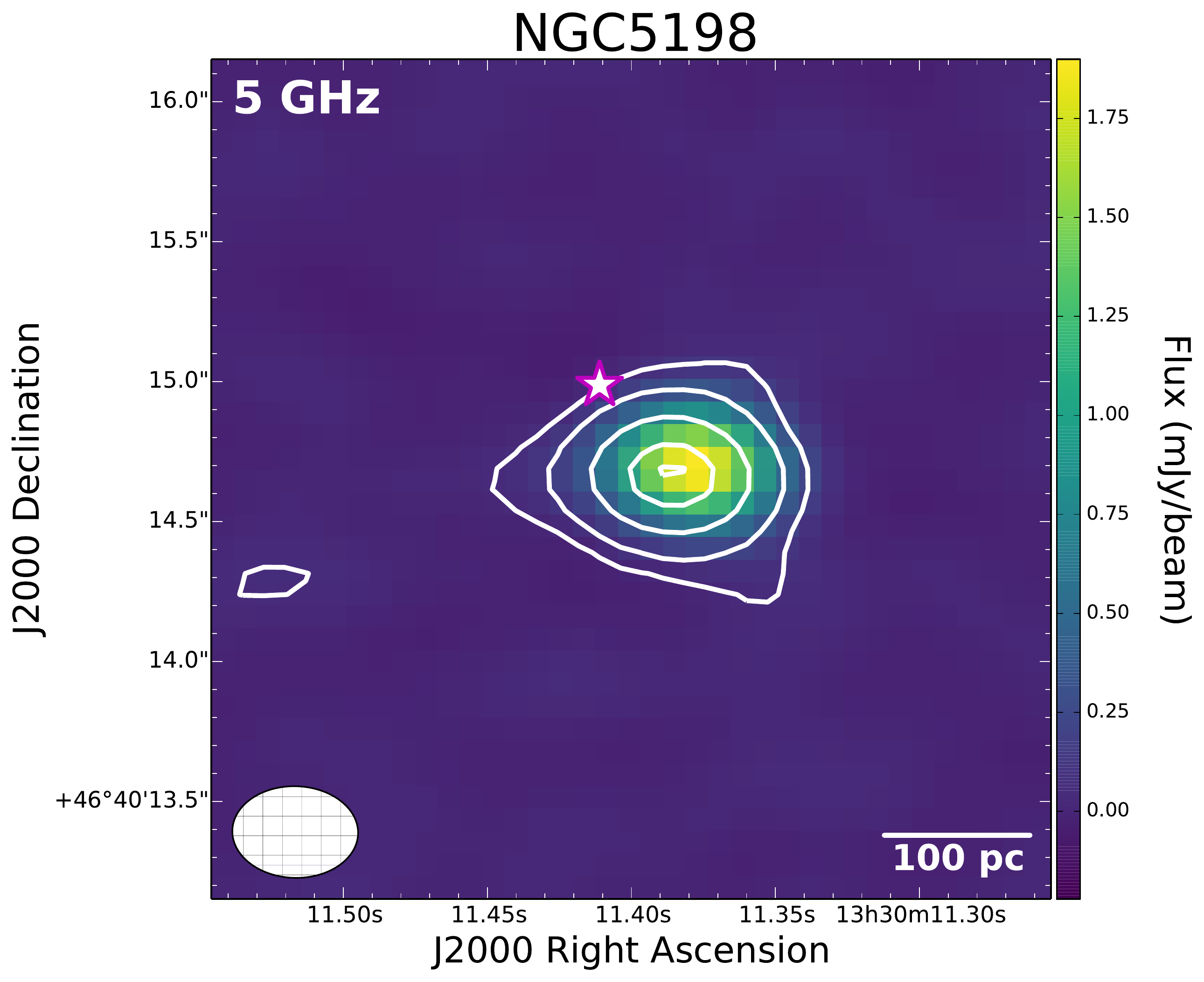}}
{\label{fig:sub:NGC5273}\includegraphics[clip=True, trim=0cm 0cm 0cm 0cm, scale=0.27]{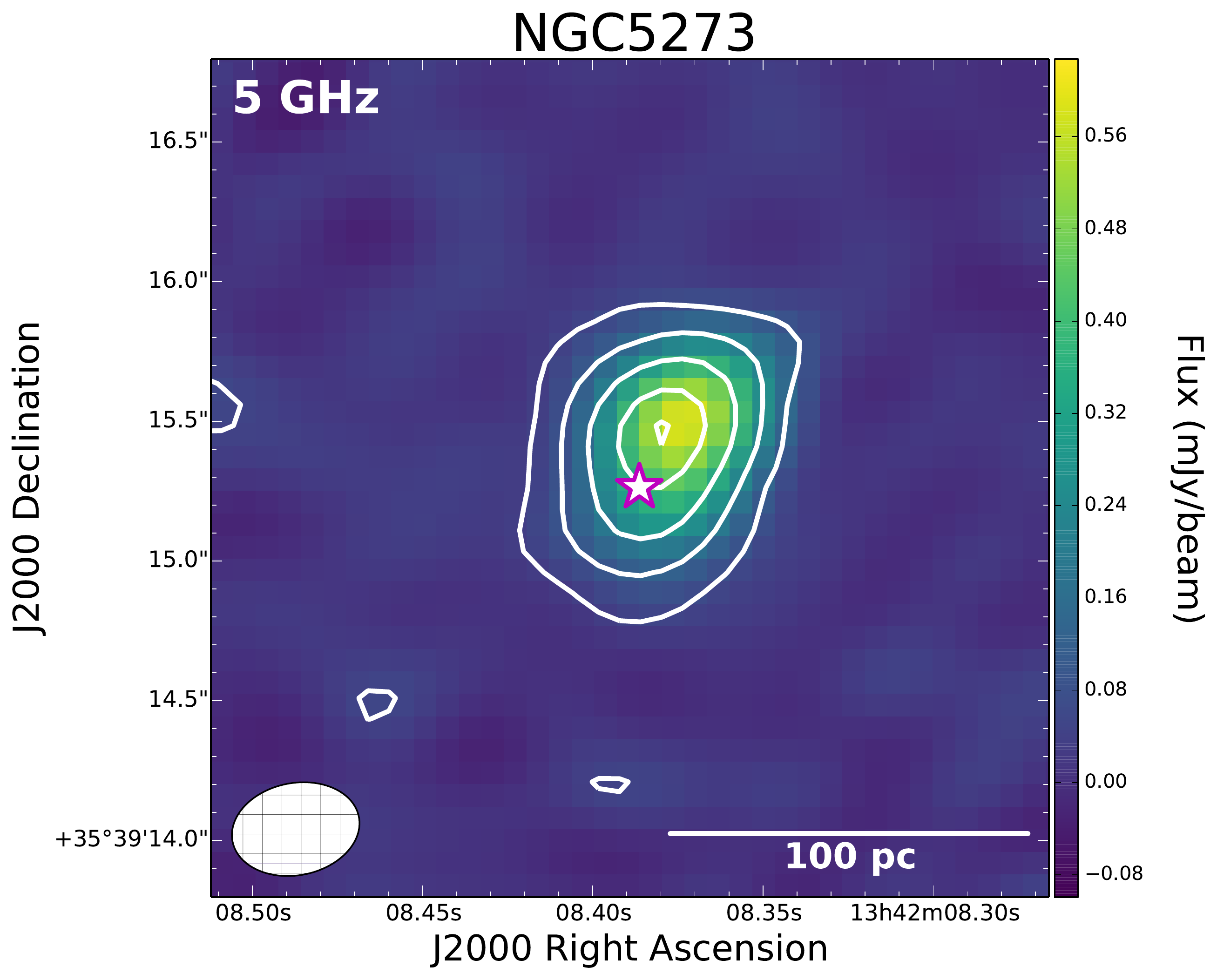}}
\end{figure*}

\begin{figure*}
{\label{fig:sub:NGC5379}\includegraphics[clip=True, trim=0cm 0cm 0cm 0cm, scale=0.27]{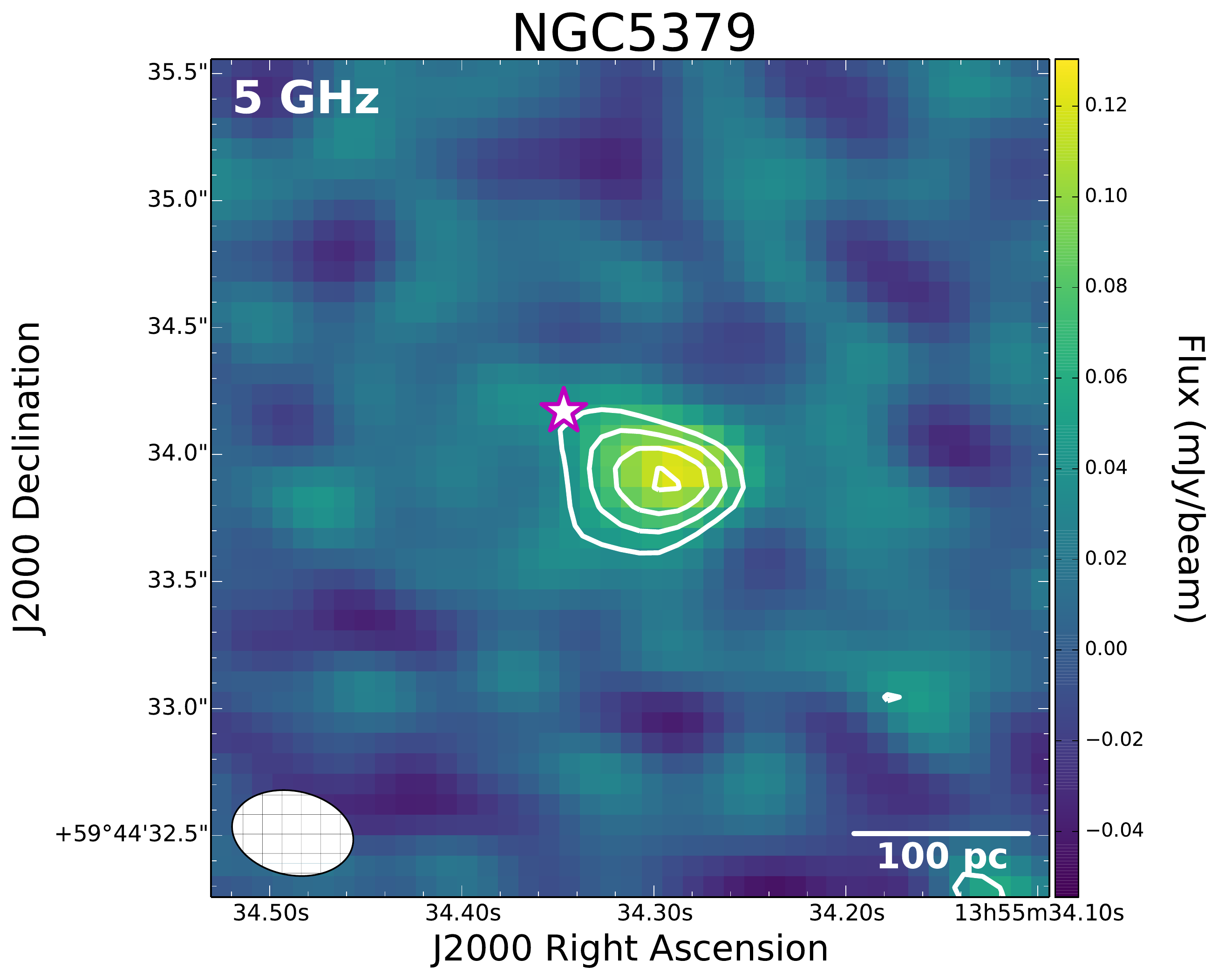}}
{\label{fig:sub:NGC5475}\includegraphics[clip=True, trim=0cm 0cm 0cm 0cm, scale=0.27]{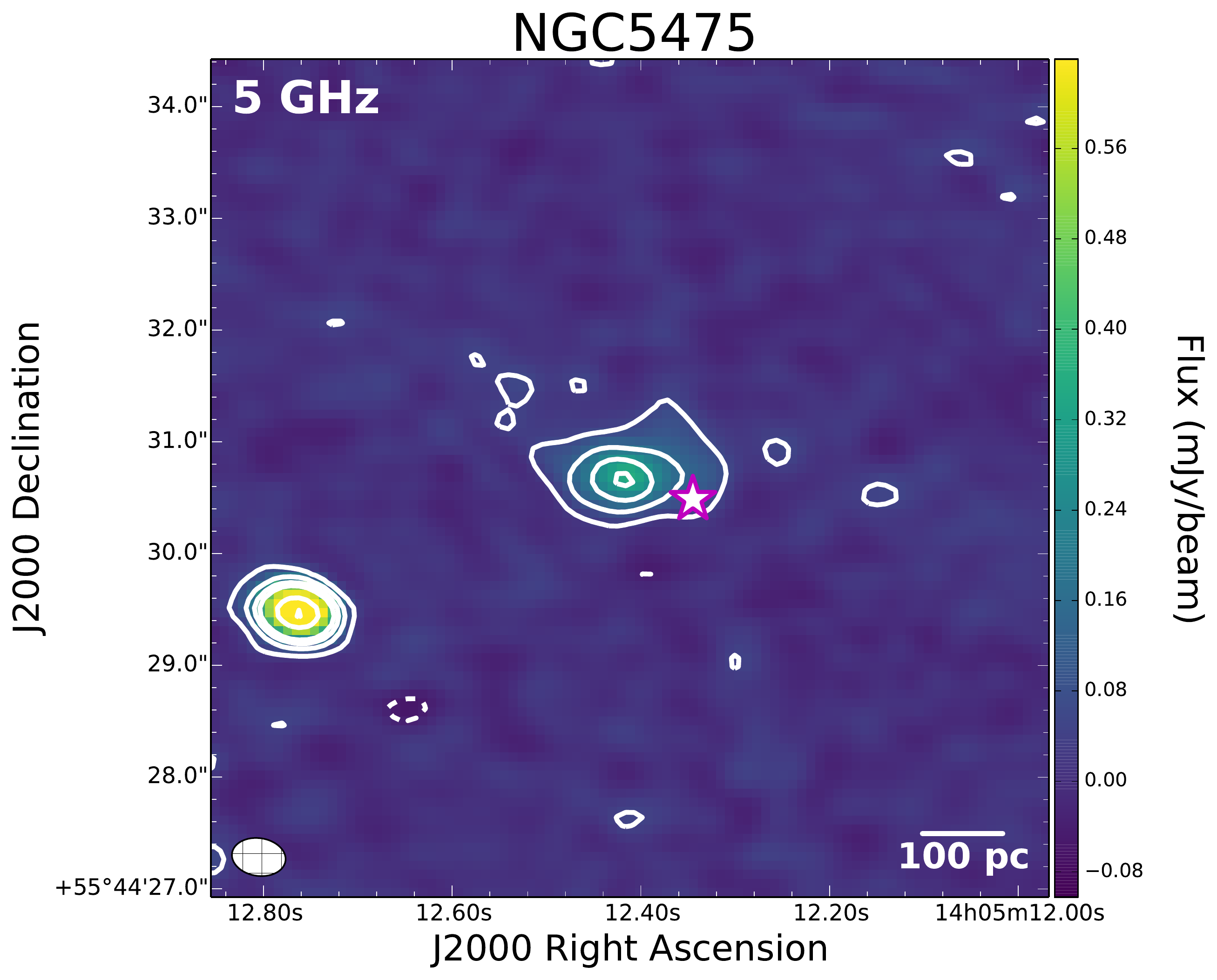}}
{\label{fig:sub:NGC5485}\includegraphics[clip=True, trim=0cm 0cm 0cm 0cm, scale=0.27]{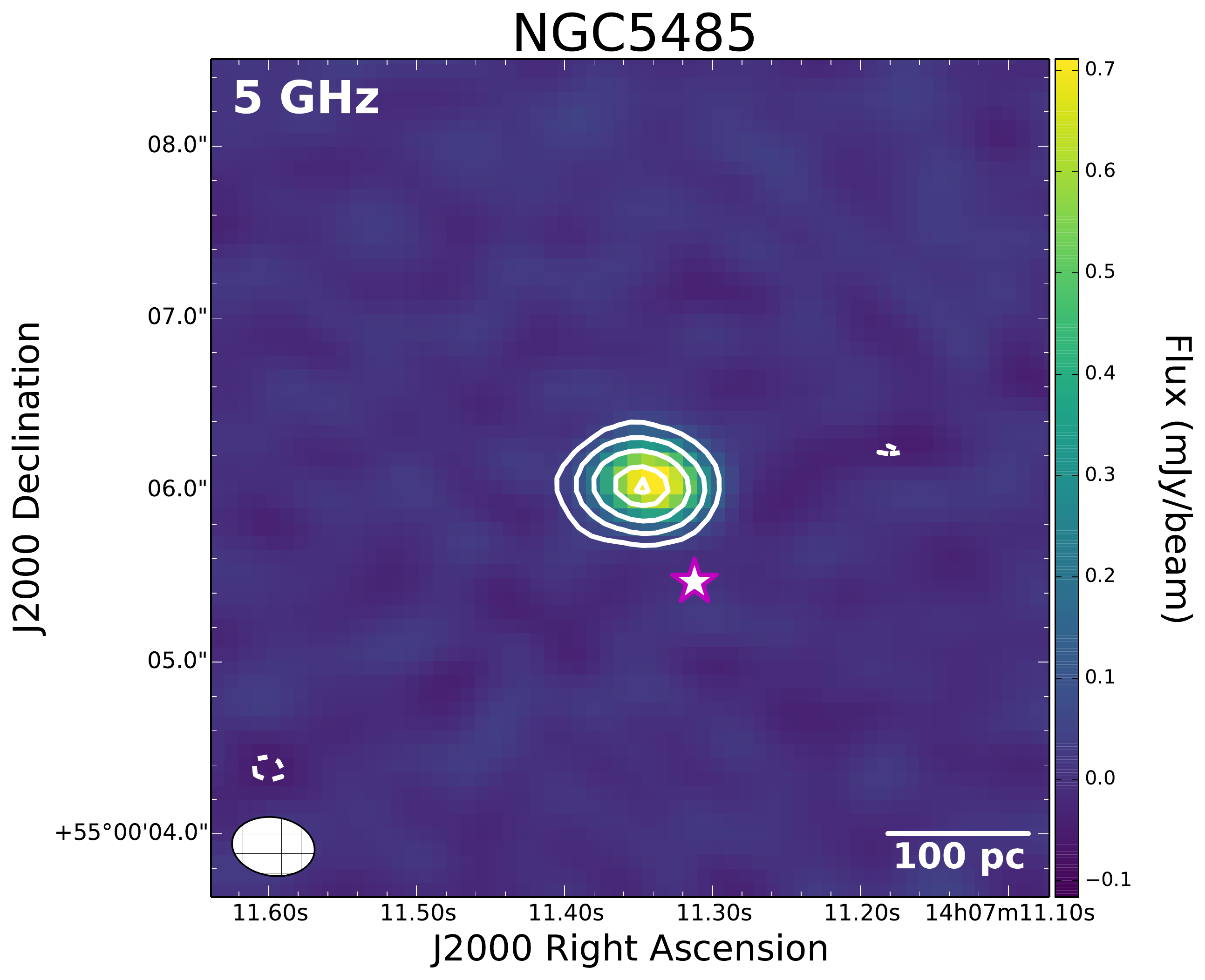}}
{\label{fig:sub:NGC5631}\includegraphics[clip=True, trim=0cm 0cm 0cm 0cm, scale=0.27]{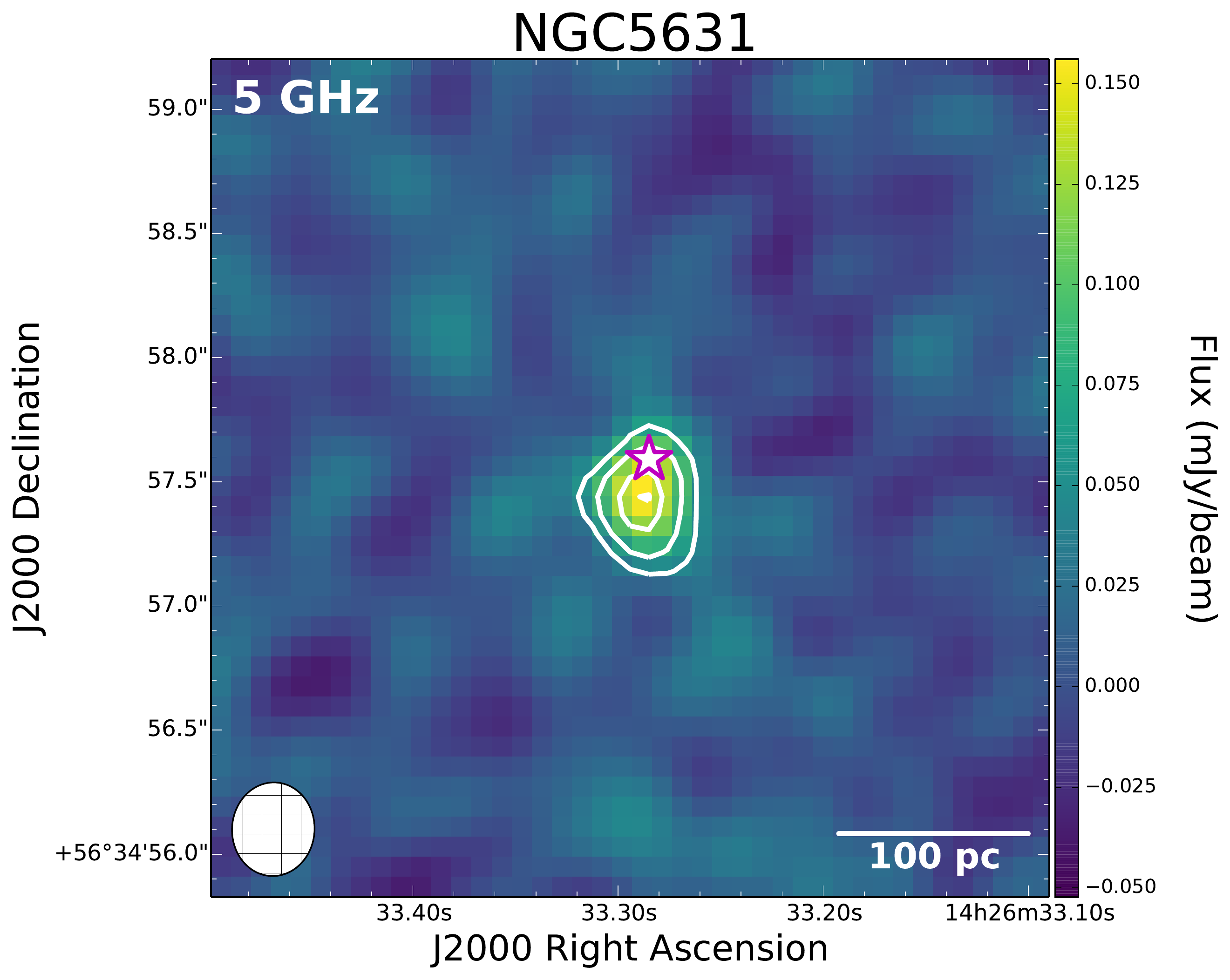}}
{\label{fig:sub:NGC6014}\includegraphics[clip=True, trim=0cm 0cm 0cm 0cm, scale=0.27]{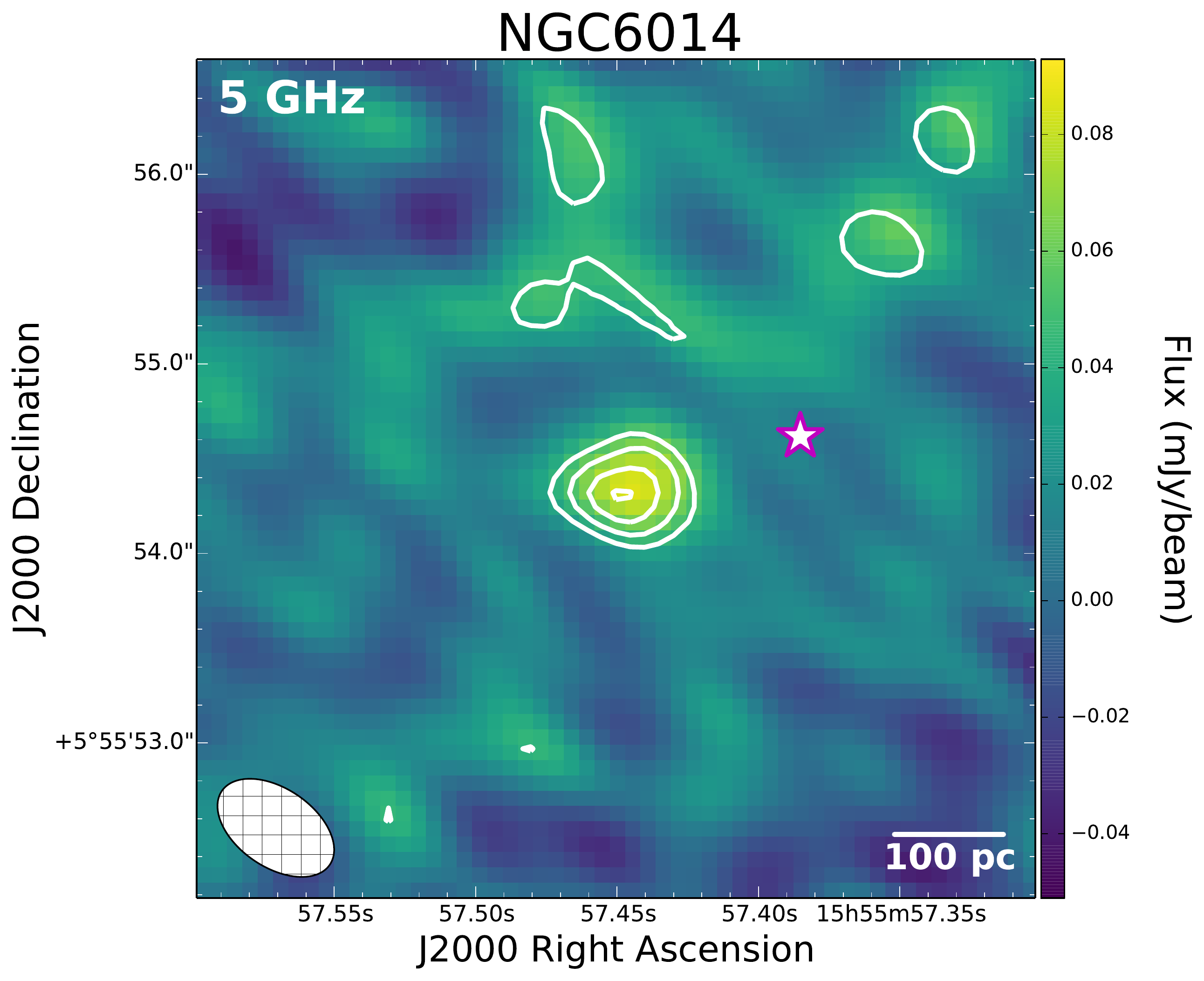}}
{\label{fig:sub:NGC6703}\includegraphics[clip=True, trim=0cm 0cm 0cm 0cm, scale=0.27]{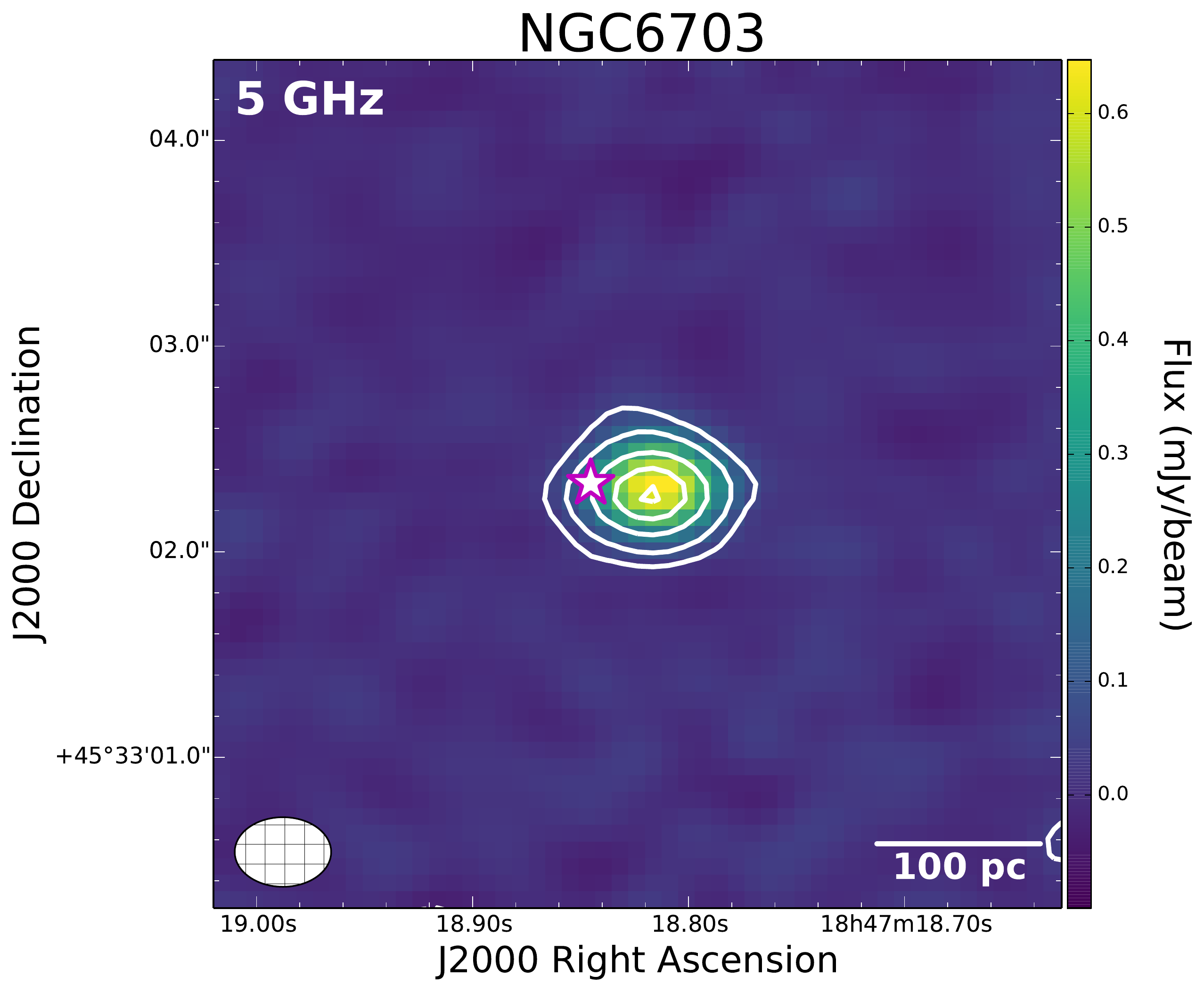}}
\end{figure*}

\begin{figure*}
{\label{fig:sub:NGC6798}\includegraphics[clip=True, trim=0cm 0cm 0cm 0cm, scale=0.27]{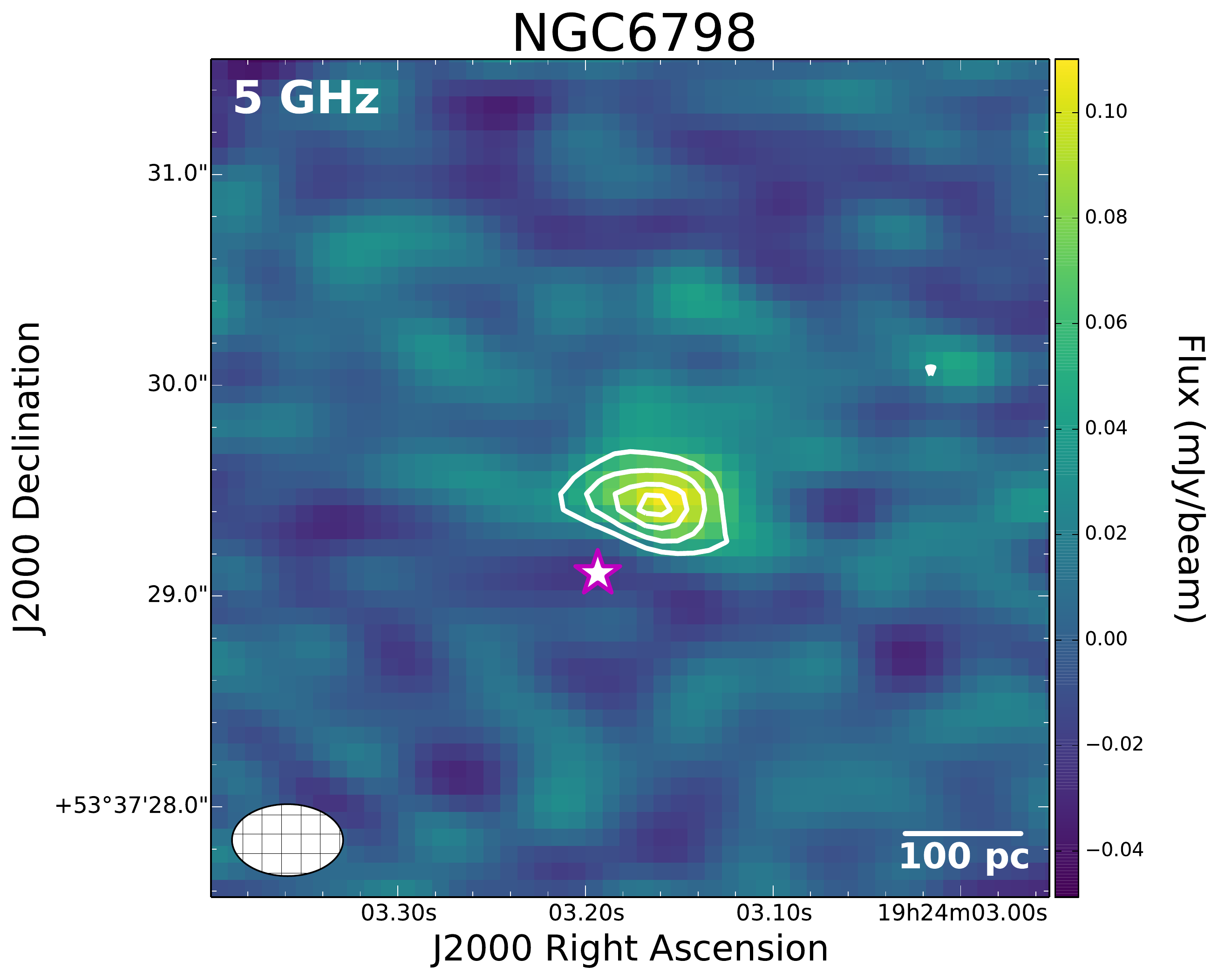}}
{\label{fig:sub:NGC7465}\includegraphics[clip=True, trim=0cm 0cm 0cm 0cm, scale=0.27]{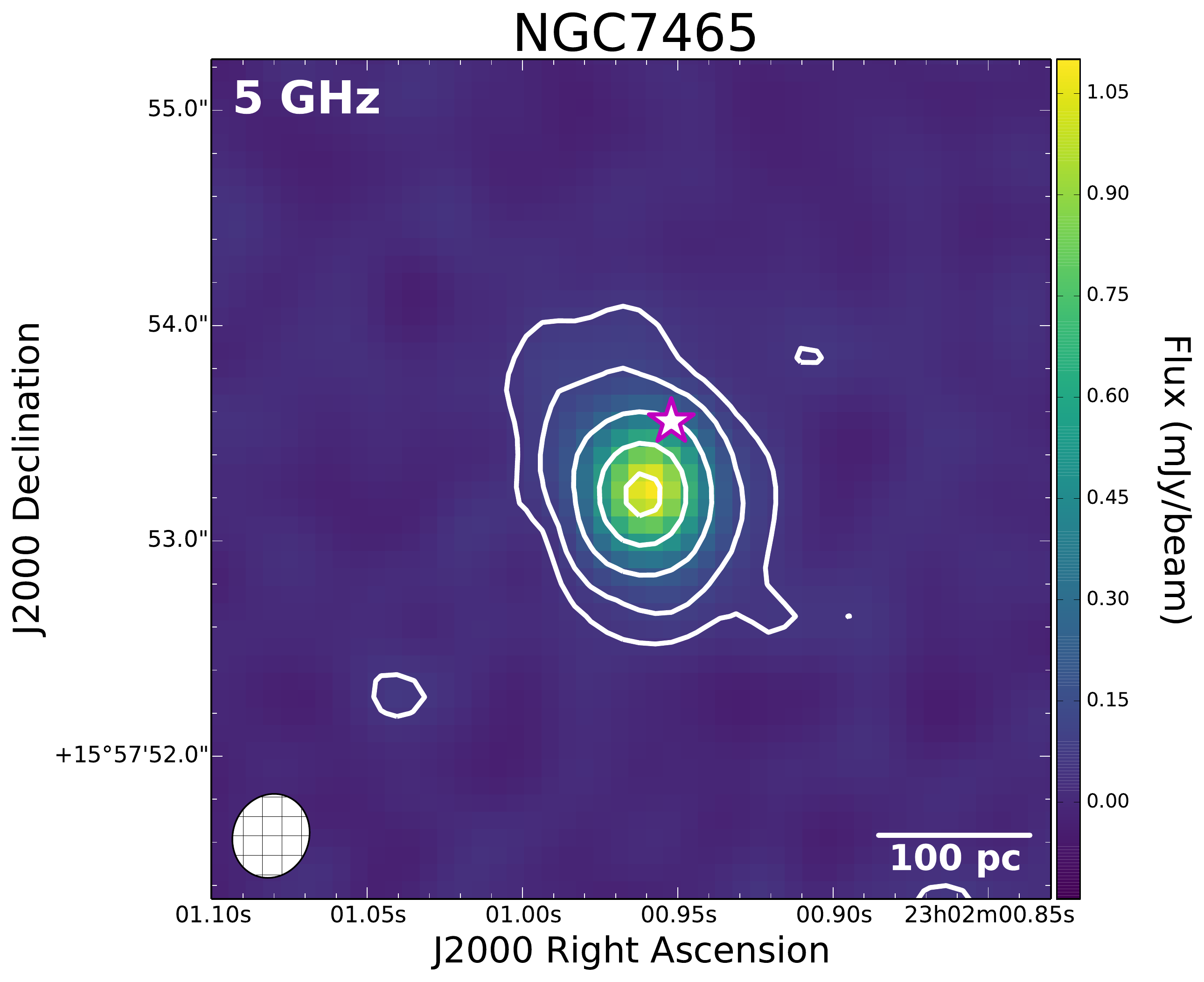}}
{\label{fig:sub:PGC029321}\includegraphics[clip=True, trim=0cm 0cm 0cm 0cm, scale=0.27]{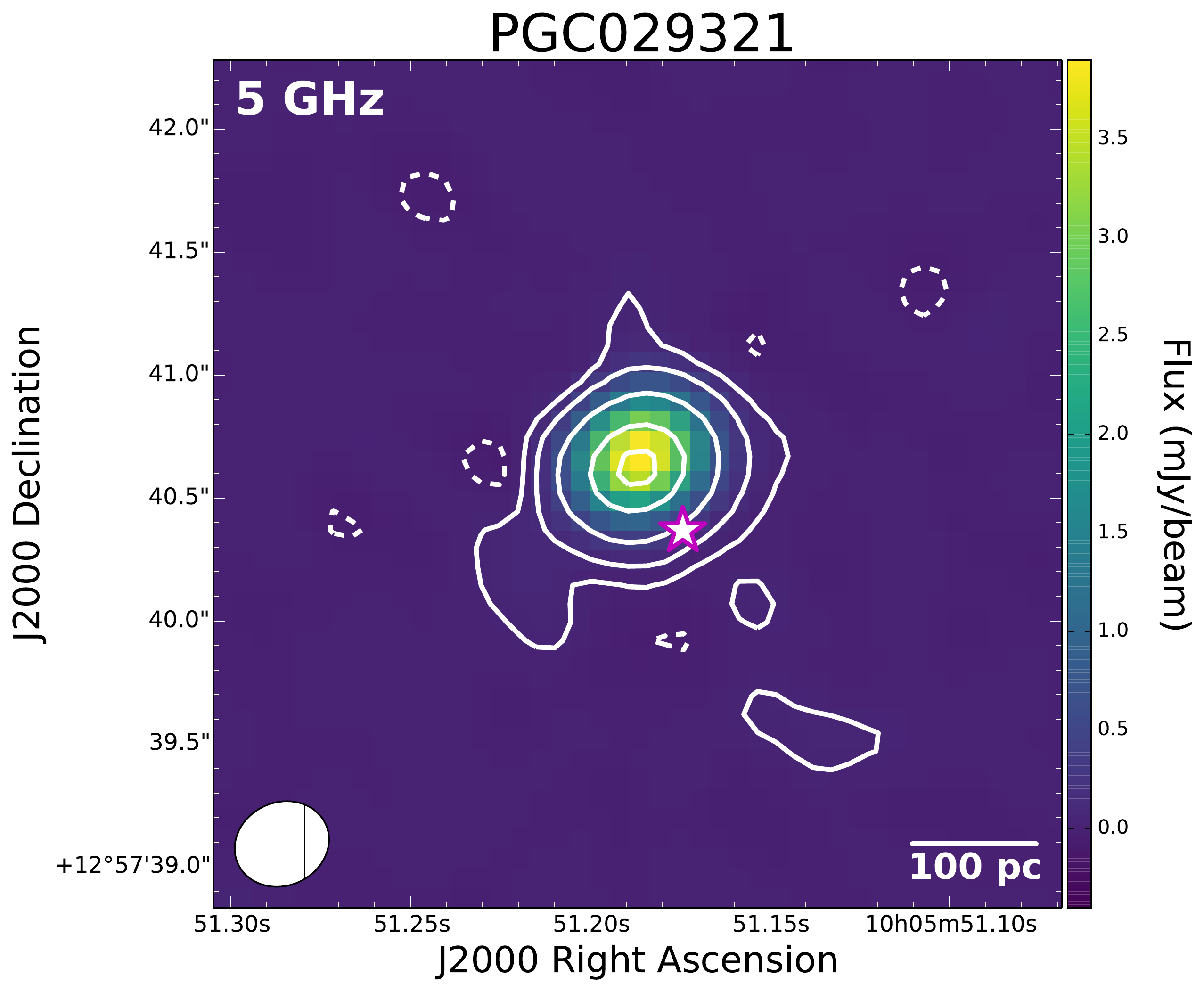}}
{\label{fig:sub:UGC05408}\includegraphics[clip=True, trim=0cm 0cm 0cm 0cm, scale=0.27]{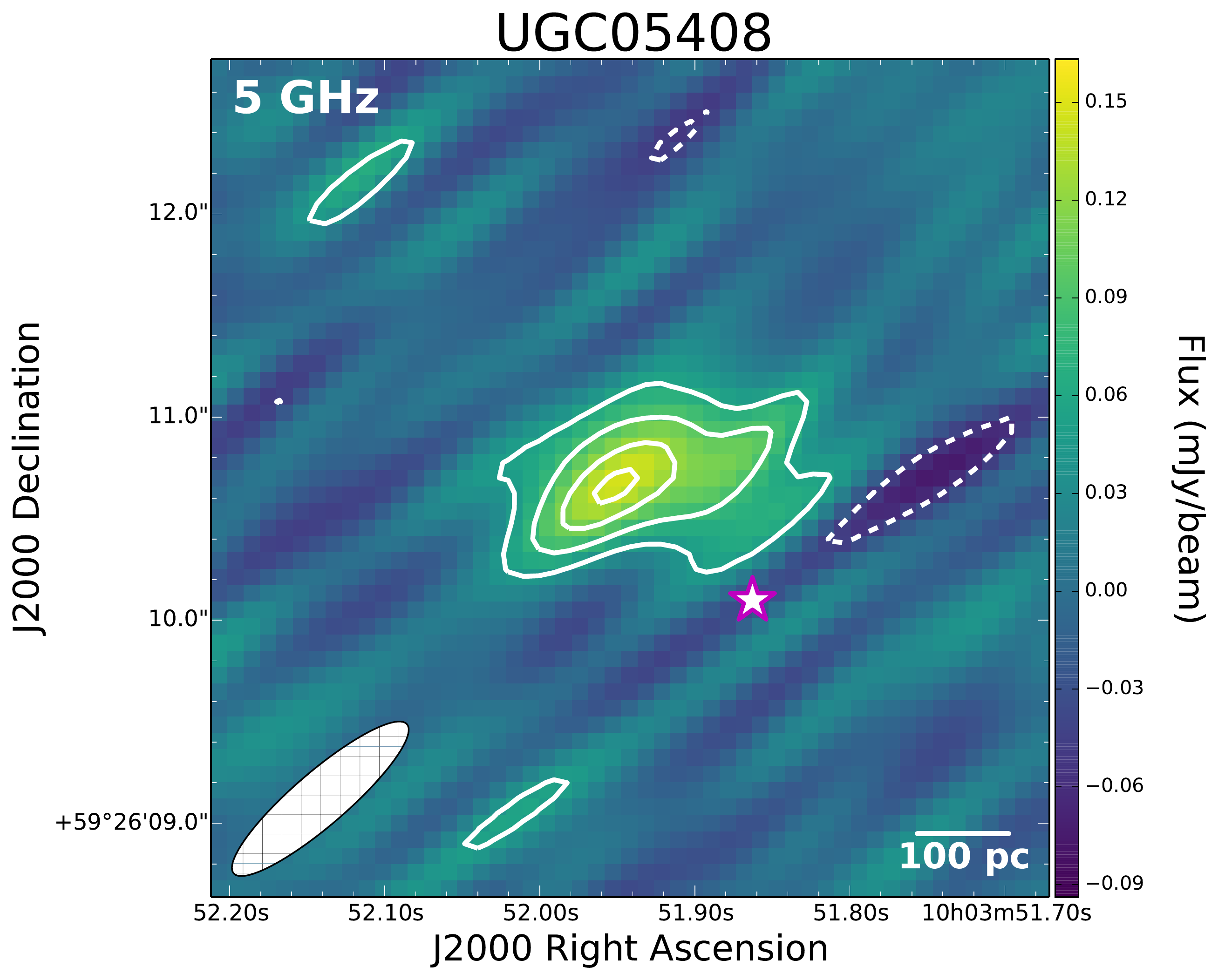}}
{\label{fig:sub:UGC06176}\includegraphics[clip=True, trim=0cm 0cm 0cm 0cm, scale=0.27]{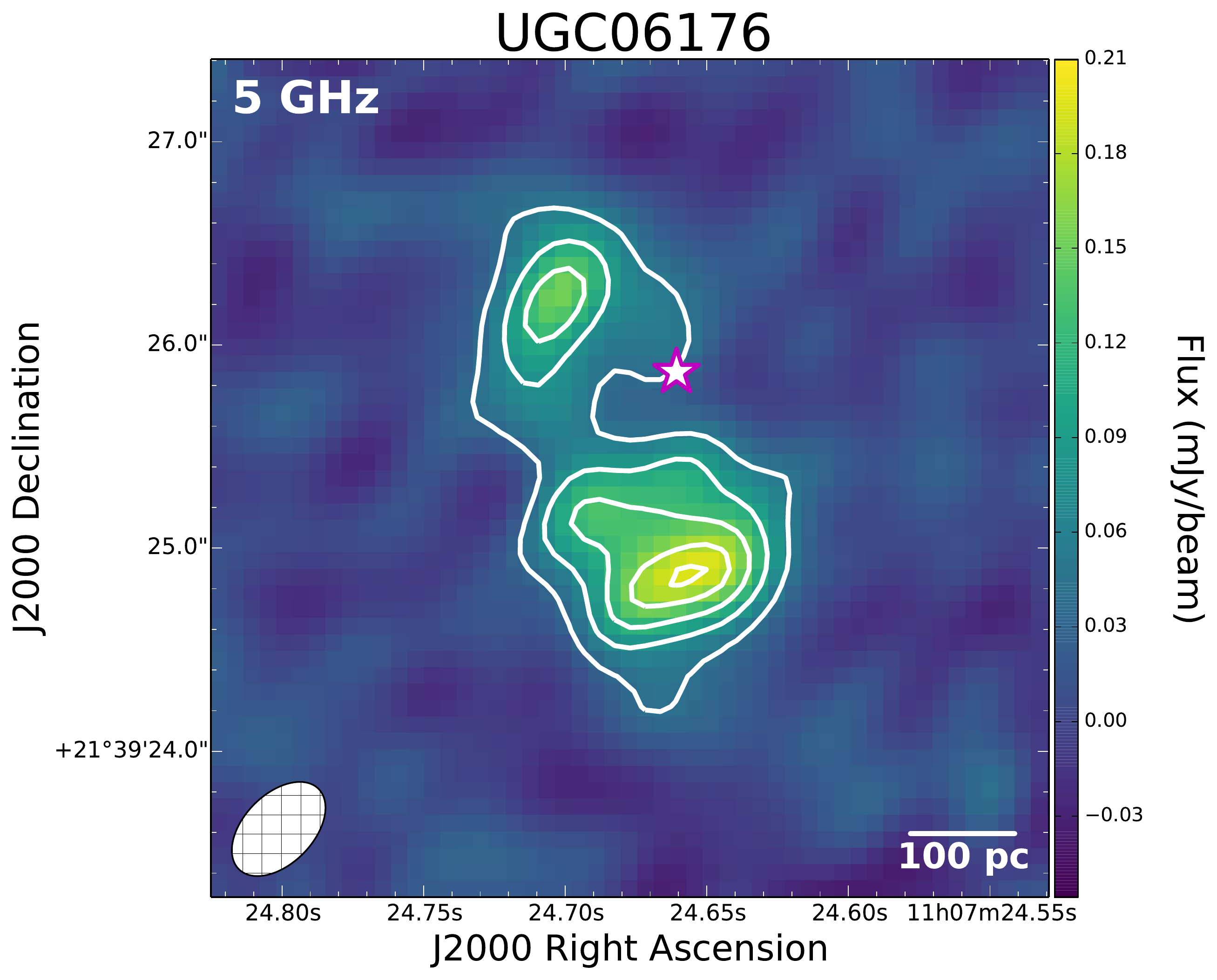}}
\caption{5~GHz continuum images with contours.  Negative contours are dashed.  The contour levels are spaced as multiples of the rms noise in each image.  Relative contour levels and rms noises are listed in Table~\ref{tab:contours}.  The synthesized beam is shown as a hatched white ellipse in the lower left corner of each image.  A white star outlined in magenta denotes the official optical position in the ATLAS$^{\mathrm{3D}}$ survey \citepalias{cappellari+11a}.  A scale bar denoting 100~pc is shown in the lower right corner of each image.
}
\label{fig:radio_images}
\end{figure*}



\bsp	
\label{lastpage}
\end{document}